\documentclass[10pt,a4paper,twoside,openright]{report}

\input{thesis-template/config/my-commands.tex}

\usepackage[utf8]{inputenc} 			
\usepackage[english]{babel} 			
\usepackage{csquotes}                   

\usepackage{amsmath}					
\usepackage{amsfonts}					
\usepackage{amssymb}					
\usepackage{mathtools}					

\usepackage{graphicx}					
\usepackage[labelfont=bf,font=small]{caption}		
\usepackage[labelfont=bf,font=small]{subcaption}	
\usepackage{sidecap}					

\usepackage{emptypage}					
\usepackage{fancyhdr}					
\usepackage{geometry}					
\usepackage{setspace}
\usepackage[bottom,para]{footmisc}      

\usepackage{tabularx}					
\usepackage{booktabs}					
\usepackage{enumitem}					
\usepackage{arydshln}					

\usepackage{siunitx}					
\usepackage{physics}					

\usepackage{lipsum}						
\usepackage{comment}					
\usepackage{xcolor} 					
\usepackage[dvipsnames]{xcolor} 		
\usepackage{framed}						

\usepackage{nccmath}                    
\usepackage{bbold}                      
\usepackage{multirow}                   
\usepackage{environ}                    
\usepackage{caption}
\usepackage{subcaption}

\usepackage{hyperref}					

\usepackage{amssymb} 

\usepackage{float} 

\usepackage{dcolumn}
\newcolumntype{d}{D{.}{.}{-1}}

\colorlet{url-blue}{blue!70!black}

\hypersetup{
	pdfpagemode			= {UseOutlines}	,
	bookmarksopen		= true 			,
	bookmarksopenlevel	= 0				,
	hypertexnames		= false			,
	colorlinks			= true			, 
	citecolor			= blue			, 
	linkcolor			= blue			, 
	urlcolor			= url-blue		, 
	pdfstartview		= {FitV}			,
	breaklinks			= true,
	unicode								,
}

\sidecaptionvpos{figure}{t}
\usepackage{cmbright}	
\usepackage[T1]{fontenc}

\newcommand{\fontnamestring}{cmbr}	

\def\FontLb{
	\usefont{T1}{\fontnamestring}{b}{n}\fontsize{16pt}{16pt}\selectfont}

\def\FontMb{
	\usefont{T1}{\fontnamestring}{b}{n}\fontsize{14pt}{14pt}\selectfont}
\def\FontSn{
	\usefont{T1}{\fontnamestring}{m}{n}\fontsize{12pt}{12pt}\selectfont}

\makeatletter
\DeclareFontEncoding{LS1}{}{}
\DeclareFontEncoding{LS2}{}{\noaccents@}
\DeclareFontSubstitution{LS1}{stix}{m}{n}
\DeclareFontSubstitution{LS2}{stix}{m}{n}
\makeatother

\DeclareMathAlphabet\mathscr{LS1}{stixscr}{m}{n}
\SetMathAlphabet\mathscr{bold}{LS1}{stixscr}{b}{n}

\DeclareFontFamily{OT1}{cmbr}{\hyphenchar\font45 }
\DeclareFontShape{OT1}{cmbr}{m}{n}{%
	<-9>cmbr8
	<9-10>cmbr9
	<10-17>cmbr10
	<17->cmbr17
}{}
\DeclareFontShape{OT1}{cmbr}{m}{sl}{%
	<-9>cmbrsl8
	<9-10>cmbrsl9
	<10-17>cmbrsl10
	<17->cmbrsl17
}{}
\DeclareFontShape{OT1}{cmbr}{m}{it}{%
	<->ssub*cmbr/m/sl
}{}
\DeclareFontShape{OT1}{cmbr}{b}{n}{%
	<->ssub*cmbr/bx/n
}{}
\DeclareFontShape{OT1}{cmbr}{bx}{n}{%
	<->cmbrbx10
}{}

\makeatletter
\g@addto@macro\bfseries{\boldmath}
\makeatother

\newcommand{\coverThesis}{@undefined} 

\newcommand{\coverExaminationCommittee}{@undefined}
\newcommand{\coverChairperson}{@undefined} 
\newcommand{\coverSupervisor}{@undefined}

\addto\captionsenglish{\renewcommand{\coverThesis}{Thesis to obtain the Master of Science Degree in}}
\addto\captionsenglish{}
\addto\captionsenglish{\renewcommand{\coverExaminationCommittee}{Examination Committee}}
\addto\captionsenglish{\renewcommand{\coverChairperson}{Chairperson}}
\addto\captionsenglish{\renewcommand{\coverSupervisor}{Supervisor}}
\addto\captionsenglish{}

\newcommand{\acknowledgments}{@undefined} 
\addto\captionsenglish{\renewcommand{\acknowledgments}{Acknowledgements}}

\usepackage{nomencl}
\makenomenclature

\RequirePackage{ifthen}

\renewcommand{\nomgroup}[1]{%
	\ifthenelse{	\equal{#1}{R}	}{	\item[\textbf{Roman symbols}]	}{%
	\ifthenelse{	\equal{#1}{G}	}{	\item[\textbf{Greek symbols}]	}{%
   	\ifthenelse{	\equal{#1}{S}	}{	\item[\textbf{Subscripts}]		}{%
  	\ifthenelse{	\equal{#1}{T}	}{	\item[\textbf{Superscripts}]	}{%
  	}}}}}

\usepackage[nomain,acronym,nogroupskip,nopostdot,nonumberlist,indexonlyfirst]{glossaries}
\makeglossaries

\usepackage[backend=biber, style=ieee, citestyle=numeric-comp, dashed=false, giveninits=true, doi=false, url=true, isbn=false, sorting=none, maxbibnames=5]{biblatex}

\DefineBibliographyExtras{english}{\stdpunctuation}

\AtEveryBibitem{%
	\clearfield{note}%
	\clearname{editor}%
}

\newbibmacro{string+doiurlisbn}[1]{%
	\iffieldundef{url}{%
		\iffieldundef{doi}{%
			\iffieldundef{isbn}{%
				\iffieldundef{issn}{%
					#1%
				}{%
					\href{http://books.google.com/books?vid=ISSN\thefield{issn}}{#1}%
				}%
			}{%
				\href{http://books.google.com/books?vid=ISBN\thefield{isbn}}{#1}%
			}%
		}{%
			\href{http://dx.doi.org/\thefield{doi}}{#1}%
		}%
	}{%
		\href{\thefield{url}}{#1}%
	}%
}

\DeclareFieldFormat{title}{\usebibmacro{string+doiurlisbn}{\mkbibemph{#1}}}

\DeclareFieldFormat[article,inbook,incollection,inproceedings,patent,thesis,unpublished]{title}{\usebibmacro{string+doiurlisbn}{\mkbibquote{#1}}}

\DeclareFieldFormat[suppbook,suppcollection,suppperiodical]{title}{\usebibmacro{string+doiurlisbn}{#1}}

\hypersetup{%
pdftitle={},%
}

\begin{document}






\thispagestyle{empty}

\includegraphics[viewport = 9.5cm 11cm 0cm 0cm ,scale = 0.29]{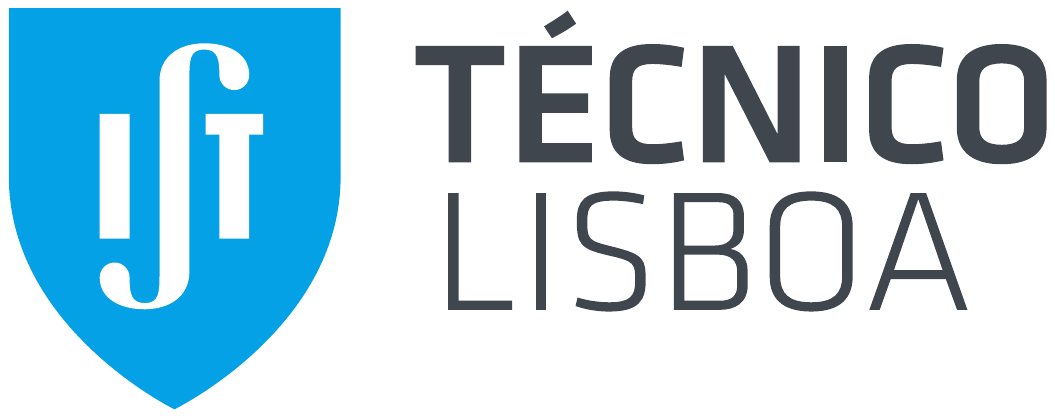}

\begin{center}

\vspace{.025\textheight}
\begin{center}
\includegraphics[height=.35\textheight]{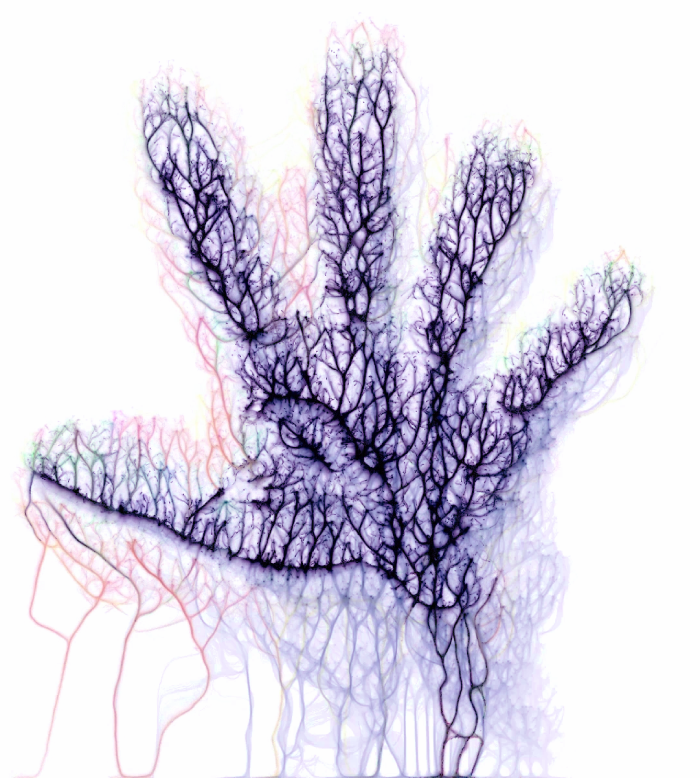}
\end{center}


\vspace{0.5cm}
{\FontLb Applications to Biological Networks of Adaptive Hagen-Poiseuille Flow on Graphs} \\ 
\vspace{0.9cm}

\vspace{0.6cm}
{\FontMb Ana Filipa Martinho Valente} \\ 

\vspace{1.5cm}
{\FontSn \coverThesis} \\
\vspace{0.3cm}
{\FontLb Engineering Physics}  \\

\vspace{1.0cm}
{\FontSn %
\begin{tabular}{ll}
	\coverSupervisor: & Rui Manuel Agostinho Dilão
\end{tabular} } \\

\vspace{1.0cm}
{\FontMb \coverExaminationCommittee} \\
\vspace{0.3cm}
{\FontSn %
\begin{tabular}{rl}
	  \coverChairperson:    & Luís Humberto Viseu Melo  \\
	  \coverSupervisor:     & Rui Manuel Agostinho Dilão   \\
	Member of the Committee:	& Ana Maria Ribeiro Ferreira Nunes
\end{tabular} } \\

\vspace{1.5cm}
{\FontMb June 2023} \\

\end{center}
\null\newpage
\thispagestyle{empty} \null

\setcounter{page}{1}
\pagenumbering{roman}


\thispagestyle{empty}

\null
\vskip5cm
\begin{flushright}
     "No matter where you are, \\
     everyone is always connected."
\end{flushright}
\vfill 
\cleardoublepage 

\chapter*{\acknowledgments}

\addcontentsline{toc}{chapter}{\acknowledgments}

I would first like to express my heartfelt gratitude to my supervisor, Professor Rui Dilão, for his insurmountable patience and availability, for the compelling and thought-provoking conversations, for the rigorous counseling and invaluable feedback, for the kindness, compassion and encouragement he has shown me, and for all the expertise and wisdom he so kindly shared with me. Thanks to his continuous support, to everything he has taught me, and to his commitment to quality academic work, I have a newfound love for research. His guidance has most definitely not only enriched my academic journey but also encouraged my personal growth. 

I'm forever indebted to my wonderful dear boyfriend, André Pereira, for all the unwavering love and support he has shown me throughout not only this difficult, jarring time but ever since we first met. Thank you so much for all your gentle and kindhearted words, for the endless motivation you provide, for all the time we spent relaxing, going on adventures and celebrating, and for helping me flourish into a person I can be ever more proud of. Being able to love you is an enormous privilege.

I would like to deeply thank all my dearest friends, in particular Inês Ferreira, Rita Silva, Bernardo Barbosa, Tomás Lopes and José Maria Nina Carreira Ferreira da Cruz. The conversations we had, the games we played, the music we shared, the cats and the funny images we laughed at together during this intense time all helped me immensely to stay sane and energetic enough to fight through life. I wish you all the very best, peaceful, happy times in your life.

Last but not least, I'm eternally grateful for my family, in particular my parents, for all their unconditional support and caring questions. Despite having to go through some of the toughest times in their life, they have shown me nothing but love and patience and support throughout these last few years, even when I proved difficult and stubborn.

I appreciate and cherish every single one of you. I wish you all the very best in your life.

\vfill
{\centering 
\small{\it I'd like to extend my gratitude to Georgios Cherouvim for allowing his mesmerizing artwork to be displayed on the cover page of this thesis. Cover page image adapted from \cite{image:cover}.}
}
\clearpage


\chapter*{Resumo}

\addcontentsline{toc}{chapter}{Resumo}

\textit{Physarum polycephalum} é um protista unicelular com numerosos núcleos cujo corpo é constituído por uma rede de veias. À medida que explora o seu meio, adapta-se e otimiza a sua rede, tendo em conta estímulos externos. Foi demonstrado que exibe comportamentos complexos, como resolver labirintos, encontrar o caminho mais curto e criar redes robustas, eficientes e económicas. Vários modelos foram desenvolvidos para tentar simular a adaptação da sua rede para compreender os mecanismos por detrás do seu comportamento e desenvolver redes eficientes. 
Esta tese pretende estudar um modelo recentemente desenvolvido e fisicamente consistente baseado em fluxos de Hagen-Poiseuille adaptativos; para tal, irá determinar propriedades das árvores produzidas pelo modelo e irá examiná-las para determinar se são realistas e consistentes com a experiência. Esta tese também pretende usar o mesmo modelo para produzir redes curtas e eficientes, aplicando-o a uma rede de transporte real.
Observámos que o modelo é capaz de criar redes que são consistentes com outras redes biológicas: seguem a lei de Murray no estado estacionário, mostram estruturas semelhantes às presentes nas redes do \textit{Physarum} e ainda exibem peristalse (oscilação dos raios das veias) e \textit{shuttle streaming} (o movimento de trás para a frente do citoplasma do \textit{Physarum}) em algumas partes das redes.
Usámos também o modelo em conjunto com diferentes algoritmos estocásticos para produzir redes curtas e eficientes; quando comparadas com a rede ferroviária de Portugal continental, todos os algoritmos produziram redes mais eficientes que a rede real e alguns produziram redes com melhor relação custo-benefício.

\vfill
{\centering 
\begin{tabular}{p{0.25\linewidth} p{0.65\linewidth}}
	\textbf{\Large Palavras-chave:} & \textit{Physarum polycephalum}; Fluxo de Hagen-Poiseuille; Rede adaptativa; \\%
									& Optimisação de redes; Eficiência de transporte; Árvore mínima de Steiner    %
\end{tabular}
}   
\clearpage

\chapter*{Abstract}

\addcontentsline{toc}{chapter}{Abstract}

\textit{Physarum polycephalum} is a single-celled, multi-nucleated slime mold whose body constitutes a network of veins. As it explores its environment, it adapts and optimizes its network to external stimuli. It has been shown to exhibit complex behavior, like solving mazes, finding the shortest path, and creating cost-efficient and robust networks. 
Several models have been developed to attempt to mimic its network's adaptation in order to try to understand the mechanisms behind its behavior as well as to be able to create efficient networks. 
This thesis aims to study a recently developed, physically-consistent model based on adaptive Hagen-Poiseuille flows on graphs, determining the properties of the trees it creates and probing them to understand if they are realistic and consistent with experiment. 
It also intends to use said model to produce short and efficient networks, applying it to a real-life transport network example. 
We have found that the model is able to create networks that are consistent with biological networks: they follow Murray's law at steady state, exhibit structures similar to \textit{Physarum}'s networks, and even present peristalsis (oscillations of the vein radii) and shuttle streaming (the back-and-forth movement of cytoplasm inside \textit{Physarum}'s veins) in some parts of the networks.
We have also used the model paired with different stochastic algorithms to produce efficient, short, and cost-efficient networks; when compared to a real transport network, mainland Portugal's railway system, all algorithms proved to be more efficient and some proved to be more cost-efficient.

\vfill
{\centering 
\begin{tabular}{p{0.25\linewidth} p{0.65\linewidth}}
	\textbf{\Large Keywords:} 	& \textit{Physarum polycephalum}; Hagen-Poiseuille flow; Adaptive network; \\ %
								& Network optimisation; Transport efficiency; Steiner minimal tree%
\end{tabular}
}
\clearpage



\phantomsection
\renewcommand{\contentsname}{Table of Contents}
\addcontentsline{toc}{chapter}{\contentsname}
{\hypersetup{linkcolor=black}
\tableofcontents%
}
\null\newpage

\phantomsection
\addcontentsline{toc}{chapter}{\listtablename}
{\hypersetup{linkcolor=black}
\listoftables%
}

\phantomsection
\addcontentsline{toc}{chapter}{\listfigurename}
{\hypersetup{linkcolor=black}
\listoffigures%
}

\renewcommand{\nomname}{List of Symbols}
\phantomsection
\renewcommand{\nomgroup}[1]{%
	\ifthenelse{	\equal{#1}{R}	}{	\item[\textbf{Variables}]	}{%
	\ifthenelse{	\equal{#1}{G}	}{	\item[\textbf{Greek symbols}]	}{%
	\ifthenelse{	\equal{#1}{S}	}{	\item[\textbf{Subscripts}]		}{%
	\ifthenelse{	\equal{#1}{T}	}{	\item[\textbf{Superscripts}]	}{%
	\ifthenelse{	\equal{#1}{Z}	}{	\item[\textbf{Other}]			}{%
}}}}}}

\setlength{\nomlabelwidth}{1.5cm}

%
%
%
%
%


\nomenclature[r]{$(i,j)$}{Edge connecting nodes $i$ and $j$.}

\nomenclature[r]{$L_{ij}$}{Length of edge $(i,j)$.}

\nomenclature[r]{$L$}{Total length of graph tree.}

\nomenclature[r]{$r_{ij}$}{Radius of edge $(i,j)$.}

\nomenclature[r]{$Q_{ij}$}{Volumetric flux flowing through edge $(i,j)$.}

\nomenclature[r]{$D_{ij}$}{Conductivity of edge $(i,j)$.}

\nomenclature[r]{$p_{i}$}{Fluid pressure at node $i$.}

\nomenclature[r]{$p_{ij}$}{Pressure difference between nodes $i$ and $j$.}

\nomenclature[r]{$\mathcal{G}$}{Graph (set of nodes connected by a set of edges).}

\nomenclature[r]{$\mathcal{V}$}{Set of nodes of a graph.}

\nomenclature[r]{$E$}{Set of edges of a graph.}

\nomenclature[r]{$x_i^n$}{Spatial coordinate of node $i$ along the $n$th spatial direction.}

\nomenclature[r]{$\eta$}{Dynamic viscosity of fluid.}

\nomenclature[r]{$S_i$}{Inward/outward volumetric flux of fluid to/from the network at node $i$, making node $i$ a source/sink, if $S_i > 0$/$S_i < 0$, respectively.}

\nomenclature[r]{$V_{ij}$}{Volume of edge $(i,j)$.}

\nomenclature[r]{$V$}{Total volume of fluid of the network.}


\nomenclature[r]{$\beta$}{$8\pi\eta$.}

\nomenclature[r]{$\alpha$}{$V/\beta$.}

\nomenclature[r]{$\mathcal{P}$}{Total dissipated energy per unit time of the network.}

\nomenclature[r]{$\mathcal{L}$}{Lagrangian.}

\nomenclature[r]{$\tau$}{Re-scaled time variable: $ct$, with $c$ a positive constant.}

\nomenclature[r]{$I_0$}{Total inward volumetric flux given by all the source nodes to the network.}

\nomenclature[r]{$\nu$}{Average length between all possible pairs of sites.}

\nomenclature[r]{$E'$}{Set of effectively conducting edges of a graph (edges whose conductivity value is above a certain threshold).}

\nomenclature[r]{$L_{s_i s_j}$}{Length of shortest path in tree between sites $s_i$ and $s_j$.}

\nomenclature[r]{$\sigma$}{Standard deviation.}

\nomenclature[r]{$\overline{a}$}{Mean value of variable $a$.}

\nomenclature[r]{$\Delta\tau$}{Time step used in the adaptive conductivities network algorithm.}

\nomenclature[r]{$P$}{Perimeter of a polygon.}

\nomenclature[r]{$L_{\text{Stei}}$}{Theoretical length for the minimum Steiner tree for a polygon.}






\nomenclature[s]{$i,j,k,m$}{Indices pertaining to nodes (single index) or edges (double index) of the network.}




\printnomenclature
\addcontentsline{toc}{chapter}{\nomname}

\renewcommand{\glossaryname}{List of Abbreviations}
\phantomsection
\glsaddall
\printglossary[type=\acronymtype, title=\glossaryname]   



%
\newacronym{H-P}{H-P}{Hagen-Poiseuille.}
\newacronym{Physarum}{Physarum}{\textit{Physarum polycephalum}.}
\newacronym{hr(s)}{hr(s)}{Hour(s).}
\newacronym{FS}{FS}{Food source(s).}
\newacronym{SMT}{SMT}{Steiner minimum tree.}
\newacronym{MST}{MST}{Minimum spanning tree.}
\newacronym{FT}{FT}{Fault tolerance.}
\newacronym{TL}{TL}{Total length.}
\newacronym{Freq. inv.}{Freq. inv.}{Frequency of inversion.}
\newacronym{Rel. str.}{Rel. str.}{Relative strength.}
\newacronym{CE}{CE}{Cost-efficiency.} 
\addcontentsline{toc}{chapter}{\glossaryname}

\cleardoublepage 


\setcounter{page}{1}
\pagenumbering{arabic}


\chapter{Introduction}
\label{chapter:intro}

\section{Motivation}

There is a multitude of dynamical systems (biological, physical, electrical) that can be described using networks and graphs. Among them, transport networks are prevalent. Transport networks are essential, as they must allocate information and resources as efficiently and quickly as possible throughout a system. Some examples of transport networks whose study remains relevant today are the phloem and xylem of plants, the network of channels that transport nutrients in their leaves \cite{leaves}, the vein irrigation network of a tumor \cite{rubinow_book}, human transit networks, such as road or railway systems \cite{transport1, transport2}, various electrical networks \cite{electric} and organisms like \textit{Physarum polycephalum}, more commonly referred to as slime mold (described in section \ref{sec:physarum}), or \textit{Dictyostelium discoideum} \cite{dictyostelium}, which display highly adaptive vessel networks. There are also a number of mathematical problems and formal systems that constitute relevant and difficult network optimisation problems, such as Steiner tree geometries \cite{steiner_tero}, the traveling salesman problem \cite{travelling_salesman} and the first passage percolation \cite{percolation}, whose results still have important applications in physics, engineering and biology today. The study of these real-world networks and these theoretical problems can lead to results that are often applicable to many other different transport networks and systems.

Key aspects of these networks can be described using hydrodynamics. While some of the transport networks do in fact transport a fluid of some sort, such as the cardiovascular system, the xylem and phloem, or the adaptive network of veins of organisms like \textit{Physarum polycephalum}, many other transport networks can be approximately described using fluid dynamics. Specifically, many relevant systems can be described by the Hagen-Poiseuille formalism (which is described in section \ref{sec:hp-flow}), which describes the laminar flow of viscous fluids through a cylindrical vessel of constant cross section. Some biological systems of relevance that can be described this way are blood vascular systems and \textit{Physarum polycephalum}.

\textit{Physarum polycephalum} specifically has recently become a prolific topic of discussion, as this acellular protist displays high-level behaviours, such as solving mazes \cite{maze}, despite not having centralized control or a neurological system. The network formed by its veins adapts dynamically to its environment; this adaptation consists of the formation of new veins, the destruction of old, less important paths, and the modification of vein radii over time. Due to these adaptations, in the presence of several different food sources it can create networks with a comparable efficiency, fault tolerance and cost to real human-made networks (such as the Tokyo rail network) \cite{tero_tokyo}. By studying this brainless, simple organism, one could understand the mechanisms behind its obtainment and processing of information and decision-making. After all, \textit{Physarum} and all living beings have gone through the process of evolution for millions of years, and have thus optimised their transport networks greatly; they can provide great inspiration for the development of efficient problem-solving algorithms.

Modeling this type of Hagen-Poiseuille adaptive flow in these biological networks is important and can provide knowledge about many diverse important topics. Namely, it may grant more understanding regarding efficient, robust and cost-effective network formation, growth and adaptation. This understanding could then be applied to several different areas and challenging problems; for example, it could provide insight into angiogenic processes like cancer development.

\section{Objectives}

This thesis has two main intentions: first, to study the adaptive Hagen-Poiseuille flows model, analyzing several properties and consequences of the adaptive flow on graph structures and flow behavior, and understand how reliable it is at predicting real physical results; second, to apply said model to a specific graph theory problem and determine how effective the model is at solving it.

The first objective comprises several additional goals. Said goals are 1) determining how the model behaves when the mesh it is applied to reaches the continuous limit, 2) establishing whether or not Murray's law is verified dynamically for this model, 3) studying properties of simple networks and how they compare to real results, 4) comparing the results of the model to the patterns exhibited by \textit{Physarum polycephalum} and 5) ascertaining whether or not the model exhibits key physical phenomena shown by \textit{Physarum}: shuttle streaming and peristalsis. 

The second objective, which essentially is utilizing the model to find contrasting trees of different efficiencies and extremal lengths, will be achieved by applying the model to two different types of configurations (geometric regular polygons and a real-world transport network case) and using dynamic algorithms to obtain said different trees.

\section{Thesis outline}

This thesis is organized into five chapters. Chapter \ref{chapter:intro} starts by introducing the work's subject matter and objectives and the motivation behind the thesis.

Chapter \ref{chapter:motivation} describes into detail \textit{Physarum polycephalum}'s physical characteristics and its relevant intelligent behaviors, as well as the state-of-the-art network models used to mimic its behavior. It also introduces and explains the adaptive Hagen-Poiseuille flows model (that aims to describe \textit{Physarum}'s network dynamics) that will be the basis of the work of this thesis, and this chapter concludes with some notes regarding some graph theory problems that will be addressed later.

Chapter \ref{chapter:properties} delineates the algorithms and methods used to put said adaptive flows model into practice and tests the model regarding some relevant physical properties that are observed in the simulations. It concludes with a direct application of the model to the comparison of obtained trees with \textit{Physarum}'s real network attributes.

Chapter \ref{chapter:Steiner} illustrates another application of the model in question: providing stochastic solutions to the Steiner minimal tree problem in graph theory. The stochastic algorithms used to create the steady-state trees and the parameters used to evaluate them are first introduced and are later applied to two different cases: first, to simple geometric polygons, and then, to a real communication network (the case of mainland Portugal).

To complete the work, chapter \ref{chapter:conclusion} provides a broad synoptic analysis of the results presented in this thesis and the conclusions that are possible to derive from them, as well as suggests several additional work topics and improvements to the model that could derive from this work.
\chapter{Physarum polycephalum and its network dynamics}
\label{chapter:motivation}

\section{Physarum polycephalum} \label{sec:physarum}

\textit{Physarum polycephalum}, also known as slime mold or "the blob", has been the subject of many biological and biochemical studies throughout the 20th century and, in recent years, of physical studies. Some of the main goals of the modern studies are to shine a light on mechanisms of information processing, along with studying self-regulated networks and their topology.

\textit{Physarum} has been studied for many reasons, namely for its simplicity regarding culturing and handling in the lab, its macroscopic size (which facilitates observation), and its unique "intelligent" behavior.

I will start by describing slime mold physically in section \ref{sec:physarum_physical}; then, I will discuss certain relevant oscillatory phenomena theorized to be at the source of \textit{Physarum}'s motion and intelligent behavior in section \ref{sec:physarum_periodic}; finally, I will describe said intelligent behavior that \textit{Physarum} is reported of showing in section \ref{sec:physarum_intelligent} and discuss its implications.

\subsection{Physical characteristics} \label{sec:physarum_physical}

\textit{Physarum polycephalum} is a unicellular amoeboid protist with multiple nuclei. 

Despite being called "slime mold", it is not a fungus, but rather it displays characteristics shared by fungi, animals and plants. Its complex life cycle allows it to have strong versatility and survivability; for example, \textit{Physarum} can enter a dormant state for up to several years in non-favorable environmental conditions, and said state can be reversed easily by providing it with water and nutrients \cite{andrew_physarum_machines}.

\begin{figure}
    \centering
    \begin{subfigure}[b]{0.53\textwidth}
        \centering
        \includegraphics[width = \textwidth]{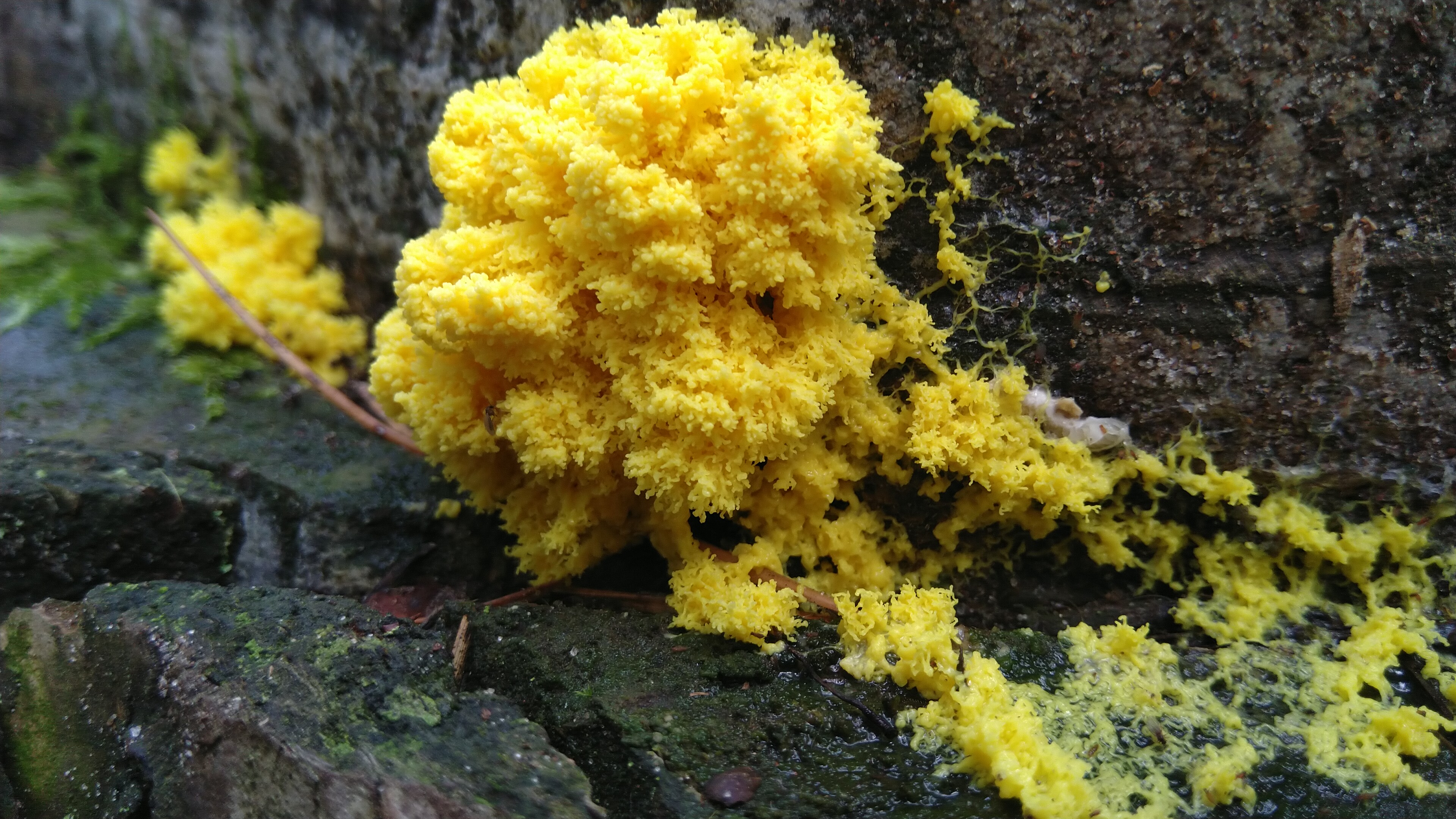}
        \caption{}
        \label{fig:physarum_tree}
    \end{subfigure}
    \begin{subfigure}[b]{0.4\textwidth}
        \centering
        \includegraphics[width = \textwidth]{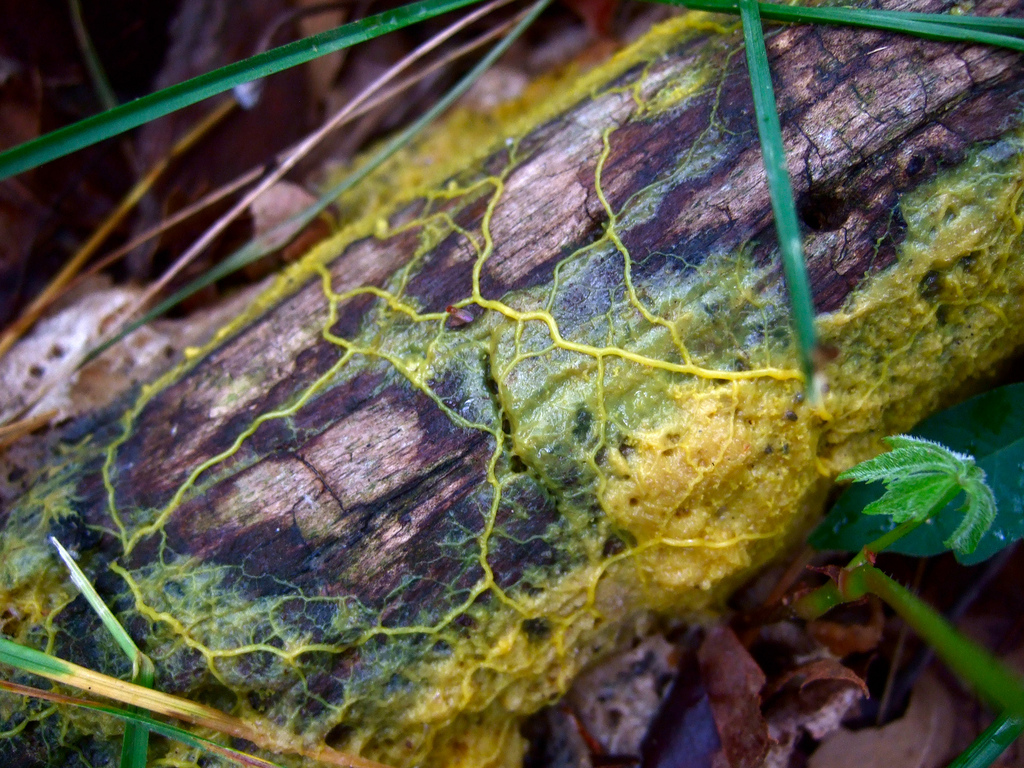}
        \caption{}
        \label{fig:physarum_tree_net}
    \end{subfigure}
    \caption{\textit{Physarum polycephalum}'s plasmodium in its natural environment, growing on tree bark. \textbf{a)} An amorphous yellow mass forms after some time of growth. \textbf{b)} The food foraging and growth process for \textit{Physarum} involves the creation of a network of veins of protoplasm, which adapts over time. Images obtained from a) \cite{image:physarum_tree_big} and b) \cite{image:physarum_tree}.}
    \label{fig:physarum}
\end{figure}

The most frequently found form of \textit{Physarum} is the plasmodium. In the plasmodial stage of its life cycle, it is a single extremely large multinucleate cell containing millions of nuclei. It appears as a macroscopic, bright yellow amorphous mass that can grow up to a size of 10 m$^2$. It prefers to live in damp and dim habitats and is often seen growing on the sides of trees (see figure \ref{fig:physarum}) \cite{Oettmeier_blob}.

\begin{figure}
    \centering
    \begin{subfigure}[b]{0.4\textwidth}
        \centering
        \includegraphics[width = \textwidth]{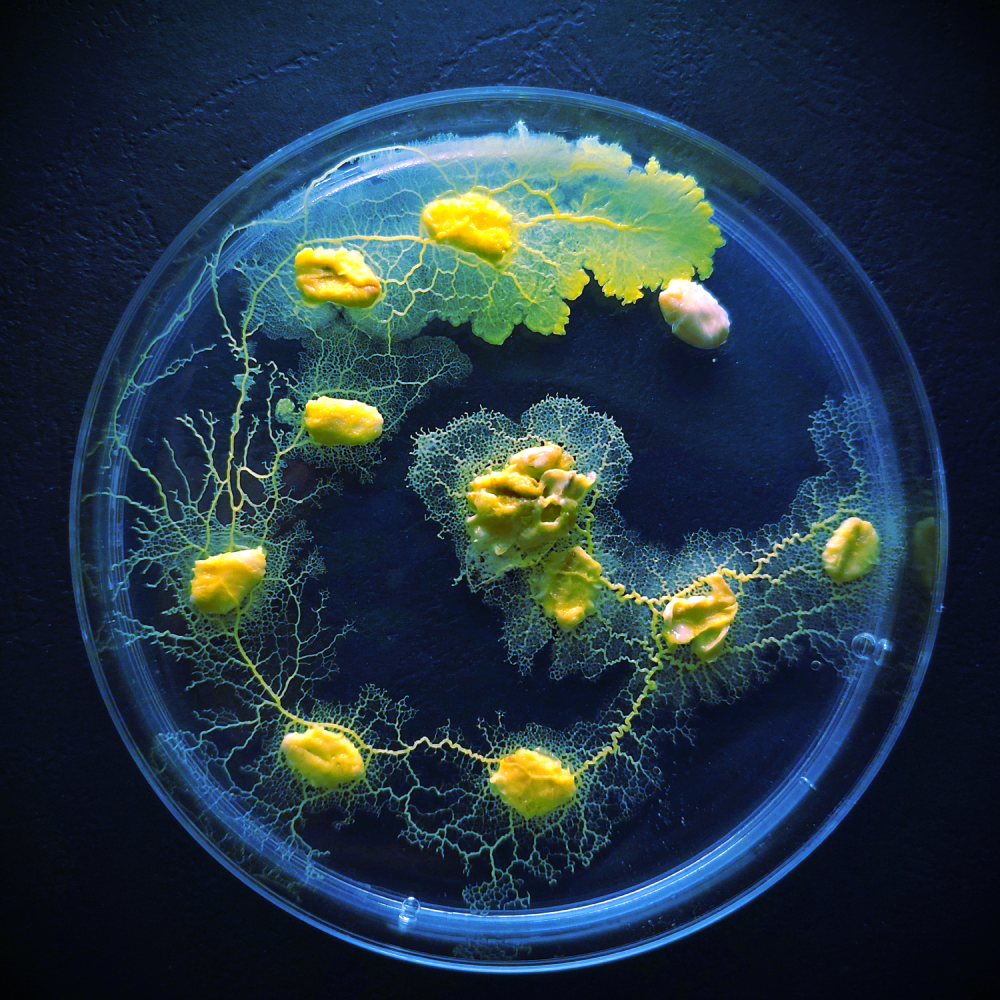}
        \caption{}
        \label{fig:physarum_lab_network}
    \end{subfigure}
    \begin{subfigure}[b]{0.538\textwidth}
        \centering
        \includegraphics[width = \textwidth]{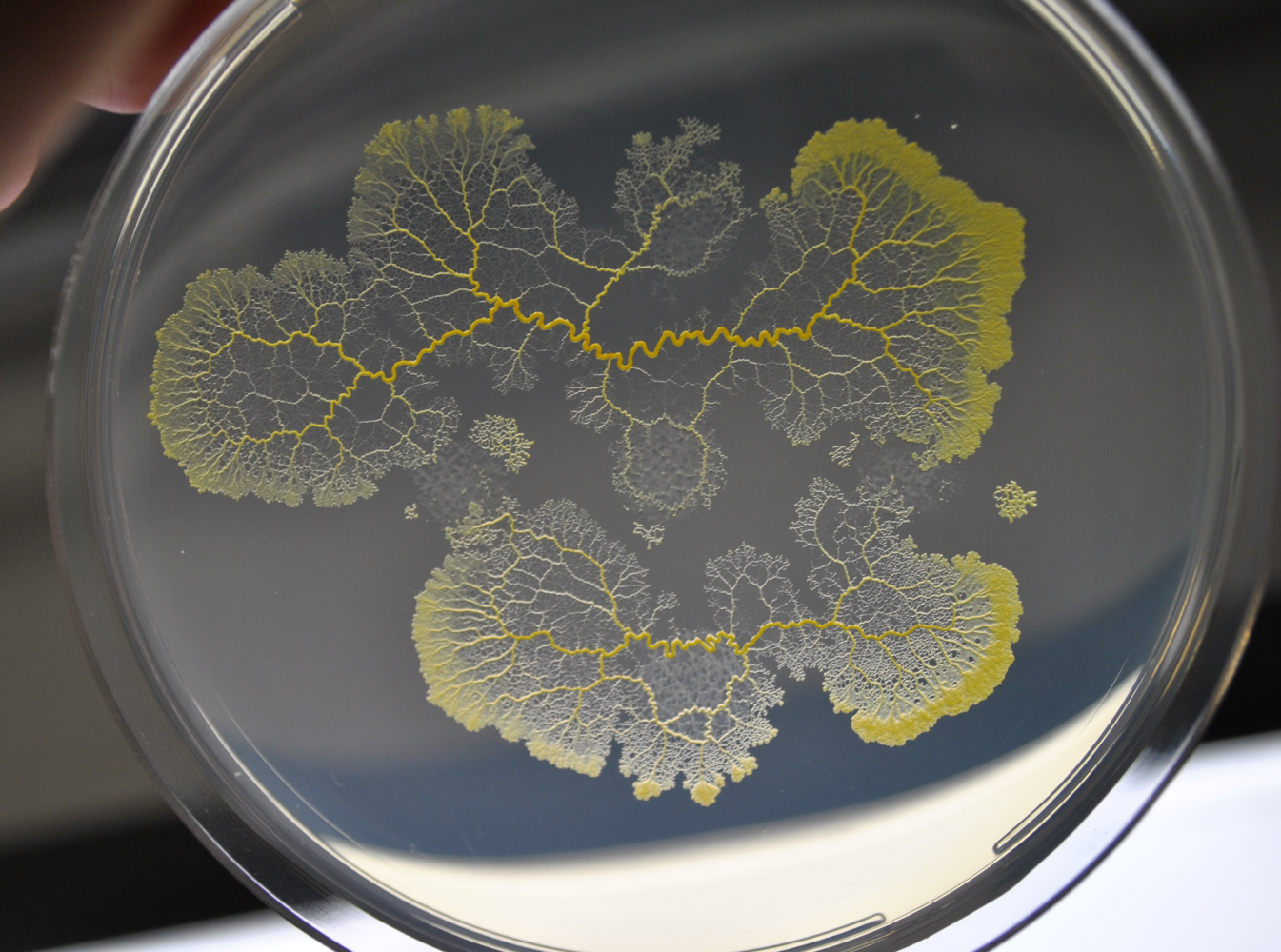}
        \caption{}
        \label{fig:physarum_lab_growth}
    \end{subfigure}
    \caption{\textit{Physarum polycephalum}'s plasmodium in a laboratory environment. \textbf{a)} Notice the network created between different food sources (the yellow blobs): a single, emphasized vein forms between the food sources, surrounded by a number of tiny, thin veins; at the top of the image, \textit{Physarum}'s network grows indiscriminately in the direction it has yet to explore. \textbf{b)} \textit{Physarum's} new growth is signaled by numerous, thin veins growing outwards in fan-like patterns (see the edges of the network), while the older part of the network shows optimised, few, thick veins connecting to the edges of the network. Images obtained from a) \cite{image:physarum_lab_network} and b) \cite{image:physarum_lab_growth}.}
    \label{fig:physarum_lab}
\end{figure}

It feeds on various organic materials, such as bacteria, mushrooms, fungal spores, and decaying matter (as well as cornflakes when fed in a laboratory environment), by covering the food with its protoplasm, which is later digested in the body by enzymes. To procure food sources, this living being can move at speeds of about 1-4 cm/h \cite{andrew_physarum_machines}; it spreads its protoplasm radially outward, in the shape of a network of tubular veins. As \textit{Physarum} grows, its network's topology is modified and optimised (see figure \ref{fig:physarum_lab}).

\textit{Physarum}'s cell is not only made up of an extremely large number of nuclei, but also of different forms of cytoplasm (that is, the fluid inside the cell that is located outside of the nuclei): endoplasm, which is the fluid cytoplasm that moves throughout the cell, and ectoplasm, which is the gel-like rigid cytoplasm that makes up the outer membrane of the cell and contains the cytoskeleton of the cell. This cytoskeleton is made up of a system of proteins, notably actin and myosin. While the filaments of actin are responsible for the structural support of the vein walls, myosin is the motor protein involved. \cite{Oettmeier_blob}

\subsection{Dynamic, periodic phenomena} \label{sec:physarum_periodic}

A solid grasp on \textit{Physarum}'s locomotion is necessary to construct a proper toy model of this organism. \textit{Physarum polycephalum} locomotion is ensured by two periodic phenomena: shuttle streaming and peristalsis, and these two are theorized to be connected.

Peristalsis is a wave of cross-sectional contractions across a tubular vein of the network. The interactions between the proteins that make up the cytoskeleton of the cell, actin and myosin, create relaxation-contraction cycles of the vein walls, which propels the movement of the endoplasm.

Shuttle streaming describes the movement of the endoplasm through the network veins. Due to peristalsis, the fluid moves in a back-and-forth pattern throughout the veins of the entire organism (with a period of about 100s). This allows for the distribution of resources throughout an entire cell.

These rhythmic oscillations and their timely coordination are what allow the organism to move in a certain direction. A gradient of pressure is produced by these waves, which propels the movement of endoplasm towards the leading edges (where the growth of ectoplasmic proteins occurs at the same time). \cite{Oettmeier_blob}

\textit{Physarum} individuals are also able to maximize their internal endoplasmic ﬂows by adapting these contraction waves to their size, thus optimizing transport. 
It was shown that the transport is optimal when the wavelength of the peristaltic wave is of the order of magnitude of the size of the network. Thus, \textit{Physarum}, despite not having a nervous system, is able to coordinate its growth and adapt the flow to its size. 
These adaptive patterns come about due to the interaction of a global mass conservation constraint with a local constraint: minimizing the phase difference between neighboring vein sections. 
These contractions are also driven by external stimuli, as the amplitude and frequency of the contractions are altered across the whole organism. Attractive stimuli, like food sources, usually increase the amplitude and frequency of the contractions, while repulsive stimuli, like excessive lighting, decrease said amplitude and frequency. \cite{alim_peristalsis}
Even though it is not yet well understood how all these oscillations are coordinated throughout the entire organism and how they arise from stimuli, it is theorized that a signaling molecule could be involved: this molecule could cause local increases in contraction amplitude as it was transported along with the endoplasm and could initiate a feedback loop to propel its own motion \cite{alim_signal_prop}.

These biochemical and hydrodynamic oscillations thus allow \textit{Physarum} to grow its network and to dynamically adapt its morphology to external stimuli.

\subsection{Intelligent behaviour observed} \label{sec:physarum_intelligent}

Even though \textit{Physarum} is a comparatively simple protistic organism, with no nervous system or centralized control, it has been shown to exhibit high-level intelligent behavior and create highly optimized networks. These behaviors are now described.

\begin{figure}
    \centering
    \begin{subfigure}[b]{0.75\textwidth}
        \centering
        \includegraphics[width = \textwidth]{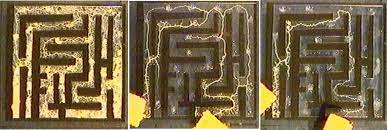}
        \caption{}
        \label{fig:physarum_maze}
    \end{subfigure}
    
    \begin{subfigure}[b]{0.75\textwidth}
        \centering
        \includegraphics[width = \textwidth]{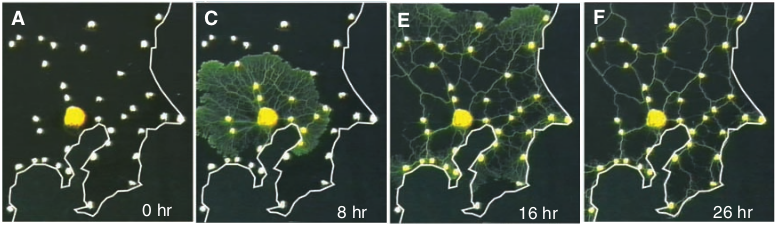}
        \caption{}
        \label{fig:physarum_tokyo}
    \end{subfigure}
    \caption{\textit{Physarum polycephalum}'s intelligent behaviour experiments. \textbf{a)} \textit{Physarum} solving a maze over time. First, the specimen is made to cover the entire maze and food sources are placed at each end of the maze (left); over time, it adapts its network (middle) until it reaches the shortest length network possible connecting both food sources/ends of the maze, which is the solution to the maze (right). \textbf{b)} \textit{Physarum} constructing a network resembling the Tokyo rail system over time. The specimen is placed in a central location, while food sources are placed in the geographical location of major cities of the Tokyo region (A - 0hr); \textit{Physarum} starts by growing radially from its original location (C - 8hr) until it reaches all food sources (E - 16hr); as it grows, it adapts its network, eliminating unnecessary links and reinforcing central ones, creating a cost-efficient and robust network (F - 26hr). Images adapted from a) \cite{maze} and b) \cite{tero_tokyo}.}
    \label{fig:physarum_maze_tokyo}
\end{figure}

A famous experiment was conducted to determine whether or not \textit{Physarum} was able to solve a maze, and the protist did so successfully \cite{maze}. A \textit{Physarum} specimen was made to cover an entire maze with agar substrate and plastic film walls (agar acts as food for the specimen, while the plastic film is a dry surface with the specimen avoids). After an operator placed food (agar blocks) at the entrance and exit of the maze, \textit{Physarum} rearranged itself: it eliminated useless veins and enlarged the most favorable ones. It created the most efficient and shortest path between the two food sources every time, thus solving the maze (figure \ref{fig:physarum_maze}). 

Later experiments had \textit{Physarum} recreate different man-made transportation systems; most notably, \textit{Physarum} recreated the Tokyo region's railway system. Food sources (oat flakes) were placed in such a way that they mimicked the location of major cities in the region and a \textit{Physarum} specimen was placed on a central food source. Topographical limitations like mountainous terrain or bodies of water were also taken into account, and were replicated using light sources, as \textit{Physarum} avoids luminous locations. As \textit{Physarum} grew out of the central location (radially outward) and it found the food sites surrounding it, it adjusted its veins and the morphology of its network (figure \ref{fig:physarum_tokyo}). This network was found to have "comparable efficiency, fault tolerance and cost" to the Tokyo railroad network. Note that "cost" refers to the total length of the network, "efficiency" means the average minimum distance between each pair of food sources, and "fault tolerance" means the probability of disconnecting a part of the network by removing a certain link. \cite{tero_tokyo} 

These experiments showcase one of the important applications of studying \textit{Physarum}. Designing transport networks that take into account both robustness and cost-efficiency is a modern, ever-present problem. \textit{Physarum} is able to create such networks naturally as part of its growth process, and efficient algorithms can be developed inspired by it.

There are even more intelligent behaviours associated with \textit{Physarum}. For instance, slime mold has also been found to exhibit memory, both internal \cite{Saigusa_internal_memory} and external \cite{Reid_external_memory}. As \textit{Physarum} moves around its environment, it leaves behind a trail of mucus and, when the organism comes across its trail on its search for sustenance, it avoids the already explored areas - this is what one means by external memory. Additionally, \textit{Physarum} can also correctly choose the most appropriate and efficient diet for itself when provided with several options \cite{Reid_correct_food}.

Even though several protistic or fungic species (in addition to \textit{Physarum polycephalum}) show intelligent behaviour despite not having a nervous system, the mechanisms behind this intelligence are not yet comprehended. The goal of studying this brainless, simple organism is to understand the mechanisms behind the obtainment and processing of information and decision-making.

\section{\textit{Physarum} network models} \label{sec:physarum-models}

\textit{Physarum polycephalum}'s plasmodium stage is quite complex, and there exists no one model yet to describe it fully. Models that describe \textit{Physarum} usually focus on one of three connected aspects that are ultimately responsible for \textit{Physarum}'s behaviour: its network's adaptation, its growth or the oscillatory patterns described in section \ref{sec:physarum_periodic}. This thesis shall focus on \textit{Physarum}'s network adaptation, and on a particular recently-developed model.

\textit{Physarum}'s network optimization models are used to not only better understand the mechanisms behind this organism's behavior, but to also solve different graph theory problems or related real-world issues \cite{sun_review, gao_survey}.

There are two main models described in the literature: the multi-agent model proposed by Jones \cite{jones_multi-agent} and the \textit{Physarum Solver} proposed by Tero et al. \cite{tero_shortest_path}. 

The multi-agent system is a phenomenological model that states that the macroscopic complex network modifications observed in \textit{Physarum} networks are a product of simple microscopic interactions between small parts of the plasmodium that acted as particle-like agents. While this model was able to obtain shortest path-like patterns and Steiner minimum tree-like solutions, it is a biologically unrealistic model as agents do not act like a continuous network and always move towards areas of positive stimuli, which is not accurate to \textit{Physarum}'s food searching process.

The \textit{Physarum solver}, on the other hand, is a model based on the hydrodynamics of the endoplasm, in which network optimization happens through a feedback loop between the flows of the network and the size of the tubular veins that make it up. This model proved to be more biologically realistic than the multi-agent system and it showed resulting networks similar to those created by \textit{Physarum}. However, the model isn't physically consistent, as it does not conserve the total volume of fluid of the network.

The model used to describe \textit{Physarum}'s network optimization dynamics in this thesis was developed by Almeida and Dilão \cite{rodrigo_artigo}. It is based on Tero's \textit{Physarum Solver} but solves its biggest physical inconsistency: the fact that volume needs to be conserved. It also uses dissipated energy minimization considerations to obtain its adaptation equations. It is described in the following section \ref{sec:adaptive-model}.

\section{Adaptive Hagen-Poiseuille flow on graphs} \label{sec:adaptive-model}

In this section, the model derived by Almeida and Dilão \cite{rodrigo_artigo} used in this thesis to illustrate the \textit{Physarum} network optimization process is described. First, an introduction to the Hagen-Poiseuille law, which is essentially the physical basis of the model, is done in section \ref{sec:hp-flow}. Then, an overview of the derivation of the model is done in section \ref{sec:formalism}. For a full derivation, please refer to \cite{rodrigo_tese}. Finally, a biologically relevant law illustrated in this model called Murray's law is discussed in section \ref{sec:murray-law}.

\subsection{Hagen-Poiseuille flow} \label{sec:hp-flow}

\par The Hagen-Poiseuille (H-P) flow refers to the steady laminar flow of an incompressible, viscous fluid (that is, a fluid with a low Reynolds number) through a cylindrical channel with slippery boundary conditions.
\par The formalism is obtained by applying several simplifications to the Navier-Stokes equation for a cylindrical channel, namely symmetry considerations, assuming steady-state conditions and incompressible fluid considerations, and also taking into account mass conservation. 
\par It is often used to describe blood flow \cite{rubinow, murray_og} and \textit{Physarum polycephalum}'s endoplasm flow \cite{tero_tokyo, alim_peristalsis}.
\par For a channel $(i,j)$ (that is, between points/nodes $i$ and $j$), with length $L_{ij}$ and radius $r_{ij}$, the Hagen-Poiseuille law gives the flux $Q_{ij}$ in the channel at steady state:

\begin{equation} \label{eq:HP}
    Q_{ij} = D_{ij}\frac{p_i - p_j}{L_{ij}}
\end{equation}

\noindent
in which $p_i$ is the pressure at node $i$ and $D_{ij}$ is the conductivity of the channel $(i,j)$ given by:

\begin{equation} \label{eq:Dij}
    D_{ij} = \frac{\pi r_{ij}^4}{8\eta}
\end{equation}

\noindent
in which $\eta$ is the dynamic viscosity of the fluid. Note that $r_{ij} = r_{ji}$, and thus $D_{ij} = D_{ji}$.
\par If $Q_{ij} > 0$, then the fluid flows from $i$ to $j$; if $Q_{ij} < 0$, then the fluid flows in the opposite direction, from $j$ to $i$. As $D_{ij} = D_{ji}$ and $L_{ij} = L_{ji}$, then the direction of the flow (and the sign of $Q_{ij}$) is determined by the sign of $p_{ij} = p_i - p_j$.

\par H-P flow is the base of the mathematical formalism that will be used in this thesis (see section \ref{sec:formalism}).

\subsection{Adaptive Hagen-Poiseuille model derivation} \label{sec:formalism}

\par The model used to describe the adaptive Hagen-Poiseuille flow on graphs will be the one described in \cite{rodrigo_artigo, rodrigo_tese}. This model consists on a class of equations that describe the flow of a viscous fluid through a network of straight channels (which can be described by a graph) with several static sources and sinks and adaptive channel conductivities. 

This class of equations was derived as a way to model the dynamics and optimisation processes of \textit{Physarum polycephalum} vein networks, and to correct previously derived models. Over the last twenty years, many other models were created aiming to describe \textit{Physarum}'s network's adaptive behaviour, namely vein formation and network optimisation over time. Some of these models also used the Hagen-Pouiseuille law to describe \textit{Physarum} flow on graphs, and, to be able to achieve channel conductivity adaptation, they used local, specific evolution equations and stochastic update methods. The models converged to networks with steady paths, sometimes close to Steiner tree-like networks (see \cite{tero_steiner, caleffi_steiner}) or shortest path solutions (see \cite{zhu_shortest_path, tero_shortest_path}). Despite their successes, these models were not consistent with some key physical notions: they did not preserve the volume of fluid of the network and, as such, the process of updating the conductivities on one site of the network was independent of the rest of the network. This is not a realistic mechanic as, for an incompressible fluid (which the H-P flow equation describes), for constant volume of fluid, any local change in conductivity must be followed by a global response. Thus, the class of equations used in this thesis preserves the volume of fluid and uses an update law that takes into consideration the entire network.

The class of equations used in this thesis and its derivation is now described. To describe a network of veins, one uses undirected graphs. Consider simple graphs $\mathcal{G}=(\mathcal{V},E)$ embedded in a $n$ dimensional Euclidean space. These graphs are made up of a set of $N$ nodes, $\mathcal{V}$, with spatial coordinates given by $ (x^1_i,...,x^n_N), \quad i = 1,...,N$, connected by a set of $M$ straight edges, $E$, described by $(i,j)$ (this describes the edge connecting nodes $i$ and $j$). The edges represent cylindrical, straight channels with elastic walls (that can expand or contract transversally to the flow of fluid). Thus, edge $(i,j)$ has a length $L_{ij}$ and radius $r_{ij}$. Each edge can also be described by its conductivity $D_{ij}$ (see eq. \eqref{eq:Dij}). The flow can be described using the Hagen-Poiseuille law (eq. \eqref{eq:HP}).
\par The network has $K$ sources and $R$ sinks, which are located at fixed nodes. At these nodes, there is a inward flux of fluid to the network (in the case of a source) or an outward flux of fluid from the network (in the case of a sink). The inward/outward flux of fluid at node $j$ is represented by $S_j$. If $S_j > 0$, node $j$ is a source; if $S_j < 0$, node $j$ is a sink; if $S_j = 0$, node $j$ is a regular node.
\par As the fluid is incompressible,

\begin{equation} \label{eq:sources_sinks}
    \sum_{j=1}^N S_j = \sum_{j:\text{sources}} S_j + \sum_{j:\text{sinks}} S_j = 0
\end{equation}

\par This shows the volume of fluid in the network should be constant. The volume of the cylindrical channel $(i,j)$ is given by $V_{ij} = \pi r_{ij}^2 L_{ij} = L_{ij}\sqrt{8\pi\eta}\sqrt{D_{ij}}$. As such, the (constant) volume of fluid in the network is given by:

\begin{equation} \label{eq:volume}
    V = \sum_{(i,j)\in E} V_{ij} = \beta \sum_{(i,j)\in E} L_{ij}\sqrt{D_{ij}}
\end{equation}

\noindent
with $\beta = \sqrt{8\pi\eta}$. Thus, the total volume of fluid $V$ is determined by the channel conductivities and the channel lengths. As the volume is constant, this volume is thus set by the initial conductivities of the system (see sec. \ref{sec:renormalization_volume} for more information).
\par The steady state is determined by the Kirchhoff conservation laws, which state that flux is conserved at each node $j$:

\begin{equation} \label{eq:kirch}
    \sum_{(i,j)\in E} Q_{ij} = S_i \quad,\quad i = 1, ... , N
\end{equation}

\par However, the fluxes and pressures determined by equations \eqref{eq:kirch} are a steady state solution obtained for fixed values of conductivities $D_{ij}$ and lengths $L_{ij}$. One wants to obtain an adaptive solution over time as, for living organisms, the flow of the veins is shown to alter the radii/conductivities of the channels \cite{wohlfarth_physarum, rubinow}.

\par The adaptation law for the rate of change of the conductivities of the network $D_{ij}$ chosen was:

\begin{equation} \label{eq:Dij_ansatz}
    \frac{d}{dt}\sqrt{D_{ij}} = f(\mathbf{Q}) - c\sqrt{D_{ij}}
\end{equation}

\noindent
where $f(\mathbf{Q})$ is a function of the global flux of the network (with $f(\mathbf{0}) = 0$) and $c$ is some positive constant. Notice that $\sqrt{D_{ij}}$ is proportional to the transverse sectional area of channel $(i,j)$.

Using the condition of conservation of volume and eq. \eqref{eq:volume}, one gets:

$$\frac{d}{dt}V = \beta \sum_{(i,j)\in E}L_{ij}\frac{d}{dt}\sqrt{D_{ij}} = 0$$

\noindent
and, using eq. \eqref{eq:Dij_ansatz}, one gets:

$$f(\mathbf{Q}) \sum_{(i,j)\in E}L_{ij} = c \sum_{(i,j)\in E} L_{ij} \sqrt{D_{ij}} = \frac{c}{\beta}V$$

\noindent
which is constant.

One can define a new function $g$ by:

\begin{equation} \label{eq:f}
    f(\mathbf{Q}) = V\frac{c}{\beta}\frac{g(Q_{ij})}{\sum_{(k,m)\in E} L_{km}g(Q_{km})}
\end{equation}

\par Substituting the function $f$ defined in \eqref{eq:f} in eq. \eqref{eq:Dij_ansatz}, one obtains:

\begin{equation} \label{eq:Dij_adapt}
    \frac{d}{d\tau}\sqrt{D_{ij}} = \alpha\frac{g(Q_{ij})}{\sum_{(k,m)\in E}L_{km}g(Q_{km})} - \sqrt{D_{ij}} \quad,\quad (i,j)\in E
\end{equation}

\noindent
where $\tau = ct$ and $\alpha = V/\beta = V/\sqrt{8\pi\eta}$.
\par For any choice of the function $g$, the adaptation equation \eqref{eq:Dij_adapt} ensures that the volume of fluid $V$ remains constant, and thus it contains a term that depends on the conductivities of all the edges of the network. 
\par To be able to compute the temporal evolution of the system given by eq. \eqref{eq:Dij_adapt} and observe the dynamic changes in network shape and conductivity values, one needs to choose a $g$ function. This function was obtained by introducing a criterion based on minimizing dissipated energy, keeping the volume of fluid constant.
\par The dissipated power of the H-P steady flow $\mathcal{P}$ is given by:

\begin{equation} \label{eq:dissipated-power}
   \mathcal{P} = \sum_{(i,j)\in E} \frac{Q_{ij}^2}{D_{ij}}L_{ij}
\end{equation}

\par Thus, using eqs. \eqref{eq:dissipated-power} and \eqref{eq:volume}, one can construct the Lagrangian:

$$\mathcal{L} = \mathcal{P} - \lambda\left(V - \beta\sum_{(k,m)\in E}L_{km}\sqrt{D_{km}}\right)$$

\noindent
where $\lambda$ is a Lagrange multiplier. Minimising $\mathcal{L}$ with respect to $D_{ij}$ and $\lambda$, and using eq. \eqref{eq:dissipated-power}, one gets:

$$\begin{dcases}
\frac{\partial \mathcal{L}}{\partial \sqrt{D_{ij}}} & = \left(-\frac{Q^2_{ij}}{D^2_{ij}}L_{ij} + 2 \sum_{(k,m)\in E} \frac{Q_{km}}{D_{km}}\frac{\partial Q_{km}}{\partial D_{ij}}L_{km} + \lambda\beta\frac{L_{ij}}{2\sqrt{D_{ij}}}\right)2\sqrt{D_{ij}} = 0\\
\frac{\partial \mathcal{L}}{\partial \lambda} & = V - \beta\sum_{(k,m)\in E}L_{km}\sqrt{D_{km}} = 
\end{dcases}$$

The second term on the right-hand side of the first equation is zero (compare to lemma 2.1 of \cite{Haskovec_2019}). As such, solving the equations in respect to $D_{ij}$ and $\lambda$, the conductivity values that minimise the dissipated power are:

\begin{equation} \label{eq:solution_lagrangian}
    D^*_{ij} = \alpha^2\frac{Q^{4/3}_{ij}}{\sum_{(k,m)\in E}L_{km}Q^{2/3}_{km}} \qquad,\qquad D^{**}_{ij} = 0
\end{equation}

Comparing $D^*_{ij}$ of eq. \eqref{eq:solution_lagrangian} to eq. \eqref{eq:Dij_adapt}, one obtains the form of $g$ that minimises dissipated energy per unit time:

\begin{equation} \label{eq:g_23}
    g(Q_{ij}) = Q_{ij}^{2/3}
\end{equation}

\par Thus, substituting eq. \eqref{eq:g_23} in eq. \eqref{eq:Dij_adapt} one obtains the adaptation equation used in this thesis:

\begin{equation} \label{eq:adaptation_eq}
    \frac{d}{d\tau}\sqrt{D_{ij}} = \alpha\frac{|Q_{ij}|^{2/3}}{\sum_{(k,m)\in E}L_{km}|Q_{km}|^{2/3}} - \sqrt{D_{ij}} \quad,\quad (i,j)\in E
\end{equation}

\noindent
where $\tau = ct$ and $\alpha = V/\beta$.

\subsection{Murray's law} \label{sec:murray-law}

Murray's law is a theoretical law for an idealised network structure used in biophysical fluid dynamics that relates the radii of fluid-transporting, cylindrical veins at bifurcations in a network. It states that:

\begin{equation} \label{eq:murray_theo}
    R^3 = \sum_i r_i^3
\end{equation}

\noindent
in which, at a bifurcation in a network, $R$ is the radius of the parent branch and $r_i$ are the radii of the children branches.

Murray states that the total work involved in operating a section of a vessel should be minimum for an ideal network. This energy required to maintain the flow is given by the work required to overcome viscous drag forces (for a fluid described by the Hagen-Poiseuille law \eqref{eq:HP}, this work is given by $Q_{ij}\left(p_i - p_j\right)$) and the metabolic cost of maintaining the network structure, which is said to be proportional to its volume \cite{murray}.

Murray believed this idealised law would work on biological systems because living beings have undergone millions of years of Darwinian evolution, and have thus optimized their transport networks to an idealised state. Murray’s law has, in fact, been supported by measurements of small vessel branching relationships of arteries, in which the exponents obtained were in the range [2.7, 3.0] \cite{sherman_murray}, and by measurements of \textit{Physarum} networks, where the measured exponents were in the range [2.53, 3.29] \cite{akita_murray} (by "exponents", one means the exponents of equation \eqref{eq:murray_theo}).

\par The class of equations used in this thesis (described in section \ref{sec:formalism}) uses similar energy minimisation considerations to the original Murray's law analysis. A significant difference between the two is that the formalism described takes the metabolic cost as a constraint (constant volume). Nevertheless,  a cubic law relationship between the flow rates and the radii of the channels was found ($Q_{ij} \propto D^{3/4}_{ij} \propto r^3_{ij}$, as seen in eq. \eqref{eq:solution_lagrangian}). As the flux is conserved at each node (eq. \eqref{eq:kirch}), one obtains the generalised Murray's law \cite{rodrigo_artigo}:

\begin{equation} \label{eq:generalised_murray}
    \sum_{j:(i,j)\in E, Q_{ij} > 0} r_{ij}^3 = \sum_{j:(i,j)\in E, Q_{ij} < 0} r_{ij}^3 \quad,\quad i = 1, ... ,\text{N}
\end{equation}

Murray's law was verified for steady state flow using the class of equations described in sec. \ref{sec:formalism} in \cite{rodrigo_artigo}.

In this thesis, using the base formalism mentioned previously, I intend to verify if Murray's law is verified dynamically for adaptive flow with fixed sources and sinks.

\section{Steiner tree problem and \textit{Physarum}} \label{sec:steiner_intro}


The Steiner minimum tree problem is a combinatorial optimisation problem. In its most general form, it asks for, given a set of objects, the optimal interconnect of all said objects according to a predefined objective function. When applied to undirected graphs, the Steiner tree problem seeks to find the tree that connects all nodes of interest (being able to contain nodes additional to the nodes of interest) and that also minimizes the total weight of the tree edges.

It's important to distinguish Steiner trees from minimum spanning trees (MST). For undirected graphs, the minimum spanning tree also seeks to find the tree that connects all nodes of interest and that also minimizes the total weight of the tree edges, but it \textit{cannot} contain additional vertices besides the nodes of interest, unlike the Steiner minimum trees.

For graphs that symbolize paths between physical, geographical locations, like the graphs in this thesis, the "weight" that is to be minimized by the minimum Steiner tree is the length of said tree. Thus, for this thesis, the Steiner minimum tree (SMT) symbolizes the shortest tree that connects all nodes of interest.

\begin{figure}
    \centering
    \begin{subfigure}[b]{0.3\textwidth}
        \centering
        \includegraphics[width = \textwidth]{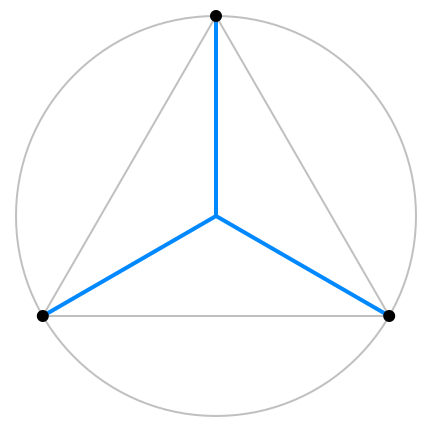}
        \caption{}
        \label{fig:steiner_geometric_triangle}
    \end{subfigure}
    \hfill
    \begin{subfigure}[b]{0.3\textwidth}
        \centering
        \includegraphics[width = \textwidth]{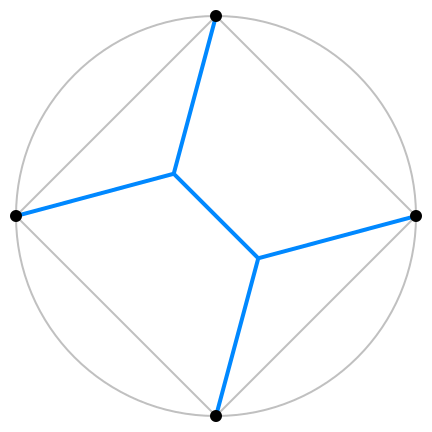}
        \caption{}
        \label{fig:steiner_geometric_square}
    \end{subfigure}
    \hfill
    \begin{subfigure}[b]{0.3\textwidth}
        \centering
        \includegraphics[width = \textwidth]{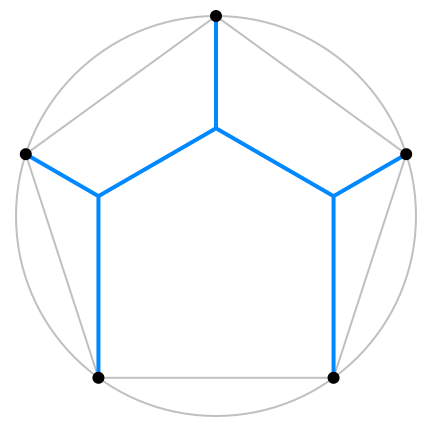}
        \caption{}
        \label{fig:steiner_geometric_penta}
    \end{subfigure}
    \caption{Steiner minimum trees (in blue) for three regular polygons: a) a triangle, b) a square and c) a pentagon. Images courtesy of Martin Janecke, obtained from \cite{image:steiner_geometric}.}
    \label{fig:steiner_geometric}
\end{figure}

As an example, figure \ref{fig:steiner_geometric} shows the SMT (in blue) for three regular polygons. This figure also shows a key feature of SMTs: Steiner points, which are nodes that arise in the tree that are not part of the original group of nodes of interest. In figure \ref{fig:steiner_geometric}, the Steiner points are the unmarked nodes that connect three (or more) edges of the blue SMT. For the triangle, there is a single Steiner point that corresponds to the center of said triangle; for the square, there are two Steiner points in the SMT and, for the pentagon, there are three Steiner points in the SMT.

Finding the Steiner minimum tree of a graph has a great deal of benefits and is a frequent problem in areas that involve building a cost-efficient and robust network of paths, like transportation, circuitry and telecommunications. After all, as a Steiner tree has a minimum length, it also has a minimum building and maintenance cost. Finding the SMT of a problem can also help increase the cost-efficiency of the network or can make it more robust (by providing minimum cost redundant paths).

Most versions of the Steiner problem in graphs are NP-hard \cite{karp_np-hard} (this means it is a much harder problem than problems that can be solved by a nondeterministic Turing machine in polynomial time). Essentially, this means that it is unlikely that one is able to find an efficient algorithm to solve the Steiner tree problem for large graphs. As such, this is still an open problem. Biologically-inspired algorithms have thus tried to provide efficient ways to solve this very pertinent problem.

\begin{figure}
    \centering
    \begin{subfigure}[b]{0.65\textwidth}
        \centering
        \includegraphics[width = \textwidth]{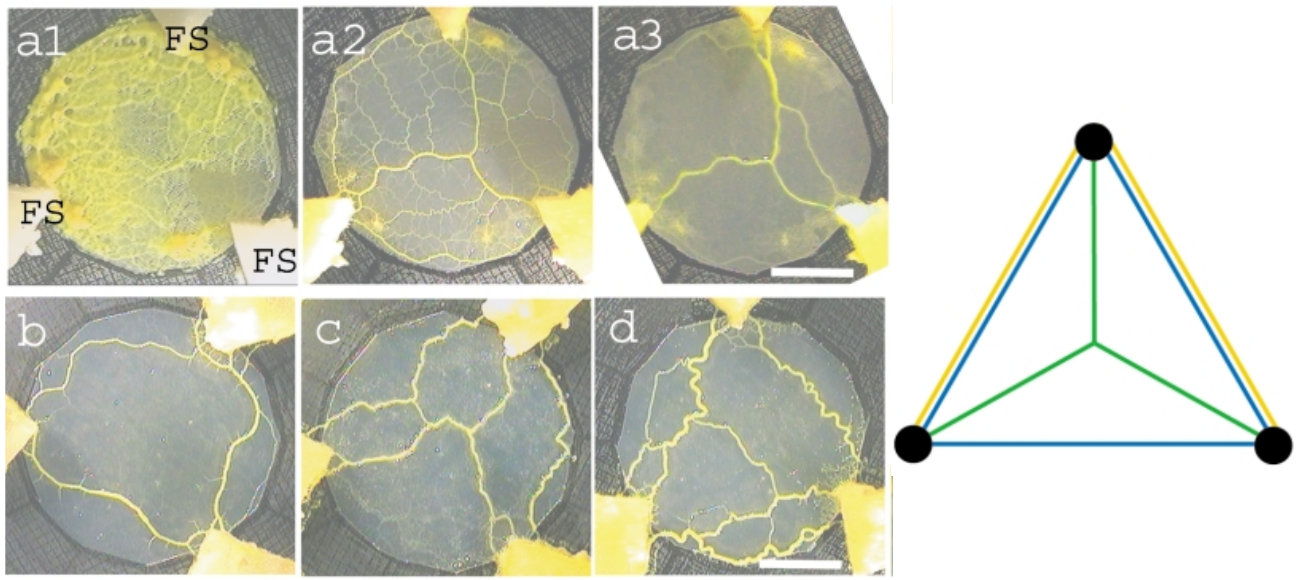}
        \caption{}
        \label{fig:Nakagaki_physarum_steiner_3FS}
    \end{subfigure}
    \hfill
    \begin{subfigure}[b]{0.27\textwidth}
        \centering
        \includegraphics[width = \textwidth]{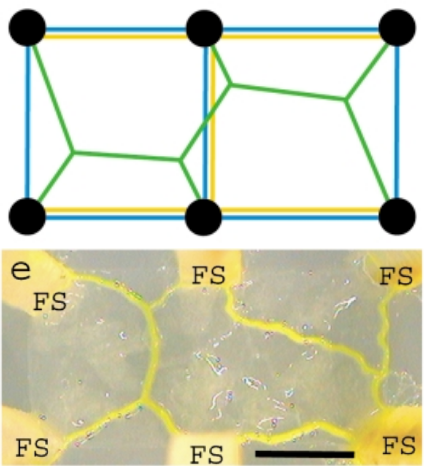}
        \caption{}
        \label{fig:Nakagaki_physarum_steiner_6FS}
    \end{subfigure}
    \caption{Comparison between \textit{Physarum polycephalum} networks and their theoretical Steiner minimum tree counterparts. \textbf{a)} For 3 food sources (FS) located at the edges of a plate, the \textit{Physarum} specimen is made to cover the whole plate. The images show the network configuration after 0hr (a1), 6hrs (a2) and 36hrs (a3) of the start of the experiment. Images b-d show three different typical networks obtained. The graphic on the right shows the food sources locations as black circles and the SMT in green. \textbf{b)} For 6 FS, the image shows \textit{Physarum}'s network after some time had passed. The graphic above shows the food sources locations as black circles and the SMT in green. Images adapted from \cite{Nakagaki_physarum_smart_network}.}
    \label{fig:Nakagaki_physarum_steiner}
\end{figure}

Real \textit{Physarum}'s network trees have already been shown to be able to converge to Steiner minimum trees, thus estimating solutions close to that of Steiner tree problems (figure \ref{fig:Nakagaki_physarum_steiner}) \cite{Nakagaki_physarum_smart_network}. In figure \ref{fig:Nakagaki_physarum_steiner_3FS}, notice the resemblance between figures a3 and the SMT of the triangle; in figure \ref{fig:Nakagaki_physarum_steiner_6FS}, \textit{Physarum}'s network also shows a resemblance to the theoretical Steiner tree of the system. \textit{Physarum}'s networks aren't perfect: the paths aren't perfectly straight and the Steiner points aren't located at the most efficient locations. However, the \textit{Physarum} networks shown in figure \ref{fig:Nakagaki_physarum_steiner} showed larger average values of FT/TL (fault tolerance / total length) than the the theoretical SMT values \cite{Nakagaki_physarum_smart_network}. Fault tolerance is the probability of the network becoming disconnected when one of the links is removed, while the total length of the network is a measure of the cost of the network. Thus, as the \textit{Physarum} network has a larger FT/TL value than the Steiner minimimum tree, it is a more robust and cheaper network than the SMT. An important note is that, while \textit{Physarum}'s network can converge to Steiner-like trees, they do not converge to a MST every time (see figures b-d of figure \ref{fig:Nakagaki_physarum_steiner_3FS}; figure b of fig. \ref{fig:Nakagaki_physarum_steiner_3FS} shows \textit{Physarum}'s network converging to the triangle's perimeter, for example).

As \textit{Physarum polycephalum}'s networks were shown to be able to converge to Steiner tree-like configurations, biologically-inspired algorithms were developed to mimic \textit{Physarum}'s behavior with the objective to reproduce these short networks. The \textit{Physarum} network models described in section \ref{sec:physarum-models} have been shown to converge to Steiner tree-like configurations, namely the hydrodynamics-based model \textit{Physarum solver} \cite{tero_steiner}. Several hydrodynamics-based models have been used to solve real-world problems, like the minimal exposure problem in wireless sensor networks \cite{song_wireless_sensor}, finding subnetworks for drug repositioning \cite{sun_drug_repositioning}, designing communication networks \cite{sun_fast_algorithms} and identifying elements of cancer-related signaling pathways \cite{sun_steiner_cancer}. The possible applications for these algorithms are vast and important.
\chapter{Properties of adaptive Hagen-Poiseuille flows}
\label{chapter:properties}

In order to determine the value and usefulness of a model, one must test it to see if it can accurately reproduce real physical results of what it is trying to portray. That is what we want to aim to do in this chapter of this thesis with the model described in section \ref{sec:adaptive-model}. First, we describe the algorithms used to implement said model (sec. \ref{sec:algorithms}). Then, we study the length distributions of several steady-state graphs obtained (sec. \ref{sec:length_distributions}) and attempt to dynamically validate Murray's law as described in section \ref{sec:murray-law} (sec. \ref{sec:validation_murrays_law}). Next, we look at the properties of some simple networks (sec. \ref{sec:simple_networks}). Finally, we compare the results obtained here with \textit{Physarum polycephalum}'s networks (sec. \ref{sec:comparison_physarum}). 

\section{Methods and algorithms} \label{sec:algorithms}

As mentioned previously, the \textit{Physarum polycephalum} network adaptation model used in this thesis is the one described in section \ref{sec:adaptive-model}. This model was implemented with the algorithms described below.

\subsection{Initializing the simulation}

First, the network graph $\mathcal{G}$ is initialized. $\mathcal{G}$ is a planar graph embedded in the two-dimensional Euclidean space. The graph is constructed as follows: first, a square lattice is created, and nodes are placed on the vertices of the lattice; the lattice is made to have side length $1$ and contains $N' \times N'$ nodes, $N'$ being the number of nodes on each side of the square. Then, the nodes' positions are randomly perturbed with Gaussian noise with a standard deviation $\sigma = 0.5$. Finally, the edges of the network result from a Delaunay triangulation between the nodes. Note that a planar Delaunay triangulation connects a set of nodes using triangular shapes in a way that no node is inside the circumscribed circle of any triangle of the triangulation.
The lengths of each edge $L_{ij}$ are calculated based on the positions of nodes $i$ and $j$ and are fixed for each different $\mathcal{G}$. In this thesis, only two different $\mathcal{G}$ of this kind will be taken into account, with $N' = \{25, 40\}$. 

Some of the nodes of the network $j$ may be assigned as sources or sinks, having $S_j \neq 0$, and providing that condition of eq. \eqref{eq:sources_sinks} is verified for the entire network.

The flow passing through the edges of the network $Q_{ij}$ is initialized by the starting conductivities $D_{ij}(t=0)$; these starting conductivities are initialized as positive random numbers with a homogeneous distribution.

Some parameters are kept constant for all runs, namely $\beta = 1$, $\Delta\tau = 0.1$, $V = 100$ and $\sum_{j\in\{\text{sources}\}}S_j = 1$ (unless stated otherwise).

\subsubsection{Renormalization of volume} \label{sec:renormalization_volume}

In order to be able to compare different runs of a simulation, some baseline conditions must be met for all the runs; namely, the volume of fluid should be the same for all the runs. While the model used in this thesis keeps the volume of the fluid in the network constant over time \cite{rodrigo_artigo}, no measures were put in place to actually determine that the volume of fluid in each simulation was the same. Thus, we introduce a volume renormalization in this work to bridge this gap present in the previous description of the method.

As mentioned previously, for a given distribution of sources and sinks in a Delaunay graph, with prescribed input flows in the sources, the flow is initialized by the starting conductivities $D_{ij}(t=0)$. As eq. \eqref{eq:volume} states that $V = \beta \sum_{(i,j)\in E} L_{ij}\sqrt{D_{ij}}$, the total volume of fluid in the graph is determined by $\beta$ (which is kept the same for all simulations), $L_{ij}$ (which is the same for simulations using the same baseline graph $\mathcal{G}$) and $D_{ij}$. Thus, the initial distribution of conductivities determines the total volume of fluid in the graph, and therefore the value of the parameter $\alpha$ in eq. \eqref{eq:adaptation_eq}. The steady states obtained will depend on $\alpha$. Thus, for a different choice of initial conductivities, a different volume will be used and a different steady-state will be obtained.

As such, to analyze different steady states for the same volume of fluid, the initial conductivities must be renormalized. Consider some initial distribution of conductivities $\{D_{ij}(t=0) , (i,j)\in E\}$, which correspond to volume $V_0 = \beta \sum_{(i,j)\in E} L_{ij}\sqrt{D_{ij}}$. By multiplying each $D_{ij}(t=0)$ by a constant, that is, by rescaling the initial conductivities of the network, one can change the total volume of fluid in the network from $V_0$ to the desired volume $V$. The conductivities that correspond to volume $V$ are:

\begin{equation} \label{eq:renormalization_dij}
    D_{ij}'(t=0) = \left(\frac{V}{V_0}\right)^2 D_{ij}(t=0)
\end{equation}

As such, to obtain the desired volume of fluid in the network $V$, one must, after initializing the conductivities with some random positive numbers, first calculate the current volume of fluid in the network $V_0$ and, afterward, multiply all $D_{ij}$ values by $(V/V_0)^2$.

With this renormalization, it is possible to compare different steady states for different initial conductivities and the same fluxes of sources and sinks.

\subsection{Computing the temporal evolution of flows and conductivities}

To describe the temporal evolution of the radii of the channels and the flow of fluid, two steps are repeated until a steady state is reached: first, the flows of the network are calculated by solving the linear equations described in eq. \eqref{eq:kirch}, and then all the channel conductivities are adapted according to eq. \eqref{eq:adaptation_eq}.

\subsubsection{Calculation of the pressures and flows}

Given a set of edge conductivities $D_{ij}$ and inward/outward fluxes of fluid at each node $S_j$, the linear equations described in eq. \eqref{eq:kirch} are solved to obtain the pressures at each node $p_i$ and the flows at each edge $Q_{ij}$. The method behind solving these equations will be the same as used in \cite{rodrigo_tese}. It is described below.

Substituting the H-P definition of flow of eq. \eqref{eq:HP}, one can rewrite each of the $i$ eqs. \eqref{eq:kirch} as: 

\begin{equation} \label{eq:Cij}
\sum_{(i,j)\in E} \frac{D_{ij}}{L_{ij}}(p_i - p_j) = S_i \Leftrightarrow \sum_{(i,j)\in E} C_{ij}(p_i - p_j) = S_i
\end{equation}

\noindent
in which $C_{ij} = D_{ij} / L_{ij}$ can be seen as weights of each of the edges of the network $(i,j)$ that measure the efficiency of each edge at carrying flow. One can re-write the left-hand side of eq. \eqref{eq:Cij} as:

\begin{equation}
\begin{aligned}
        \sum_{(i,j)\in E} C_{ij}(p_i - p_j) & = \left(\sum_{(i,j)\in E} C_{ij}\right)p_i - \sum_{(i,j)\in E} C_{ij} p_j = \\
        & = \left(\sum_{(i,k)\in E} C_{ik}\right)\sum_{(i,j)\in E}\delta_{ij}p_j - \sum_{(i,j)\in E} C_{ij} p_j = \\
        & = \sum_{(i,j)\in E}\left[\left(\sum_{(i,j\in E)} C_{ik}\right)\delta_{ij} - C_{ij}\right]p_j = \\
        & = \sum_{(i,j)\in E} w_{ij} p_j
\end{aligned}
\end{equation}

\noindent
where $\delta_{ij}$ is the Kronecker delta. One can define a $N \times N$ symmetric matrix $\mathbf{W}$ with entries $w_{ij} = \left(\sum_{(i,j\in E)} C_{ik}\right)\delta_{ij} - C_{ij}$; this is the generalized Laplacian matrix of the network, as the network is a weighted graph with weights $C_{ij}$. As our network's graph is simple and undirected, this Laplacian matrix can be easily calculated by $\mathbf{W} = \mathbf{D} - \mathbf{A}$, where $\mathbf{D}$ is the degree matrix (a diagonal matrix whose $i$th diagonal entry is the sum of $C_{ij}$ for all the nodes $j$ connected to node $i$) and $\mathbf{A}$ is the adjacency matrix (a symmetric matrix with a null diagonal whose $ij$th entry is $C_{ij}$) of the network.
If $\mathbf{S}$ is the N-dimensional vector whose $i$th entry is $S_i$ (vector that describes the sources and sinks of the network) and $\mathbf{p}$ is the N-dimensional vector whose $i$th entry is $p_i$ (vector that describes the pressures of the nodes of the network), then one can rewrite the system of equations \eqref{eq:kirch} as:

\begin{equation}
    \mathbf{W} \mathbf{p} = \mathbf{S}
\end{equation}

It's important to note that $\mathbf{W}$ is a singular matrix has all its rows sum up to zero, which means that it has an eigenvector with a null eigenvalue ($\mathbf{W}\mathbf{1} = \mathbf{0}$). Thus, $\text{rank}(\mathbf{W}) = N - 1$. This relates to the fact that, in the system of linear eqs. \eqref{eq:kirch}, pressures are defined up to an additive constant (physically, one can only measure pressure differences). To be able to invert the matrix and then solve the system of equations, one adds a small arbitrary constant to one diagonal element of $\mathbf{W}$ (see section 2.4 of \cite{katifori_constant}). In the case of this thesis, each time equation \eqref{eq:kirch} is solved, one diagonal entry pertaining to a sink node is randomly perturbed. \cite{rodrigo_tese} 

The pressure values are obtained by solving equation $\mathbf{p} = \mathbf{W}^{-1}\mathbf{S}$ using the Cholesky decomposition method, and the flow values are obtained by then using the H-P law \eqref{eq:HP}; these flow values are well-defined.

\subsubsection{Dynamic adaptation of conductivities}

After obtaining the $Q_{ij}$ flow values, all the channel conductivities are adapted according to eq. \eqref{eq:adaptation_eq}.

In this thesis, the simple Euler method is used to numerically solve eq. \eqref{eq:adaptation_eq} using a time step $\Delta\tau = 0.1$. As such, the equation used to obtain the conductivities at time $t + \Delta\tau$ is:

\begin{equation} \label{eq:euler_method}
    \sqrt{D_{ij}(t + \Delta\tau)} = \sqrt{D_{ij}(t)} + \Delta\tau\left(\frac{V}{\beta}\frac{\left|Q_{ij}(t)\right|^{2/3}}{\sum_{(k,m)\in E}L_{km}\left|Q_{km}(t)\right|^{2/3}} - \sqrt{D_{ij}(t)}\right) \quad,\quad (i,j)\in E
\end{equation}

\noindent
where $Q_{ij}(t)$ denotes the flow value of edge $(i,j)$ at time $t$ (and similarly for $D_{ij}(t)$). Note that the time since the beginning of the simulation is given by $t = n\Delta\tau$, where $n$ is the number of iterations that occurred up to that point.

It's important to determine whether or not this method conserves the volume of fluid in the network over time. The volume of fluid at time $t + \Delta\tau$ is: (according to eq. \eqref{eq:volume} and using eq. \eqref{eq:euler_method})

\begin{equation}
\begin{aligned}
    V(t+\Delta\tau) & = \beta\sum_{(i,j\in E)}L_{ij}\sqrt{D_{ij}(t+\Delta\tau)} = \\
    & = \beta\sum_{(i,j\in E)}L_{ij}\sqrt{D_{ij}(t)} + \tau\left(V - \beta\sum_{(i,j)\in E}L_{ij}\sqrt{D_{ij}(t)}\right) = \\
    & = V(t)
\end{aligned}
\end{equation}

Thus, equation \eqref{eq:euler_method} preserves the volume of fluid of the network while also dynamically adapting the network's conductivity values.

\subsubsection{Stopping criteria} \label{sec:stopping_criteria}

For a static configuration of sources and sinks, the two steps described above (calculating the flows and then adapting the conductivities) are repeated over time until a steady state of the channel conductivities is reached. The numerical condition for reaching the steady state is:

\begin{equation} \label{eq:stop_condition}
\text{max}_{(i,j)\in E} \left|D_{ij}(t) - D_{ij}(t - \Delta\tau)\right| \le D_{\text{thresh}}
\end{equation}

\noindent
where $t$ is the first time instance for which the condition is verified and $D_{\text{thresh}}$ symbolizes the desired precision value, being a positive small number. For this work, $D_{\text{thresh}} = 5\times10^{-6}$.

For non-static configurations of sources and sinks, the stopping criteria is different. By non-static configuration, one means that the sources and/or sinks locations change as time passes (at every time step of the simulation in the case of this work, unless stated otherwise). As the active sites change so frequently, the values of the conductivities $D_{ij}$ never get to a stable value, and change greatly every time step. However, for the cases considered in this thesis, the shape of the tree obtained does reach a constant form. As the topology of the network is what one wishes to study for these types of simulations, the stopping criteria was defined as follows: if the length of the network (defined by equation \eqref{eq:total_length}) remains unchanged for $N_{\text{iter}}$ iterations, then it is considered that the shape of the network was kept constant for that number of iterations and it is considered that the simulation reached its steady state. $N_{\text{iter}}$ is a parameter that is adjusted depending on the configuration of sources and sinks and the graph $\mathcal{G}$.

The simulations present in this work were carried out using \textit{Python}, with the help of packages like \texttt{NetworkX} (for graph representation and analysis), \texttt{SciPy} and \texttt{NumPy} (for numerical computations) and \texttt{Matplotlib} (for graphical representations and plots). The code used was originally developed by Almeida \cite{rodrigo_tese} and was adapted for this work.

\section{Length distribution of the steady state graphs} \label{sec:length_distributions}

As mentioned previously (in sec. \ref{sec:renormalization_volume}), different choices of $\{D_{ij}(t=0) , (i,j)\in E\}$ may lead to trees with different shapes connecting sources and sinks. These trees are steady states of the H-P flow. 

In some cases, we may have several apparently disconnected trees. However, the graph tree has only one connected component at all times, with conductivities partitioned in two sets of high and low values. This allows for one to identify the channels that are effectively conducing flow in the tree. \cite{rodrigo_artigo}

To be able to evaluate, characterize and quantify the different possible steady states, different parameters are used. The first parameter studied will be the \textbf{length} of the tree, measured at steady state. The total length of the graph is given by:

\begin{equation} \label{eq:total_length}
    L = \sum_{(i,j)\in E'} L_{ij}
\end{equation}

\noindent
where $E'$ is the set of effectively conducting edges, which are edges with conductivity values above a certain threshold, that is, $E' = \{(i,j) \in E: D_{ij} > D_{\text{thr}}\}$. In the case of this work, $D_{\text{thr}} = 5 \times 10^{-4}$.

\begin{figure}
     \centering
     \begin{subfigure}[b]{\textwidth}
         \centering
         {\includegraphics[width=0.3\textwidth]{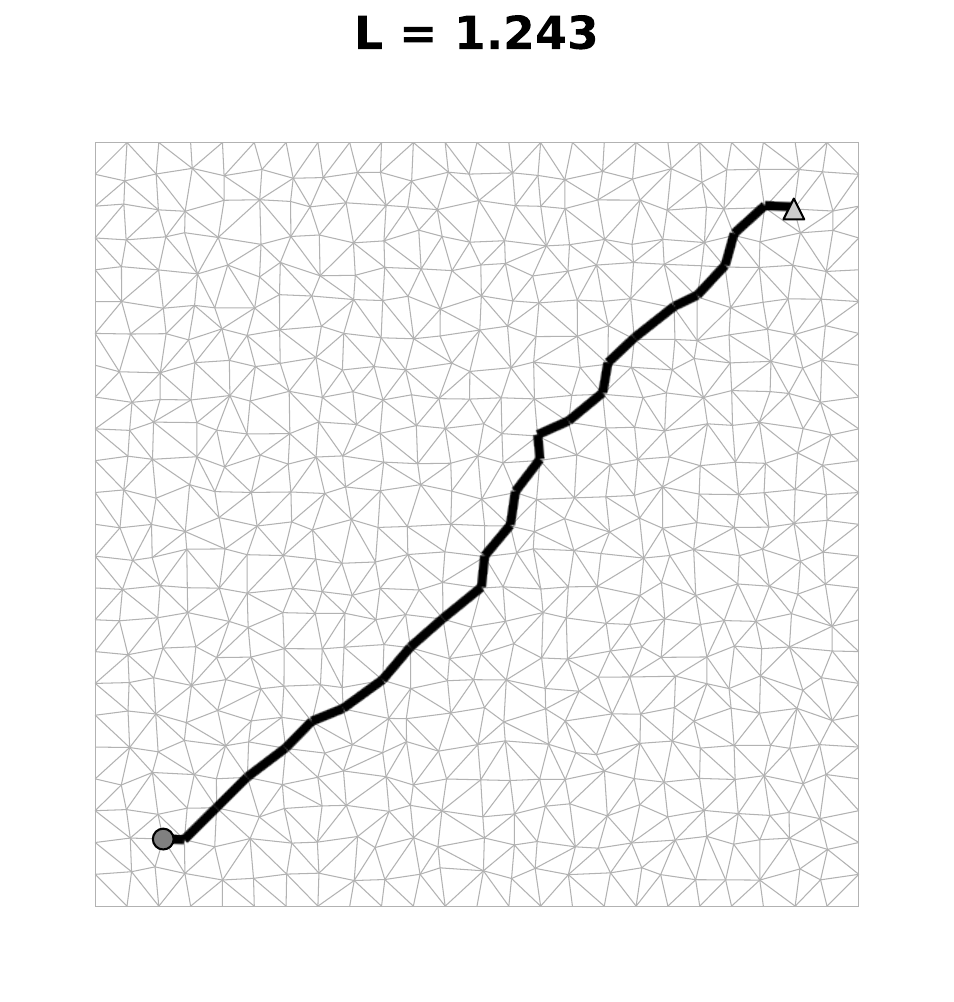}\hfill
         \includegraphics[width=0.3\textwidth]{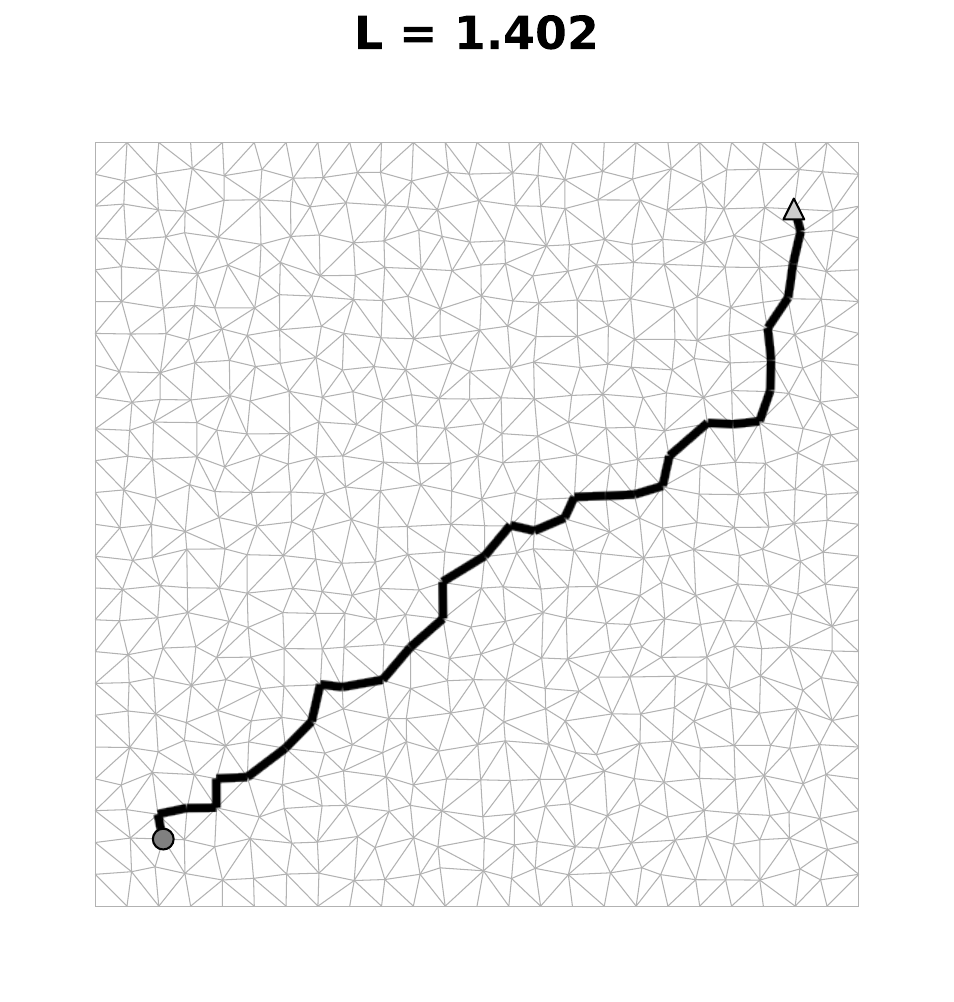}\hfill
         \includegraphics[width=0.37\textwidth]{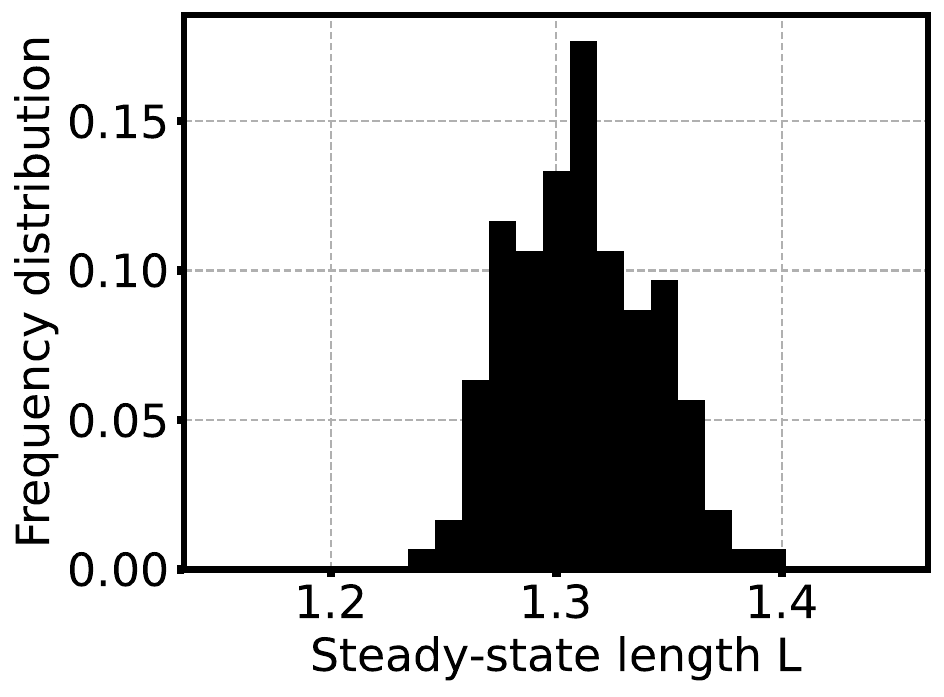}}
         \caption{$25\times25$ lattice. The mean of the length distribution is $\overline{L} = 1.311$ and the standard deviation is $\sigma = 0.030$.}
         \label{fig:l_dist_1_25}
     \end{subfigure}
     
     \begin{subfigure}[b]{\textwidth}
         \centering
         {\includegraphics[width=0.3\textwidth]{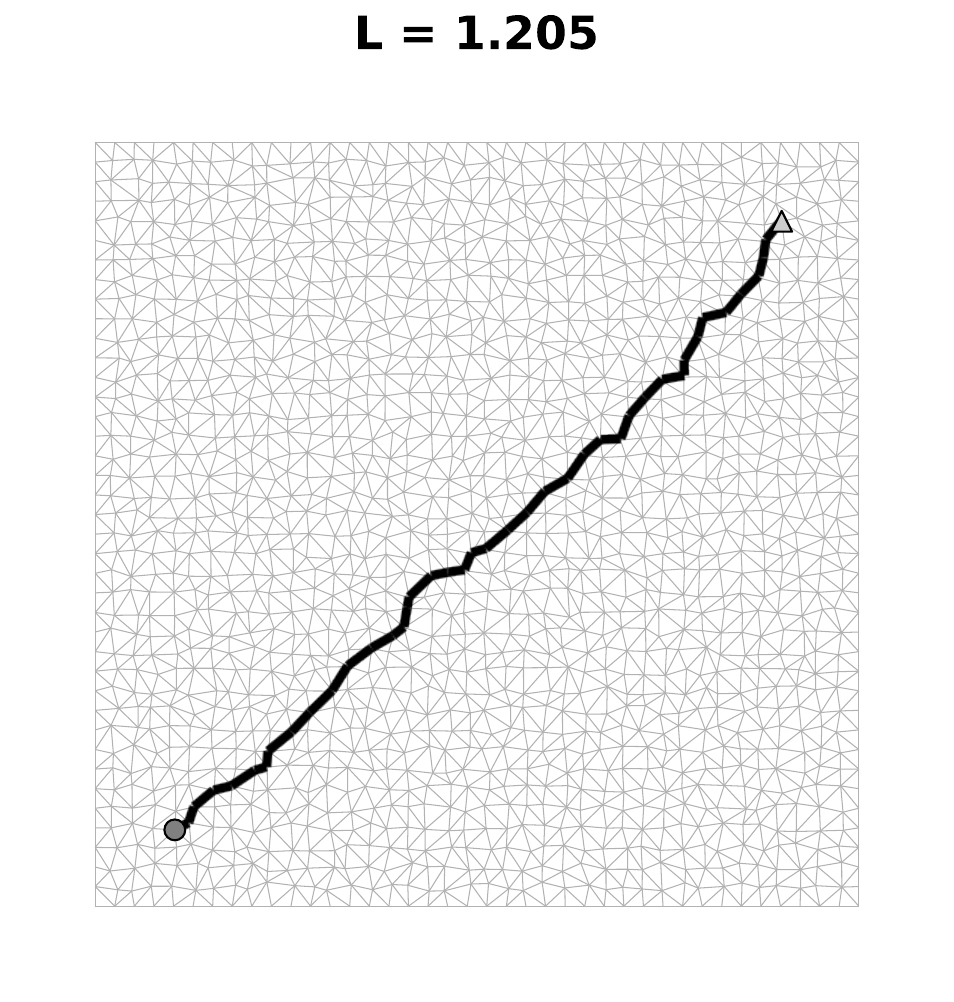}\hfill
         \includegraphics[width=0.3\textwidth]{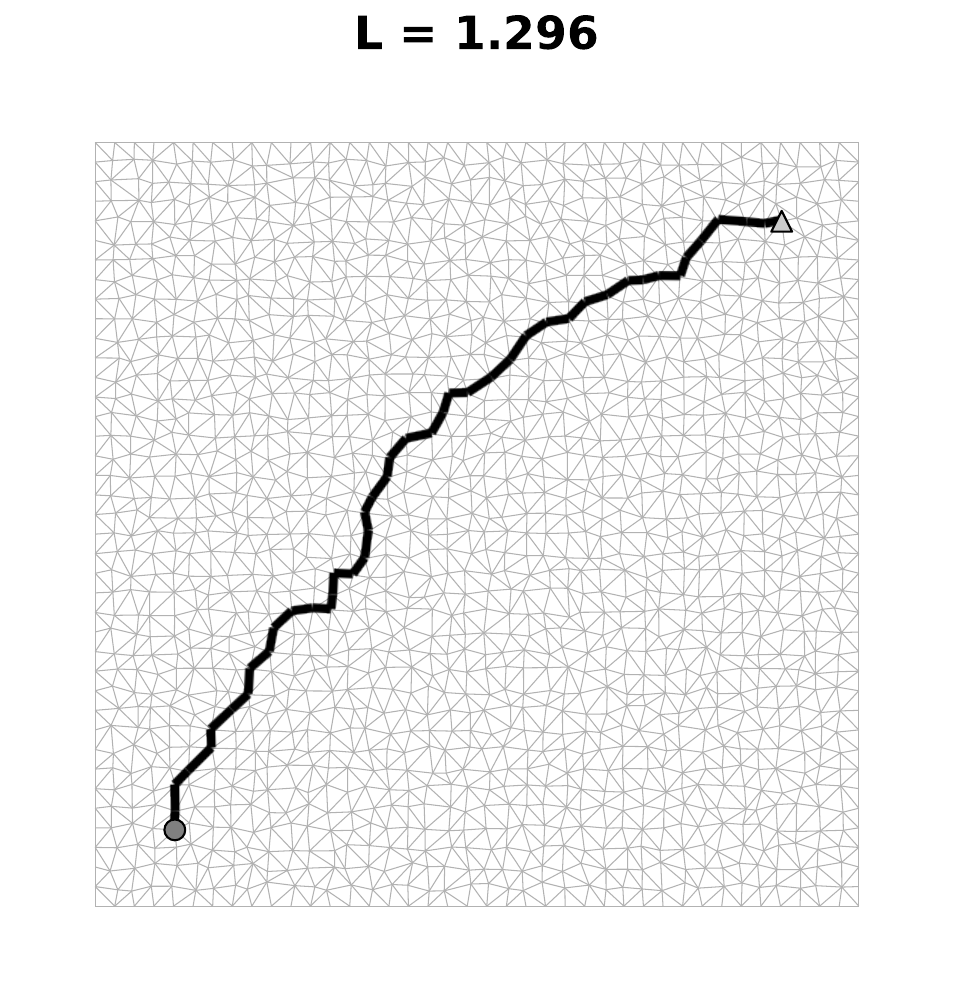}\hfill
         \includegraphics[width=0.37\textwidth]{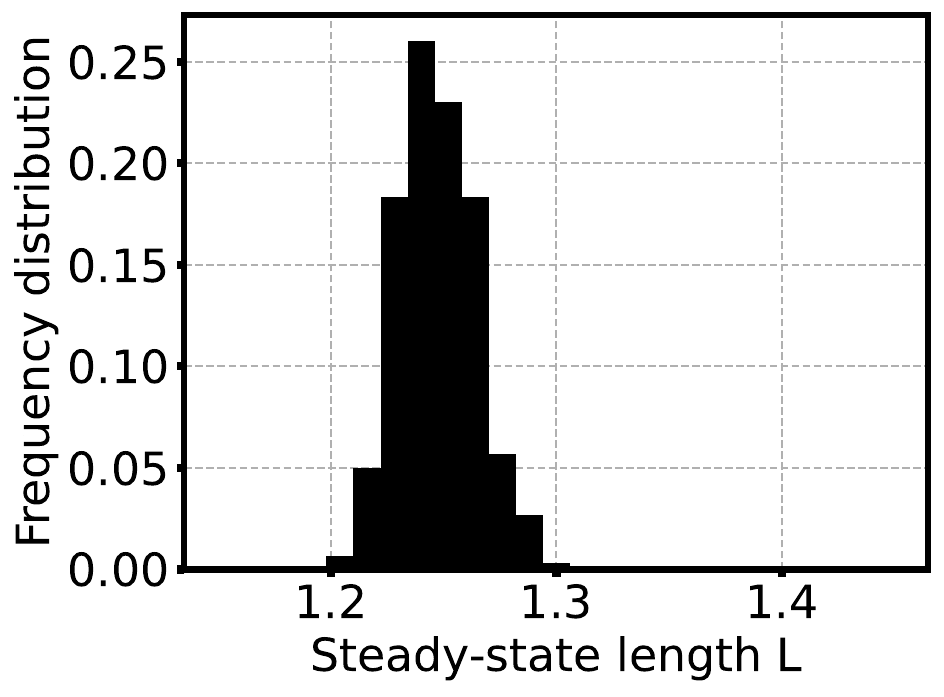}}
         \caption{$40\times40$ lattice. The mean of the length distribution is $\overline{L} = 1.247$ and the standard deviation is $\sigma = 0.017$.}
         \label{fig:l_dist_1_40}
     \end{subfigure}
     
    \caption{Length distribution of the steady state trees (right), shortest steady state tree obtained (left) and longest steady state tree obtained (middle), for one source and one sink, for different $\mathcal{G}$. The length of the steady state tree is written above its image. The linear distance between source and sink is $L_{\text{lin}} = 4\sqrt{2}/5 \approx 1.131$. In principle, in the limit $N' \to \infty$, $L\to L_{\text{lin}}$.}
    \label{fig:length_distribution_one}
\end{figure}

\begin{figure}
     \centering
     \begin{subfigure}[b]{\textwidth}
         \centering
         {\includegraphics[width=0.3\textwidth]{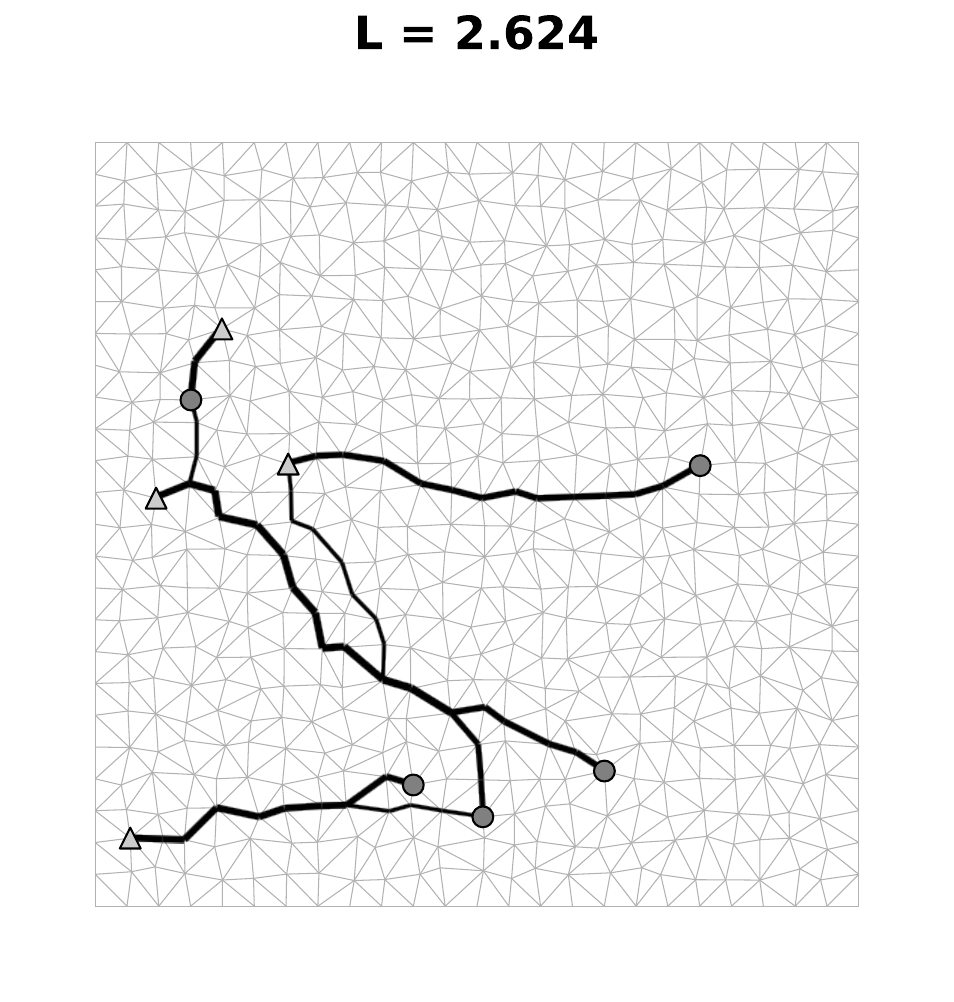}\hfill
         \includegraphics[width=0.3\textwidth]{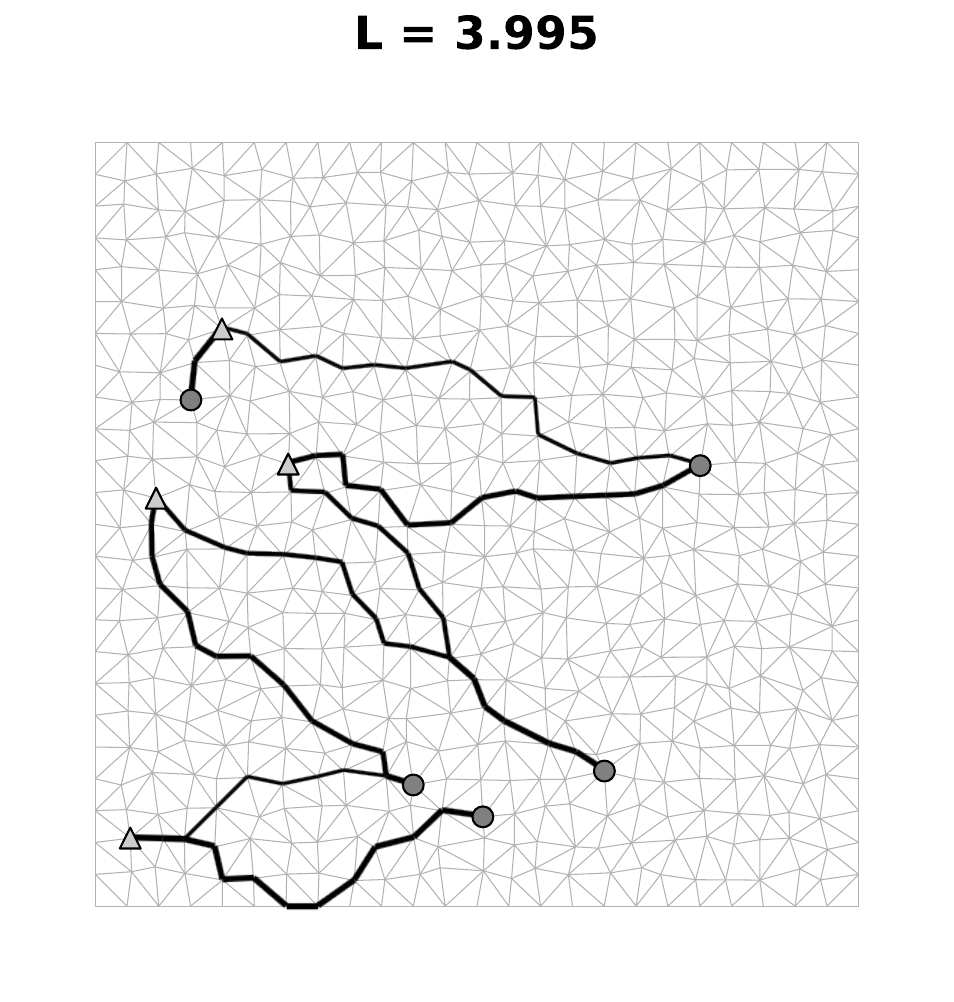}\hfill
         \includegraphics[width=0.37\textwidth]{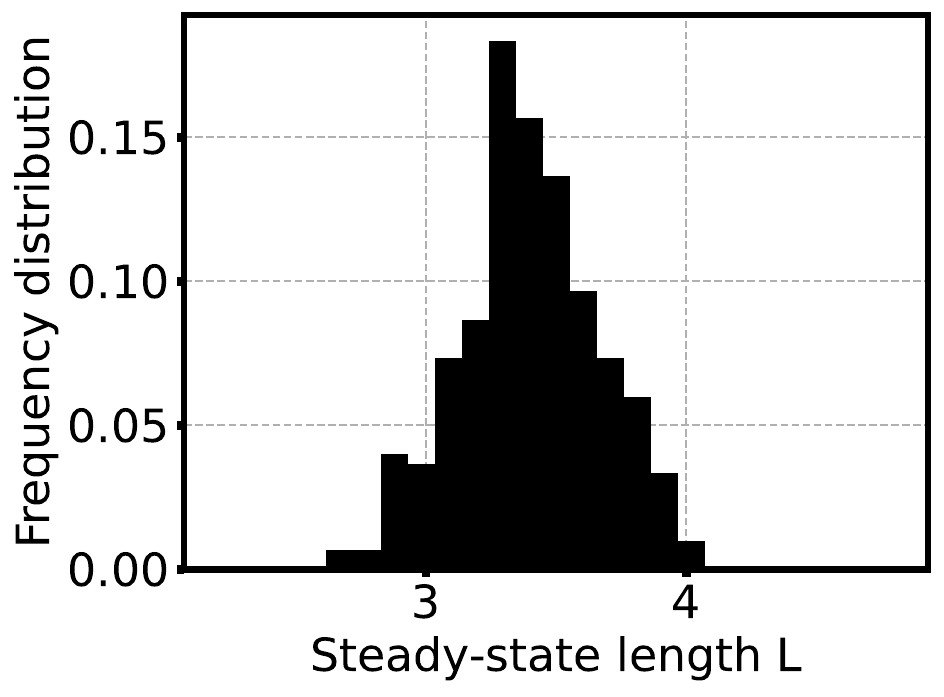}}
         \caption{$25\times25$ lattice. The mean of the length distribution is $\overline{L} = 3.40$ and the standard deviation is $\sigma = 0.27$.}
         \label{fig:l_dist_more_25}
     \end{subfigure}

     \begin{subfigure}[b]{\textwidth}
         \centering
         {\includegraphics[width=0.3\textwidth]{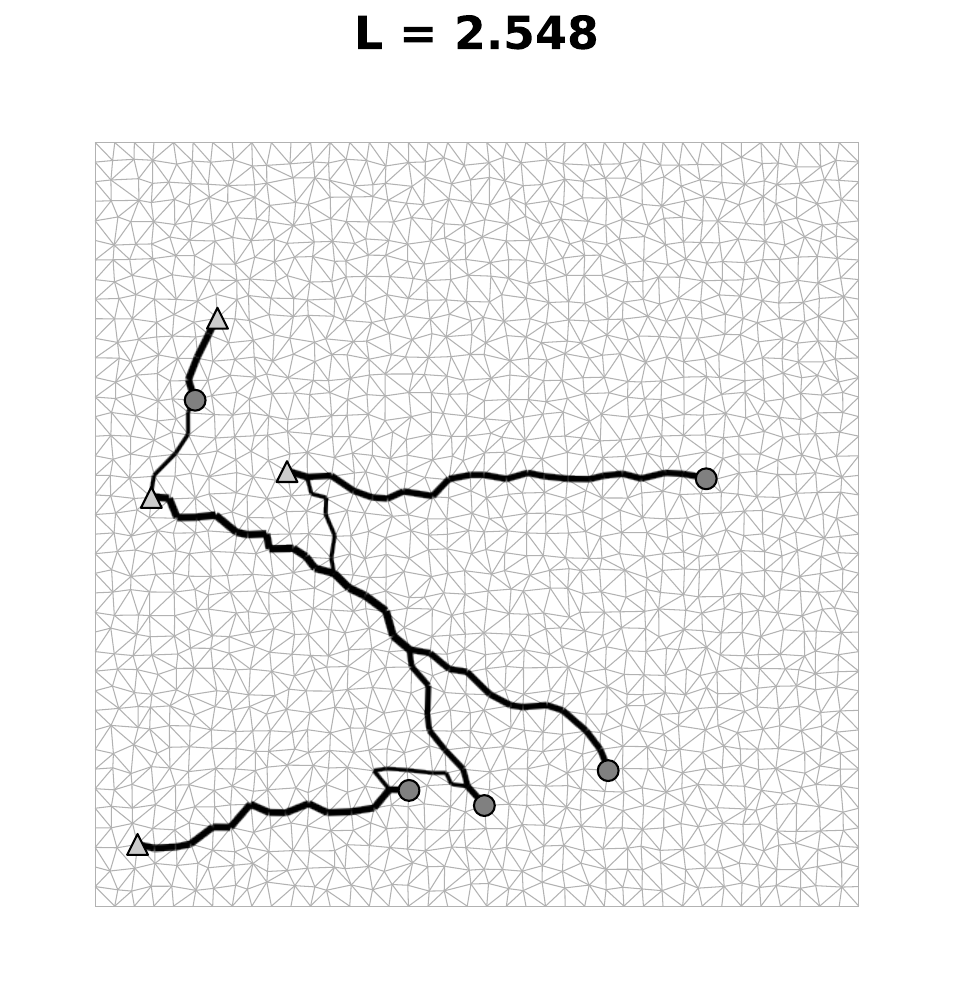}\hfill
         \includegraphics[width=0.3\textwidth]{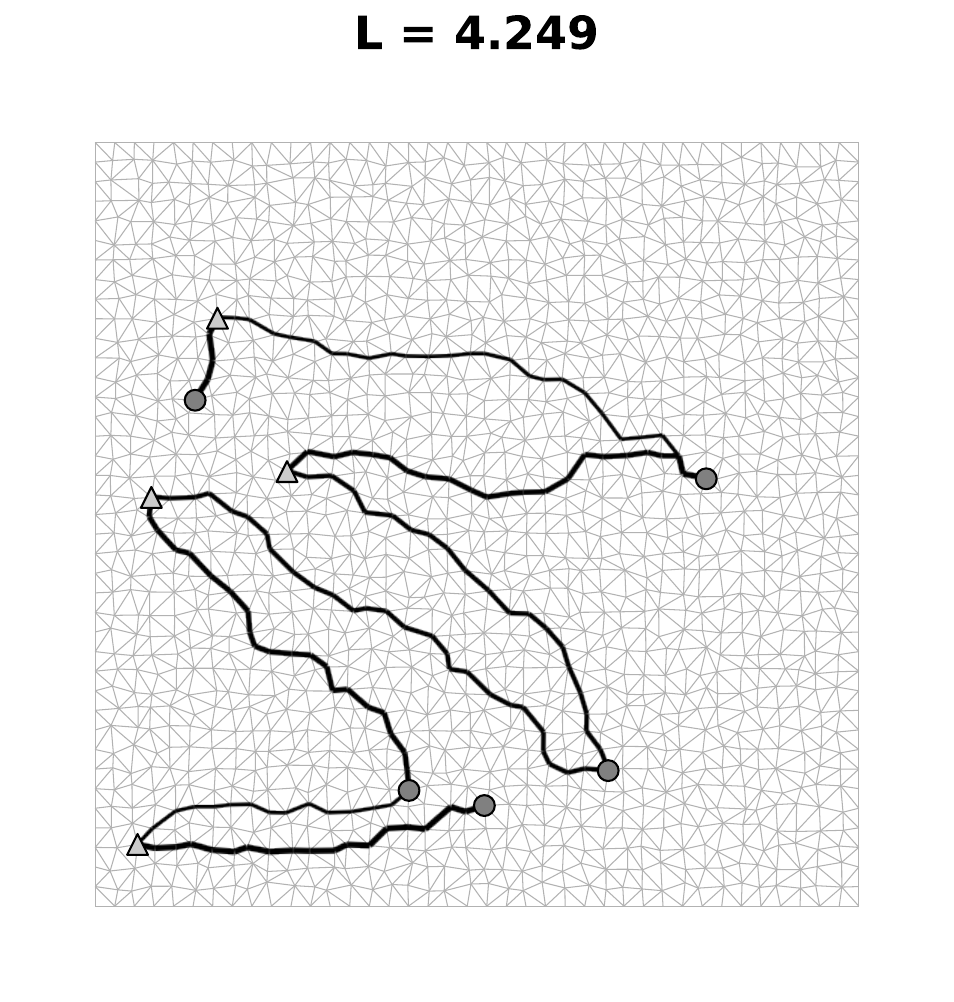}\hfill
         \includegraphics[width=0.37\textwidth]{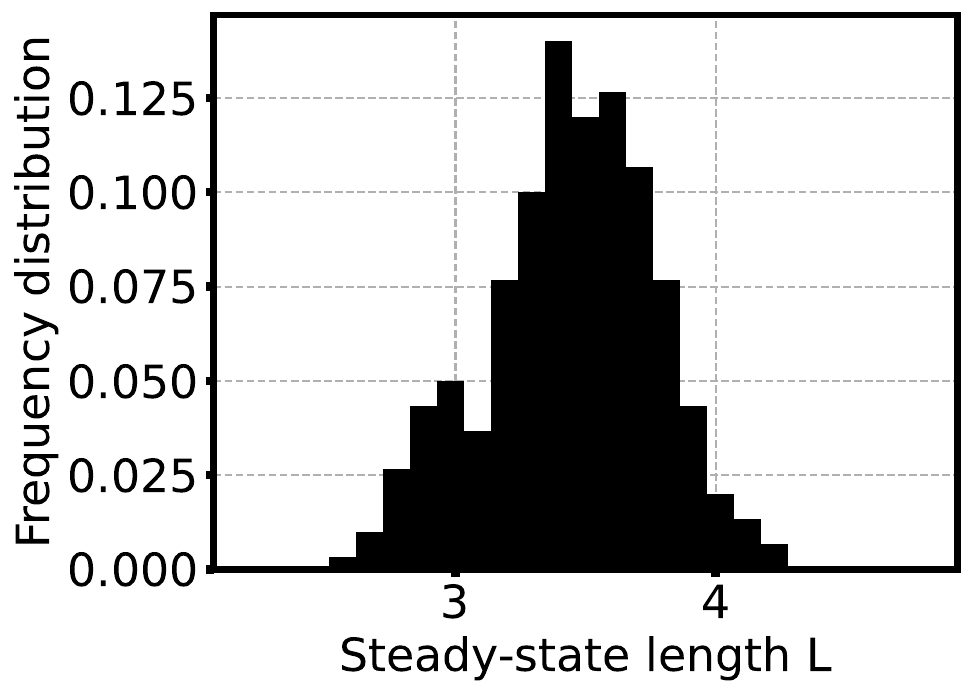}}
         \caption{$40\times40$ lattice. The mean of the length distribution is $\overline{L} = 3.44$ and the standard deviation is $\sigma = 0.33$.}
         \label{fig:l_dist_more_40}
     \end{subfigure}
     
    \caption{Length distribution of the steady state trees (right), shortest steady state tree obtained (left) and longest steady state tree obtained (middle), for five sources and four sinks, for different lattices. The length of the steady state tree is written above its image.}
        \label{fig:length_distribution_more}
\end{figure}

A study of the length distribution of the steady state graphs for different sets of sources and sinks can be seen in figures \ref{fig:length_distribution_one} and \ref{fig:length_distribution_more} for one source and one sink and five sources and four sinks, respectively. The trees were generated on the square lattice graph described previously of side length $1$ with $N = N'\times N'$ nodes, with $V = 100$ and $\sum_{j\in\text{sources}}S_j = 1$. Each figure contains the results for a graph $\mathcal{G}$ with $N' = 25$ and $N' = 40$. For each set of sources and sinks and for each $\mathcal{G}$, 300 trees were generated, each with different random initial conductivity conditions.

For all the steady state figures presented in this paper, the initial Delaunay triangulation $\mathcal{G}$ is shown in grey, and the thickness of the black line is proportional to the radius of the edges $\sim D_{ij}^{1/4}$. Additionally, source nodes are represented as circles, and sink nodes are represented as triangles.

The length distributions of both figures \ref{fig:length_distribution_one} and \ref{fig:length_distribution_more} resemble normal distributions. These images confirm the statement that different initial conductivity conditions lead to several different steady states.

As the size of the lattice $N'$ increases, the continuous limit is approached. For the case of one single source and one single sink (figure \ref{fig:length_distribution_one}), the shortest graph connecting them tends to reach a straight line between the source and the sink (compare figs. \ref{fig:l_dist_1_25} to \ref{fig:l_dist_1_40}). This is corroborated by the fact that the length of the shortest tree for the $N' = 40$ lattice is smaller than the length of the shortest tree for $N' = 25$, while also being closer to the linear distance between source and sink ($L_{25} > L_{40} > L_{\text{lin}}$). This is also true for the mean value of the length distribution of both lattices ($\overline{L_{25}} > \overline{L_{40}} > L_{\text{lin}}$).

The tendency of the simulation to take the shortest path possible between source and sink can also be shown by the fact that the standard deviation of the length distribution of the $N' = 40$ case is smaller than that of the $N' = 25$ case. This is because, when the number of nodes increases, so does the number of edges and therefore the number of possible paths the fluid can take; a smaller standard deviation shows the fluid didn't create very long trees and stayed as close to the shortest path as possible.

For the case of five sources and four sinks (figure \ref{fig:length_distribution_more}), we once again observe the tendency of larger node number leading to shorter tree lengths, as once again the length of the shortest tree for the $N' = 40$ lattice is smaller than the length of the shortest tree for $N' = 25$. However, this time both the mean and the standard deviation of the length distribution for the $N' = 40$ lattice are larger than that of the $N' = 25$ case. This might be because of the fact that with more sources and sinks come a lot more possibilities of topologies of the tree, that can be further expressed with a larger node number.

\begin{figure}
     \centering
     \begin{subfigure}[b]{0.45\textwidth}
         \centering
         \includegraphics[width=0.85\textwidth]{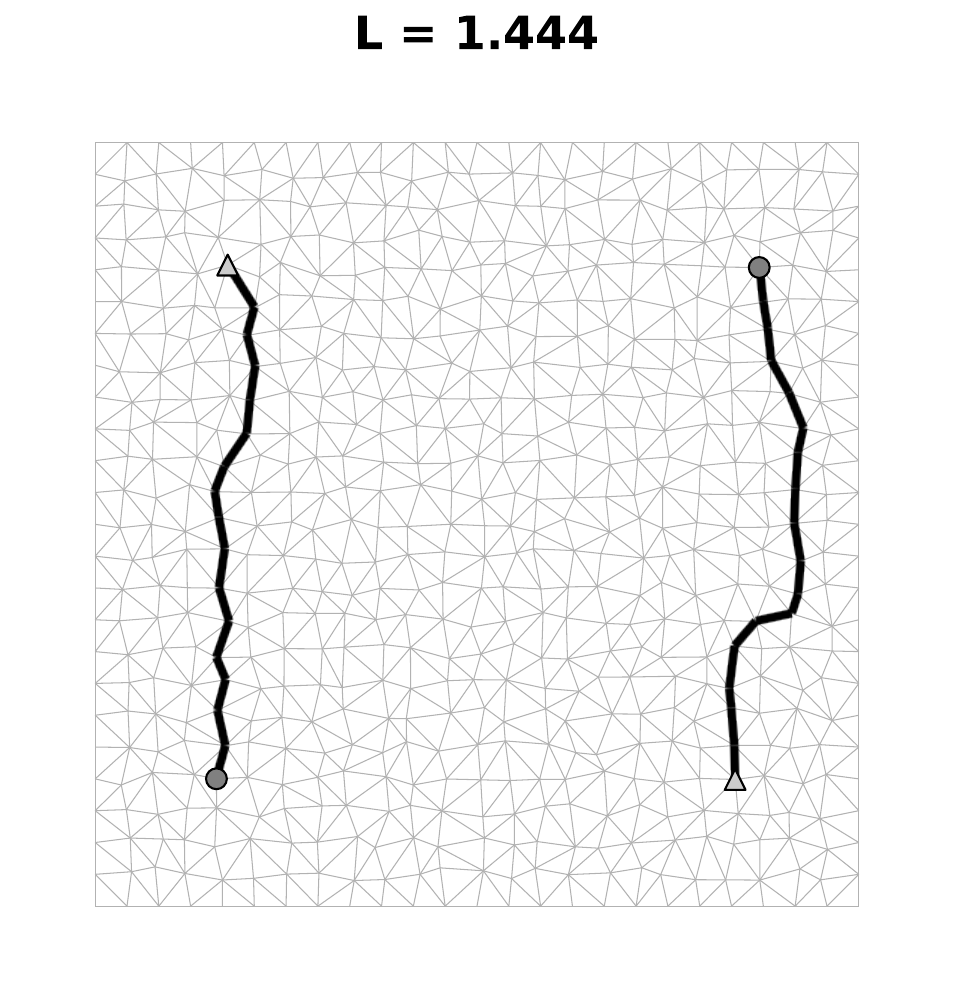}
         \caption{$S_{\text{left source}} = S_{\text{right source}} = 0.5 = - S_{\text{left sink}} = - S_{\text{right sink}}$.}
         \label{fig:2_disconnected}
     \end{subfigure}
     \hfill
     \begin{subfigure}[b]{0.45\textwidth}
         \centering
         \includegraphics[width=0.85\textwidth]{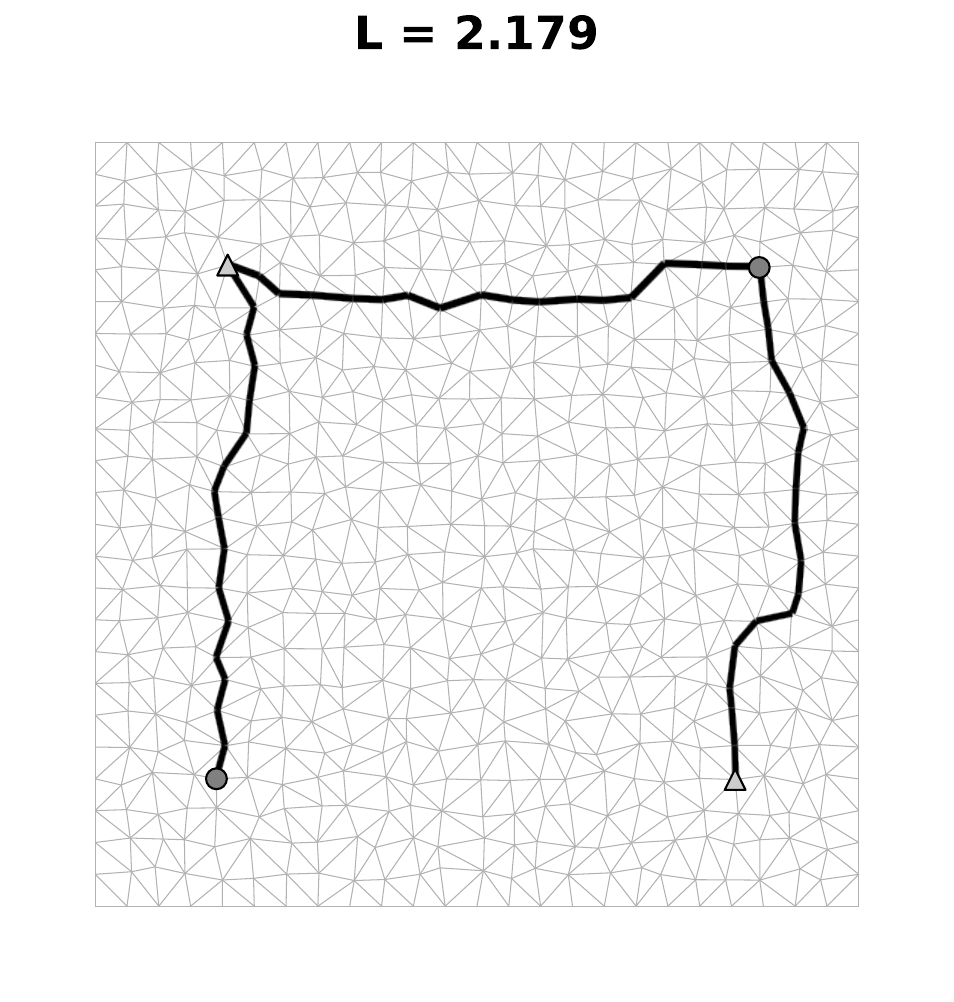}
         \caption{$S_{\text{left source}} = - S_{\text{right sink}} = 0.3$, and $S_{\text{right source}} = - S_{\text{left sink}} = 0.7$.}
         \label{fig:2_connected}
     \end{subfigure}
    \caption{Two different possible steady state configurations for two sources and two sinks in a $25\times25$ lattice.}
    \label{fig:2_sources_sinks}
\end{figure}

The case of two sources and two sinks was also studied and, to illustrate it, two different trees were generated and are presented in figure \ref{fig:2_sources_sinks}. Even though the source and sink nodes were the same, each tree was generated with different values of incoming/outgoing flow of fluid at each source/sink node $j$, $S_j$. The trees were generated using the same initial conditions, and were generated on the square lattice graph described previously of side length $1$ with $N = 25\times 25$ nodes, with $V = 100$ and $\sum_{j\in\text{sources}}S_j = 1$.

\par Figure \ref{fig:2_sources_sinks} shows that not all steady state trees are necessarily visibly connected. Figure \ref{fig:2_disconnected} seems like it contains two disconnected graphs. However, as stated earlier, the graph tree has only one connected component at all times, with conductivities partitioned in two sets of high and low values, which allows the identification of the channels that are effectively conducing flow in the tree. 

\section{Validation of Murray's law} \label{sec:validation_murrays_law}

While the generalised Murray's law (eq. \eqref{eq:generalised_murray}) was verified for steady state mathematically, it has yet to be verified in simulation. As it is a direct consequence of the set of equations used to determine the steady state, the hypothesis that it is verified at steady state is likely, but it is unknown whether or not it is verified dynamically.

To determine if the generalised Murray's law \eqref{eq:generalised_murray} is verified, the configuration of sources and sinks of figure \ref{fig:length_distribution_more} will be used. First, the initial conductivities will be chosen to be $D_{ij}(t=0) = 1, \forall (i,j)\in E$, in order to ensure the steady state obtained will always be the same for all the simulations. Then, the steady state will be observed and two separate bifurcation points in the tree (at steady state) will be identified. By bifurcation points, one means nodes $j$ which are connected to more than two conducting edges. For these two chosen bifurcation nodes, the quantity $\left|\sum_{j:(i,j)\in E, Q_{ij} < 0} r_{ij}^3 - \sum_{j:(i,j)\in E, Q_{ij} > 0} r_{ij}^3\right| := \left|R_{\text{out}} - R_{\text{in}}\right|$ will be measured as time progresses. For Murray's law to be verified, the quantity $\left|R_{\text{out}} - R_{\text{in}}\right|$ must be zero.

\begin{figure}
     \centering
     \begin{subfigure}[b]{0.32\textwidth}
         \centering
         \includegraphics[width=\textwidth]{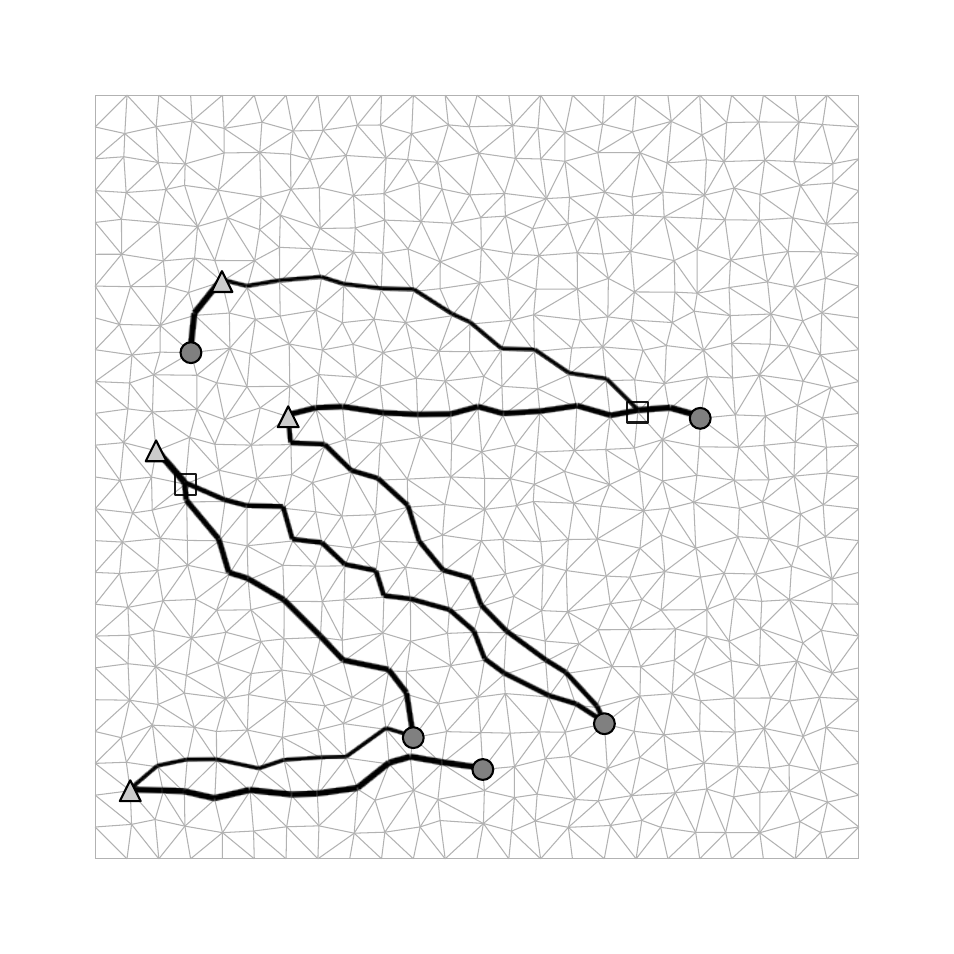}
         \caption{Steady state analyzed, with the bifurcation nodes chosen surrounded by black square boxes.}
         \label{fig:murray_ss_nodes}
     \end{subfigure}
     \hfill
     \begin{subfigure}[b]{0.32\textwidth}
         \centering
         \includegraphics[width=\textwidth]{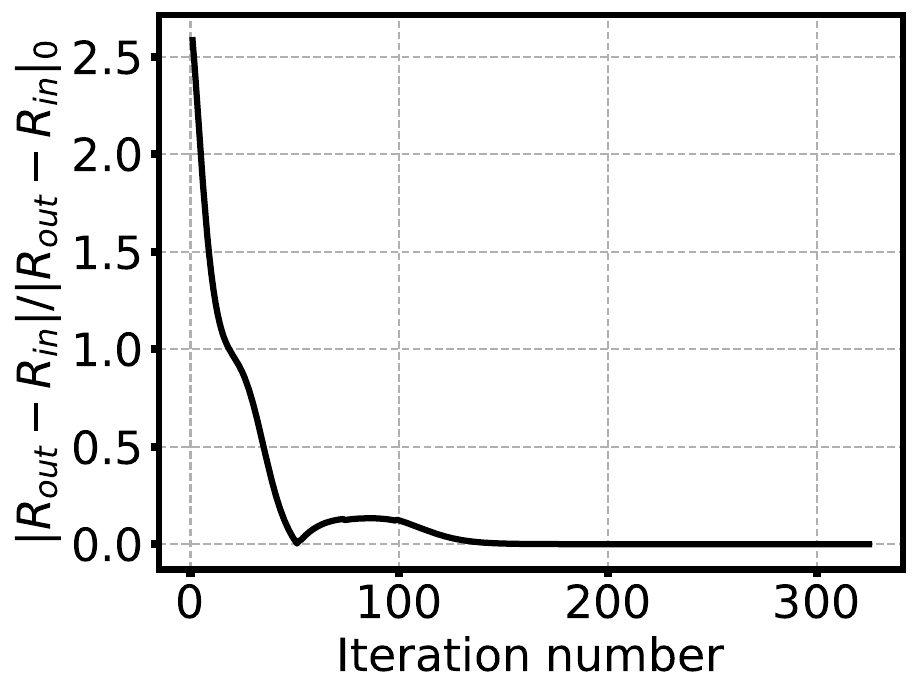}
         \caption{Variation of $\left|R_{\text{out}} - R_{\text{in}}\right|$ over time for leftmost bifurcation node.}
         \label{fig:murray_graph_left}
     \end{subfigure}
     \hfill
     \begin{subfigure}[b]{0.32\textwidth}
         \centering
         \includegraphics[width=\textwidth]{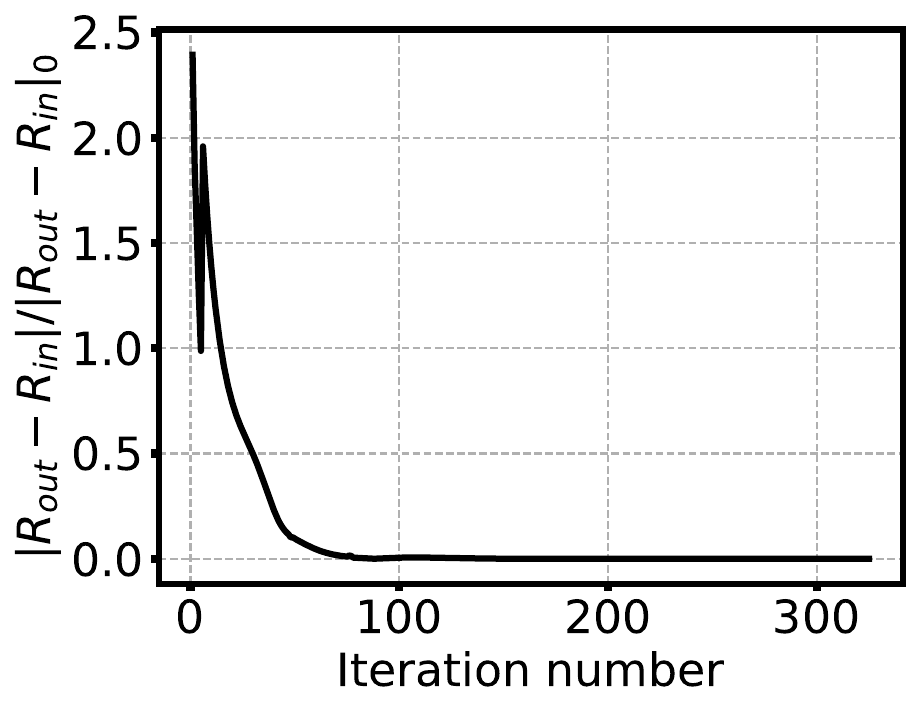}
         \caption{Variation of $\left|R_{\text{out}} - R_{\text{in}}\right|$ over time for rightmost bifurcation node.}
         \label{fig:murray_graph_right}
     \end{subfigure}
    \caption{Dynamic verification of Murray's law for two different bifurcation nodes.}
    \label{fig:murrays_law}
\end{figure}

The graph of $\left|R_{\text{out}} - R_{\text{in}}\right|$ (normalized to its initial value) as a function of time is shown for both bifurcation nodes chosen in figure \ref{fig:murrays_law}, as well as the steady state and the bifurcation nodes chosen (the bifurcation nodes chosen are surrounded by a black square box).

As the graphs of figures \ref{fig:murray_graph_left} and \ref{fig:murray_graph_right} show, $\left|R_{\text{out}} - R_{\text{in}}\right| > 0$ for some initial portion of time for both nodes; after some time has passed, at steady state, $\left|R_{\text{out}} - R_{\text{in}}\right| = 0$. Thus, the generalized Murray's law \eqref{eq:generalised_murray} is not verified dynamically but is verified at steady state.

\section{Properties of simple networks} \label{sec:simple_networks}

To obtain more insight into the equations of adaptive H-P flows (see sec. \ref{sec:adaptive-model}), we studied the analytical solutions to such equations for cases of simple networks, that is, networks comprised of one and two channels. 

\subsection{One channel networks}

We start by applying the conductivity adaptation equations to a single channel of length $L_{12}$, with one source and sink. Let us assume that the source is located at the node number $1$ and the sink at node number $2$, and that $S_1 = - S_2$. As there is only a single channel, the flux going through it will be $Q_{12} = S_1$. Thus, the adaptation equation \eqref{eq:adaptation_eq} becomes:

\begin{equation}
\frac{d}{d\tau}\sqrt{D_{12}} = \alpha\frac{|Q_{12}|^{2/3}}{L_{12}|Q_{12}|^{2/3}} - \sqrt{D_{12}} = \alpha \frac{1}{L_{12}}-\sqrt{D_{12}}
\label{eqn1}
\end{equation}

From eq. \eqref{eq:volume} and as discussed previously in sec. \ref{sec:renormalization_volume}, the volume of fluid in the network is determined by the initial conductivity values $D_{ij}(0)$. Thus, $V = \beta L_{12} \sqrt{D_{12}(0)}$. Equation \ref{eqn1} can be further simplified by substituting $\alpha=V/\beta = L_{12} \sqrt{D_{12}(0)}$:

\begin{equation}
\frac{d}{d\tau}\sqrt{D_{12}}=\sqrt{D_{12}(0)} - \sqrt{D_{12}},
\label{eqn1.1}
\end{equation}

Equation \eqref{eqn1} is linear and its general solution is:

\begin{equation}
D_{12}(\tau)=  \left(\frac{\alpha}{L_{12}}-\left(\frac{\alpha}{L_{12}}-\sqrt{D_{12}(0)}\right)e^{-\tau} \right)^2=D_{12}(0).
\label{eqn2}
\end{equation}

The steady state solution for the conductivity value is $D_{12}^*=\lim_{\tau\to \infty}D_{12}(t)= \alpha^2 / L_{12}^2=D_{12}(0)$.

As the conductivity is defined by $D_{12}=\pi r_{12}^4/(8\eta)$, and the Hagen-Poiseuille flux is given by $Q_{12}=D_{12} (p_1-p_2)/L_{12}$ (see eq. \eqref{eq:HP}), the flux at steady state (when $\tau\to\infty$) is:

\begin{equation}
Q_{12}^*=D_{12}^*\frac{1}{L_{12}} (p_1(\infty)-p_2)=\frac{\alpha^2}{L_{12}^3} p_1(\infty) = \frac{D_{12}(0)}{L_{12}} p_1(\infty) = S_1 ,
\label{eqn3}
\end{equation}
\noindent
where $p_1(\infty)$ is determined by the flux at the source $Q_{12}=S_1$, and we made the choice $p_2=0$.

The steady-state solution $Q_{12}^*$ in \eqref{eqn3} shows that, for fixed volume $V$ ($\alpha$ constant), the flux at the source increases linearly with the pressure. The radius of the channel is constant for every $\tau\ge 0$ and is determined by $D_{12}(0)$.

\begin{figure}
\begin{center}
\includegraphics[width=\textwidth]{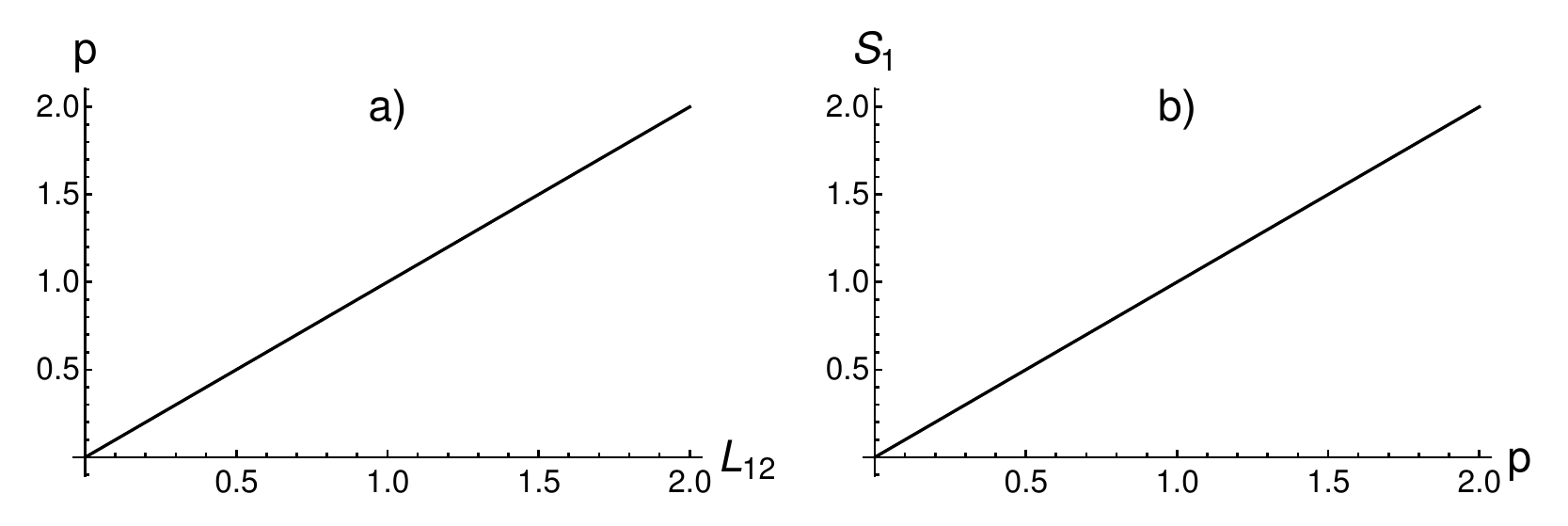}
\caption{For constant volume of the fluid, \textbf{a)} Relationship between the pressure at the source $p$ and channel length $L_{12}$ and \textbf{b)} Relationship between the pressure at the source $p$ and source flux $S_1$, at constant volume. Parameter values: $\alpha=1$, $S_1=1$, $L_{12}=1$.}
\label{fig1}
\end{center}
\end{figure}

In figure~\ref{fig1}, we show the relationship between the pressure and channel length and source fluxes. 

\subsection{Two channel networks}

Let us now study the case of a network comprised of two channels with two sinks connected to a single source with an input flux $S_1$. The adaptation equations for the Hagen-Poiseuille flow (see eq. \eqref{eq:adaptation_eq}) are:

\begin{equation}
\begin{array}{l}\displaystyle
\frac{d}{d\tau}\sqrt{D_{12}}=\alpha \frac{Q_{12}^{2/3}}{L_{12}Q_{12}^{2/3}+L_{13}Q_{13}^{2/3}}-\sqrt{D_{12}}\\ \displaystyle
\frac{d}{d\tau}\sqrt{D_{13}}=\alpha \frac{Q_{13}^{2/3}}{L_{12}Q_{12}^{2/3}+L_{13}Q_{13}^{2/3}}-\sqrt{D_{13}}
\end{array}
\label{eqn4}
\end{equation}

\noindent
with node fluxes given by eq. \eqref{eq:kirch}:

\begin{equation}
Q_{12}+Q_{13}=S_1>0,\ Q_{21}=S_2<0,\ \hbox{and} \ Q_{31}=S_3<0 
\label{eqn5}
\end{equation}
and the conservation law $S_1+S_2+S_3=0$ (see \eqref{eq:sources_sinks}).

Introducing equations \eqref{eqn5} into \eqref{eqn4}, these equations simplify and become linear. The steady states conductivities are:

\begin{equation}
\begin{array}{l}\displaystyle
D_{12}^*=\alpha^2 \frac{|S_2|^{4/3}}{(L_{12}|S_2|^{2/3}+L_{13}|S_3|^{2/3})^2}\\ \displaystyle
D_{13}^*=\alpha^2 \frac{|S_3|^{4/3}}{(L_{12}|S_2|^{2/3}+L_{13}|S_3|^{2/3})^2},
\end{array}
\label{eqn6}
\end{equation}
where $\alpha=V/\beta=L_{12} \sqrt{D_{12}(0)}+L_{13} \sqrt{D_{13}(0)}$. Therefore, as $D_{12}^*\le \alpha^2/L_{12}^2$ and $D_{13}^*\le \alpha^2/L_{13}^2$, we have:

\begin{equation}
Q_{12} + Q_{13} = S_1 \iff \left(D_{12}^*\frac{1}{L_{12}}+D_{13}^*\frac{1}{L_{13}}\right)p_1(\infty)=S_1.
\label{eqn7}
\end{equation}

Therefore, for fixed volume $V$, at steady state, the pressure at the source is linear as a function of the flux at the source. However, as time passes, the flow will adapt for the conductivity values $D_{12}^*$ and $D_{13}^*$. 

\begin{figure}
\begin{center}
\includegraphics[width=\textwidth]{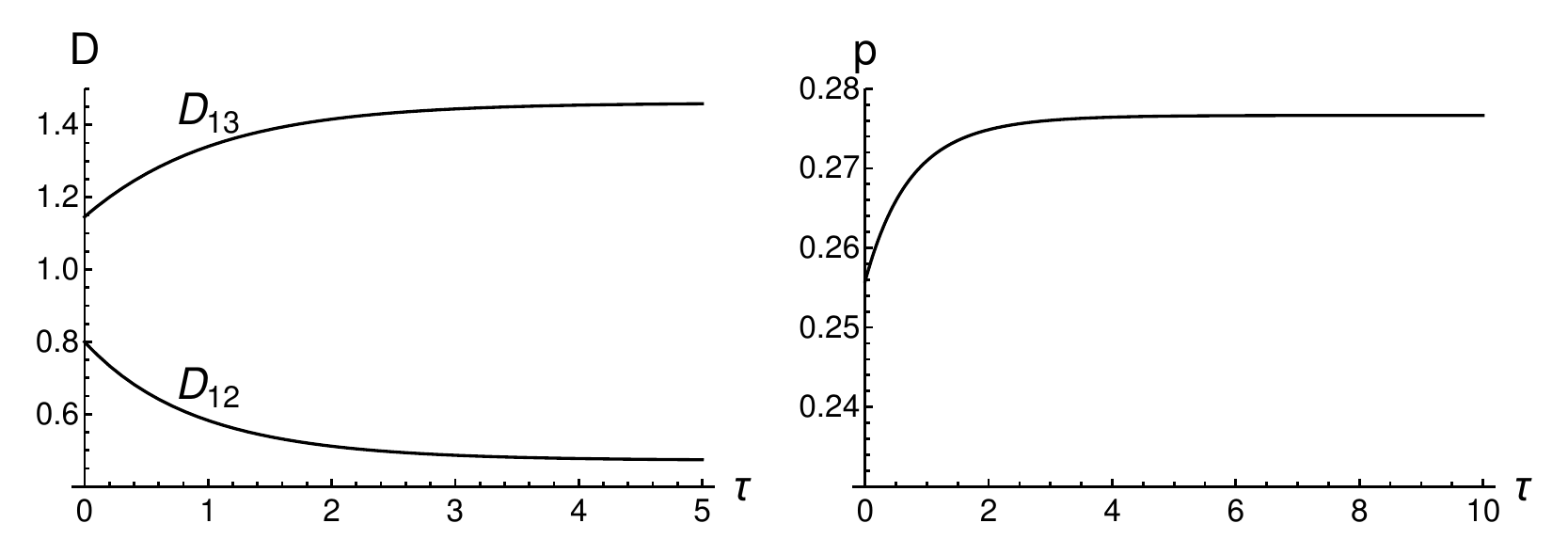}
\caption{Channel conductivities $D_{12}$ and $D_{13}$ and pressure at the source $p$ as a function of time $\tau$. Parameter values: $\alpha=1$, $S_1=1$, $S_2=-0.3$, $S_3=-0.7$, $L_{12}=0.4$, $L_{13}=0.6$, $D_{12}(0)=0.8$, $D_{13}(0)=1.14572$.}
\label{fig2}
\end{center}
\end{figure}

In figure~\ref{fig2}, we show the time evolution of the adapted conductivities and the pressure at the source for the two channels with fixed lengths and constant volume. Equation \eqref{eqn7} shows that the steady-state pressure at the source depends on the lengths of the channels.

During the adaptation process, despite the fact that the input fluxes are constant, the pressure at the sources adapts over time. While the conductivities saturate at the steady state values, the pressure at the sources increases as the length of the vein increases without an apparent limit.

\begin{figure}
\begin{center}
\begin{subfigure}[c]{0.45\textwidth}
         \centering
         \includegraphics[width=\textwidth]{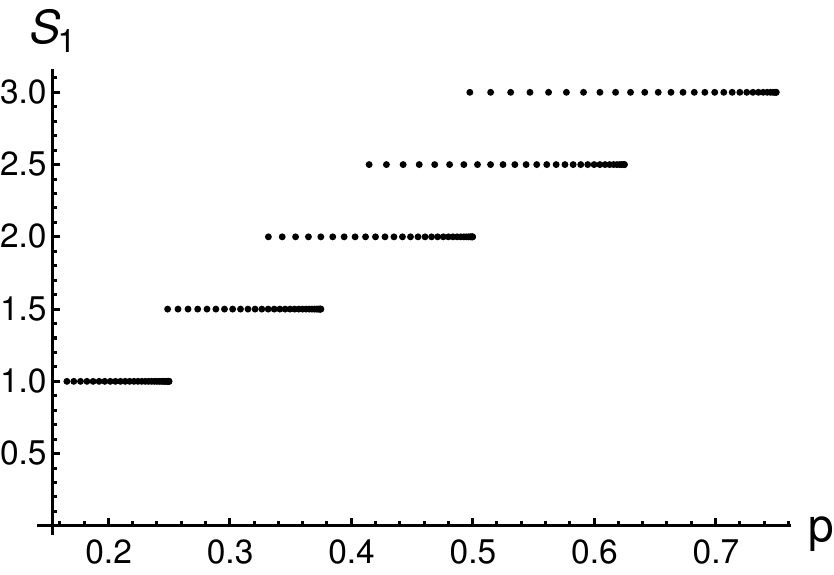}
         \caption{}
         \label{fig:3_data}
\end{subfigure}
\begin{subfigure}[c]{0.45\textwidth}
         \centering
         {\includegraphics[width=\textwidth]{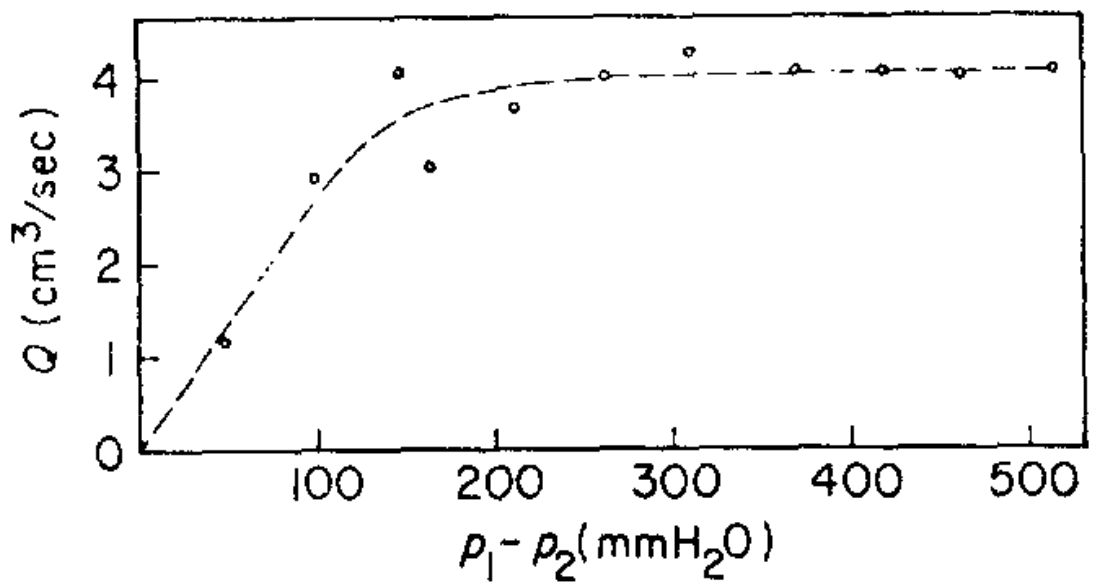}\\
         \includegraphics[width=\textwidth]{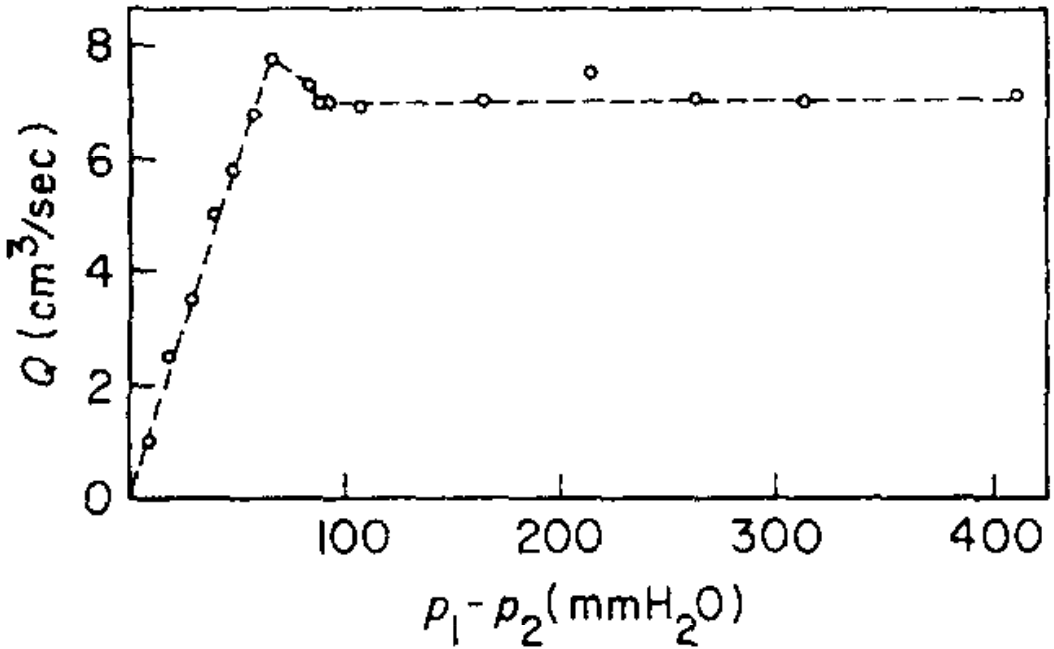}}
         \caption{}
         \label{fig:3_brecher}
     \end{subfigure}
\caption{\textbf{a)} Fluxes $S_1$ and pressures at the source $p$, for several different channel lengths and several different fluxes at the source. Parameter values: $\alpha=1$, $S_1\in [1,3]$, $S_2=-S_1/2$, $I_3=-S_1/2$, $L_{12}\in [0.5,0.8]$, $L_{12}+L_{13}=1$, $D_{12}(0)=0.8$, $D_{13}(0)=1.146$. \textbf{b)} Measurements of the steady flux of fluid through a vessel as a function of the pressure difference at the edges of said vessel, for the superior \textit{vena cava} of a dog (above) and for a collapsible rubber tube (below). Images from \cite{rubinow}.}
\label{fig3}
\end{center}
\end{figure}

We now analyse the case where the lengths of the channels vary, and the volume of the fluid remains constant. In figure~\ref{fig3}, we have calculated the 
steady-state pressures at the source $p$ for different channel lengths and several fluxes at the source. Data shows the increase of pressure for fixed fluxes at the source. This phenomenon has been observed experimentally for contractile veins, specifically for the \textit{vena cava} of a dog and a collapsible rubber tube (see figure \ref{fig3}) \cite{rubinow}.

\section{Comparison with \textit{Physarum polycephalum}} \label{sec:comparison_physarum}

\subsection{Synchronous configuration} \label{sec:physarum_sync}

The class of models that describe adaptive Hagen-Poiseuille flow on graphs studied in this thesis were developed as a way to describe \textit{Physarum polycephalum}'s network adaptation patterns. Thus, the results of this model are now compared to \textit{Physarum polycephalum}'s behavior.

\begin{figure}
    \centering
    \begin{subfigure}[b]{\textwidth}
         \centering
         {\includegraphics[width=0.3\textwidth]{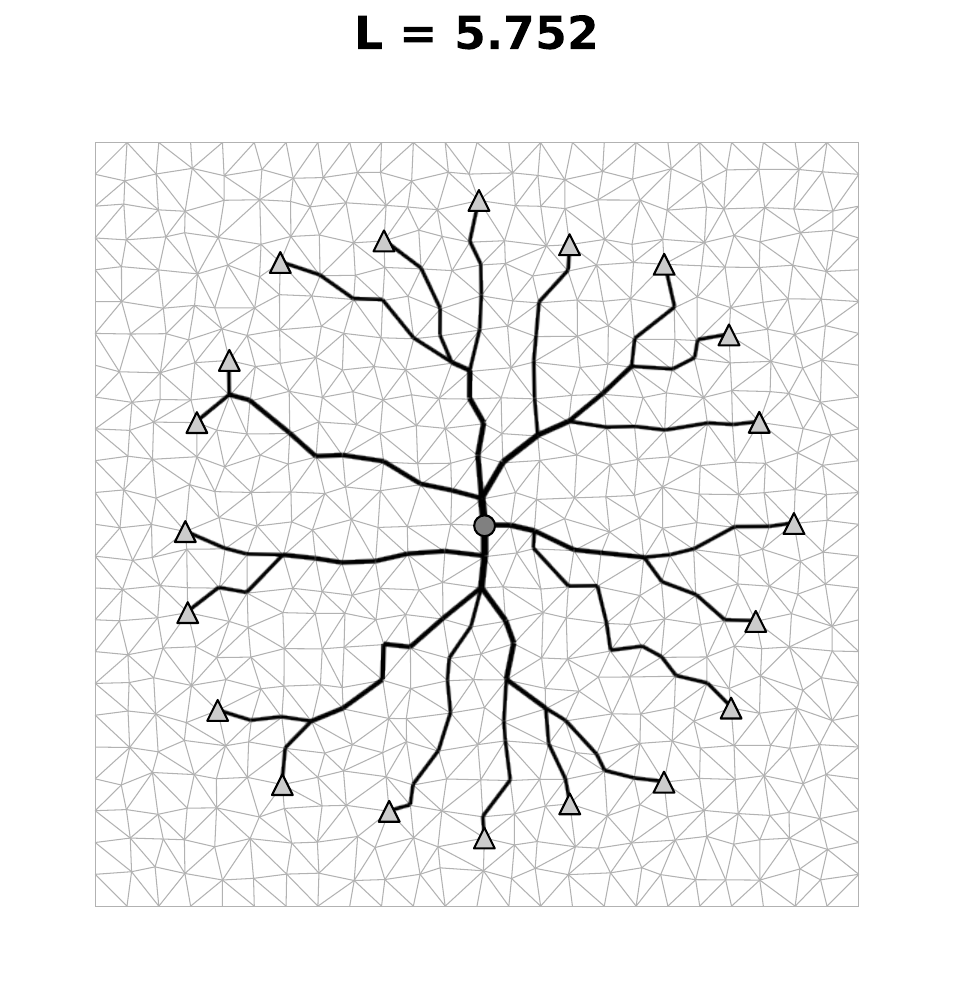}\hfill
         \includegraphics[width=0.3\textwidth]{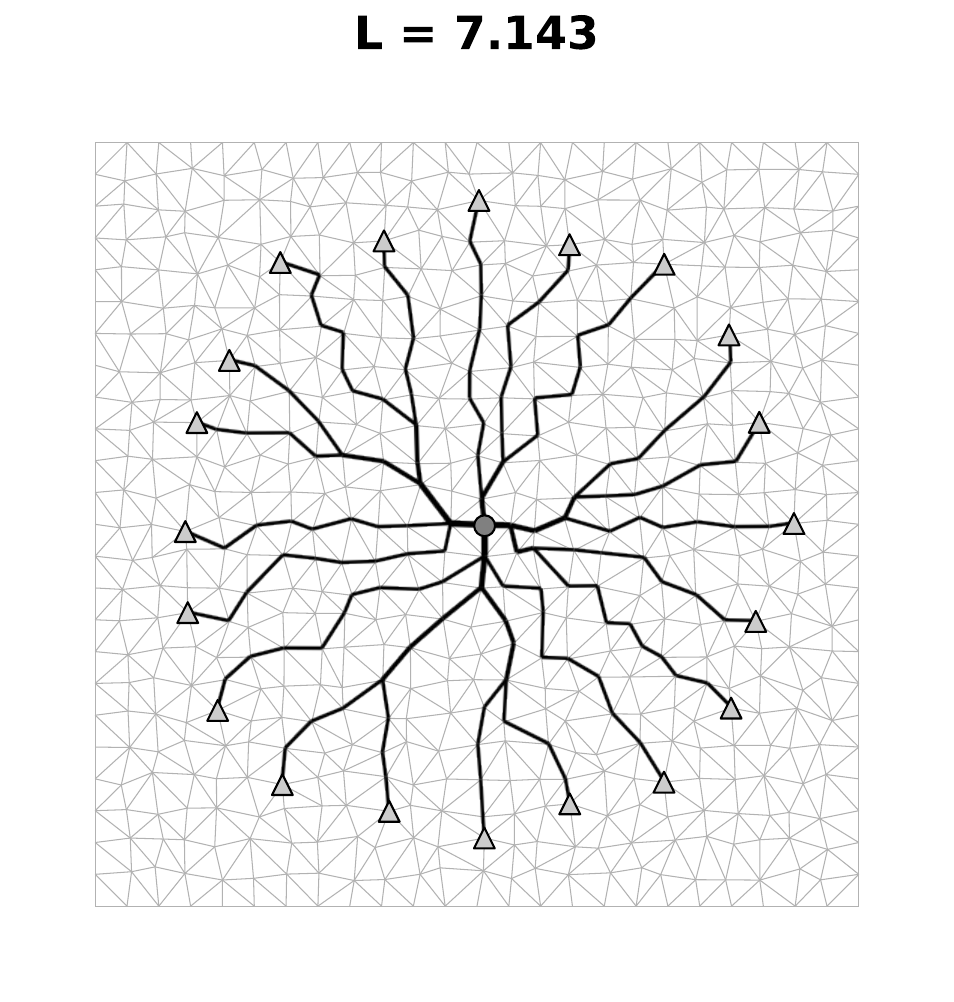}\hfill
         \includegraphics[width=0.37\textwidth]{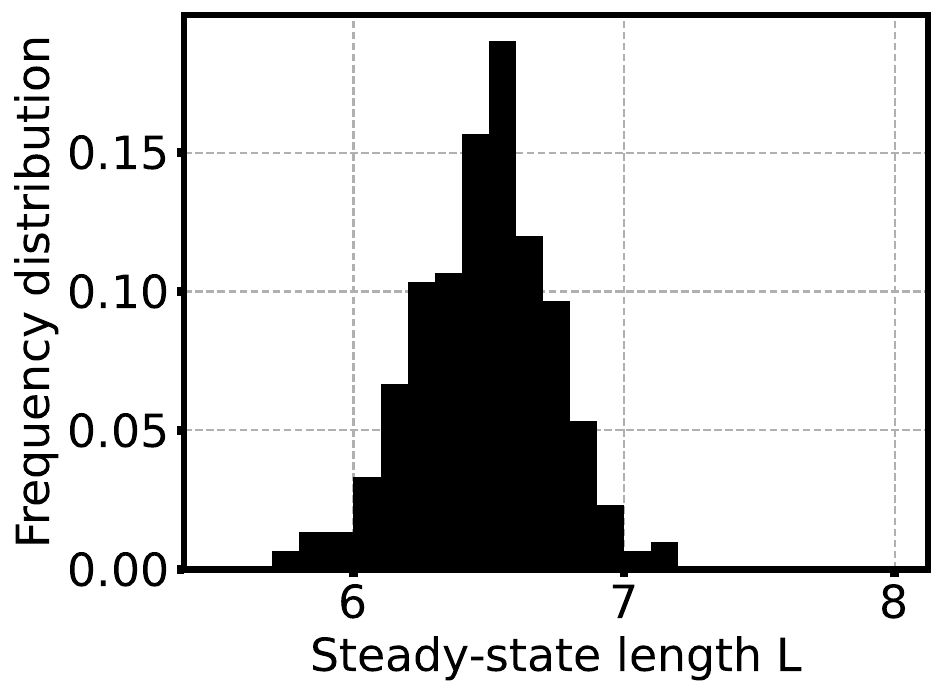}}
         \caption{Length distribution of the steady state graphs (right), shortest steady state graph obtained (left) and longest steady state graph obtained (middle), for a synchronous \textit{Physarum}-like configuration, for a $25\times25$ graph. The mean of the length distribution is $\overline{L} = 6.48$ and the standard deviation is $\sigma = 0.25$.}
         \label{fig:physarum_sync_L_dist_25}
     \end{subfigure}

     \begin{subfigure}[b]{\textwidth}
         \centering
         {\includegraphics[width=0.3\textwidth]{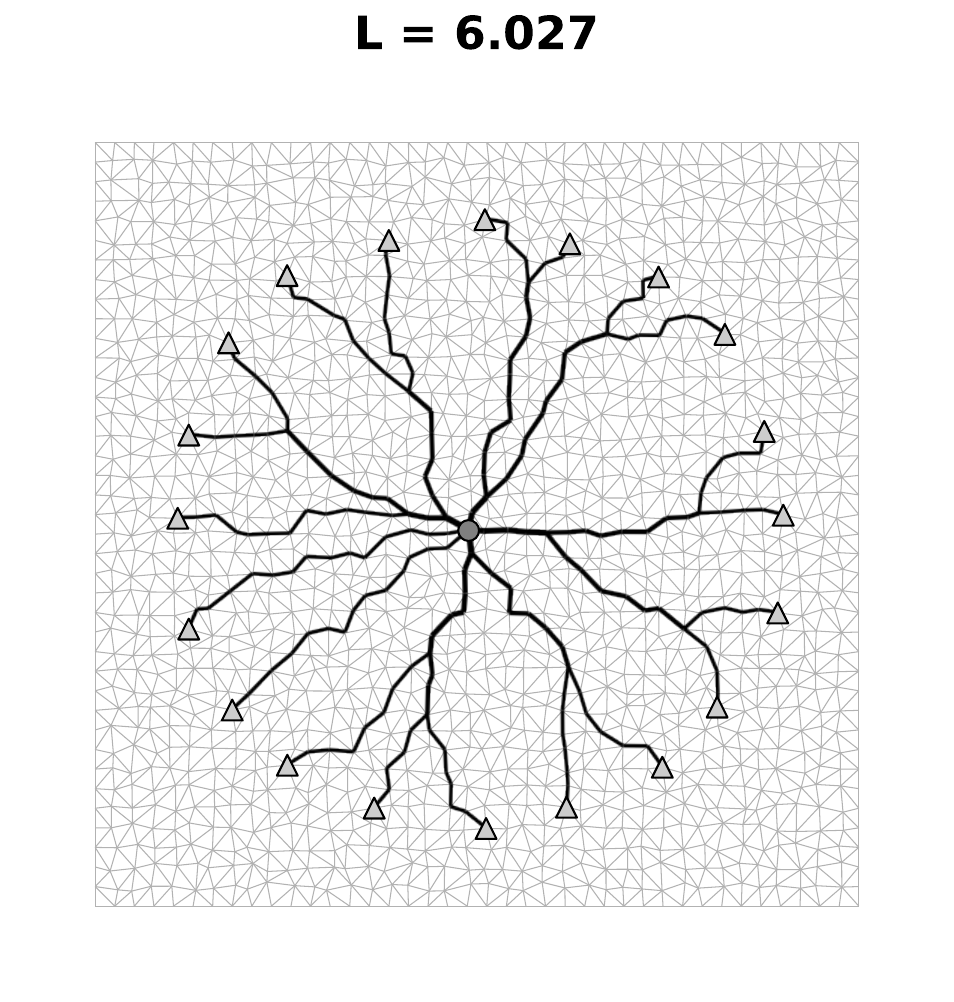}\hfill
         \includegraphics[width=0.3\textwidth]{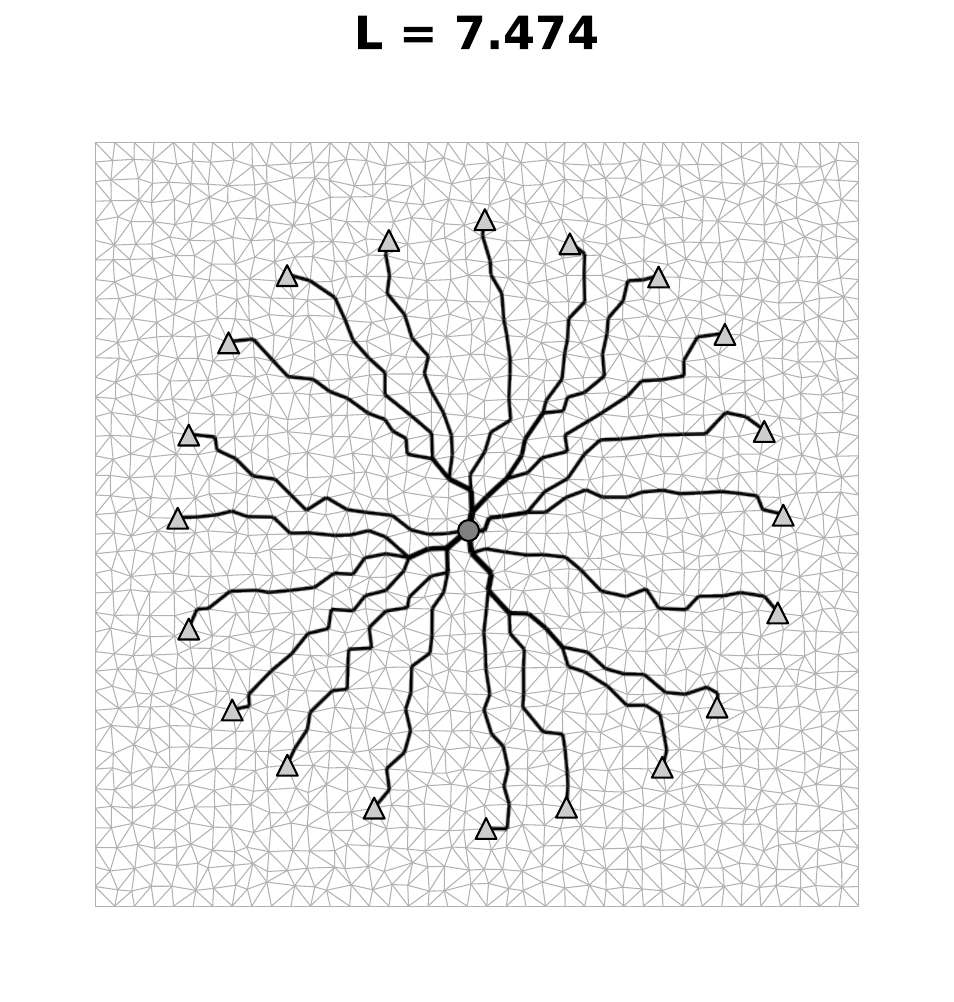}\hfill
         \includegraphics[width=0.37\textwidth]{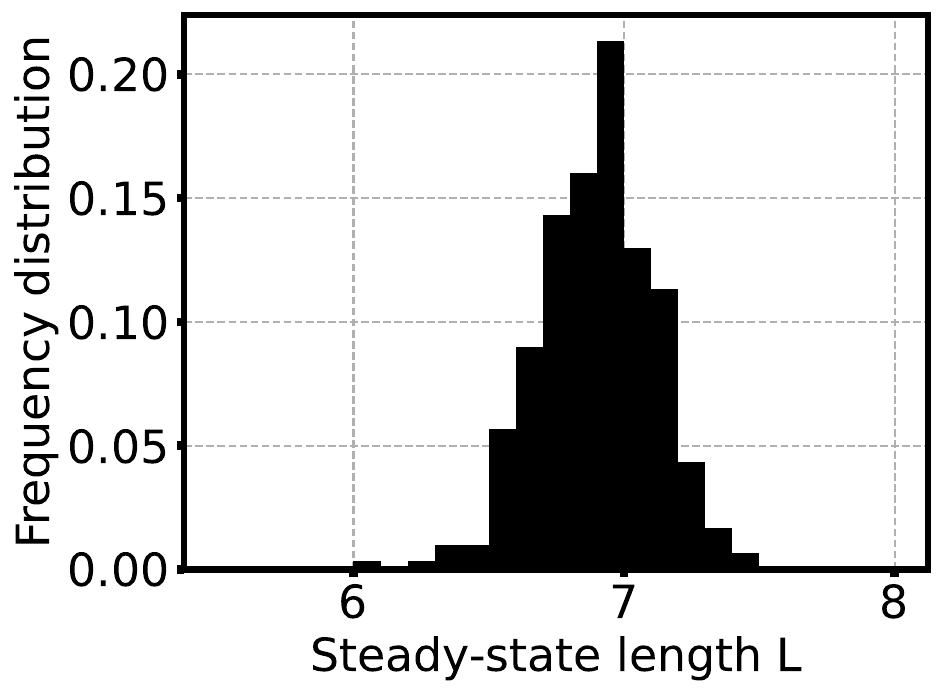}}
         \caption{Length distribution of the steady state graphs (right), shortest steady state graph obtained (left) and longest steady state graph obtained (middle), for a synchronous \textit{Physarum}-like configuration, for a $40\times40$ graph. The mean of the length distribution is $\overline{L} = 6.90$ and the standard deviation is $\sigma = 0.22$.}
         \label{fig:physarum_sync_L_dist_40}
     \end{subfigure}
     
     \begin{subfigure}[b]{\textwidth}
         \centering
         {\includegraphics[width=0.43\textwidth]{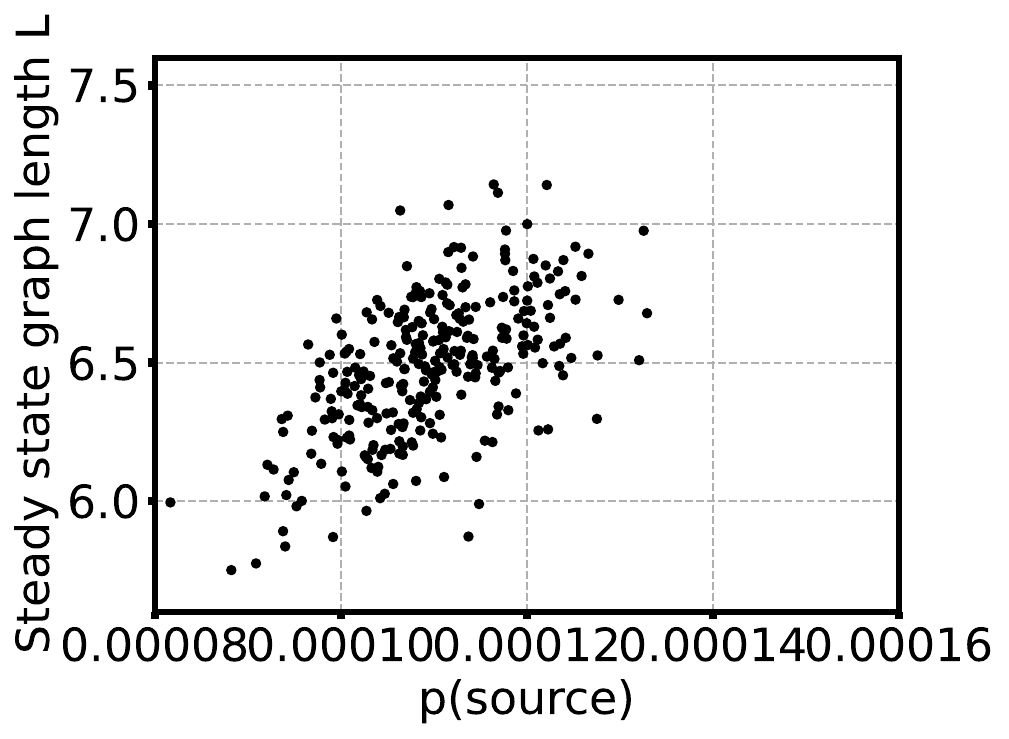}\hfill
         \includegraphics[width=0.43\textwidth]{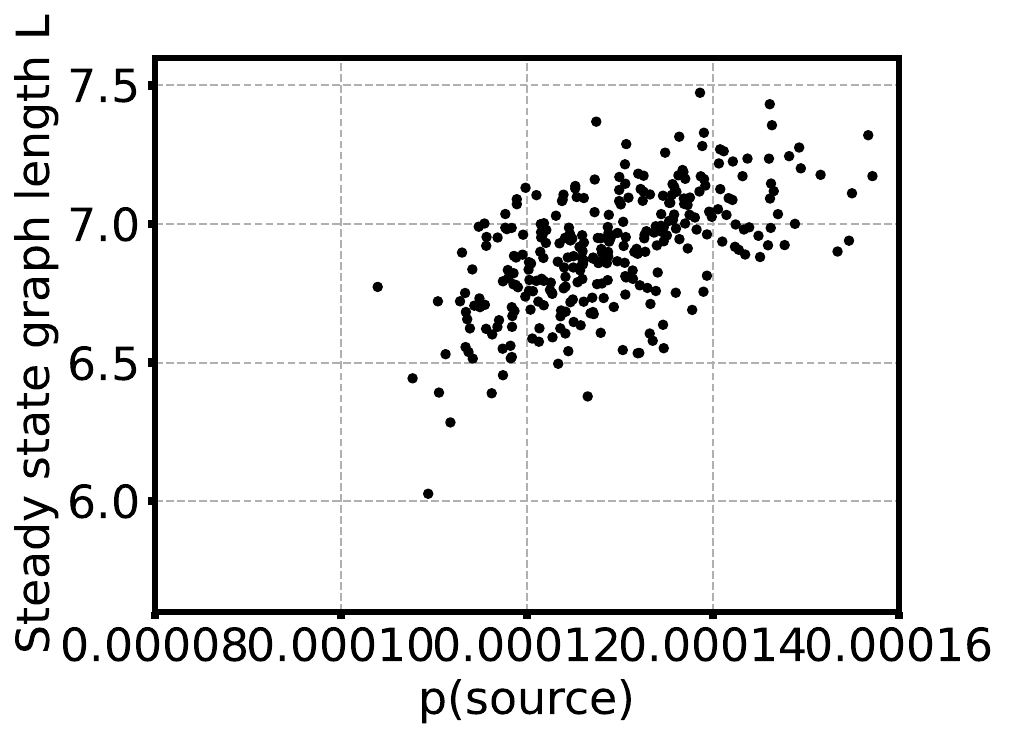}}
         \caption{Steady-state graph length $L$ as a function of the pressure at the source node $p(\text{source})$ for a synchronous \textit{Physarum}-like configuration, for a $25\times25$ graph (left) and a $40\times40$ graph (right).}
         \label{fig:physarum_sync_L_deltap}
     \end{subfigure}
     
\caption{Study for a synchronous \textit{Physarum}-like configuration of sources and sinks, for $V = 100$, $S_{\text{source}} = 1$, $S_j = -1/20, \forall j \in \{\text{sinks}\}$, for 300 runs of the simulation with random starting conductivities.}
\label{fig:physarum_sync}
\end{figure}

First, a synchronous \textit{Physarum}-like configuration of sources and sinks is tested. By synchronous, one means all sinks are active at all times (as opposed to the asynchronous configuration described ahead). This configuration consists on one source with $S_{\text{source}} = 1$ in the middle of the grid surrounded by 20 sinks evenly placed in a circumference of radius 0.4 around it, each with $S_j = -1/20, \forall j \in \{\text{sinks}\}$. All sources and sinks are active at all times. This mimics the outward radial growth pattern \textit{Physarum polycephalum} shows. The steady state graph length distribution, as well as the shortest and longest graphs at steady state for this configuration are shown in figures \ref{fig:physarum_sync_L_dist_25} and \ref{fig:physarum_sync_L_dist_40} for $N' = 25$ and $N' = 40$, respectively. Figure \ref{fig:physarum_sync_L_deltap} shows the steady state graph length $L$ as a function of the pressure at the source node $p_{\text{source}}$ for this same configuration, for both lattice sizes. Note that $p_j = 0, \forall j\in\{\text{sinks}\}$.

The steady state trees shown in figure \ref{fig:physarum_sync} exhibit paths from the source to the sinks.

Comparing figures \ref{fig:physarum_sync_L_dist_25} and \ref{fig:physarum_sync_L_dist_40}, we see that both the mean of the length distribution and the shortest steady state tree length for $N' = 40$ are larger than that of the $N' = 25$ case, while the standard deviation of the length distribution is slightly smaller for $N' = 40$.

Figure \ref{fig:physarum_sync_L_deltap} shows that, for $N' = 40$, not only are the steady state trees longer than the trees obtained with $N' = 25$, but the pressure at the source for $N' = 40$ is also generally larger for $N' = 40$.

Topologically, figures \ref{fig:physarum_sync_L_dist_25} and \ref{fig:physarum_sync_L_dist_40} show a pattern among the \textit{Physarum} simulations. As a tree of veins forms from the central source to the outward sinks, the veins are thicker and fewer near the source and can have several bifurcation points, where they split into thinner (and more numerous) veins; this splitting seems to increase in probability as the distance to the source increases. There are also no connecting paths between sinks whatsoever.

The more refined mesh ($N' = 40$) leading to longer networks is probably because of the fact that it is more likely that, for $N' = 40$, there are less bifurcations in the veins that connect the source to the sinks, and it is therefore more likely that the $N'=40$ case presents more numerous direct connections between the source and the sinks, thus increasing the tree length. This also means that the $N' = 40$ case has a source with more conducting edges connected to it. As such, this might also explain why the case of $N' = 40$ shows higher values of pressure difference between the source and sinks (see figure \ref{fig:physarum_sync_L_deltap}).

\subsection{Asynchronous configuration} \label{sec:physarum_async}

An asynchronous \textit{Physarum}-like configuration of sources and sinks is hereby tested. This configuration consists on one source in the middle of the grid surrounded by 20 sinks evenly placed in a circumference of radius 0.4 around it. By asynchronous, one means not all sinks are active at all times: at each iteration (every $\Delta\tau = 0.1$), a random number of sinks $N_{\text{active sinks}}$ (between 1 and 20) is chosen to become active with $S_j = -1/N_{\text{active sinks}}$. The inactive sinks at each iteration have $S_j = 0$. The source is always active with $S_{\text{source}} = 1$. This configuration also mimics the outward radial growth pattern \textit{Physarum polycephalum} exhibits, and similar algorithms were used in the past to simulate its adaptation \cite{tero_tokyo}. This algorithm mimics the asynchrony of resource consumption in \textit{Physarum}.

\begin{figure}
    \centering
    \begin{subfigure}[b]{0.245\textwidth}
         \centering
         \includegraphics[width=\textwidth]{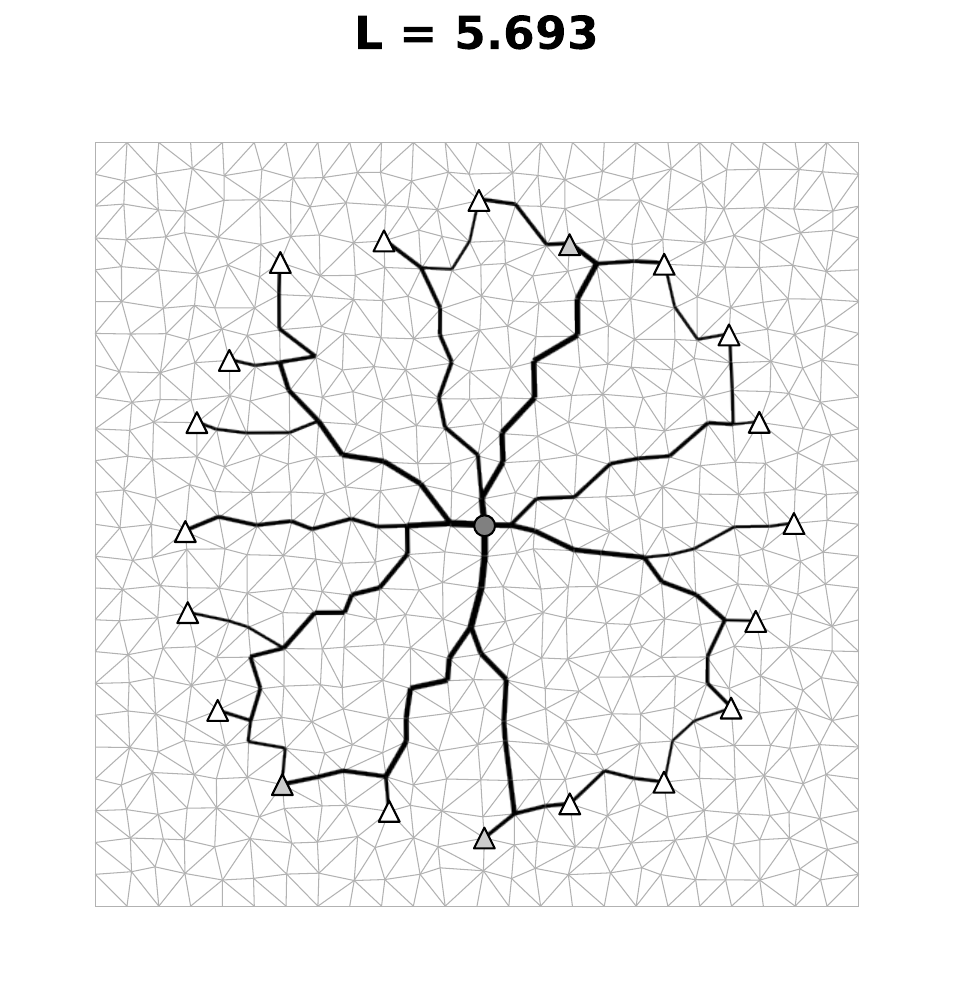}
         \caption{}
         \label{fig:physarum_async_1}
     \end{subfigure}
     \hfill
     \begin{subfigure}[b]{0.245\textwidth}
         \centering
         \includegraphics[width=\textwidth]{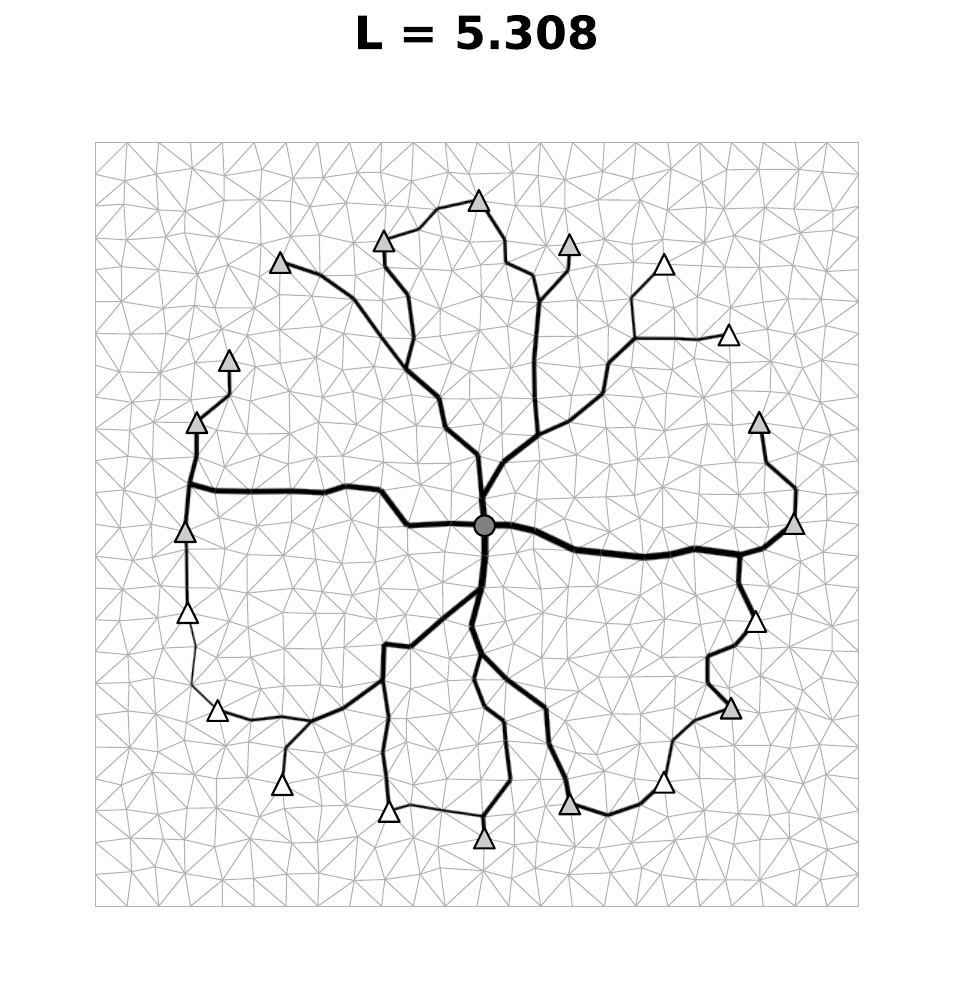}
         \caption{}
         \label{fig:physarum_async_2}
     \end{subfigure}
     \hfill
     \begin{subfigure}[b]{0.245\textwidth}
         \centering
         \includegraphics[width=\textwidth]{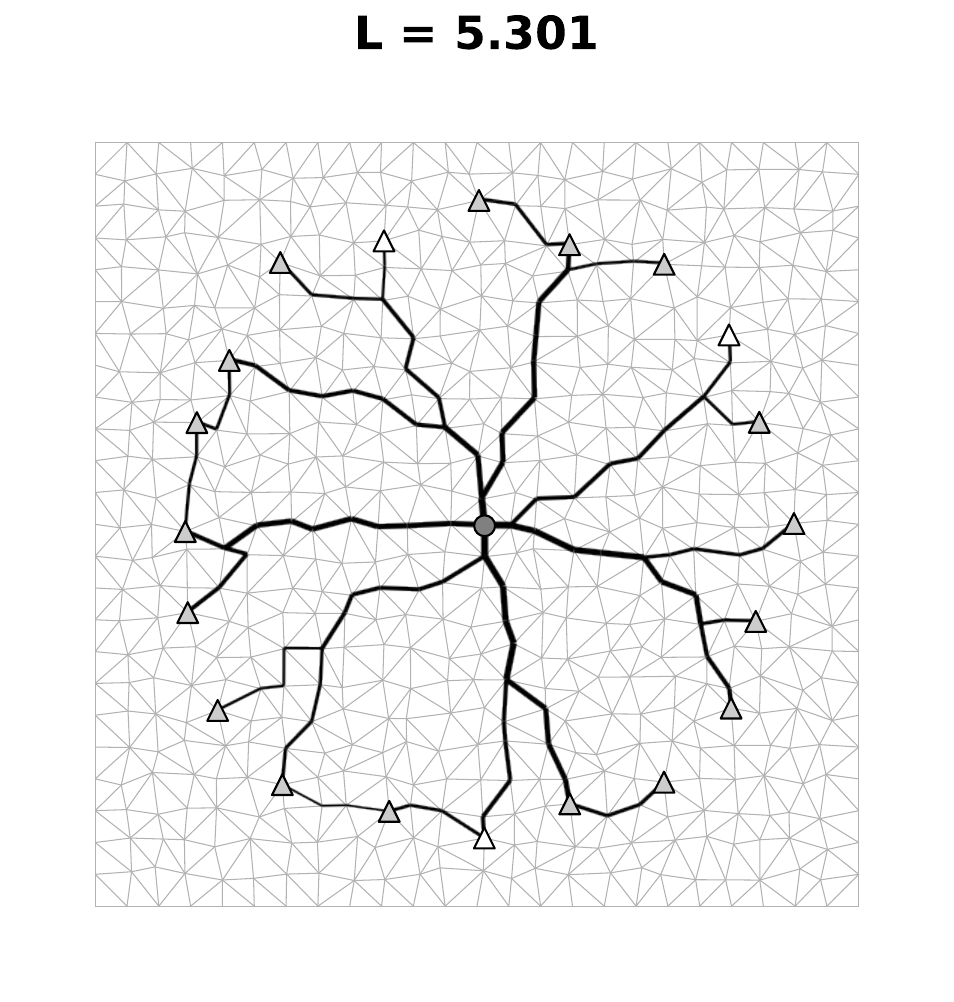}
         \caption{}
         \label{fig:physarum_async_3}
     \end{subfigure}
     \hfill
     \begin{subfigure}[b]{0.245\textwidth}
         \centering
         \includegraphics[width=\textwidth]{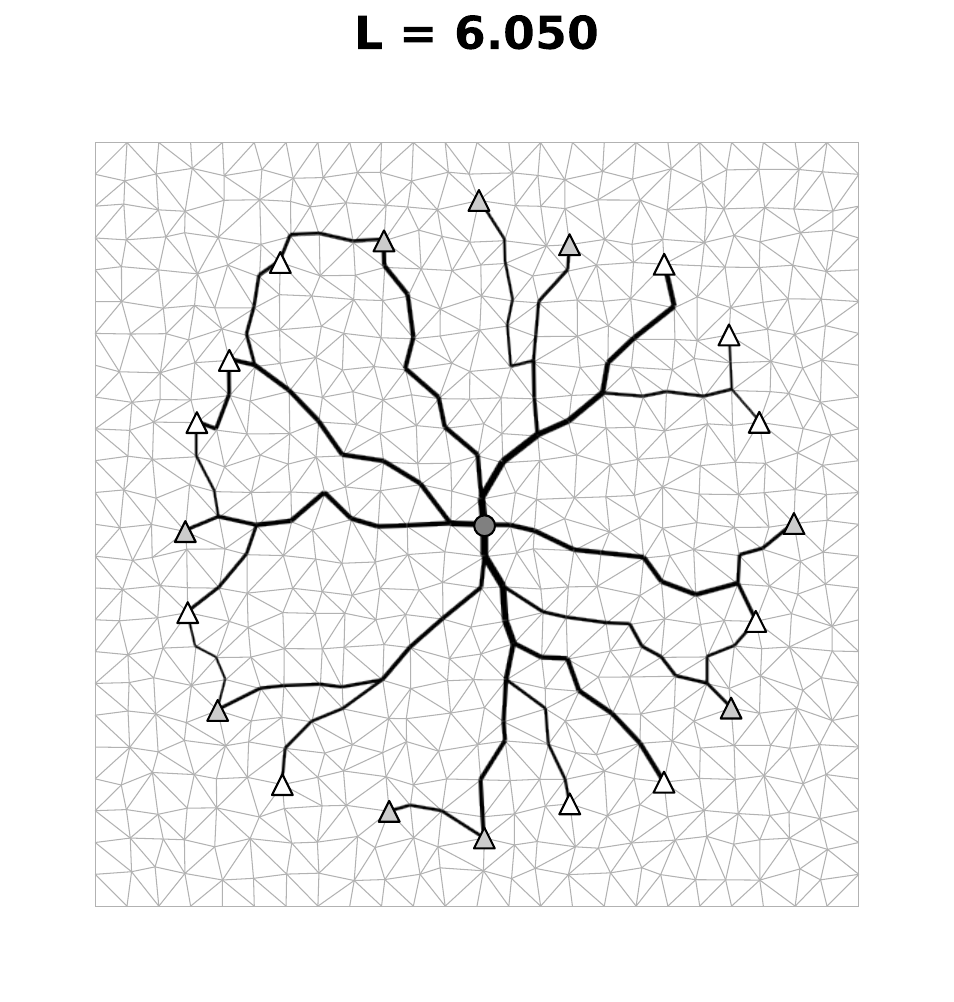}
         \caption{}
         \label{fig:physarum_async_4}
     \end{subfigure}
     \\
     \begin{subfigure}[b]{0.245\textwidth}
         \centering
         \includegraphics[width=\textwidth]{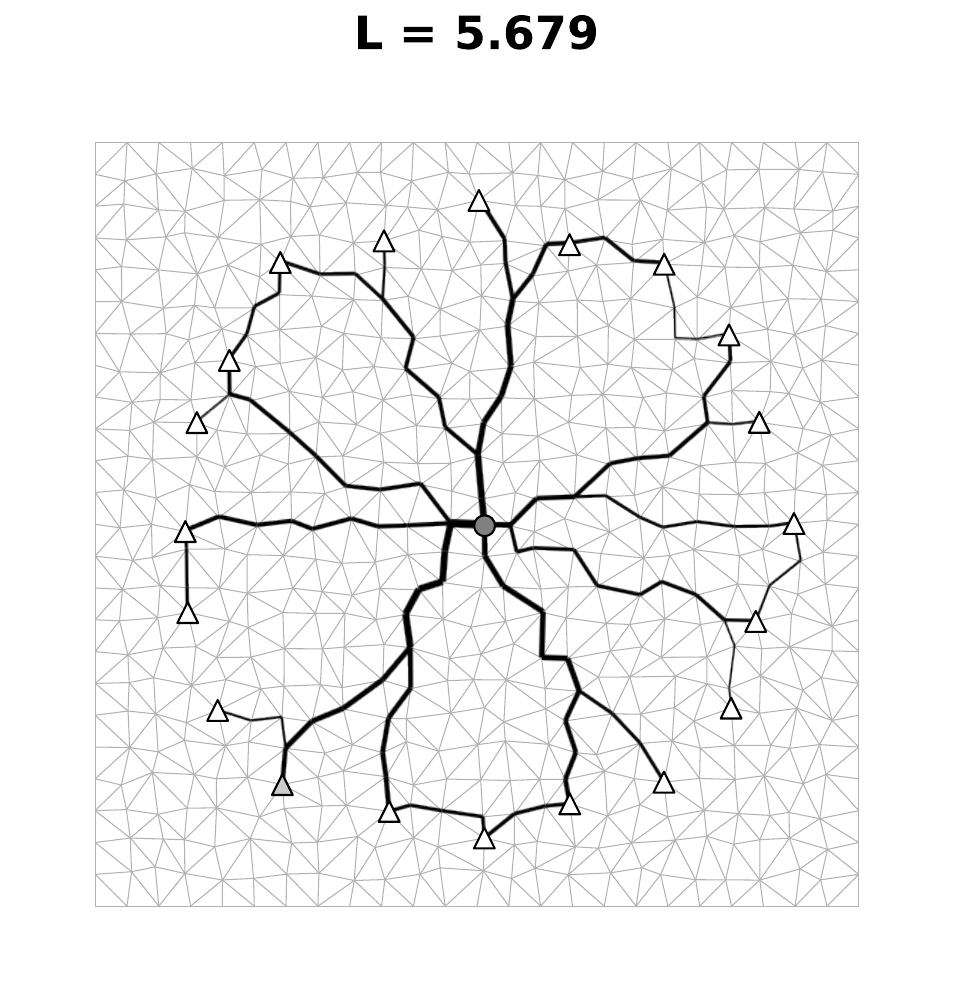}
         \caption{}
         \label{fig:physarum_async_5}
     \end{subfigure}
     \hfill
     \begin{subfigure}[b]{0.245\textwidth}
         \centering
         \includegraphics[width=\textwidth]{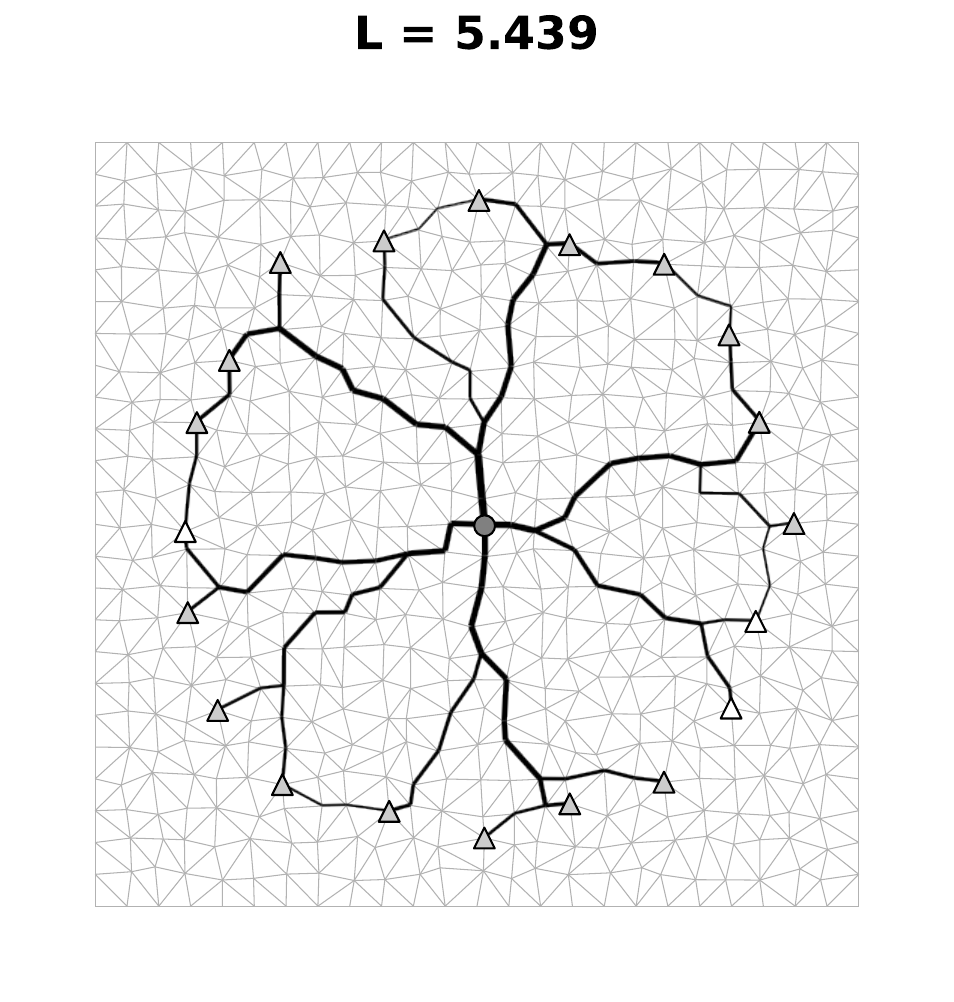}
         \caption{}
         \label{fig:physarum_async_6}
     \end{subfigure}
     \hfill
     \begin{subfigure}[b]{0.245\textwidth}
         \centering
         \includegraphics[width=\textwidth]{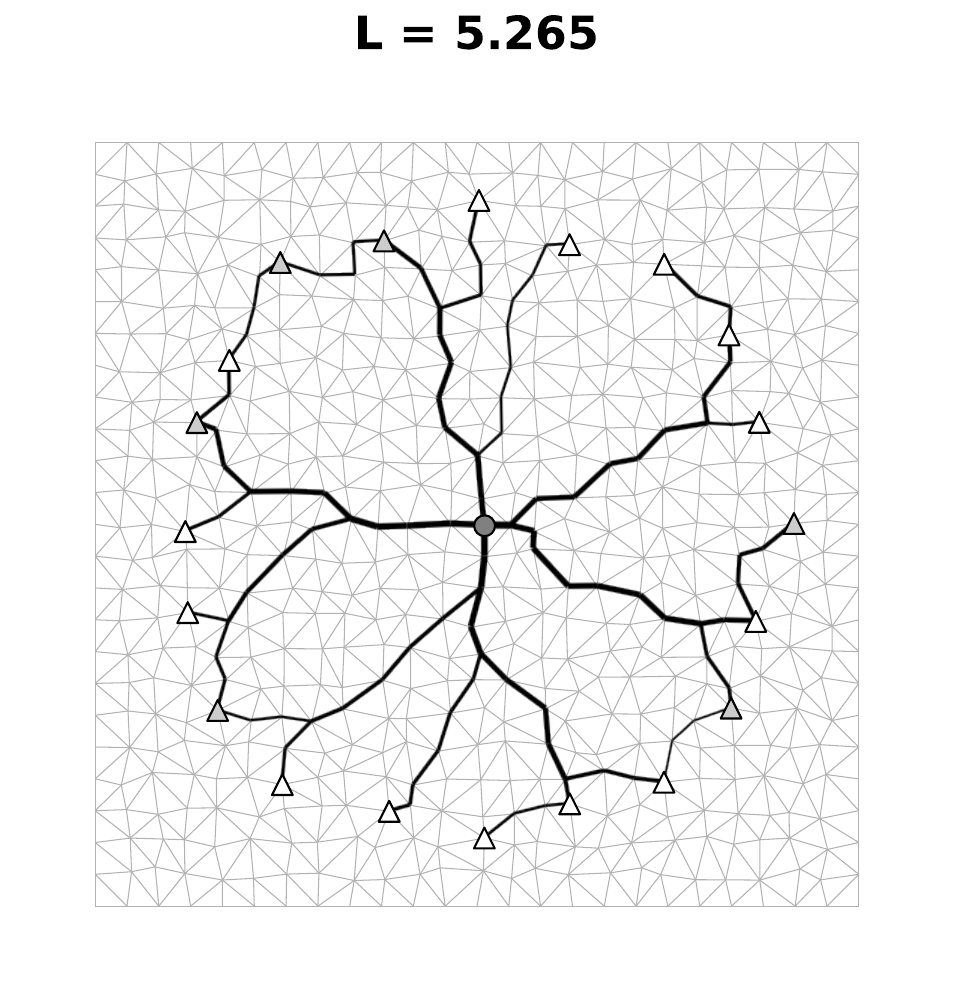}
         \caption{}
         \label{fig:physarum_async_7}
     \end{subfigure}
     \hfill
     \begin{subfigure}[b]{0.245\textwidth}
         \centering
         \includegraphics[width=\textwidth]{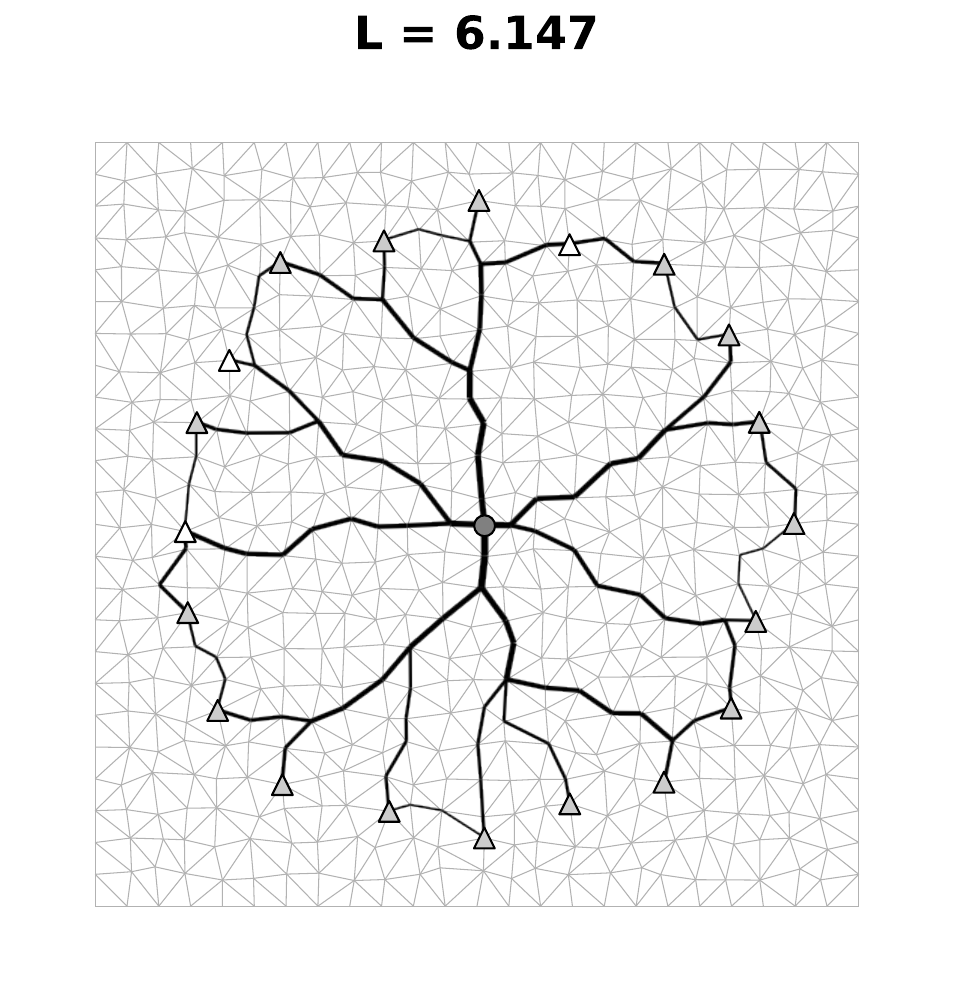}
         \caption{}
         \label{fig:physarum_async_8}
     \end{subfigure}
\caption{Eight final states for an asynchronous \textit{Physarum}-like configuration of sources and sinks, for $V = 100$, with different random starting conductivities.}
\label{fig:physarum_async}
\end{figure}

Figure \ref{fig:physarum_async} shows eight different final states for this configuration and their respective length $L$, obtained using different random initial conditions. In this case, as sources and sinks are not static, the steady state reached is not defined the same way the previous steady states were (see section \ref{sec:stopping_criteria}; in this case, $N_{\text{iter}} = 500$). For the last iteration, active sinks are shown as grey triangles, while inactive sinks are shown as white triangles.

Figure \ref{fig:physarum_async} shows a recurring vein network pattern among these simulations: veins are thicker and less numerous near the source, and they split into thinner (and more numerous) veins as the distance to the source increases (see figures \ref{fig:physarum_async_2} and \ref{fig:physarum_async_7}). Additionally, all steady states show closed loops involving the source and a few possible sinks, formed by thicker veins connecting the sinks to the source and thinner veins connecting the sinks on the outer edge of the network (see figures \ref{fig:physarum_async_1} and \ref{fig:physarum_async_3}). 

Comparing these patterns to the ones observed for the \textit{Physarum} synchronous simulations (see figures \ref{fig:physarum_sync_L_dist_25} and \ref{fig:physarum_sync_L_dist_40}), we see that the first pattern mentioned - thicker veins in the center splitting into thinner veins as the distance to the source increases - is also visible in the \textit{Physarum} synchronous simulations. However, the second pattern mentioned - paths connecting sinks to each other and possibly forming loops - is not seen at all in the \textit{Physarum} synchronous simulations.

\begin{figure}
    \centering
    \includegraphics[width=0.4\textwidth]{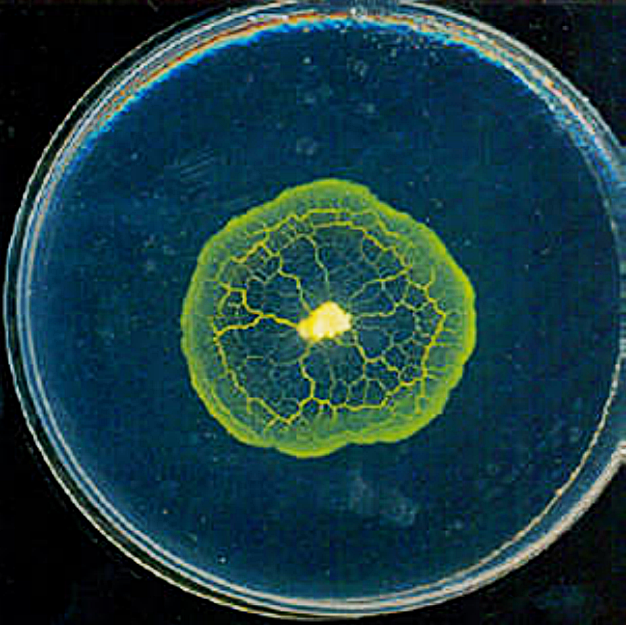}
    \caption{\textit{Physarum polycephalum}'s radial growth from a single site on a nutrient-rich surface. Image adapted from \cite{adamatzky:physarum_radial_growth}.}
    \label{fig:physarum_radial_growth}
\end{figure}

These patterns are observed in \textit{Physarum} networks. Figure \ref{fig:physarum_radial_growth} shows a \textit{Physarum} specimen growing from a single location radially outward. As figure \ref{fig:physarum_lab} also shows, \textit{Physarum}'s network's new growth occurs with many tiny veins growing outwardly at the edge of the network, while \textit{Physarum}'s older network adapts itself, reinforcing certain central veins and getting rid of unimportant paths. This results in \textit{Physarum} networks having a few thick central veins connecting the periphery of the network to the center, and the periphery of the network being heavily interconnected by a large number of thin and short veins. As such, the loops mentioned earlier arise between these thick veins and the periphery of the network.
The model being tested is able to accurately portray the adaptation of the older part of \textit{Physarum}'s network, but it was not designed to be able to mimic the growth pattern shown in the periphery of real \textit{Physarum} networks, hence why the periphery highly interconnected patterns are not visible in figure \ref{fig:physarum_sync} and not all sinks are inclosed in a loop pattern.
After all, while in figure \ref{fig:physarum_radial_growth} we see evolving patterns, the simulations of \ref{fig:physarum_async} are steady-state patterns of veins.

\subsection{Shuttle streaming} \label{sec:shuttle_streaming}

As explained in section \ref{sec:physarum_periodic}, \textit{Physarum} exhibits periodic physical behaviors that are thought to be responsible for its movement, namely shuttle streaming, which is the back-and-forth motion of fluid through the veins of the organism. This is an extremely important phenomenon, as it is thought to be responsible for \textit{Physarum}'s network's adaptation mechanisms, in conjunction with other aspects. 

As the model explored in this thesis (described in sec. \ref{sec:adaptive-model}) seeks to model those exact adaptation dynamics, it would show great consistency if it could model shuttle streaming as well. As elucidated in section \ref{sec:physarum_periodic}, shuttle streaming and peristalsis (a wave of contractions of the walls of the veins of the network) are correlated. This model seeks to adapt the conductivities of the channels of the network over time, thus adapting the radii of the veins of the network over time; if this adaptation of the radii of the network was driven by non-static sources, it could be responsible for the contraction and relaxation of the walls of the veins of the network, thus contributing to shuttle streaming.

To test this hypothesis, the \textit{Physarum polycephalum}-like asynchronous configuration and algorithm of section \ref{sec:physarum_async} were used: the source in the middle is always active with $S_{\text{source}} = 1$ and, out of all the 20 possible sinks in the periphery circling the source, at each $\Delta t$, $N_{\text{active sinks}}$ are randomly chosen to be active sinks with $S_j = -1/N_{\text{active sinks}}$ (the inactive sinks have $S_j = 0$).

\begin{figure}
\centering
\begin{subfigure}[b]{0.44\textwidth}
 \centering
 \includegraphics[width=\textwidth]{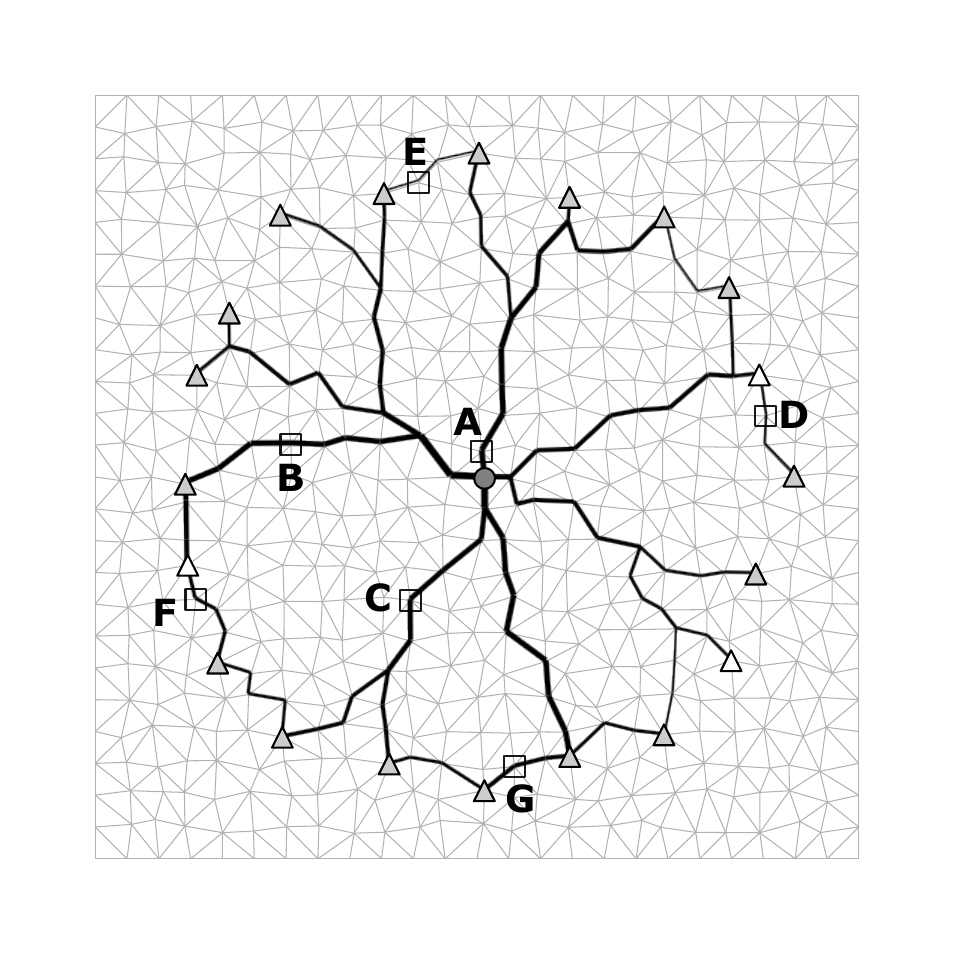}
 \caption{Steady state obtained after 955 iterations.}
 \label{fig:shuttle_streaming_ss_short}
 \end{subfigure}
 \begin{subfigure}[b]{0.44\textwidth}
 \centering
 \includegraphics[width=\textwidth]{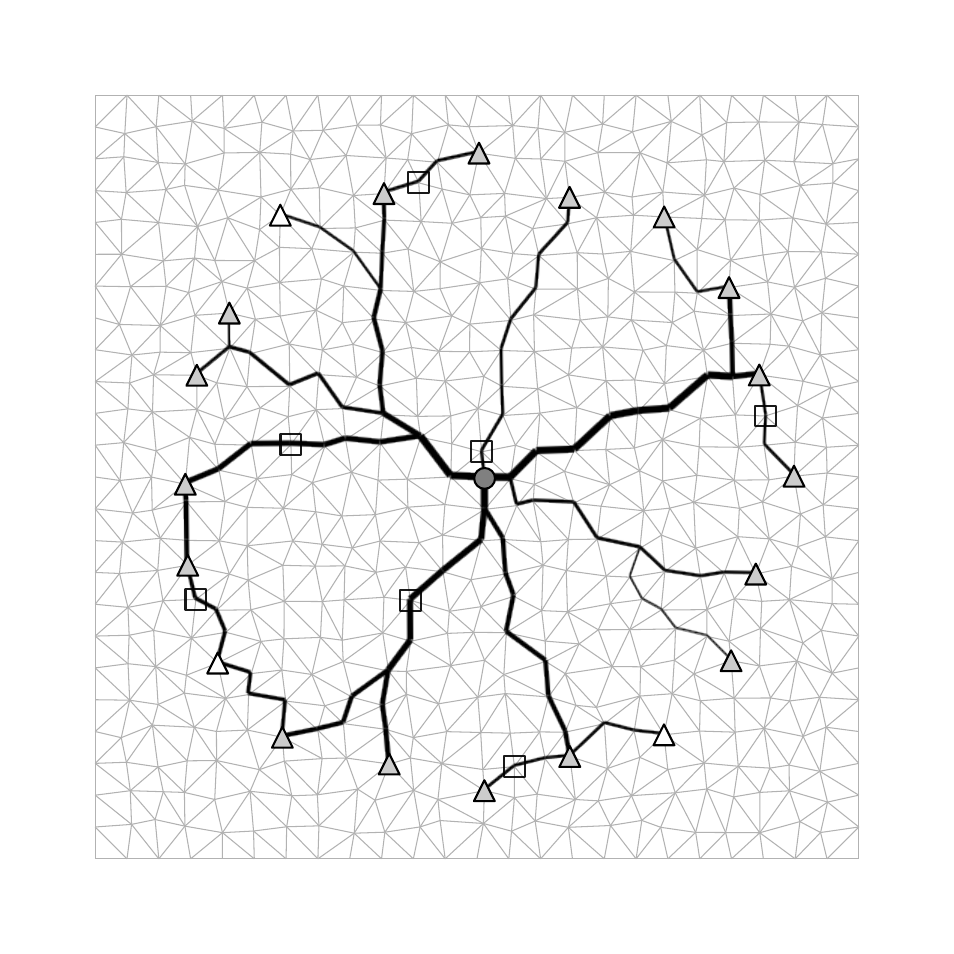}
 \caption{Steady state obtained after 10000 iterations.}
 \label{fig:shuttle_streaming_ss_long}
 \end{subfigure}
\caption{Steady states obtained. Chosen nodes (A-G) are highlighted by a box surrounding them.}
\label{fig:shuttle_streaming_ss}
\end{figure}

Out of curiosity and as a way to make sure the stopping criterion for the simulations was appropriate, we decided to run two types of simulations with two different stopping criteria: first, the stopping criteria used (described in section \ref{sec:stopping_criteria}) was that the length remained unchanged for $N_{\text{iter}}=500$ iterations, for which the simulation ran for 955 iterations; then, we ran the simulation with no concrete stopping criteria for $10000$ iterations. The steady states obtained for both these stopping criteria can be seen in figure \ref{fig:shuttle_streaming_ss}.

Steady states shown in figure \ref{fig:shuttle_streaming_ss} are quite similar, but present some key differences: some previous existing paths disappeared from figure \ref{fig:shuttle_streaming_ss_short} to figure \ref{fig:shuttle_streaming_ss_long}, which made some loops disappear as well (by loops, we mean the ones discussed in section \ref{sec:physarum_async}, formed by paths between the source and sinks and also by paths between the sinks themselves) - see the loop which included nodes A and E, and the loop which included nodes C and G, for example.

\begin{figure}
\centering
\includegraphics[width=0.6\textwidth]{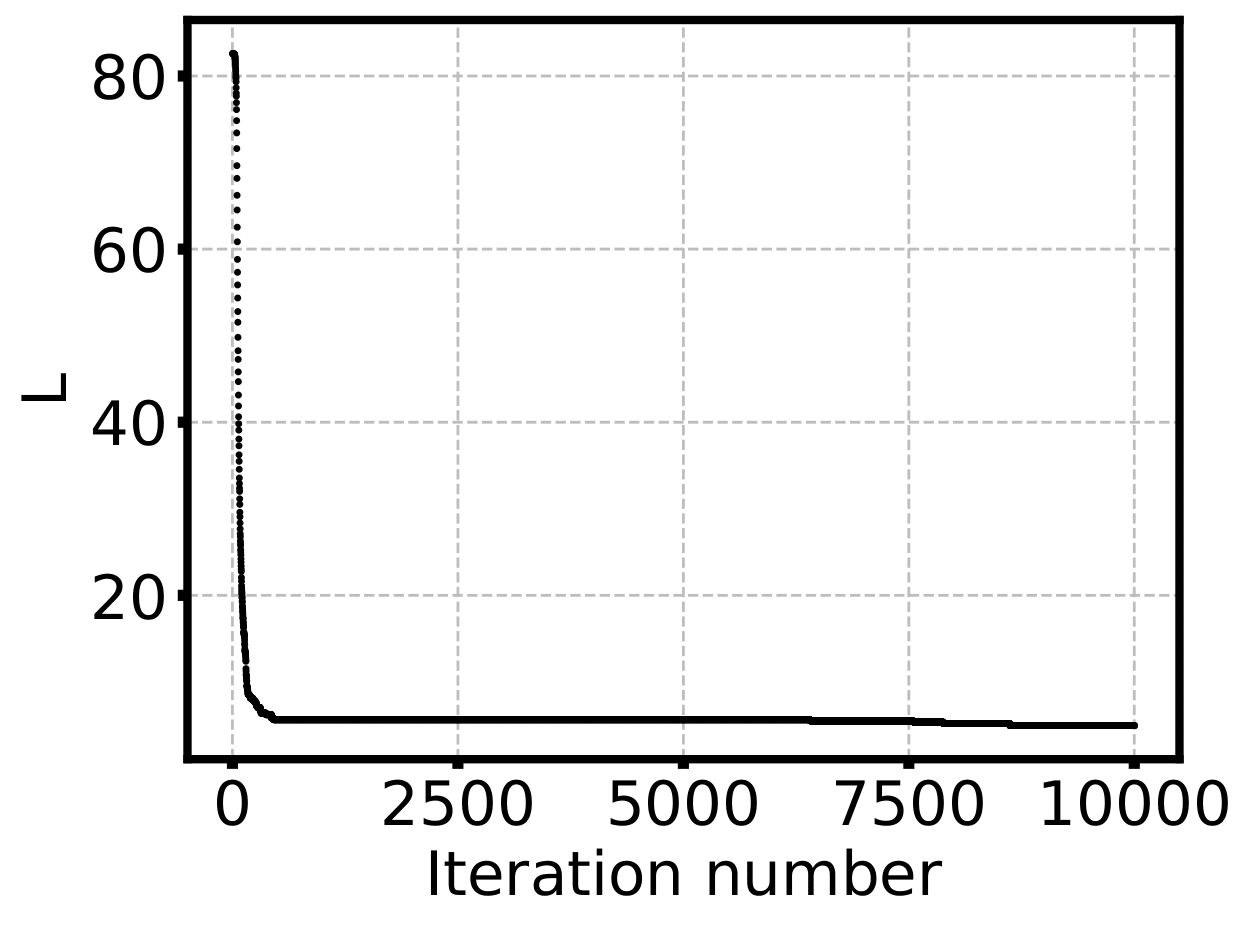}
\caption{Length of the tree obtained for every step of the simulation, for 10000 iterations.}
\label{fig:shuttle_streaming_Lt}
\end{figure}

To determine why the stopping criteria wasn't working as intended, the length of the tree was calculated for every iteration and is shown in figure \ref{fig:shuttle_streaming_Lt}. In this figure, we can see that the length of the simulation remained constant for about 5600 iterations (from iterations $\approx 625 - 6250$), then decreased at iteration number $\approx 6250$ and then proceeded to decreased once again at iteration number $\approx 8700$. This shows how difficult it is to find a steady state condition for non-static sources and sinks for this model. While the state obtained after 955 iterations may not be the true steady state, it shows some important and interesting features which will be discussed further. For the purpose of this study, both steady states will be taken into account. 

After a steady state was obtained, several different random nodes were chosen. These nodes had to be actively-conducting nodes at steady state, could not be a source or sink, and had to be connected to only 2 other actively-conducting nodes (that is, couldn't be bifurcation nodes). These nodes can be seen in figure \ref{fig:shuttle_streaming_ss_short}. 

The nodes and edges selected are in the following locations: node A is part of a path that connects the source to a bifurcation point; node B is part of a path that connects a bifurcation point (coming directly from the source) to a sink; node C is part of a path that connects two bifurcation points, one of which connects directly to a source and another that connects to two sinks; node D is part of a path that connects two sinks to each other, one of which is not connected to any other path; node E is a part of a path that connects two sinks to each other, and that makes up a larger path that creates a loop with the source; node F is a part of a path that connects two sinks as well, and makes up a path that creates a loop with two additional sinks and that includes nodes B and C; node G is once again a part of a path that connects two sinks, and makes up a path that creates a loop with one additional sink and that includes node C.

It's relevant to note that the simulation can be ran under the exact same conditions by using the same initial conductivities. Additionally, the number of sinks and the set of chosen sinks at each iteration is kept the same by using the same random seed at each $\Delta t$ (the seed chosen was a multiple of the iteration number).

\begin{figure}
    \centering
    \begin{subfigure}[b]{0.45\textwidth}
     \centering
     \includegraphics[width=\textwidth]{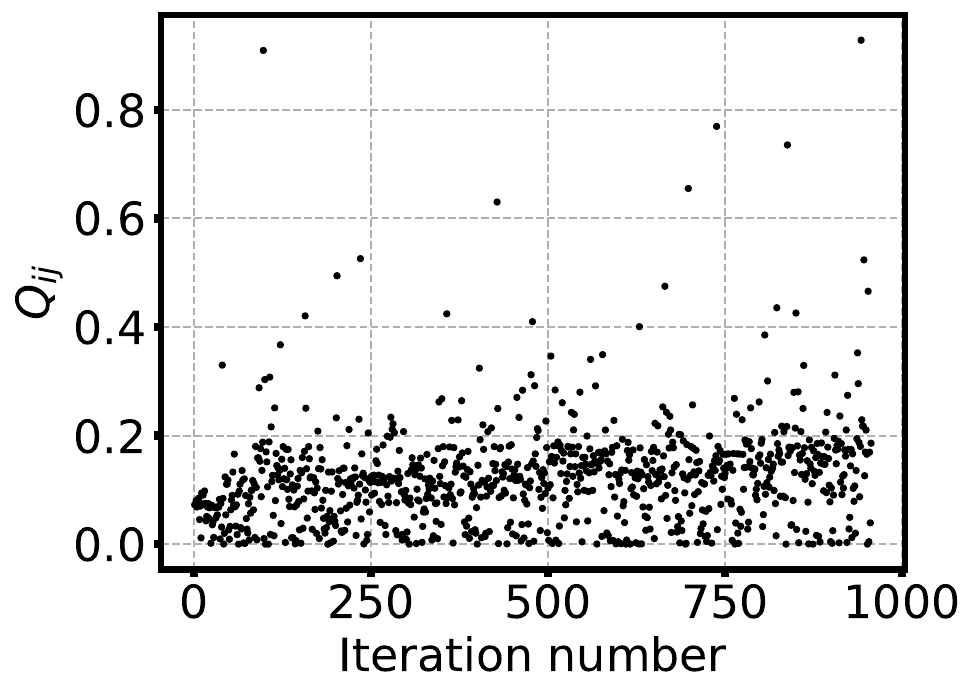}
     \caption{Node A.}
     \label{fig:nodeA}
     \end{subfigure}
     \begin{subfigure}[b]{0.45\textwidth}
     \centering
     \includegraphics[width=\textwidth]{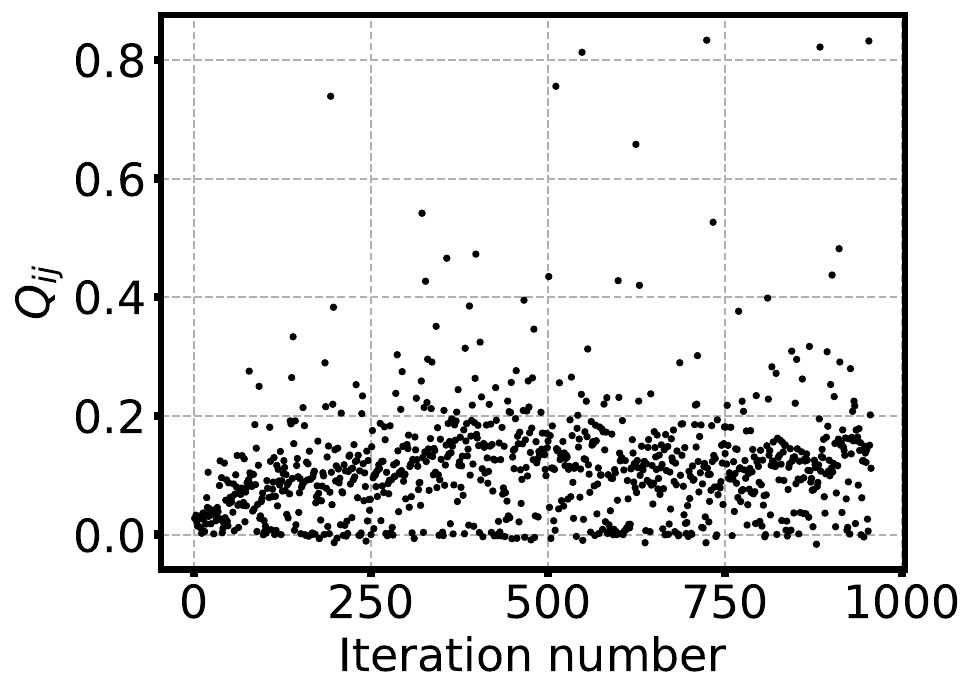}
     \caption{Node B.}
     \label{fig:nodeB}
     \end{subfigure}
     \\
     \begin{subfigure}[b]{0.45\textwidth}
     \centering
     \includegraphics[width=\textwidth]{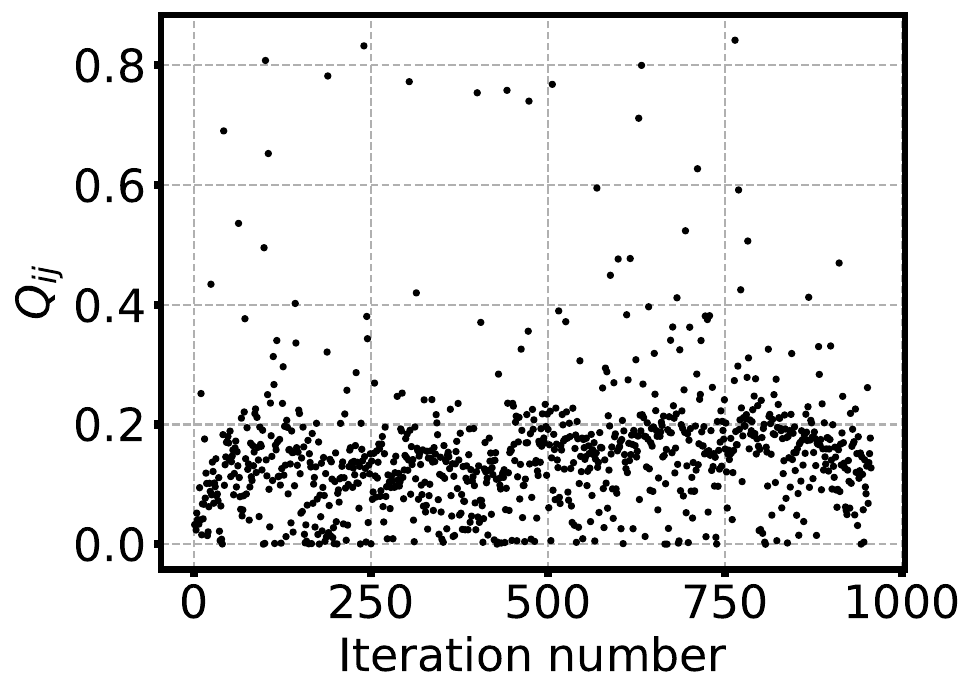}
     \caption{Node C.}
     \label{fig:nodeC}
     \end{subfigure}
     \begin{subfigure}[b]{0.45\textwidth}
     \centering
     \includegraphics[width=\textwidth]{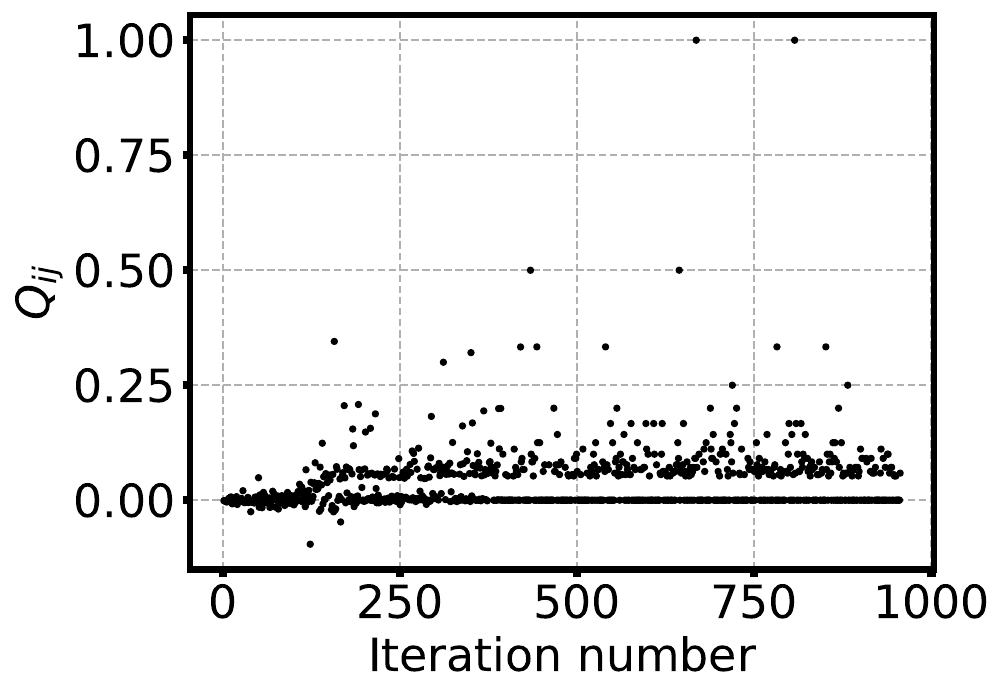}
     \caption{Node D.}
     \label{fig:nodeD}
     \end{subfigure}
     \begin{subfigure}[b]{0.45\textwidth}
     \centering
     \includegraphics[width=\textwidth]{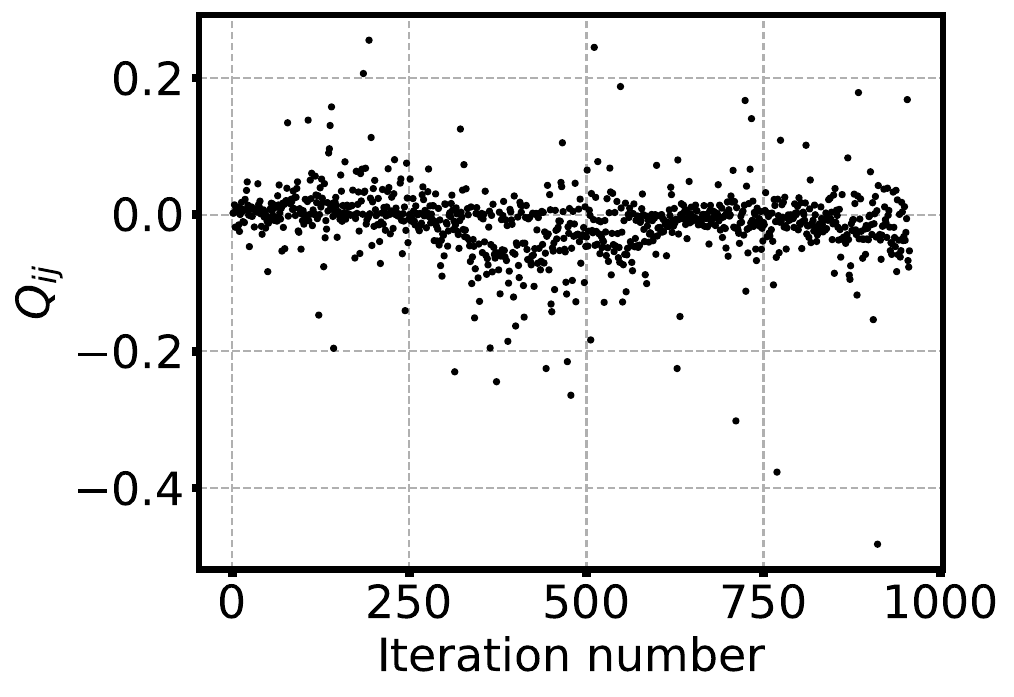}
     \caption{Node E.}
     \label{fig:nodeE}
     \end{subfigure}
     \begin{subfigure}[b]{0.45\textwidth}
     \centering
     \includegraphics[width=\textwidth]{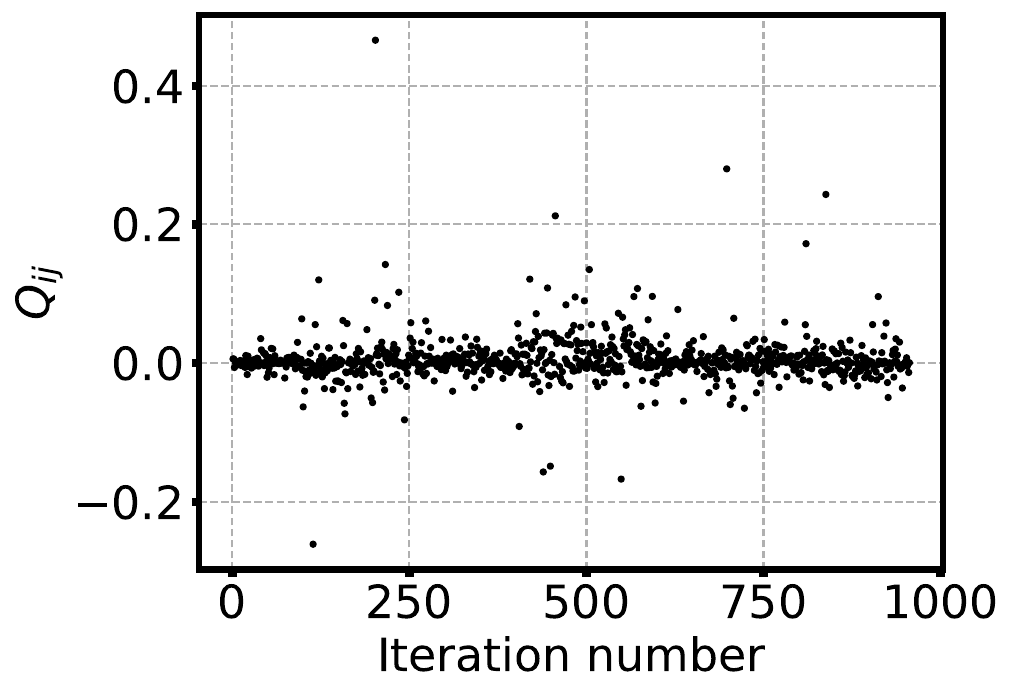}
     \caption{Node F.}
     \label{fig:nodeF}
     \end{subfigure}
     \begin{subfigure}[b]{0.45\textwidth}
     \centering
     \includegraphics[width=\textwidth]{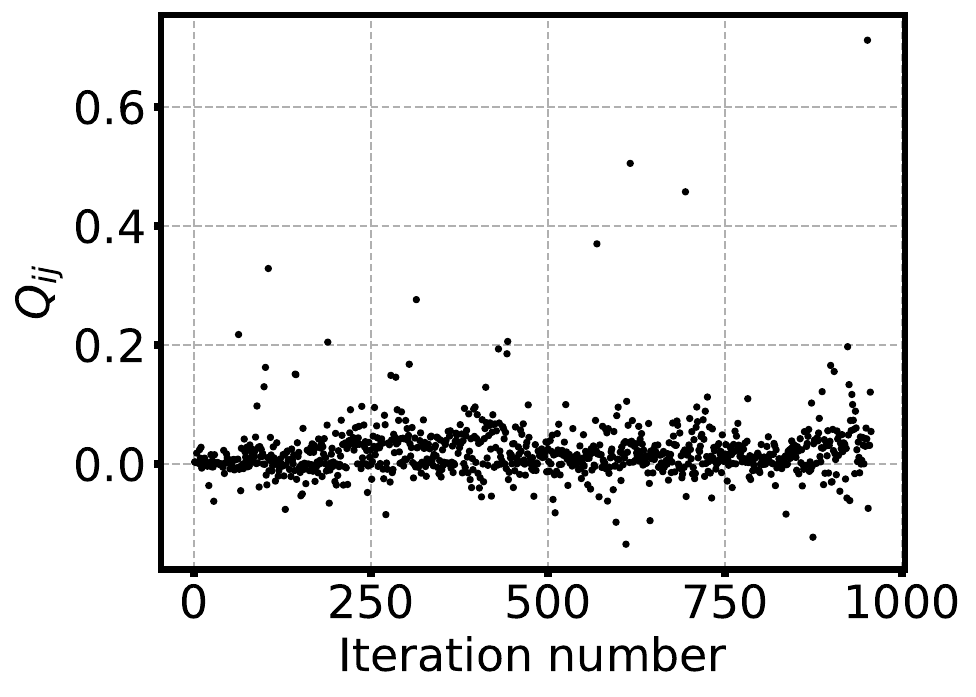}
     \caption{Node G.}
     \label{fig:nodeG}
     \end{subfigure}
    \caption{$Q_{ij}$ for all edges for 955 iterations.}
    \label{fig:shuttle_streaming_Q}
\end{figure}

\begin{figure}
    \centering
    \begin{subfigure}[b]{0.45\textwidth}
     \centering
     \includegraphics[width=\textwidth]{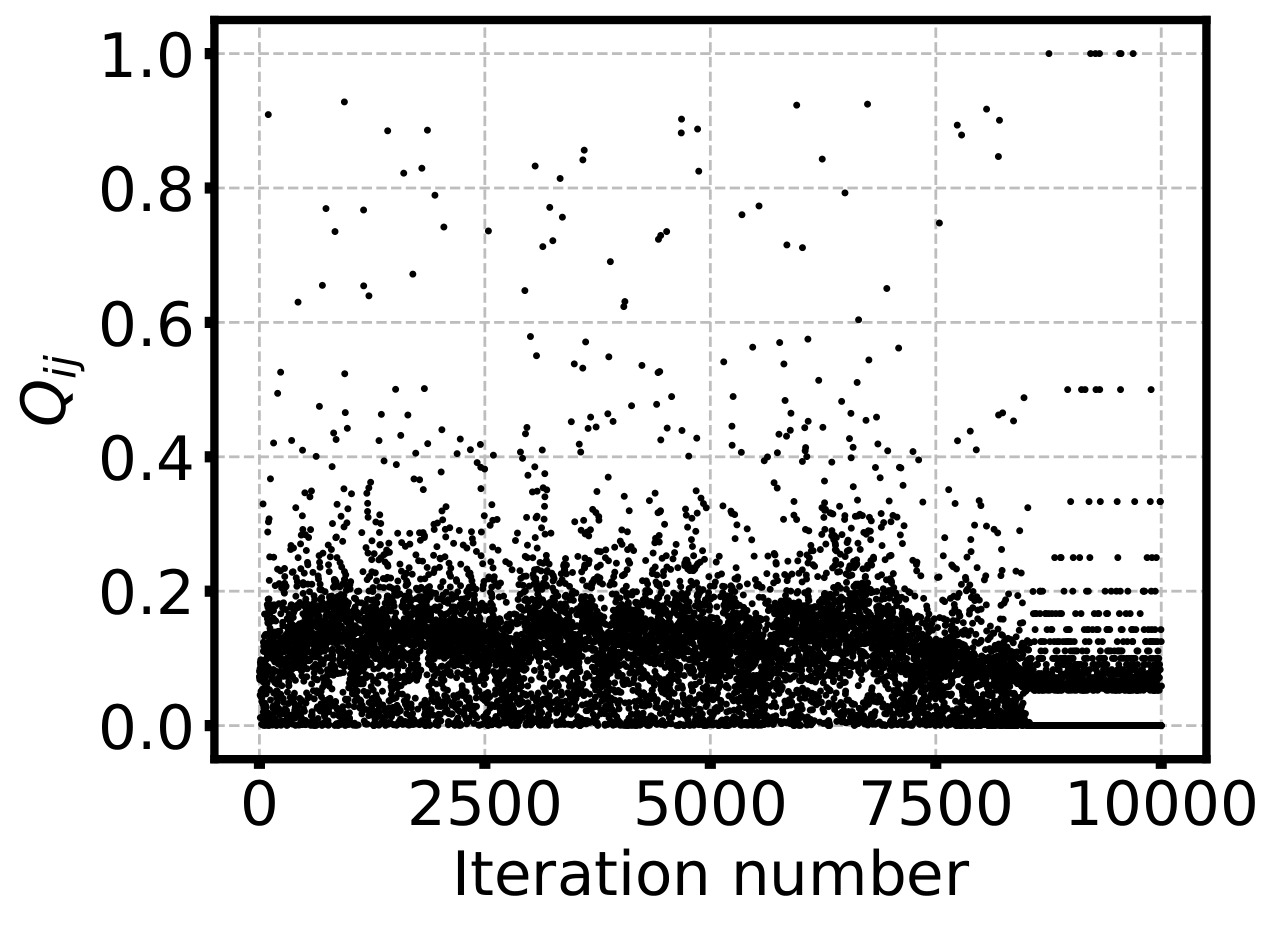}
     \caption{Node A.}
     \label{fig:nodeA_long}
     \end{subfigure}
     \begin{subfigure}[b]{0.45\textwidth}
     \centering
     \includegraphics[width=\textwidth]{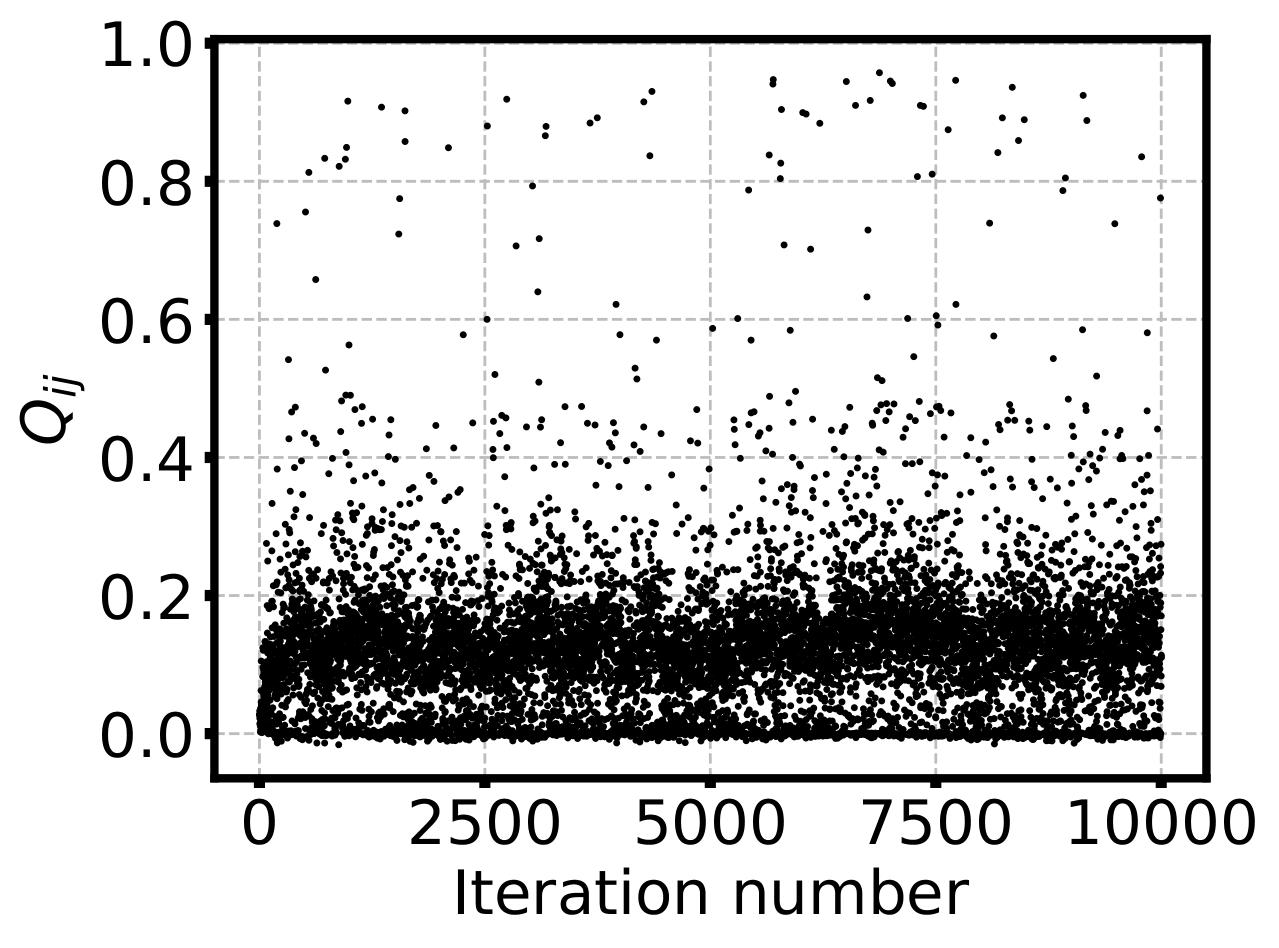}
     \caption{Node B.}
     \label{fig:nodeB_long}
     \end{subfigure}
     \\
     \begin{subfigure}[b]{0.45\textwidth}
     \centering
     \includegraphics[width=\textwidth]{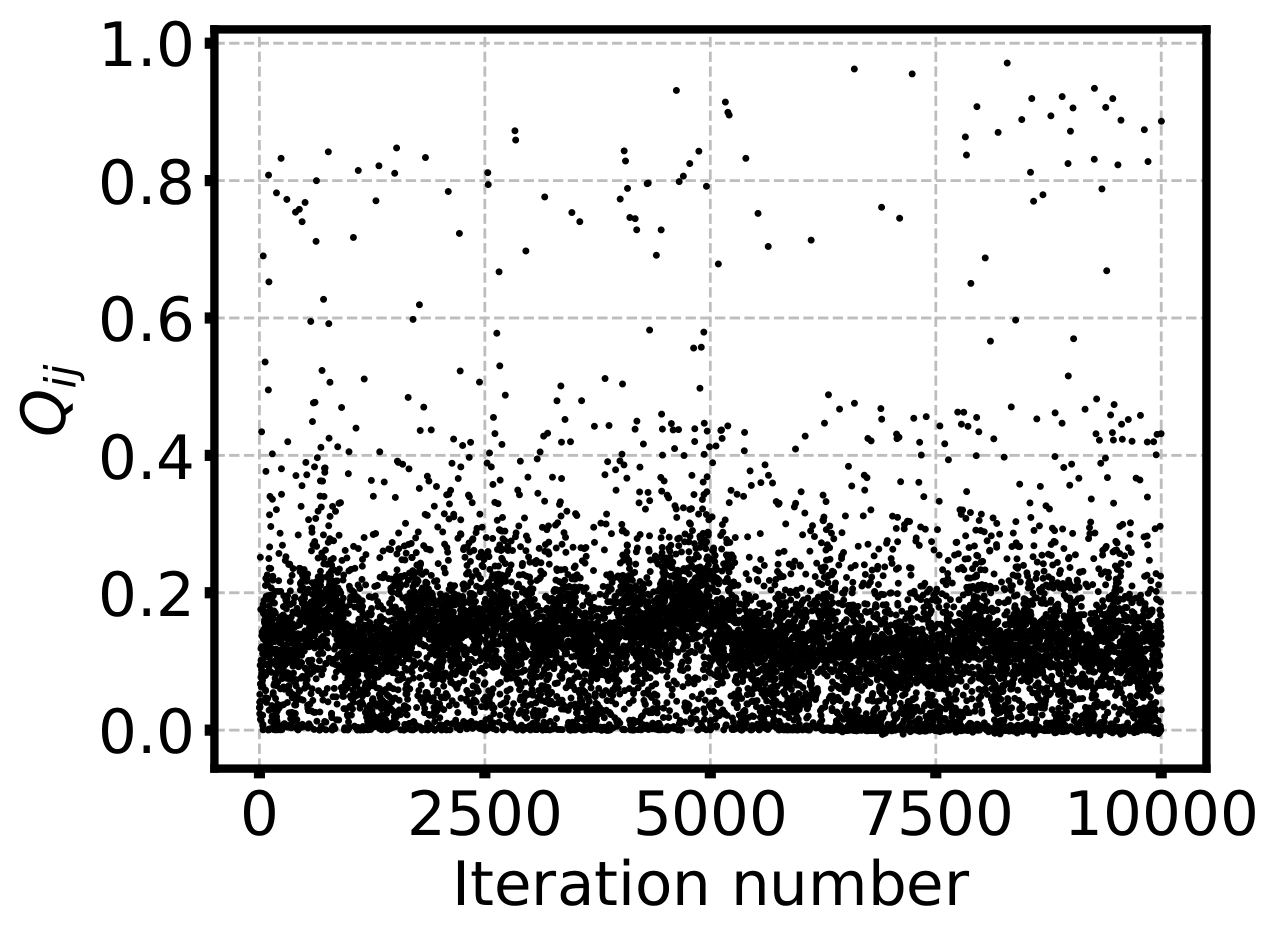}
     \caption{Node C.}
     \label{fig:nodeC_long}
     \end{subfigure}
     \begin{subfigure}[b]{0.45\textwidth}
     \centering
     \includegraphics[width=\textwidth]{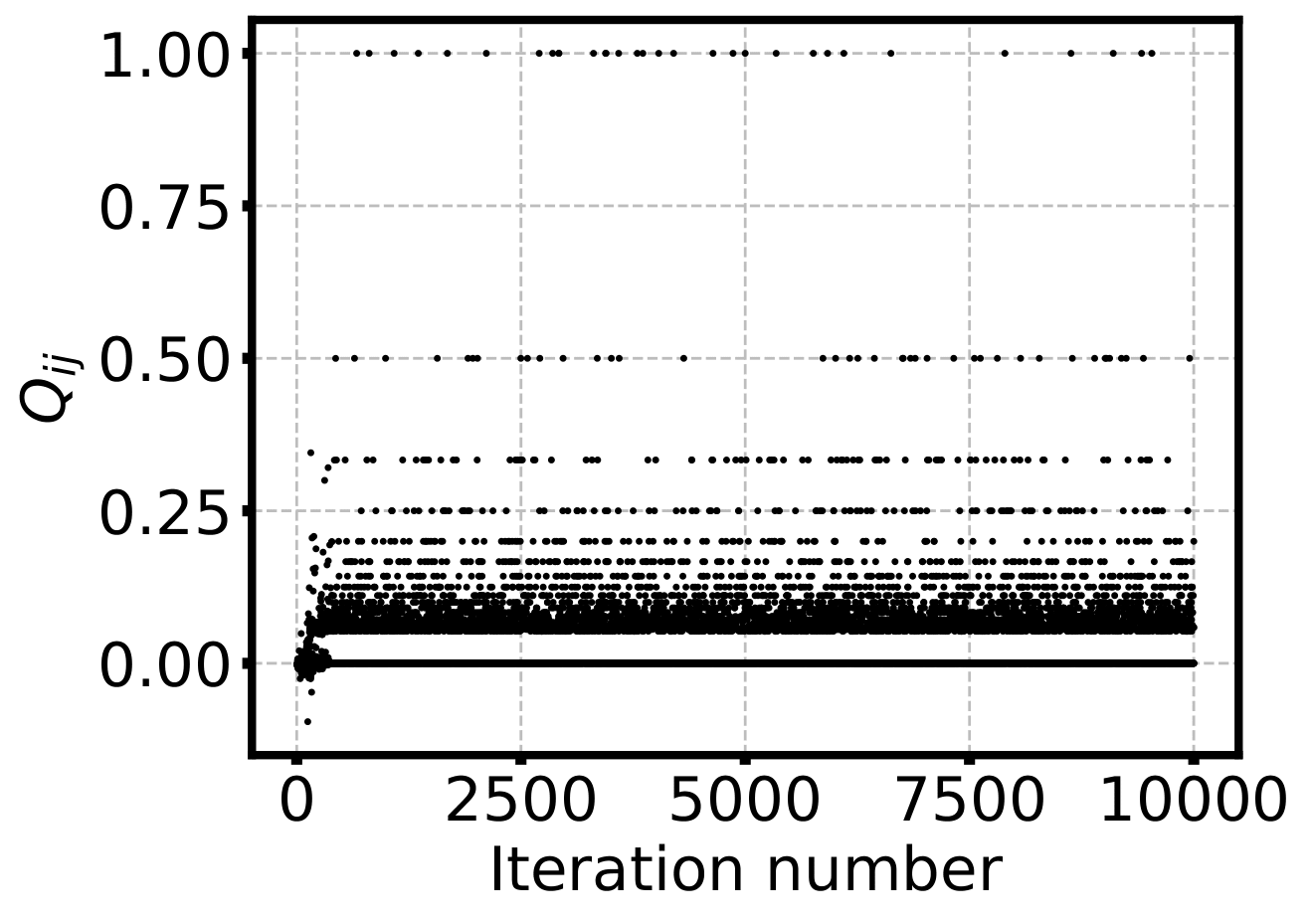}
     \caption{Node D.}
     \label{fig:nodeD_long}
     \end{subfigure}
     \begin{subfigure}[b]{0.45\textwidth}
     \centering
     \includegraphics[width=\textwidth]{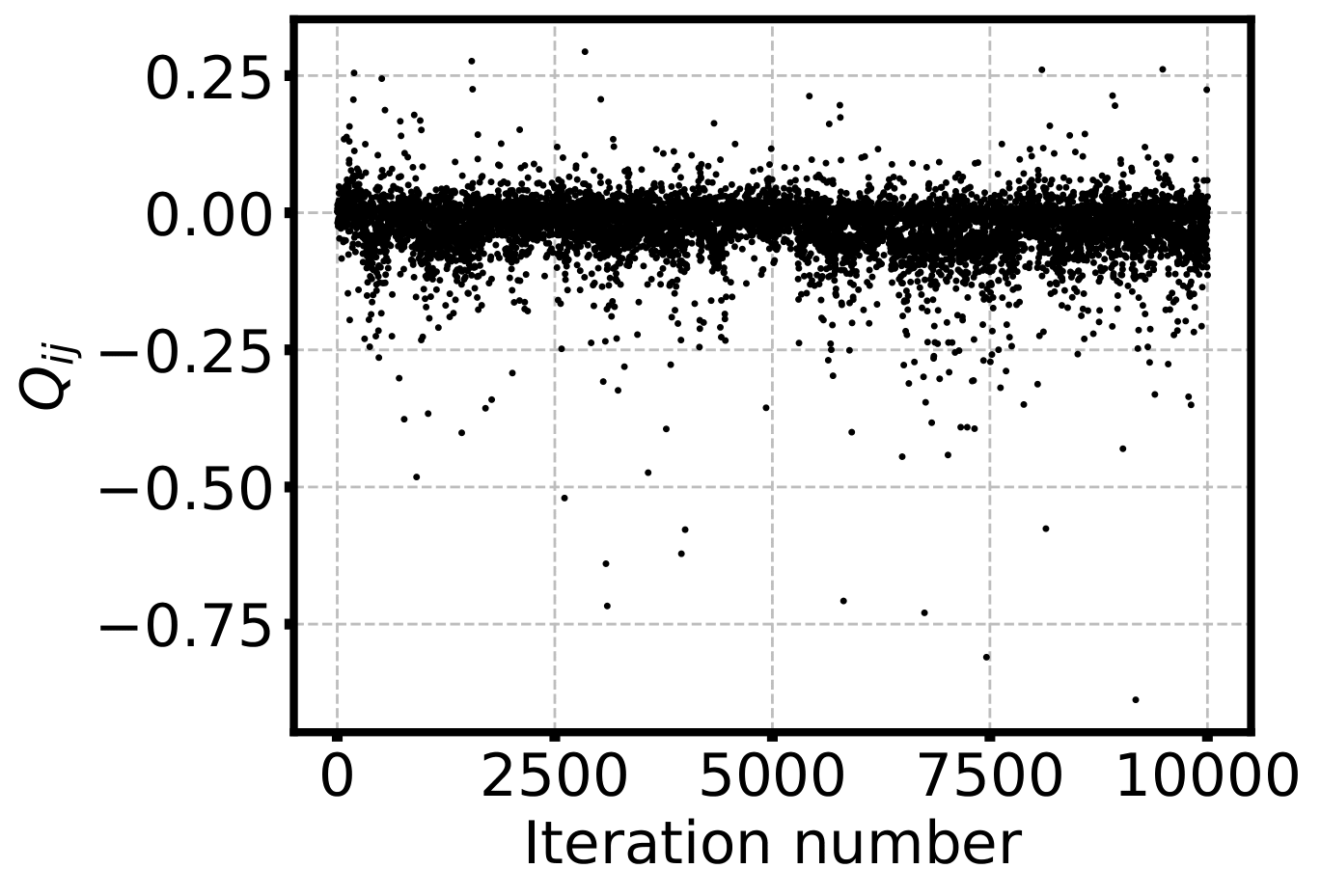}
     \caption{Node E.}
     \label{fig:nodeE_long}
     \end{subfigure}
     \begin{subfigure}[b]{0.45\textwidth}
     \centering
     \includegraphics[width=\textwidth]{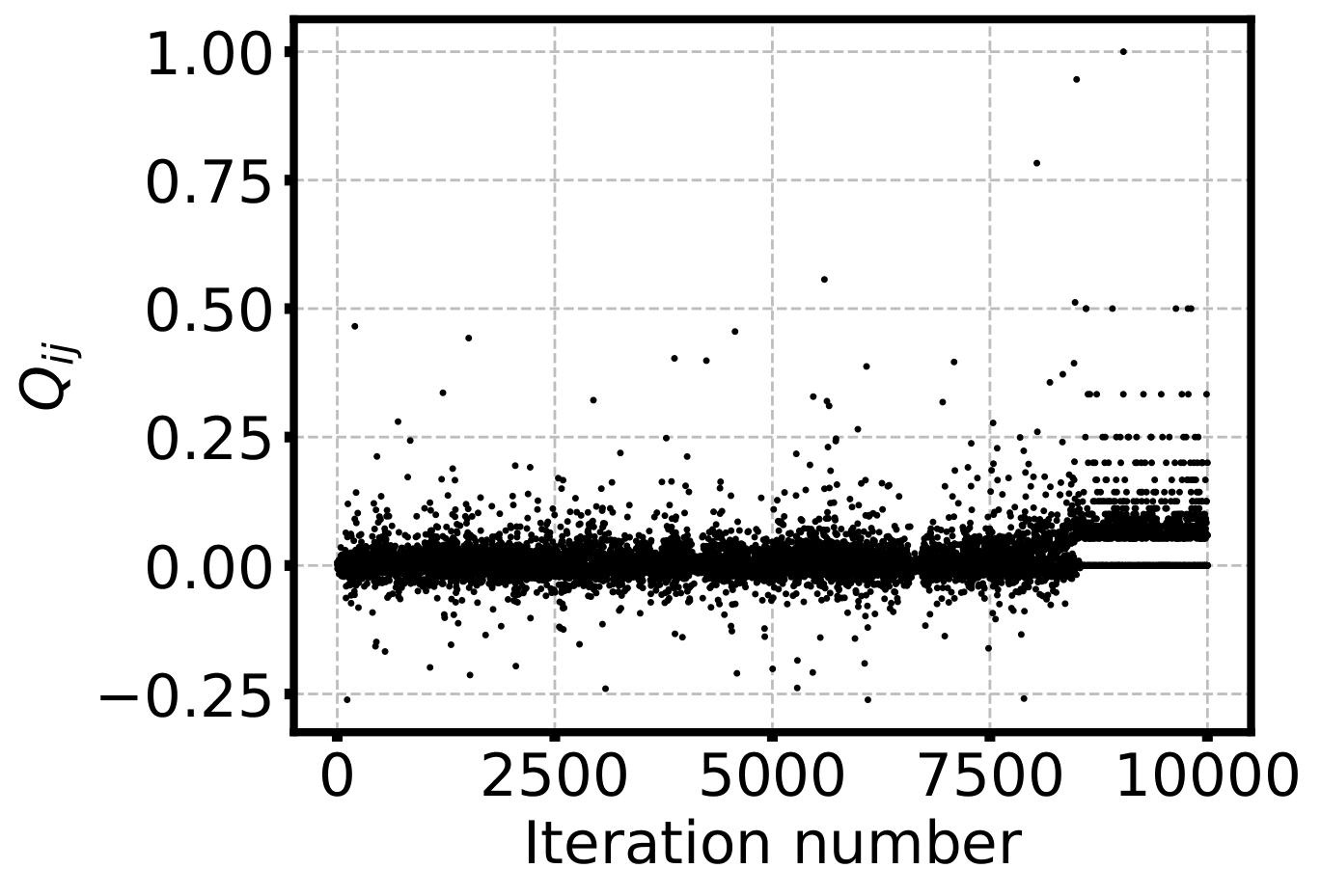}
     \caption{Node F.}
     \label{fig:nodeF_long}
     \end{subfigure}
     \begin{subfigure}[b]{0.45\textwidth}
     \centering
     \includegraphics[width=\textwidth]{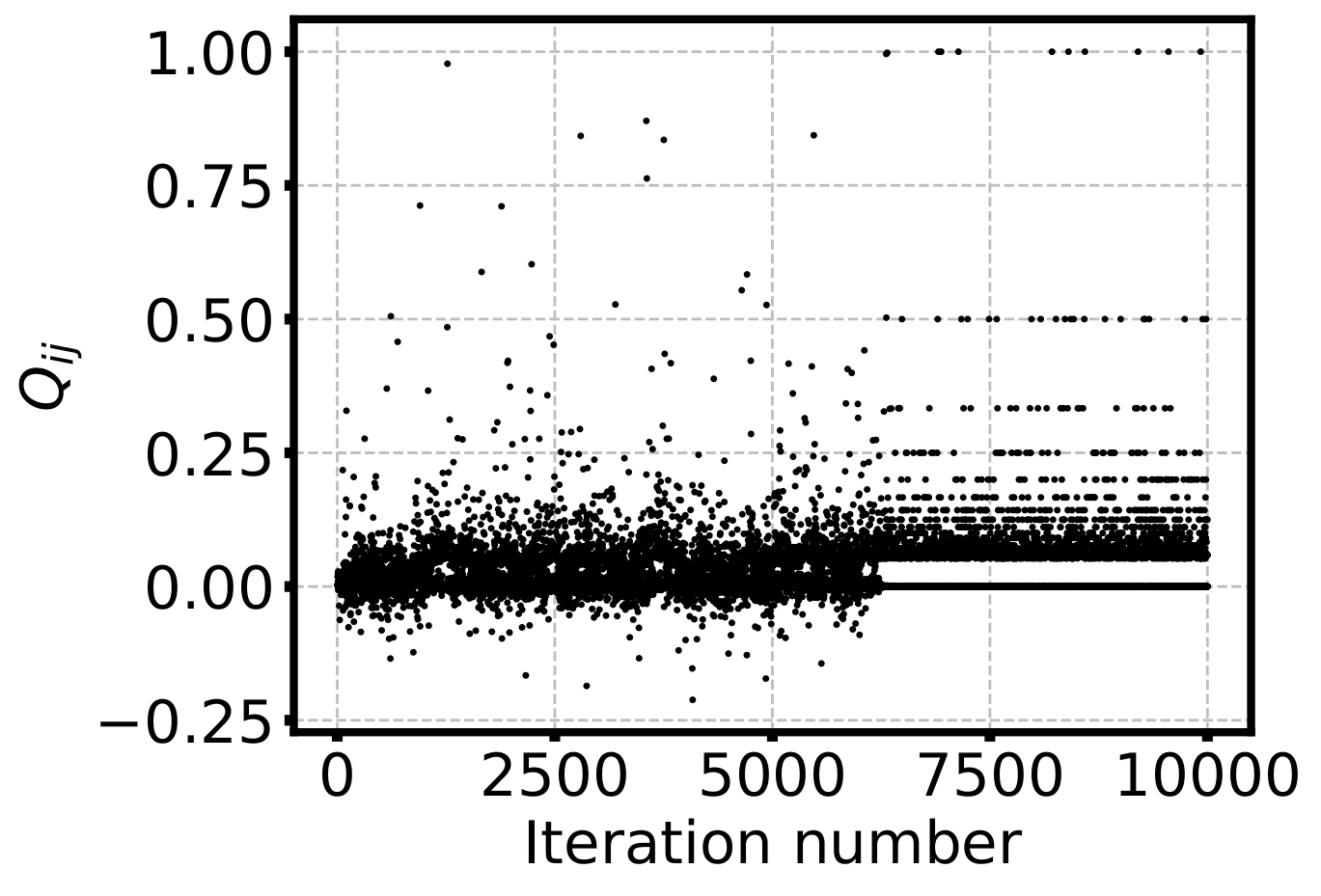}
     \caption{Node G.}
     \label{fig:nodeG_long}
     \end{subfigure}
    \caption{$Q_{ij}$ for all edges for 10000 iterations.}
    \label{fig:shuttle_streaming_Q_long}
\end{figure}

After selecting relevant nodes, one of the two edges to which that node was connected to was selected to have its $Q_{ij}$ value evaluated over time. The value of $Q_{ij}$ (that is, the value of $|Q_{ij}|$ with the sign given by $p_i - p_j$) was then saved for all those edges for all iterations. $Q_{ij}$ as a function of time for the edges concerned is shown in figure \ref{fig:shuttle_streaming_Q} for the simulation that ran for 955 iterations, and is shown in figure \ref{fig:shuttle_streaming_Q_long} for the simulation that ran for 10000 iterations. Note that positive $Q_{ij}$ direction is defined as radially outward (from source to sinks) for edges A-C. For edges D-G, positive $Q_{ij}$ direction is defined as clockwise. 

Relevant data to determine whether or not shuttle streaming was observed is shown in table \ref{tab:shuttle_streaming} for both these simulations. The data shown is: \textbf{Freq. inv.}, which stands for frequency of inversion, and is the percentage of iterations in which the sign of $Q_{ij}$ is different from the more frequent sign; $\mathbf{\text{\textbf{min}}(Q_{ij})}$, which is the smallest $Q_{ij}$ value recorded; $\mathbf{\text{\textbf{max}}(Q_{ij})}$, which is the biggest $Q_{ij}$ value recorded; and \textbf{Rel. str.}, which stands for relative strength, and, for edges in which the dominant sign for $Q_{ij}$ is positive, corresponds to $|\text{min}(Q_{ij}) / \text{max}(Q_{ij})|$, and, for edges in which the dominant sign for $Q_{ij}$ is negative, corresponds to $|\text{max}(Q_{ij}) / \text{min}(Q_{ij})|$.

\begin{table}
\centering
\begin{tabular}{cc|cc|cc|cc|cc}
\toprule
& & \multicolumn{2}{c}{\textbf{Freq. inv. (\%)}} & \multicolumn{2}{c}{$\text{\textbf{min}} \mathbf{\left(Q_{ij}\right)}$} & \multicolumn{2}{c}{$\text{\textbf{max}} \mathbf{\left(Q_{ij}\right)}$} & \multicolumn{2}{c}{\textbf{Rel. str.}} \\
\midrule
& Nb. Iterations & 955 & 10000 & 955 & 10000 & 955 & 10000 & 955 & 10000 \\
\midrule
\multirow{7}{*}{ Node } & A & 0.00 & 0.00 & 0.00 & 0.00 & 0.93 & 1.00 & 0.00 & 0.00 \\
& B & 5.24 & 7.12 & -0.02 & -0.02 & 0.83 & 0.96 & 0.02 & 0.02 \\
& C & 0.00 & 1.00 & 0.00 & -0.01 & 0.84 & 0.97 & 0.00 & 0.01 \\
& D & 13.61 & 1.30 & -0.10 & -0.10 & 1.00 & 1.00 & 0.10 & 0.10 \\
& E & 38.95 & 30.45 & -0.48 & -0.89 & 0.25 & 0.29 & 0.53 & 0.33 \\
& F & 45.97 & 35.50 & -0.26 & -0.26 & 0.47 & 1.00 & 0.56 & 0.26 \\
& G & 26.70 & 14.46 & -0.13 & -0.21 & 0.71 & 1.00 & 0.19 & 0.21\\
\bottomrule
\end{tabular}
\vspace{-0.7 em}
\caption{$Q_{ij}$ data for nodes A-G, for simulations with 955 and 10000 iterations. From left to right, the data is: frequency of inversion, which is the percentage of iterations in which the sign of $Q_{ij}$ is different from the more frequent sign; $\text{min}(Q_{ij})$, which is the smallest $Q_{ij}$ value recorded; $\text{max}(Q_{ij})$, which is the biggest $Q_{ij}$ value recorded; and relative strength, which is, for edges in which the dominant sign for $Q_{ij}$ is positive, corresponds to $|\text{min}(Q_{ij}) / \text{max}(Q_{ij})|$, and, for edges in which the dominant sign for $Q_{ij}$ is negative, corresponds to $|\text{max}(Q_{ij}) / \text{min}(Q_{ij})|$. The dominant sign for $Q_{ij}$ for all edges is positive, except for edge E, which is negative.}
\label{tab:shuttle_streaming}
\end{table}

First, on one hand, notice how edges A-D all predominantly show positive flux values. This indicates that the motion of fluid is directed radially outward, that is, from the source to the sinks, as expected.

Looking at figure \ref{fig:shuttle_streaming_Q_long} specifically, notice how a few edges show a very different $Q_{ij}$ behavior after a certain iteration: for nodes A and F, this happens at about iteration number 8700 (figs. \ref{fig:nodeA_long} and \ref{fig:nodeF_long}); for node G, this occurs at iteration number $\approx 6250$ (fig. \ref{fig:nodeG_long}). These are the iteration numbers where the length of the tree was seen to change in figure \ref{fig:shuttle_streaming_Lt}. This different behavior mentioned seems to be reaching a steady state, as the flux value for these edges seems to oscillate between the same values. However, nodes B, C or E show no such periodic $Q_{ij}$ values and no steady state. As such, it seems that the steady state is reached at different iteration numbers for different edges of the network.

Let us now analyze figures \ref{fig:shuttle_streaming_Q} and \ref{fig:shuttle_streaming_Q_long}, along with table \ref{tab:shuttle_streaming}, and determine whether shuttle streaming is observed. Shuttle streaming is the back-and-forth motion of fluid. Thus, we are looking for inversions in the direction of movement of the fluid, which is translated in a change of sign for $Q_{ij}$ values.

We start by looking at figure \ref{fig:shuttle_streaming_Q} to study the first 955 iterations of the simulation. 
Edges A and C show no negative $Q_{ij}$ values and edge B shows very small, very infrequent negative $Q_{ij}$ values (present in less than 6\% of the iterations, with absolute values less than $3\%$ of the largest $Q_{ij}$ value recorded). Thus, edges A-C are considered to show no significant shuttle streaming.
Edges E-G all show frequent and significant inversions of the direction of motion, as they show frequency of inversion values larger than 25\% and show relative strength values of at least 19\%. Edge D also shows some somewhat significant negative $Q_{ij}$ values (14\% of the time with 10\% of the largest flux value observed). As such, edges E-G all show significant shuttle streaming, and edge D shows some signs of shuttle streaming.

Let us now look at figure \ref{fig:shuttle_streaming_Q_long} to study all 10000 iterations of the simulation and see if anything changed.
Edge A continued to show no signs of flow direction inversion. Edges B-D all showed signs of minor flow direction inversion: the frequencies of inversion obtained were less than 8\% for edge B and less than 2\% for edges C and D, while the relative strength values were less than 3\% for edges B and C and 10\% for edge D once again. The edge with largest frequency (B) showed very small relative strength values, and the edge with larger relative strength values (D) showed very small frequency. Thus, it was considered that these edges continued not to show shuttle streaming.
As for edges E-G, these edges showed larger frequency of inversion values (larger than 30\% for edges E and F, and larger than 14\% for edge G) and also larger relative strength values (larger than 20\% for all edges). Thus, all these edges are considered to have significant shuttle streaming. However, notice how both the frequency of inversion and relative strength values decreased from the simulation with 955 iterations to the one with 10000 iterations. It seems that, as the simulations get closer to steady state, shuttle streaming becomes less frequent.

Thus, edges A-C showed no considerable shuttle streaming, while edges E-G showed significant shuttle streaming (and edge D was not clear). 

It's relevant to notice how edges D-G are peripheral edges, and are all part of paths that connect two sinks, as opposed to edges A-C, which are central and directly connect the source to the sinks. It seems it is more likely to observe shuttle streaming in peripheral edges than in central edges with the model and algorithm. Another relevant fact is how edges E-G are all part of loops for the first 955 iterations. Edge F stops showing shuttle streaming after the length of the network decreases, as does edge G (see figs. \ref{fig:nodeF_long} and \ref{fig:nodeG_long}), albeit at different times, while edge E shows shuttle streaming even after the network's length decreases. Edge G's loop is shown to become disconnected after 10000 iterations, as does edge E's loop; however, edge F's loop does not become disconnected (see fig. \ref{fig:shuttle_streaming_ss_long}). Also, notice that edges B, C and F are all part of the same loop that remains connected throughout the entire 10000-step simulation. However, only edge F is shown to exhibit significant shuttle streaming. One can conclude that the presence of loops may contribute to the shuttle streaming phenomenon, but cannot be the only contributing factor.

\subsection{Peristalsis} \label{sec:peristalsis}

As mentioned in section \ref{sec:shuttle_streaming}, the model used in this thesis adapts the radii of the edges of the network over time, contracting and relaxing the walls of the veins of the network. This seems to be a process quite similar to peristalsis (the periodic wave-like contraction and relaxation of the vein walls). Additionally, we observed shuttle streaming in some edges of the network generated by the model (see section \ref{sec:shuttle_streaming}). Shuttle streaming and peristalsis are highly connected processes. Thus, it is worth understanding whether the simulations generated by the model also exhibit peristalsis.

To study this, we use the steady state and edges that were previously used to study shuttle streaming, shown in figure \ref{fig:shuttle_streaming_ss_short}. Similarly to section \ref{sec:shuttle_streaming}, we evaluate the radius $r_{ij}$ of the previously-selected edges (A-G) over time. We use the same stop condition as in the previous section: if the length of the graph is unchanged for 500 iterations, it is considered that the simulation has reached steady state.

The radius of an edge can be obtained using its conductivity. If one substitutes $\beta = \sqrt{8\pi\eta}$ in equation \eqref{eq:Dij}, one obtains the radius of the edge $(i,j)$:

\begin{equation} \label{eq:radius}
    r_{ij} = \sqrt{\frac{\beta}{\pi}} \left(D_{ij}\right)^{1/4}
\end{equation}

\noindent
where we're using $\beta = 1$ as usual. 

The results of $r_{ij}$ over time (for 955 iterations) for edges A-G are shown in figure \ref{fig:peristalsis}. We observe that, after an initial period of $\approx 100$ iterations, all edges show periodic variations of their radius.

The average values of $r_{ij}$ and their standard deviation is shown in table \ref{tab:peristalsis}.
The average value of $r_{ij}$ for edges A-C is 40\% larger than for edges D-G. Edges A-C are central edges to the network, and edges D-G are peripheral edges of the network, thus presenting lower flux values and thinner veins. As mentioned in section \ref{sec:shuttle_streaming}, \textit{Physarum} also exhibits this pattern of having thicker central veins and thinner peripheral veins.

Regarding the oscillations, their amplitude seems to be very similar for all edges (with differences of up to 14\%); according to the standard deviation values of table \ref{tab:peristalsis}, the average amplitude value for all edges is $0.25$. Determining the period of the oscillations is not trivial, as they appear to be composed of a sum of oscillations with different periods. This is probably heavily dependant on the sequence of active sinks.

We have thus shown that the simulations produced by the model exhibit peristalsis on all edges of the network due to the elastic adaptation process of the channels. As such, the model has been shown to exhibit two crucial \textit{Physarum polycephalum} periodic phenomena: shuttle streaming and peristalsis. These processes are theorized to be driven by signaling molecules \cite{alim_signal_prop}. The model studied does not take into account chemical or biological mechanisms, it only considers hydrodynamic processes driven by inward/outward flux of fluid into/out of the network. The (asynchronous) outward flux of sections \ref{sec:shuttle_streaming} and \ref{sec:peristalsis} symbolizes the asynchronous resource consumption that occurs in \textit{Physarum}.

\begin{table}
\centering
\begin{tabular}{c|c}
\toprule
\textbf{Node} & $\mathbf{r_{ij}}$\\
\midrule
A & 2.77 $\pm$ 0.24 \\
B & 2.69 $\pm$ 0.26 \\
C & 2.95 $\pm$ 0.27 \\
D & 1.79 $\pm$ 0.26 \\
E & 1.71 $\pm$ 0.27 \\
F & 1.36 $\pm$ 0.23 \\
G & 1.67 $\pm$ 0.25 \\
\bottomrule
\end{tabular}
\caption{Average values of $r_{ij}$ and its standard deviation for edges A-G (of fig. \ref{fig:shuttle_streaming_ss_short}), for 955 iterations, without taking into account the first 125 iterations (using data of fig. \ref{fig:peristalsis}).}
\label{tab:peristalsis}
\end{table}

\begin{figure}
\centering
\includegraphics[width=0.5\textwidth]{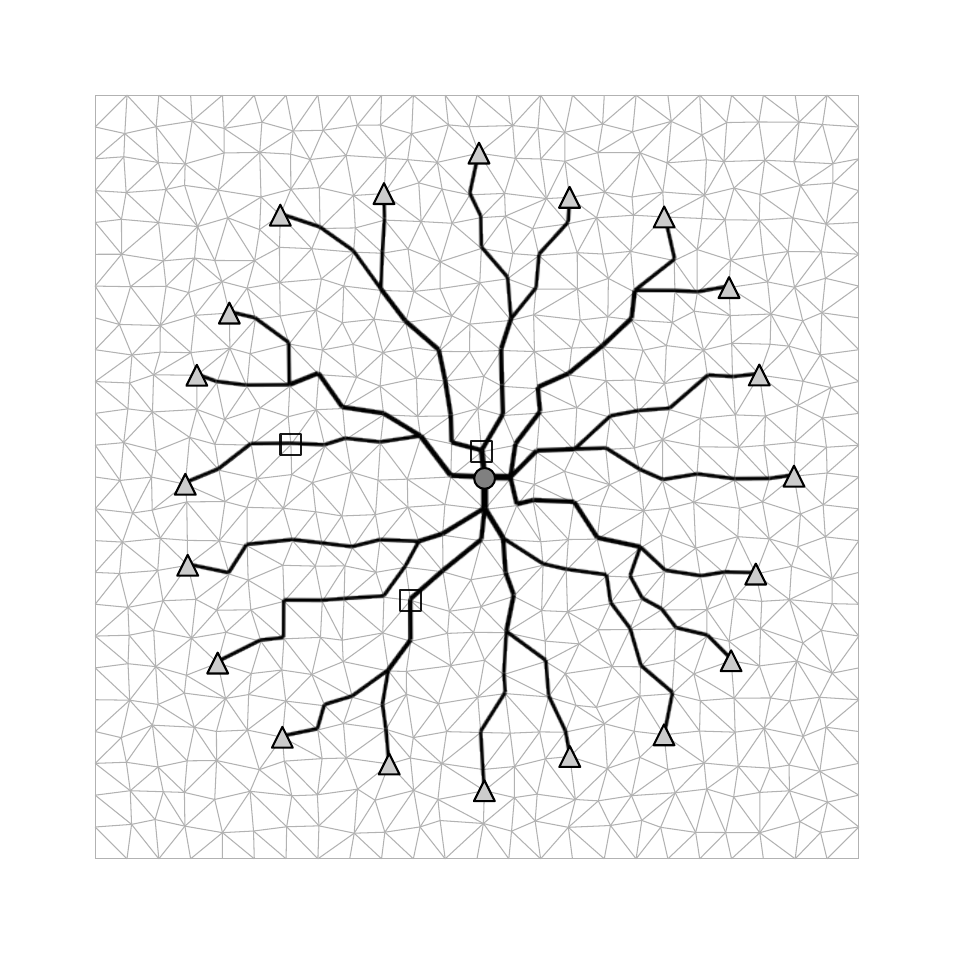}
\caption{Steady state obtained after 295 iterations for synchronous sites. Nodes A-C are highlighted by a box surrounding them (see figure \ref{fig:shuttle_streaming_ss_short}).}
\label{fig:peristalsis_ss_sync}
\end{figure}

\begin{figure}
\centering
\begin{subfigure}[b]{0.45\textwidth}
\centering
\includegraphics[width=\textwidth]{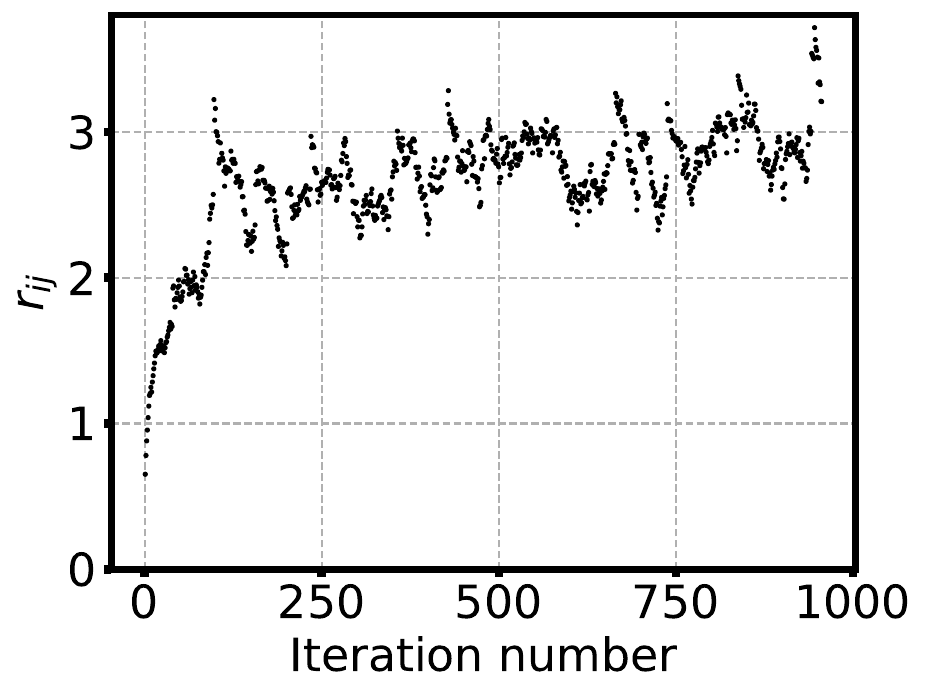}
\caption{Node A.}
\label{fig:nodeA_peri}
\end{subfigure}
\begin{subfigure}[b]{0.45\textwidth}
\centering
\includegraphics[width=\textwidth]{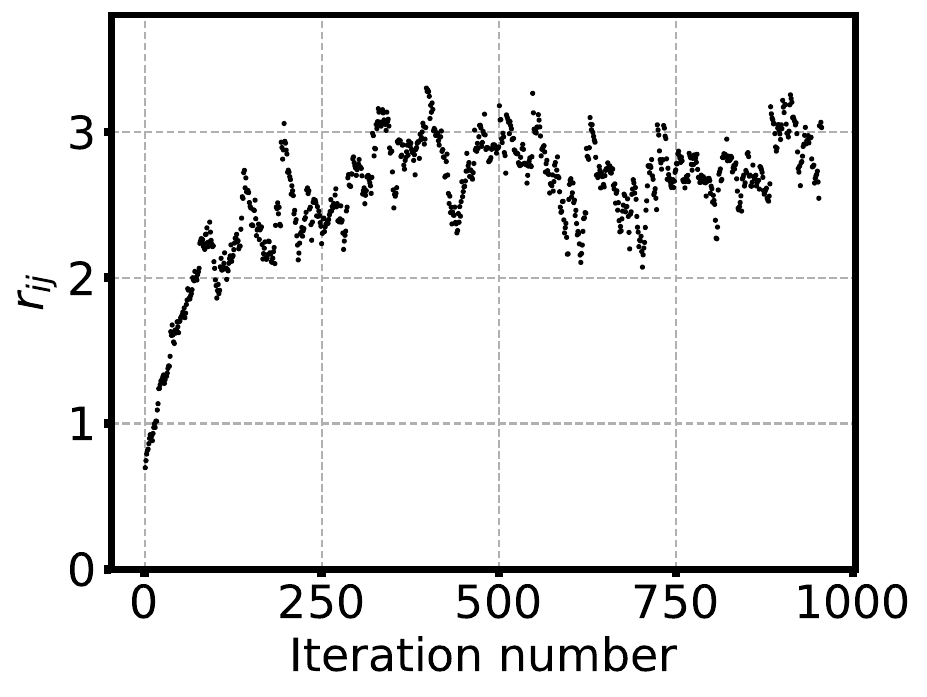}
\caption{Node B.}
\label{fig:nodeB_peri}
\end{subfigure}
\\
\begin{subfigure}[b]{0.45\textwidth}
\centering
\includegraphics[width=\textwidth]{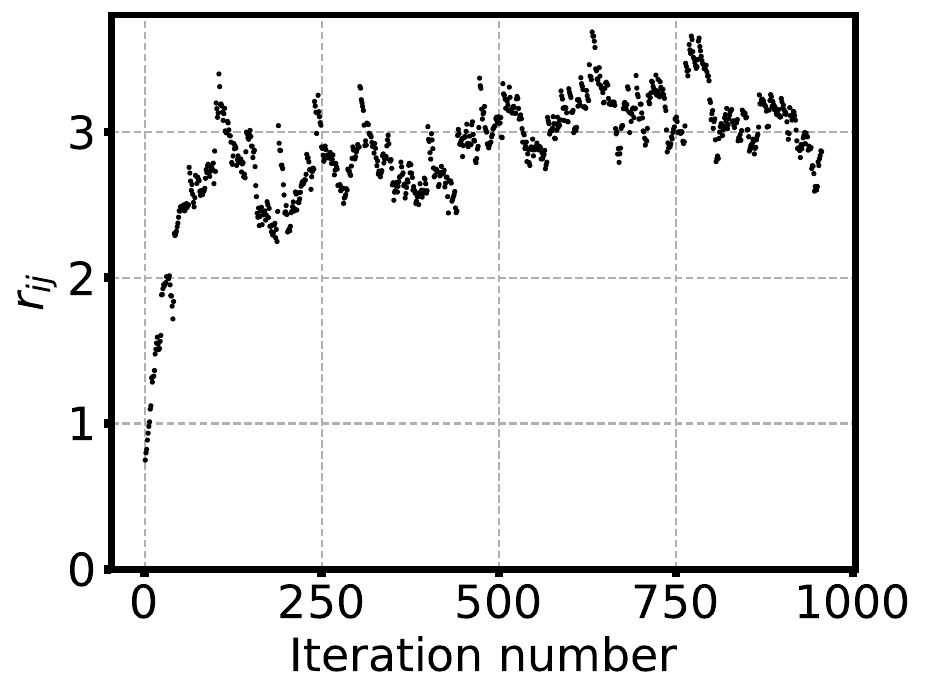}
\caption{Node C.}
\label{fig:nodeC_peri}
\end{subfigure}
\begin{subfigure}[b]{0.45\textwidth}
\centering
\includegraphics[width=\textwidth]{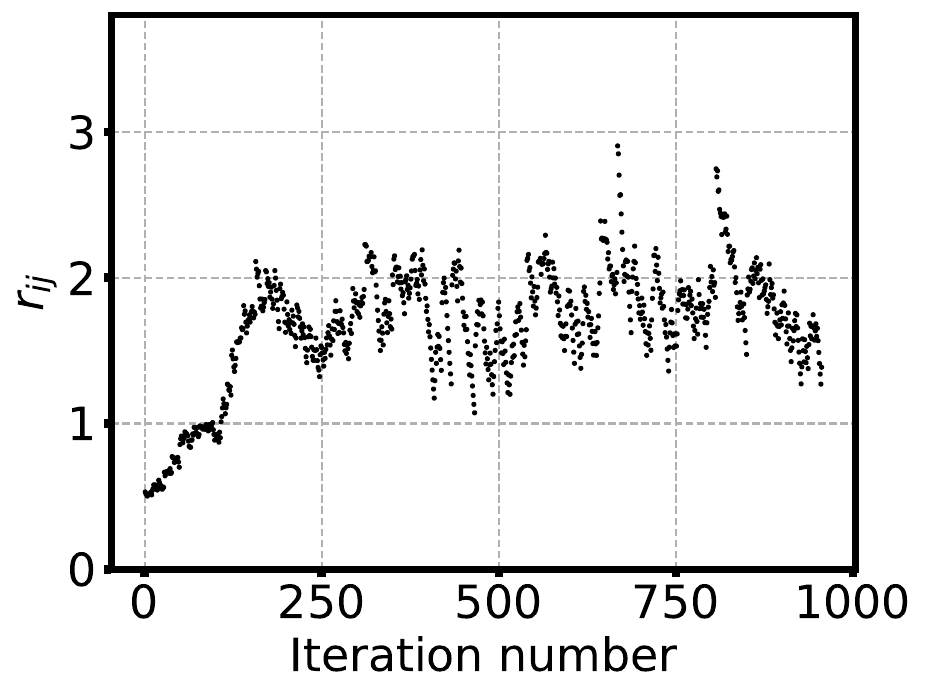}
\caption{Node D.}
\label{fig:nodeD_peri}
\end{subfigure}
\begin{subfigure}[b]{0.45\textwidth}
\centering
\includegraphics[width=\textwidth]{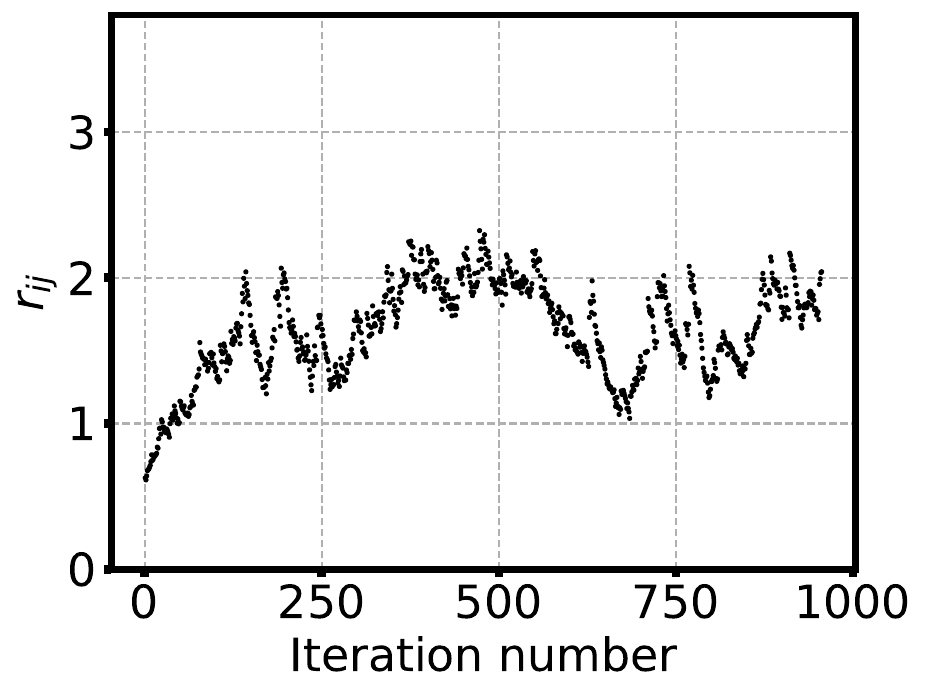}
\caption{Node E.}
\label{fig:nodeE_peri}
\end{subfigure}
\begin{subfigure}[b]{0.45\textwidth}
\centering
\includegraphics[width=\textwidth]{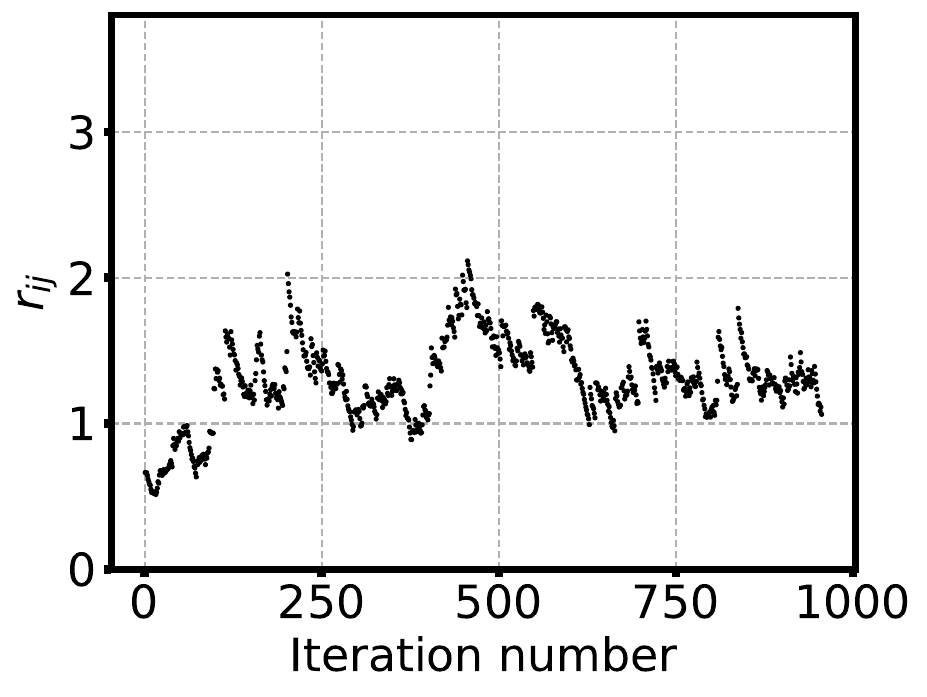}
\caption{Node F.}
\label{fig:nodeF_peri}
\end{subfigure}
\begin{subfigure}[b]{0.45\textwidth}
\centering
\includegraphics[width=\textwidth]{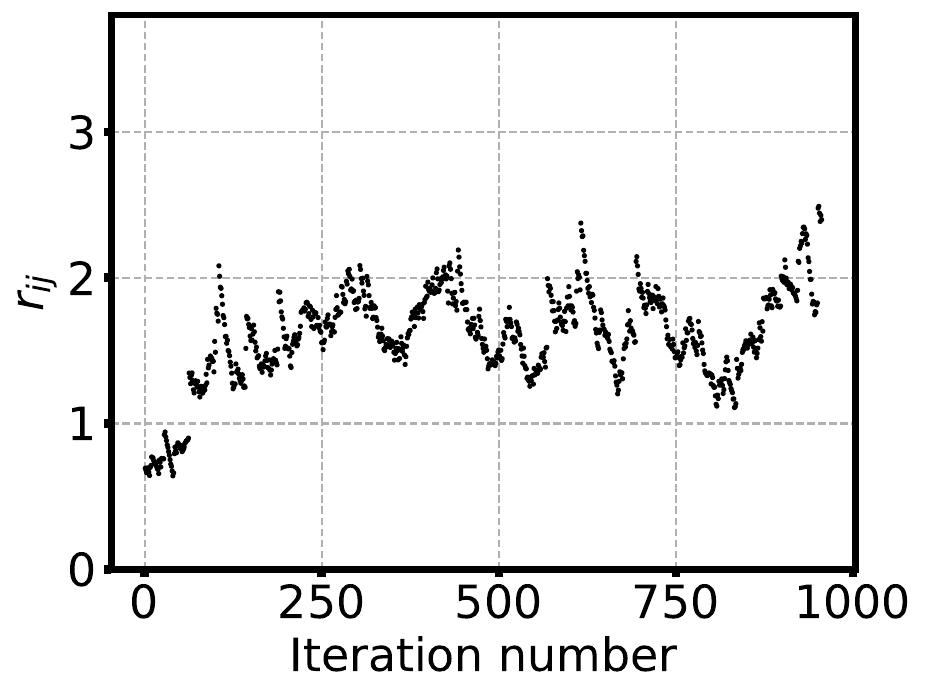}
\caption{Node G.}
\label{fig:nodeG_peri}
\end{subfigure}
\caption{$r_{ij}$ for all edges for 955 iterations.}
\label{fig:peristalsis}
\end{figure}

\begin{figure}
\centering
\begin{subfigure}[c]{0.95\textwidth}
\centering
{\includegraphics[width=0.49\textwidth]{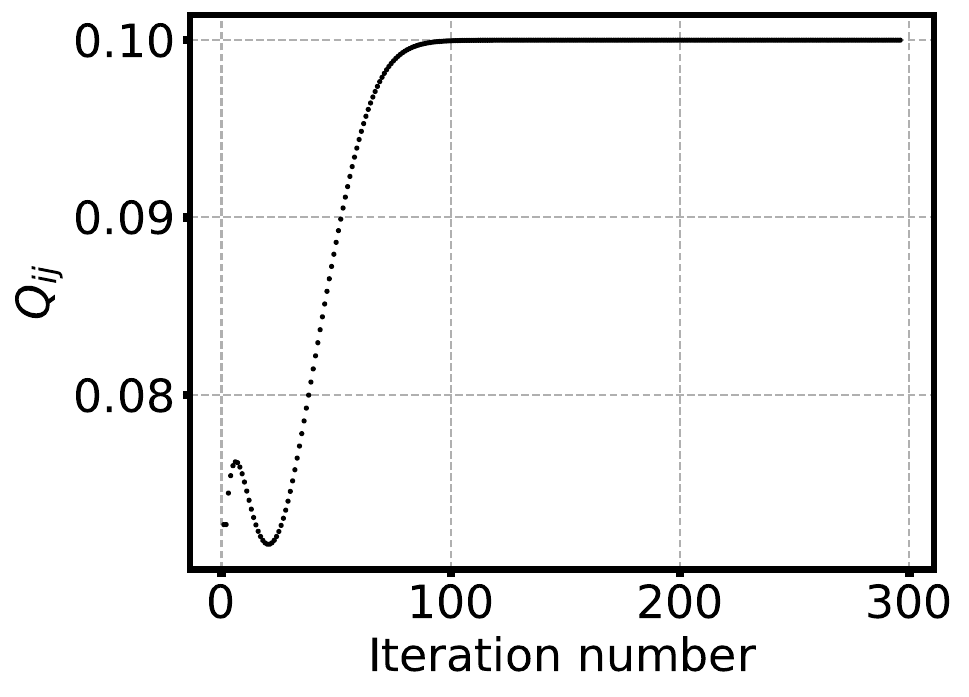}
\includegraphics[width=0.47\textwidth]{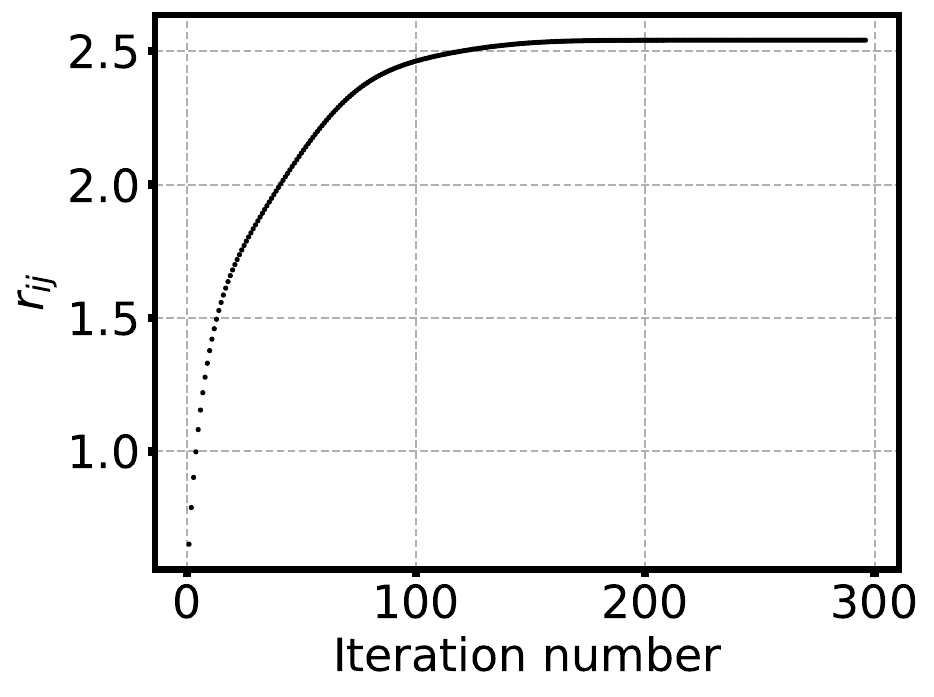}}
\caption{Node A.}
\label{fig:nodeA_sync}
\end{subfigure}
\begin{subfigure}[c]{0.95\textwidth}
\centering
{\includegraphics[width=0.49\textwidth]{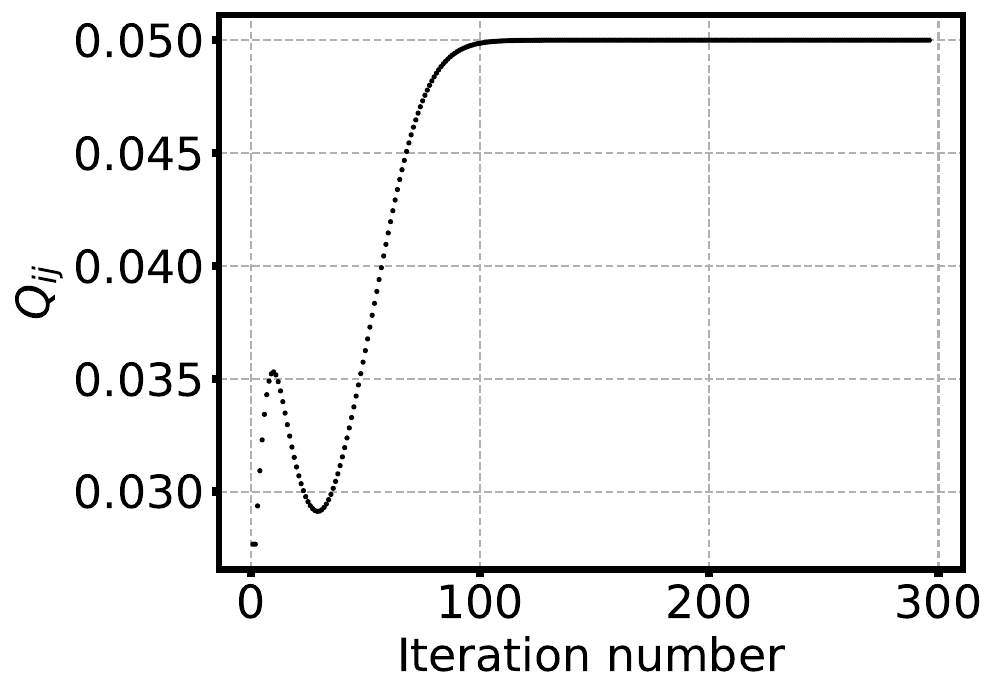}
\includegraphics[width=0.47\textwidth]{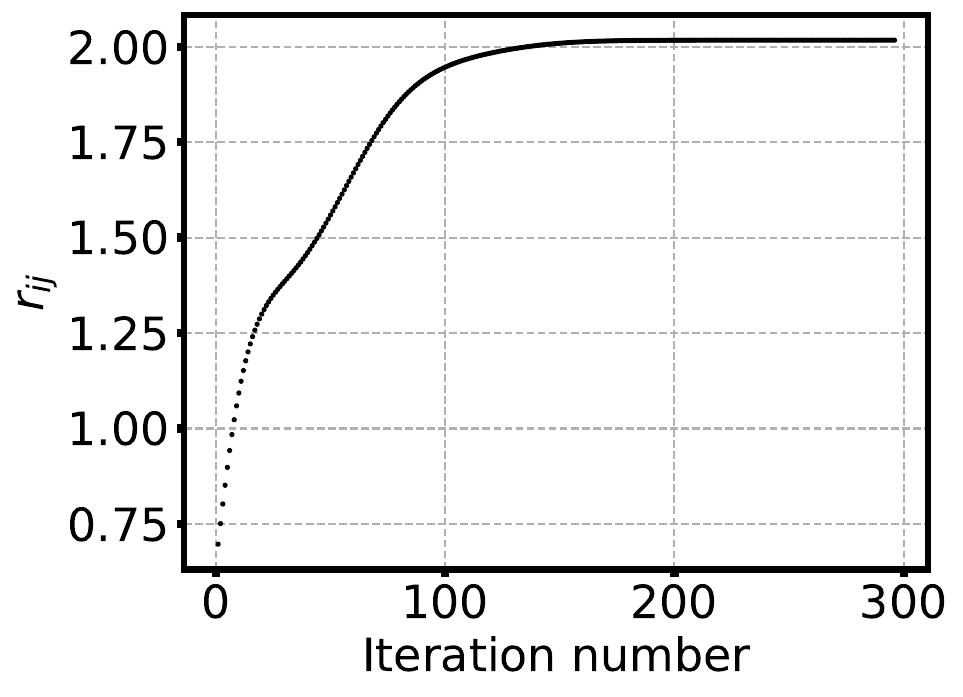}}
\caption{Node B.}
\label{fig:nodeB_sync}
\end{subfigure}
\begin{subfigure}[c]{0.95\textwidth}
\centering
{\includegraphics[width=0.49\textwidth]{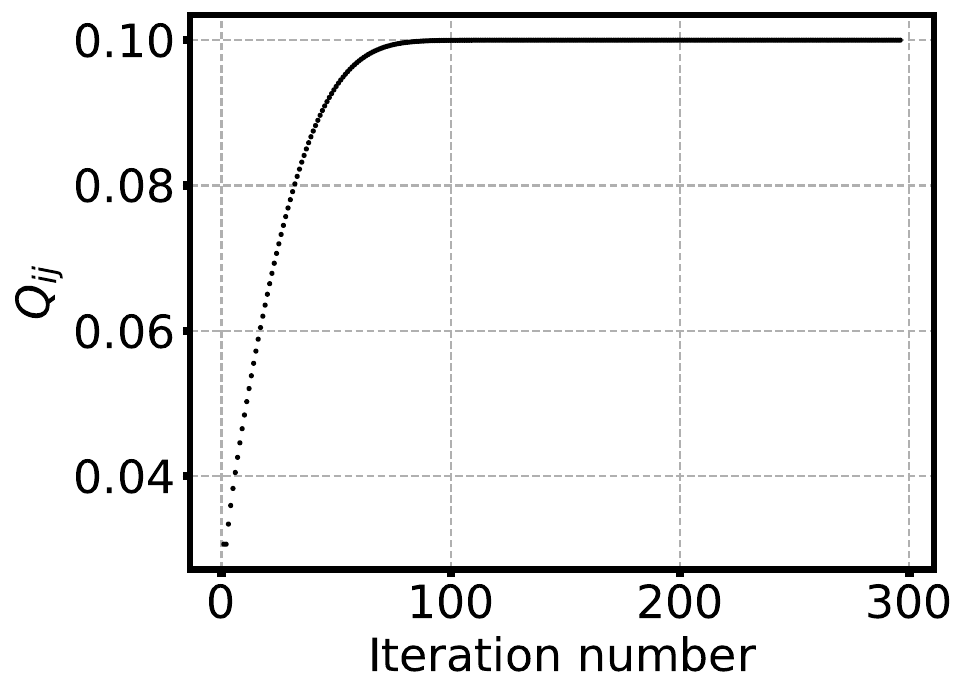}
\includegraphics[width=0.47\textwidth]{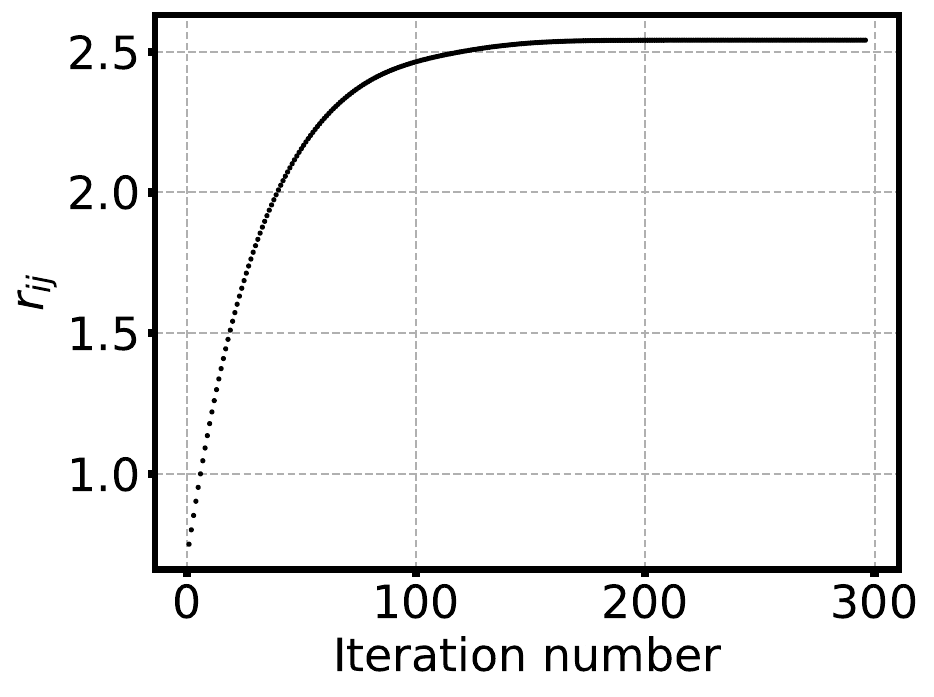}}
\caption{Node C.}
\label{fig:nodeC_sync}
\end{subfigure}
\caption{$Q_{ij}$ (left) and $r_{ij}$ (right) for all edges for the synchronous configuration of figure \ref{fig:peristalsis_ss_sync}, for 295 iterations.}
\label{fig:peristalsis_sync}
\end{figure}

It's important to note that peristalsis and shuttle streaming are only shown in simulations with asynchronous sites. To show this, we used the exact same configuration of sources and sinks and the same initial conditions, but made every sink be active at all times. The stopping condition used was the same as the one used for the synchronous configuration (see section \ref{sec:physarum_sync}), and the simulation converged after 295 iterations. The steady state obtained can be seen in figure \ref{fig:peristalsis_ss_sync}, where nodes A-C are highlighted by a black box. We can observe that the edges D-G, which were connections between sinks, are not present in this steady state. The values of $Q_{ij}$ and $r_{ij}$ for these edges were recorded over time and can be seen in figure \ref{fig:peristalsis_sync}. As we can see, the radii of the edges does not oscillate over time, and the flux does not change its sign at any point. We can conclude that shuttle streaming and peristalsis are most likely consequences of the asynchronous resource consumption that occurs in \textit{Physarum}, as mentioned previously.

\chapter{Paths of extremal lengths connecting points in an Euclidean space}
\label{chapter:Steiner}

The goal of this chapter is to analyze extremal length graphs connecting $n$ points in an Euclidean space, using the properties of adaptive Hagen-Poiseuille flows. We will explore the ability of the model studied in this work to generate approximate Steiner-type solutions in graph theory. We will develop several strategies for the design of transportation systems and we propose an algorithm to determine their efficiency.

\section{Finding paths of extremal lengths connecting the nodes of a graph}

The survival of an organism is greatly dependent on its biological network. For optimal survivability, a biological network must be efficient in transporting and distributing resources and nutrients throughout the body (by providing a short path between relevant sites). In addition to ensuring transport efficiency, a biological network should also be as short in length as possible, so as to minimise the costs of building said network. Furthermore, biological networks should be robust and redundant as well, to prevent network failure in the case of damage to some channel. In the case of \textit{Physarum}, its network's adaptability is a key factor to its survivability.

These factors are not only relevant in biological networks; in fact, they are key concepts behind the design of many man-made networks, transport networks (railroads, highways) being an example.

To first illustrate these concepts, we will present geometric examples. We consider a set of $n$ points in an Euclidean space of dimension $\mathcal{N}$. These points may be connected through a connected graph, and there is an infinity of solutions for the connected graph.

\begin{figure}[]
\centering
\begin{subfigure}[b]{0.4\textwidth}
     \centering
     \includegraphics[width=\textwidth]{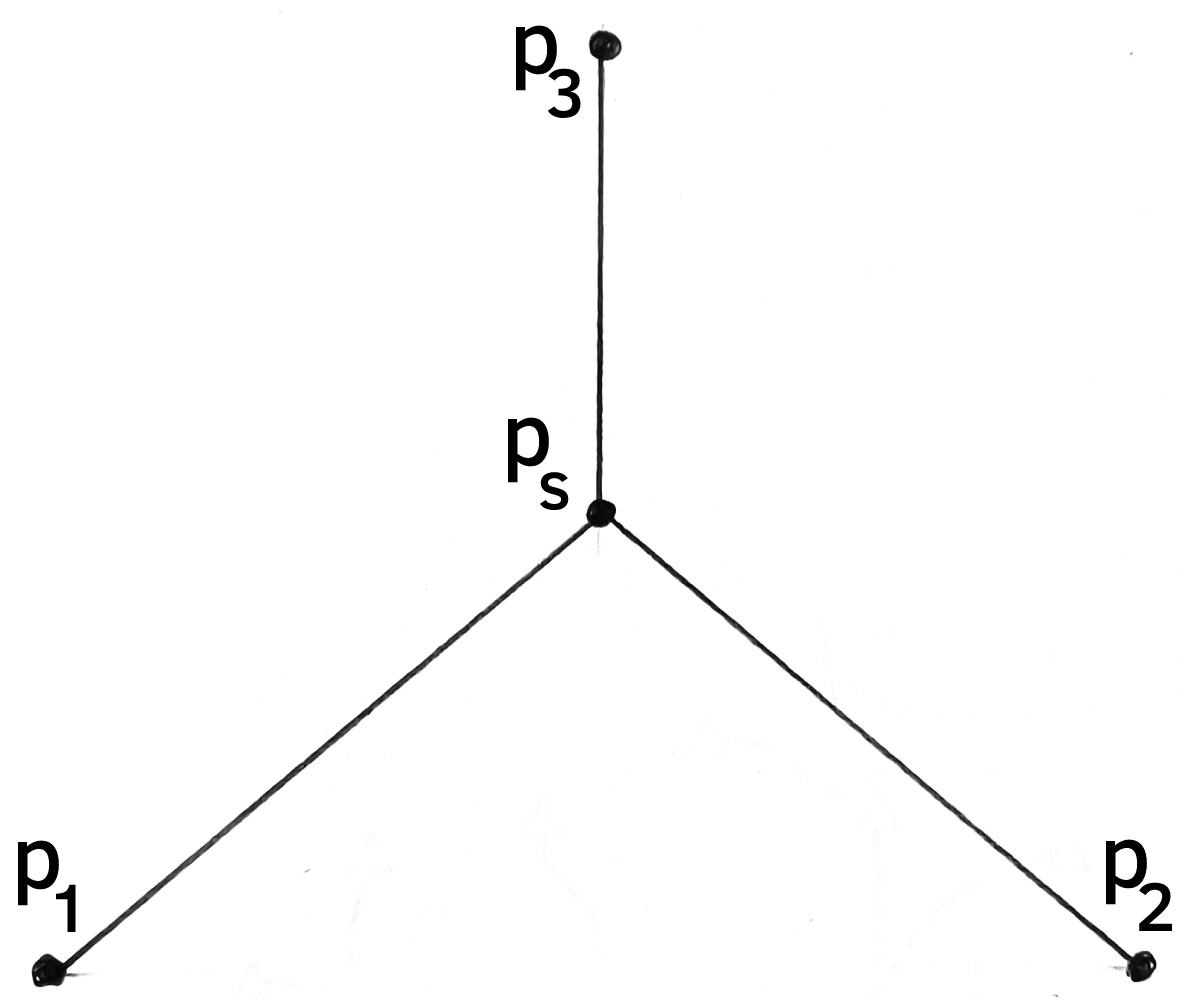}
     \caption{Minimal Steiner tree of a triangle.}
     \label{fig:triangle_theo_steiner}
 \end{subfigure}
 \hfill
 \begin{subfigure}[b]{0.4\textwidth}
     \centering
     \includegraphics[width=\textwidth]{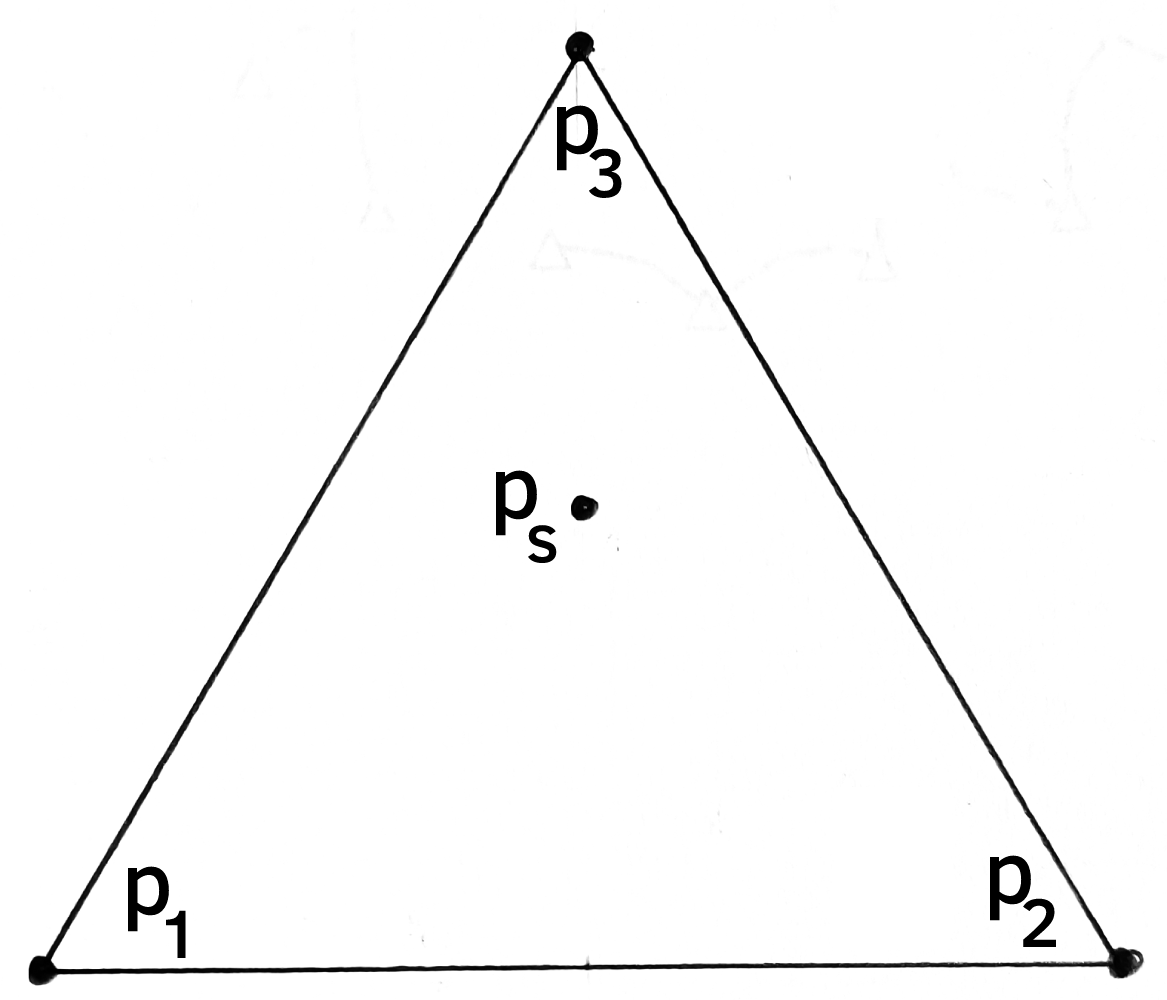}
     \caption{Perimeter of a triangle.}
     \label{fig:triangle_theo_triangle}
 \end{subfigure}
\caption{Two possible connections of 3 points in a 2D graph.}
\label{fig:triangle_theo}
\end{figure}

Let us first consider the case of $n = 3$ points in an equilateral triangular configuration on a 2D plane. Two possible trees for the graph connecting these three points are shown in figure \ref{fig:triangle_theo}.

The points $p_i, i=\{1, 2, 3\}$ make an equilateral triangle of side 1, while the point $p_S$ is the center of the same triangle. We assume the following coordinates for the 3 vertices and central point:

$$p_1 = (0,0) \qquad p_2 = (1,0) \qquad p_3 = \left(\frac{1}{2},\frac{\sqrt{3}}{2}\right) \qquad p_S = \left(\frac{1}{2},\frac{\sqrt{3}}{4}\right)$$

The length of the tree of figure \ref{fig:triangle_theo_steiner} is $L_a = 3\sqrt{7}/4\approx1.98$, while the graph length of the graph of figure \ref{fig:triangle_theo_triangle} is $L_b = 3$. The graph of figure \ref{fig:triangle_theo_steiner} is the graph of minimal lengths that connects the three points (and that is allowed to contain additional points, $p_S$), and thus is the Steiner minimum tree of the triangle vertices.

While this is an abstract system, the tree connecting the three points can also symbolize a multitude of physical, social or biological systems; for example, it can be a model of a system of roads connecting three cities. Suppose each road is used by frequent travellers that go from city $i$ to city $j$, and that the number of travelers that go from $i$ to $j$, for every $(i,j)$ pair, is asymptotically the same. For the road system described by the graph of figure \ref{fig:triangle_theo_steiner}, the mean distance between two cities is $M_a = 1.98$, and for the road system described by the graph of figure \ref{fig:triangle_theo_triangle}, this distance is $M_b = 1$.

Therefore, from the point of view of communications, the graph configuration of figure \ref{fig:triangle_theo_triangle} is more advantageous in the sense that is has lower transportation costs for the travelers. However, the graph configuration of figure \ref{fig:triangle_theo_steiner} is more advantageous in the sense that it has lower building and maintenance costs.

For a given set of configurations of points in an Euclidean space, we want to decide which graph fulfills the requirement of efficiency and low costs.

\subsection{Stochastic algorithms used}

As mentioned in \ref{sec:adaptive-model}, different initial conductivity values will lead to different steady-state tree shapes. To obtain different trees with different shapes, efficiencies, and lengths, a stochastic search will be done. The simulation will be run $N_{\text{runs}} = 300$ times, each time with different random initial conditions, for each configuration with $n$ sites.

The adaptive Hagen-Poiseuille flows equations require that there is at least one source node and one sink node for a tree to form. In this paper, for the case of a configuration with $n$ sites, each site is equal to every other site, making each site equally likely to be a source or a sink. As such, different stochastic algorithms will be used to choose which nodes are sources and which nodes are sink at each iteration step.

Each algorithm is given a name for quick identification. The algorithms are the following: at each iteration (every $\Delta t$), let $I_0 = \sum_{i\in\text{sources}} S_i$ be the total inward/outward flux of fluid to/from the network:

\begin{enumerate}
    \item \texttt{Random pair}: 1 source (with $S_i = I_0$) and 1 sink (with $S_j = -I_0$) are chosen randomly among the $n$ sites; 
    \item \texttt{Random half}: $n/2$ sources (each with $S_i = I_0/(\frac{n}{2})$) and $n/2$ sinks (each with $S_j = -I_0/(\frac{n}{2})$) are chosen randomly from the $n$ sites; if $n$ is an odd number, one of the sites is inactive;
    \item \texttt{Random source}: 1 source (with $S_i = I_0$) is chosen randomly from the $n$ sites and the remaining $n - 1$ sites are sinks (each with $S_j = - I_0 / (n-1)$);
    \item \texttt{Random random}: $1 \leq n_{so} \leq n - 1$ sources are chosen randomly from the $n$ sites (each with $S_i = I_0/n_{so}$), and the remaining $n-n_{so}$ sites are sinks (each with $S_j = -I_0/(n-n_{so})$).
\end{enumerate}

\subsection{Characterization parameters}

Three metrics will be used to characterize each steady state: its length $L$, its average length between pairs of sites $\nu$, and its cost-efficiency CE. These three parameters will be measured for all $N$ steady-state trees.

Let

\begin{equation} \label{eq:L}
    L = \sum_{(i,j)\in E'} L_{ij}
\end{equation}
\noindent
be the length $L$ of a graph, where $E'$ is the set of effectively conducting edges, that is, edges with conductivity values above a certain threshold $(E' = \{(i,j)\in E: D_{ij} > D_{\text{thresh}})\}$, with $D_{\text{thresh}} = 5\times10^{-4}$). Note that $L_{ij}$ is the length of edge $(i,j)$.

Let

\begin{equation} \label{eq:nu}
    \nu = \frac{1}{\left(^n_2\right)} \sum_{\substack{i,j\in\{\text{sites}\},\\i\neq j,\\ (i,j)\neq(j,i)}} L_{s_i s_j} = \frac{2}{n(n-1)}\sum_{\substack{i,j\in\{\text{sites}\},\\i\neq j,\\ (i,j)\neq(j,i)}} L_{s_i s_j}
\end{equation}
\noindent
be the average length between pairs of sites $\nu$, where $L_{s_i s_j}$ is the length of the path between sites $s_i$ and $s_j$. The quantity $L_{s_i s_j}$ was computed by finding the shortest path between two sites and adding up the lengths of all conducting edges on said path; the shortest path between the two sites was found using an algorithm very similar to Breadth first search \cite{cormen:breath_first_search}.

Let

\begin{equation} \label{eq:CE}
    \text{CE} = \frac{1}{L\nu}
\end{equation}
\noindent
be the cost-efficiency CE. By thinking once again of these steady-state trees as roads connecting cities, an efficient tree in terms of communication and travel would have a low $\nu$ value, while a tree with a low $L$ value would be cheap to construct or maintain. The parameter CE tries to take into account both of these facts.

\section{Optimising the Steiner tree solution} \label{sec:steiner_points_calc}

In section \ref{sec:steiner_intro}, the difference between the Steiner minimum tree and the minimum spanning tree has been explained: the STM is able to contain additional vertices other than the nodes of interest, called Steiner points, unlike the minimum spanning tree.

One of the characteristics of the steady-state solutions of the H-P adaptive flows is the existence of additional vertices other than sources and sinks (as discussed in section \ref{sec:physarum_async} and seen in figures \ref{fig:length_distribution_more}, \ref{fig:physarum_sync} and \ref{fig:physarum_async}). In some instances, the steady state of the H-P adaptive flow has the topological structure of a Steiner tree, despite not needing to necessarily be a minimum-length tree. In this case, when the topology of the graph is fixed, it is possible to adjust the coordinates of the Steiner points in such a way that the steady-state graph becomes a Steiner minimum tree.

\begin{figure}
    \centering
    \begin{subfigure}[b]{0.3\textwidth}
        \centering
        \includegraphics[width = \textwidth]{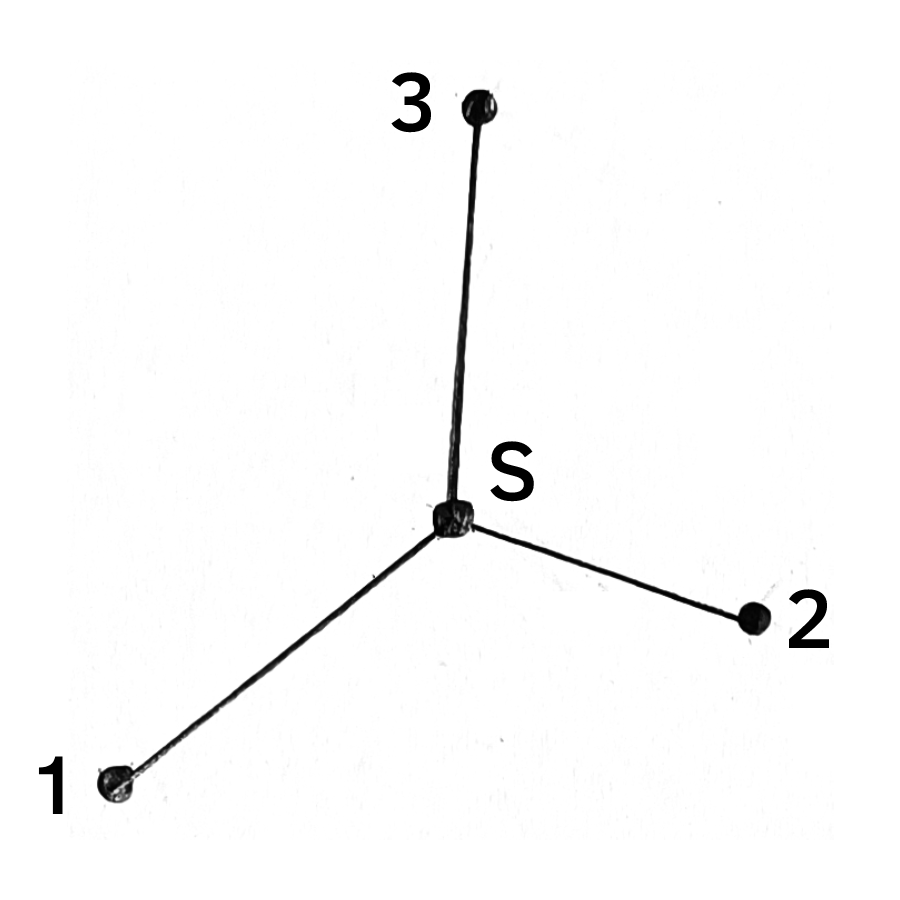}
        \caption{}
        \label{fig:wonky_steiner_triangle}
    \end{subfigure}
    \hfill
    \begin{subfigure}[b]{0.3\textwidth}
        \centering
        \includegraphics[width = \textwidth]{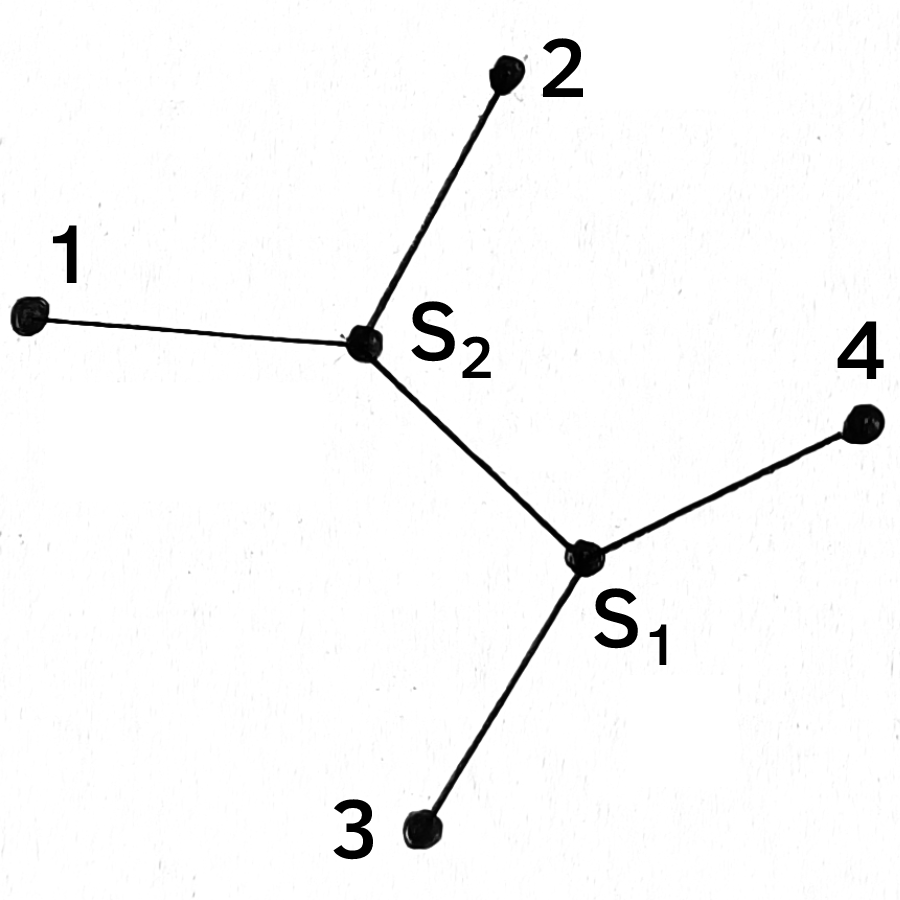}
        \caption{}
        \label{fig:wonky_steiner_square}
    \end{subfigure}
    \hfill
    \begin{subfigure}[b]{0.3\textwidth}
        \centering
        \includegraphics[width = \textwidth]{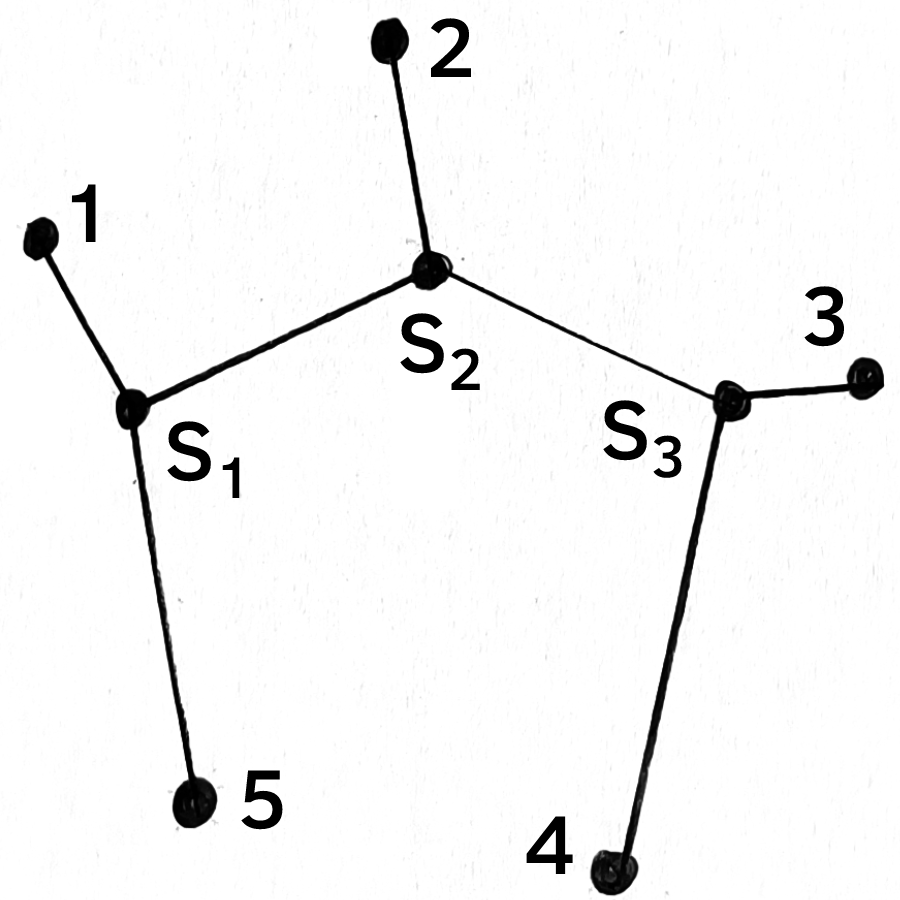}
        \caption{}
        \label{fig:wonky_steiner_penta}
    \end{subfigure}
    \caption{Trees that resemble Steiner minimum trees for three polygons: a) $n=3$, b) $n=4$ and c) $n=5$.}
    \label{fig:wonky_steiner}
\end{figure}

We exemplify this with simple examples. Consider the graphs in figure \ref{fig:wonky_steiner}. All the trees have the topology of a Steiner tree, that is, they look like they could be a Steiner minimum tree, but their Steiner points may be slightly out of place.

Let's look at the case of figure \ref{fig:wonky_steiner_triangle}, the triangle-like configuration. Consider the Steiner point is located at coordinates $(x_{S}, y_{S})$. The sum of the squares of the lengths of all the edges of the tree is

\begin{equation}
    L^2 = \sum_{i=1}^3 (x_i - x_S)^2 + (y_i - y_S)^2,
\end{equation}

Minimizing $L^2$ relative to $x_S$ and $y_S$,

\begin{equation} \label{eq:minimizing_L_steiner}
\begin{aligned} 
    \frac{\partial L^2}{\partial x_S} & = - 2 \sum_{i=1}^3 (x_i - x_S) = 0 \\
    \frac{\partial L^2}{\partial y_S} & = - 2 \sum_{i=1}^3 (y_i - y_S) = 0, \\
\end{aligned}
\end{equation}
\noindent
we obtain the solution

\begin{equation} \label{eq:steiner_points_triangle_minimum}
    x_S = \frac{1}{3} \sum_{i=1}^3 x_i \quad,\quad y_S = \frac{1}{3} \sum_{i=1}^3 y_i
\end{equation}

It's important to note that, since

\begin{equation}
    \frac{\partial^2 L^2}{\partial x_S^2} = \frac{\partial^2 L^2}{\partial y_S^2} = 2 > 0
\end{equation}
\noindent
then the solution described in eq. \eqref{eq:steiner_points_triangle_minimum} is in fact a minimum. As such, eq. \eqref{eq:steiner_points_triangle_minimum} gives the positions of the Steiner point for the tree of figure \ref{fig:wonky_steiner_triangle}. In fact, eq. \eqref{eq:steiner_points_triangle_minimum} provides the positions of the Steiner point for all trees that possess that general shape, one Steiner point and that result from configurations of three sites. 

This procedure is trivially generalised for any tree shape, number of sites and any number of Steiner points. All one must do is construct $L^2$ appropriately, taking into account all paths of the tree (paths that connect sites to Steiner points and paths that connect Steiner points themselves), and then obtain its minimum by calculating $\partial L / \partial x_{Sj}$ for all $j$ Steiner points and equating it to zero.

For example, for the tree of picture \ref{fig:wonky_steiner_square} (the square-like configuration with 2 Steiner points), if the Steiner point $i$ has coordinates $(x_{Si},y_{Si})$ and the node of interest $j$ has coordinates $(x_j, y_j)$, $L^2$ is:

\begin{equation}
    L^2 = \sum_{i=1}^4 \sum_{j=1}^2 \left((x_i - x_{Sj})^2 + (y_i - y_{Sj})^2\right) + (x_{S1} - x_{S2})^2 + (y_{S1} - y_{S2})^2
\end{equation}

After minimization, we obtain the new positions of the Steiner points:

\begin{equation}
\begin{aligned}
    x_{S1} & = \frac{1}{8}(x_1 + x_2 + 3x_3 + 3x_4) \quad,\quad y_{S1} & = \frac{1}{8}(y_1 + y_2 + 3y_3 + 3y_4) \\
    x_{S2} & = \frac{1}{8}(3x_1 + 3x_2 + x_3 + x_4) \quad,\quad y_{S2} & = \frac{1}{8}(3y_1 + 3y_2 + y_3 + y_4) 
\end{aligned}
\end{equation}

If $x_1 + x_2 = x_3 + x_4$ and $y_1 + y_2 = y_3 + y_4$, the two Steiner points coalesce into a single Steiner point.

For the tree of picture \ref{fig:wonky_steiner_penta} (the pentagonal-like configuration with 3 Steiner points), the result obtained is:

\begin{equation}
\begin{aligned}
    x_{S1} = \frac{1}{21}(8x_1 + 3x_2 + x_3 + x_4 + 8x_5) & \quad,\quad \text{and similarly for } y_{S1} \\
    x_{S2} = \frac{1}{7}(x_1 + 3x_2 + x_3 + x_4 + x_5) & \quad,\quad \text{and similarly for } y_{S2}\\
    x_{S3} = \frac{1}{21}(x_1 + 3x_2 + 8x_3 + 8x_4 + x_5) & \quad,\quad \text{and similarly for } y_{S3}\\
\end{aligned}
\end{equation}

Many times, the problem in calculating Steiner trees resides in the fact that it is unclear how many Steiner points there may be and how they could be connected to each other and to the sites. Using the algorithm of the model studied in this work, one can determine how many Steiner points the minimum tree should have (as the algorithms produce trees that tend to the shortest paths); then, using these calculations, one can determine the Steiner points' exact locations, based on the sites used and, in this case, this is a solvable linear problem.

\section{Results}

The algorithms and parameters described will first be tested and studied on networks connecting vertices of regular polygons on a planar graph embedded in the two-dimensional Euclidean space. Then, the algorithms and parameters will be used to optimize paths for communication and transport systems; as an example, we present mainland Portugal. 

\subsection{Configurations of extremal lengths} \label{sec:polygonal_studies}

To study the algorithms and parameters described, the methods described in section \ref{sec:algorithms} will be used to initialize the simulation and to compute the temporal evolution of flow and conductivities. Among the $N' \times N'$ nodes that make up the mesh, $n$ sites will be chosen as possible sources/sinks for the simulations.

The parameters used were: $\beta = 1$, $\sum_{i\in\text{active sources}}S_i = 1$, $V = 100$, $N' = 25$, $\Delta\tau = 0.1$ and $N_{\text{runs}} = 300$.

The cases of $n = \{3, 4, 5\}$ will be studied, with the sites disposed in a regular polygon configuration. The results are presented in the following sections. For additional results, see appendix \ref{appendix:steiner_geometric}.

\subsubsection{Triangular configuration}

For $n = 3$, the sites approximately form an equilateral triangle of side $0.8$. The theoretical value for the perimeter of the triangle is $P_\triangle = 2.275$ (calculated using the coordinates of the sites) and for the minimum Steiner tree length it is $L_{\text{Stei}_\triangle} = 1.313$ (calculated using the Steiner point location obtained with the equations of section \ref{sec:steiner_points_calc}).

\begin{figure}
\centering
\begin{subfigure}[b]{0.55\textwidth}
 \centering
 \includegraphics[width=0.73\textwidth]{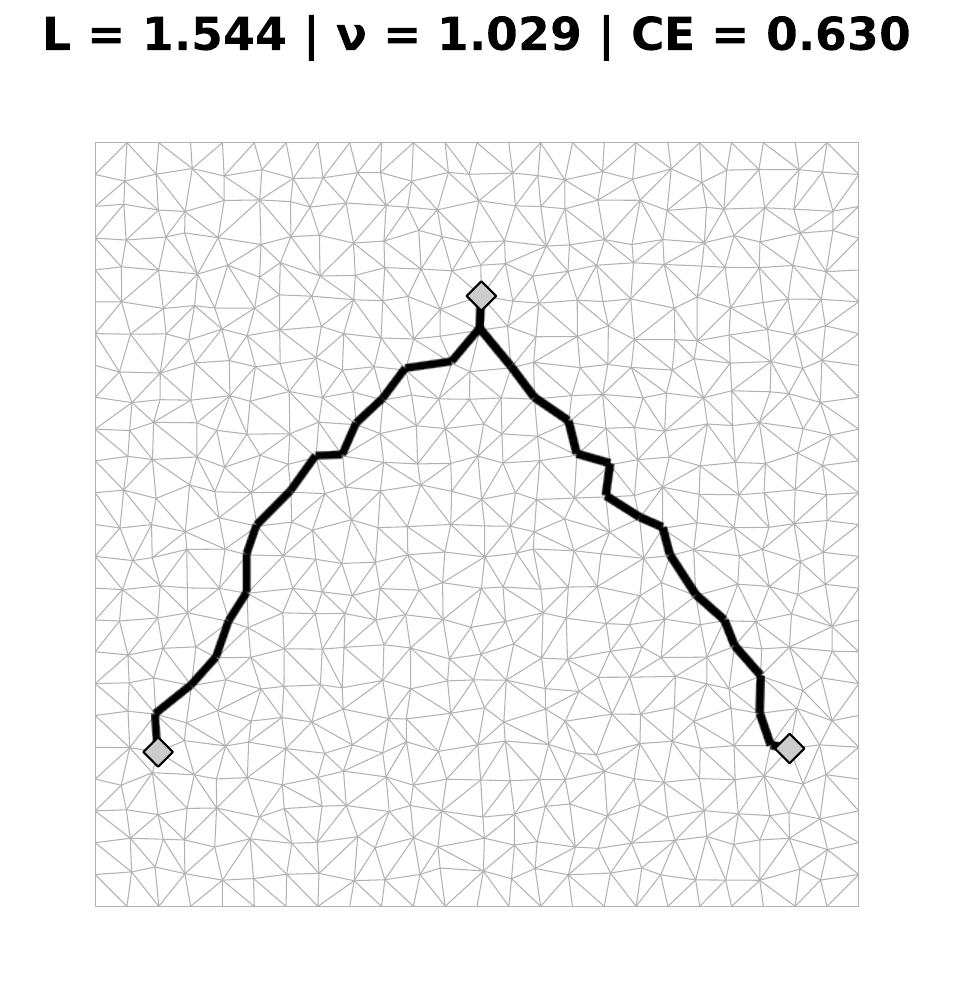}
 \caption{Steady state with smallest $L$ and largest CE.}
 \label{fig:tri_random_ss_smallest_L_largest_CE}
\end{subfigure}
\begin{subfigure}[b]{0.4\textwidth}
 \centering
 \includegraphics[width=\textwidth]{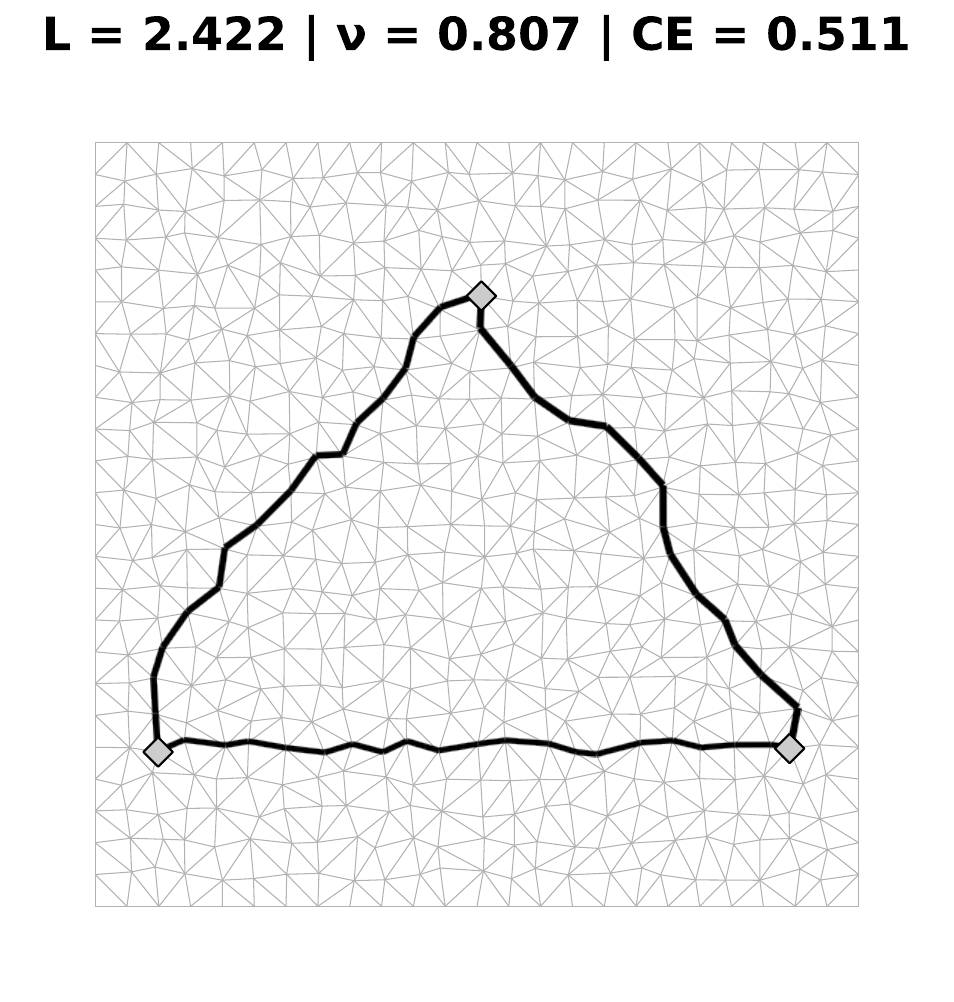}
 \caption{Steady state with smallest $\nu$.}
 \label{fig:tri_random_ss_smallest_niu}
\end{subfigure}
\caption{Best steady states for the triangular configuration, for the \textbf{Random pair} and \textbf{Random half} algorithms.}
\label{fig:tri_random}
\end{figure}

\begin{figure}
\centering
\begin{subfigure}[b]{0.6\textwidth}
 \centering
 \includegraphics[width=0.7\textwidth]{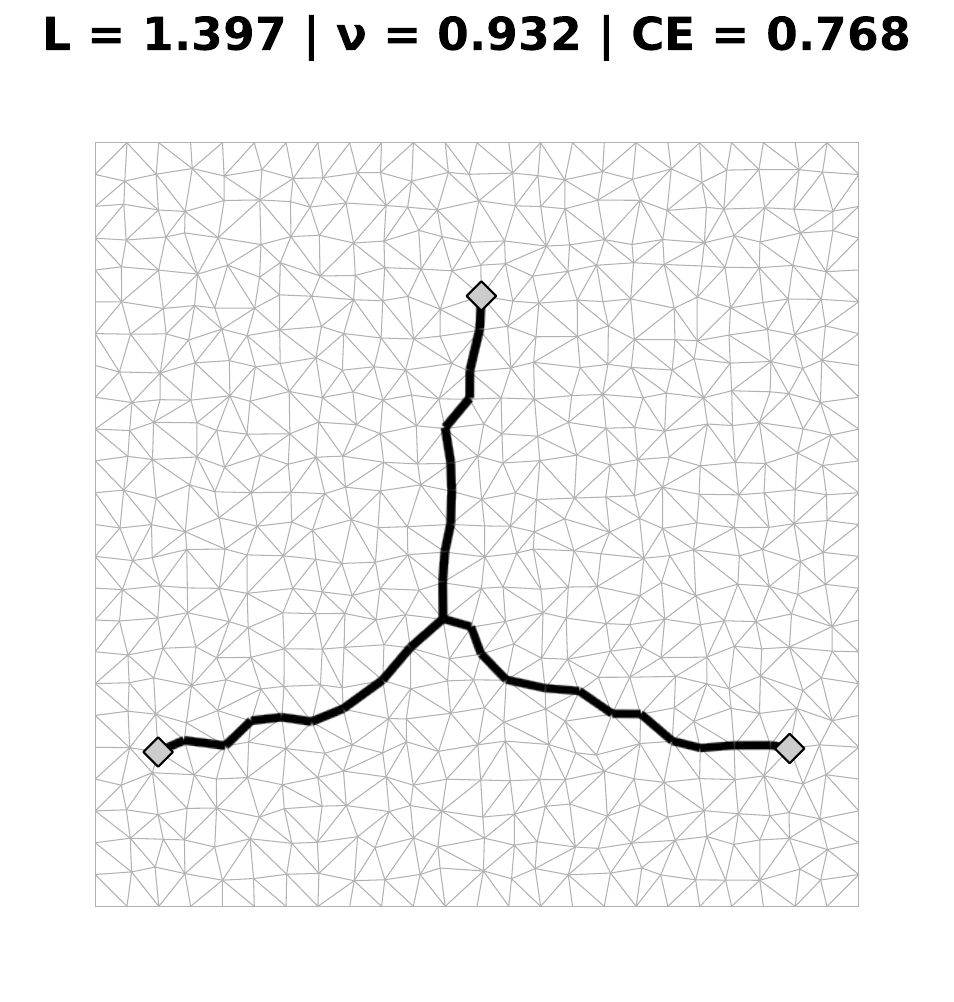}
 \caption{Steady state with smallest $L$, smallest $\nu$ and largest CE.}
 \label{fig:tri_random_source_ss_smallest_L_smallest_niu_biggest_CE}
\end{subfigure}
\caption{Best steady state for the triangular configuration, for the \textbf{Random source} algorithm.}
\label{fig:tri_random_source}
\end{figure}

\begin{figure}
\centering
\begin{subfigure}[b]{0.6\textwidth}
 \centering
 \includegraphics[width=0.7\textwidth]{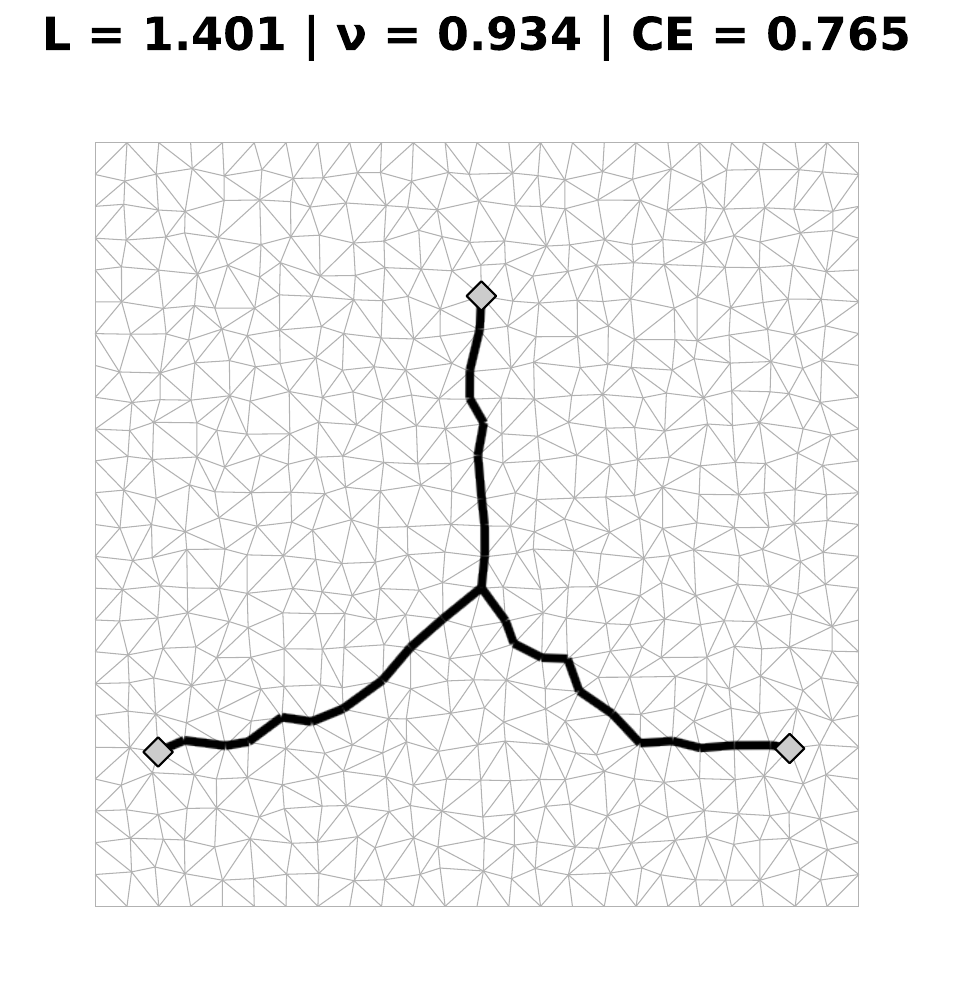}
 \caption{Steady state with smallest $L$, smallest $\nu$ and largest CE.}
 \label{fig:tri_random_fixed_I0_ss_smallest_L_smallest_niu_largest_CE}
\end{subfigure}
\caption{Best steady state for the triangular configuration, for the \textbf{Random random} algorithm.}
\label{fig:tri_random_fixed_I0}
\end{figure}

The results for algorithms \texttt{Random pair}, \texttt{Random source} and \texttt{Random random} are shown in figures \ref{fig:tri_random}, \ref{fig:tri_random_source} and \ref{fig:tri_random_fixed_I0} respectively. Note that algorithms \texttt{Random pair} and \texttt{Random half} are equivalent for the case of $n = 3$. The results shown are the steady states obtained with smallest $L$ and $\nu$ values and with the biggest CE value, and the graphs of $\nu$ as a function of $L$ for all runs and for all algorithms can be seen in figure \ref{fig:tri_L_niu}. Additional results can be viewed in appendix \ref{appendix:tri}.

\begin{figure}
\centering
\begin{subfigure}[b]{0.32\textwidth}
\centering
\includegraphics[width=\textwidth]{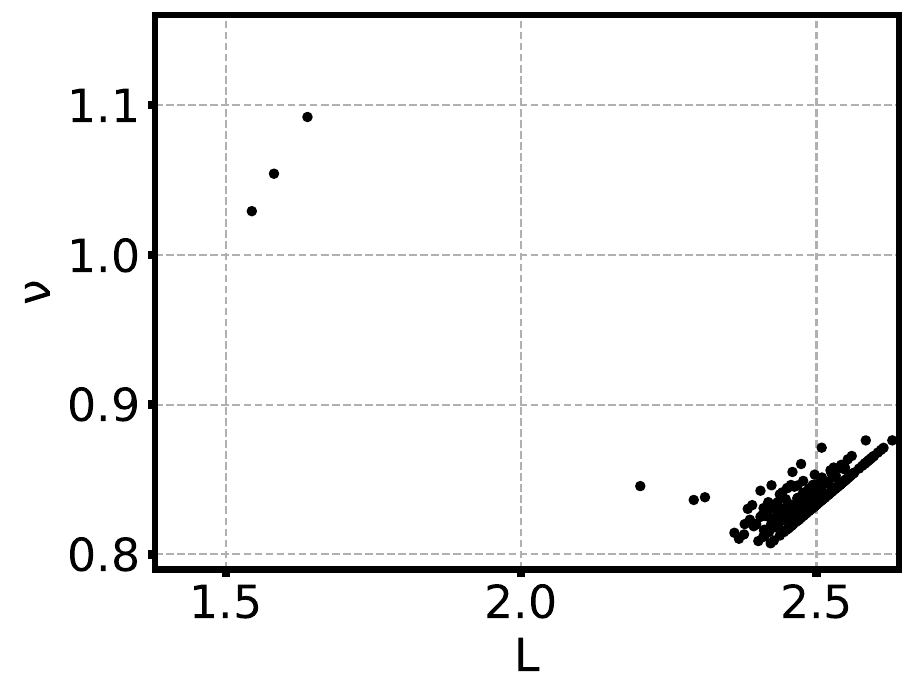}
\caption{\textbf{Random pair} and \textbf{Random half}.}
\label{fig:tri_random_L_niu}
\end{subfigure}
\hfill
\begin{subfigure}[b]{0.32\textwidth}
\centering
\includegraphics[width=\textwidth]{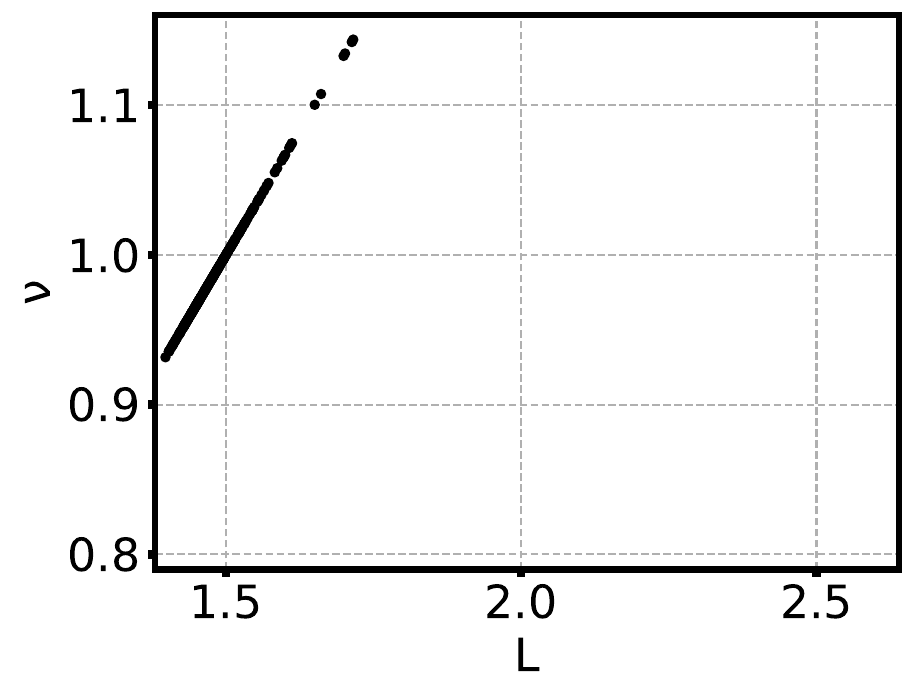}
\caption{\textbf{Random source}.}
\label{fig:tri_random_source_L_niu}
\end{subfigure}
\hfill
\begin{subfigure}[b]{0.32\textwidth}
\centering
\includegraphics[width=\textwidth]{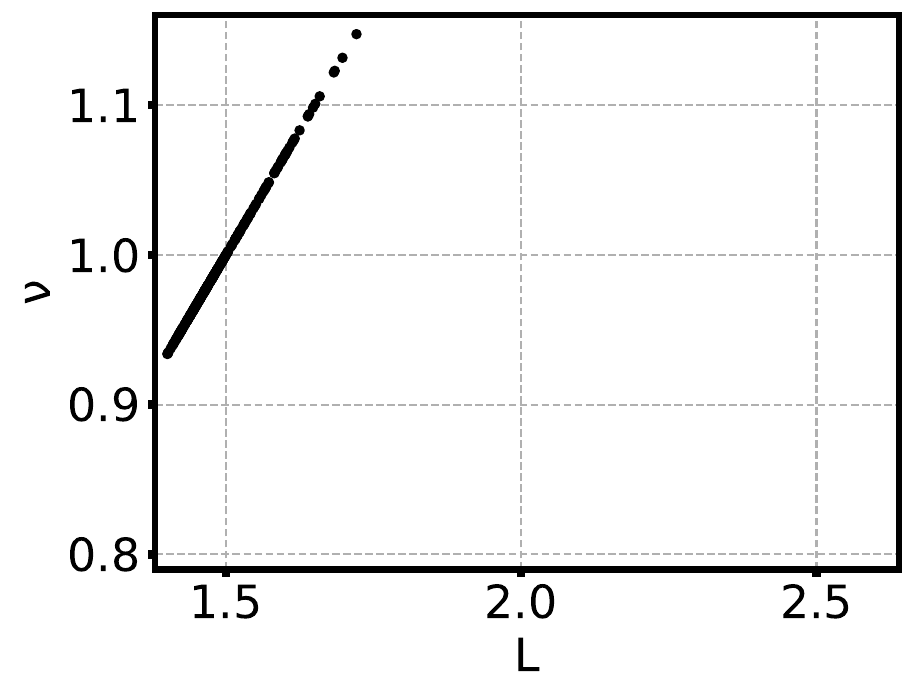}
\caption{\textbf{Random random}.}
\label{fig:tri_random_fixed_I0_L_niu}
\end{subfigure}
\caption{Steady state parameters $\nu$ as a function of $L$ for the triangular configuration, for all algorithms.}
\label{fig:tri_L_niu}
\end{figure}

The tree obtained as the minimum length, minimum $\nu$ and maximum CE tree for algorithms \texttt{Random random} and \texttt{Random source} is very similar to the minimum Steiner tree of the triangle. For these two algorithms, the graph of $\nu$ vs. $L$ shows a linear relation between the two variables: a tree with a large length would also present a large $\nu$, thus a lower efficiency.

For the algorithm \texttt{Random pair}, the minimum Steiner tree was not obtained; the minimum $\nu$ tree was the triangle with its full perimeter and the minimum $L$ and maximum CE tree was two sides of the triangle. For this algorithm, the graph of $\nu$ vs. $L$ shows two clusters: one representing trees with large $\nu$ and small $L$ (in which the two variables seem to show a linear relation) and one representing trees with small $\nu$ and large $L$ (where the lower bound of this region appears to present a linear relation).

The algorithm that produced the shortest tree and the tree with the largest CE value was \texttt{Random source}, which is also the algorithm that most frequently produced short and high CE value trees (see fig. \ref{fig:tri_random_source_L_distribution} and table \ref{tab:tri}).

The algorithm that produced the smallest $\nu$ tree was \texttt{Random pair}, which was also the algorithm that most frequently produced small $\nu$ value trees (see fig. \ref{fig:tri_random_niu_distribution} and table \ref{tab:tri}).

\subsubsection{Square configuration}

For $n = 4$, the sites approximately form a square of side $0.7$. The theoretical value for the perimeter of the square is $P_\square = 2.715$ (calculated using the coordinates of the sites) and, for the minimum Steiner tree length, it is $L_{\text{Stei}_\square} = 1.851$ (calculated using the Steiner points locations obtained with the equations of section \ref{sec:steiner_points_calc}).

\begin{figure}
\centering
\begin{subfigure}[b]{0.55\textwidth}
\centering
\includegraphics[width=0.73\textwidth]{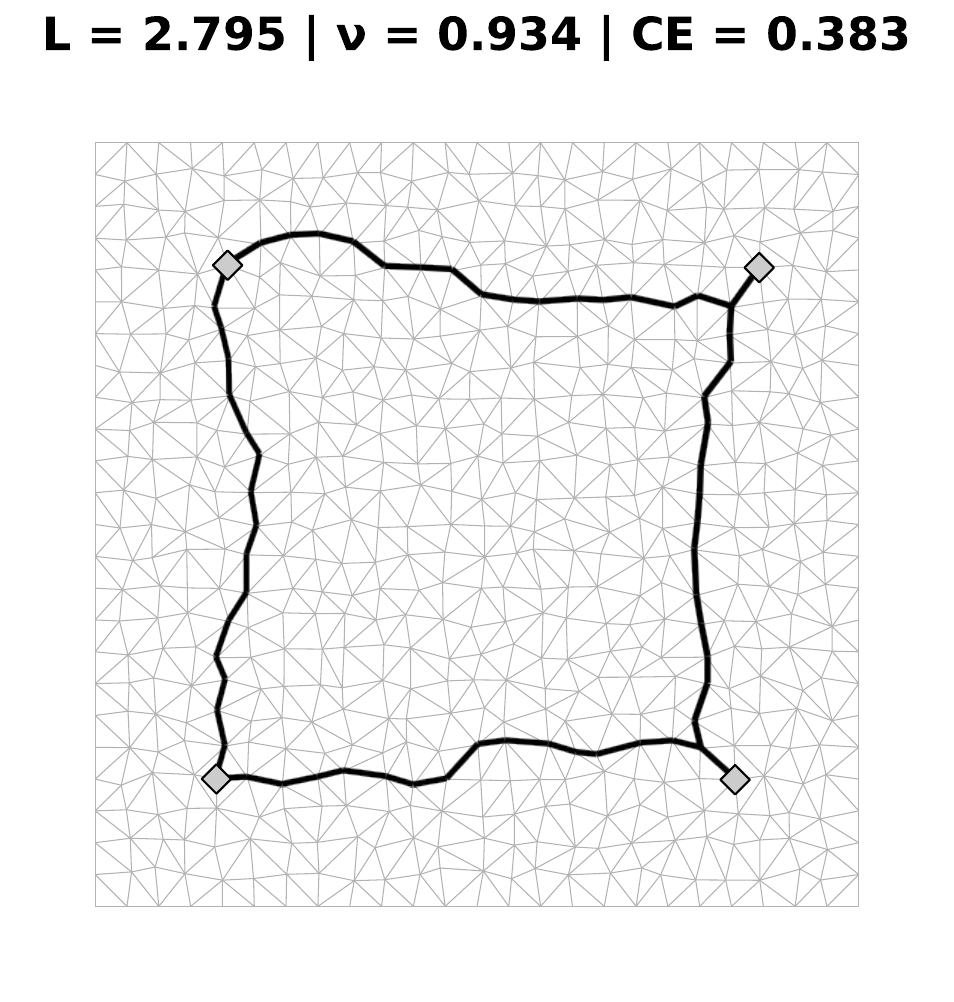}
\caption{Steady state with smallest $L$ and largest CE.}
\label{fig:square_random_ss_smallest_L_biggest_CE}
\end{subfigure}
\begin{subfigure}[b]{0.4\textwidth}
\centering
\includegraphics[width=\textwidth]{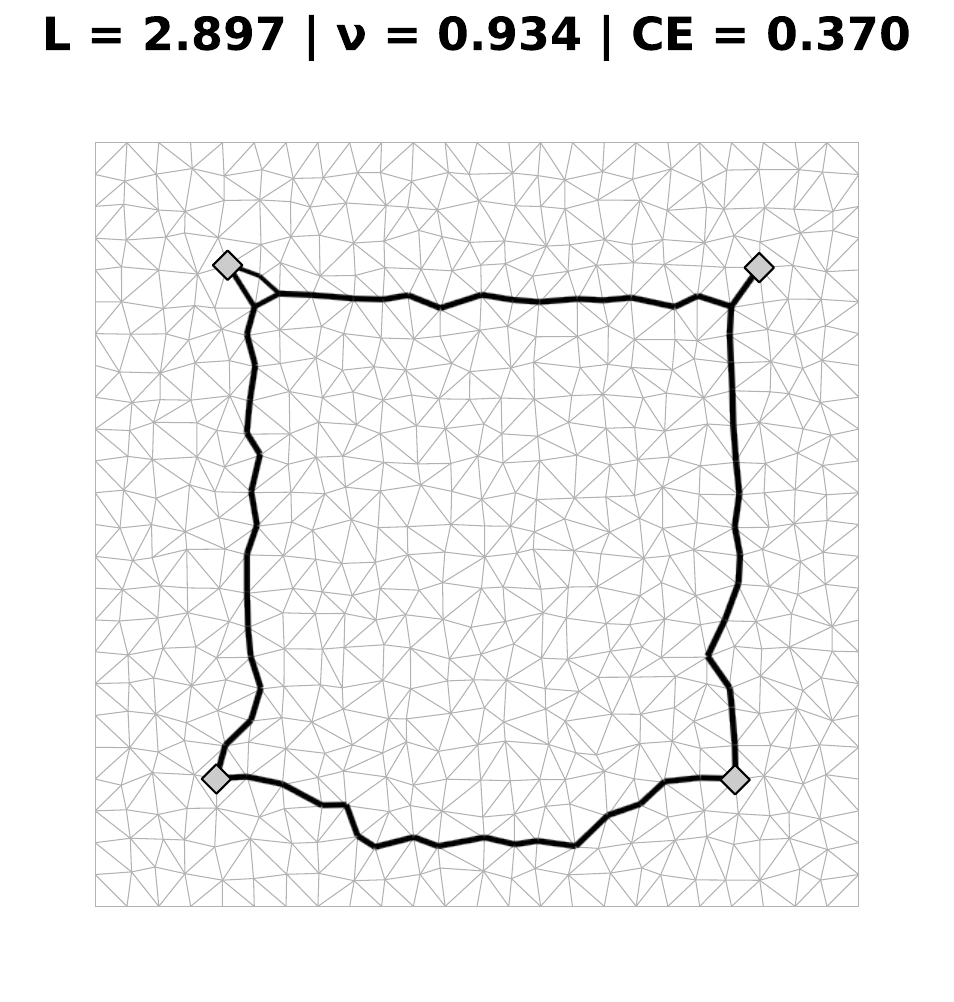}
\caption{Steady state with smallest $\nu$.}
\label{fig:square_random_ss_smallest_niu}
\end{subfigure}
\caption{Best steady states for the square configuration, for the \textbf{Random pair} algorithm.}
\label{fig:square_random}
\end{figure}

\begin{figure}
\centering
\begin{subfigure}[b]{0.32\textwidth}
\centering
\includegraphics[width=\textwidth]{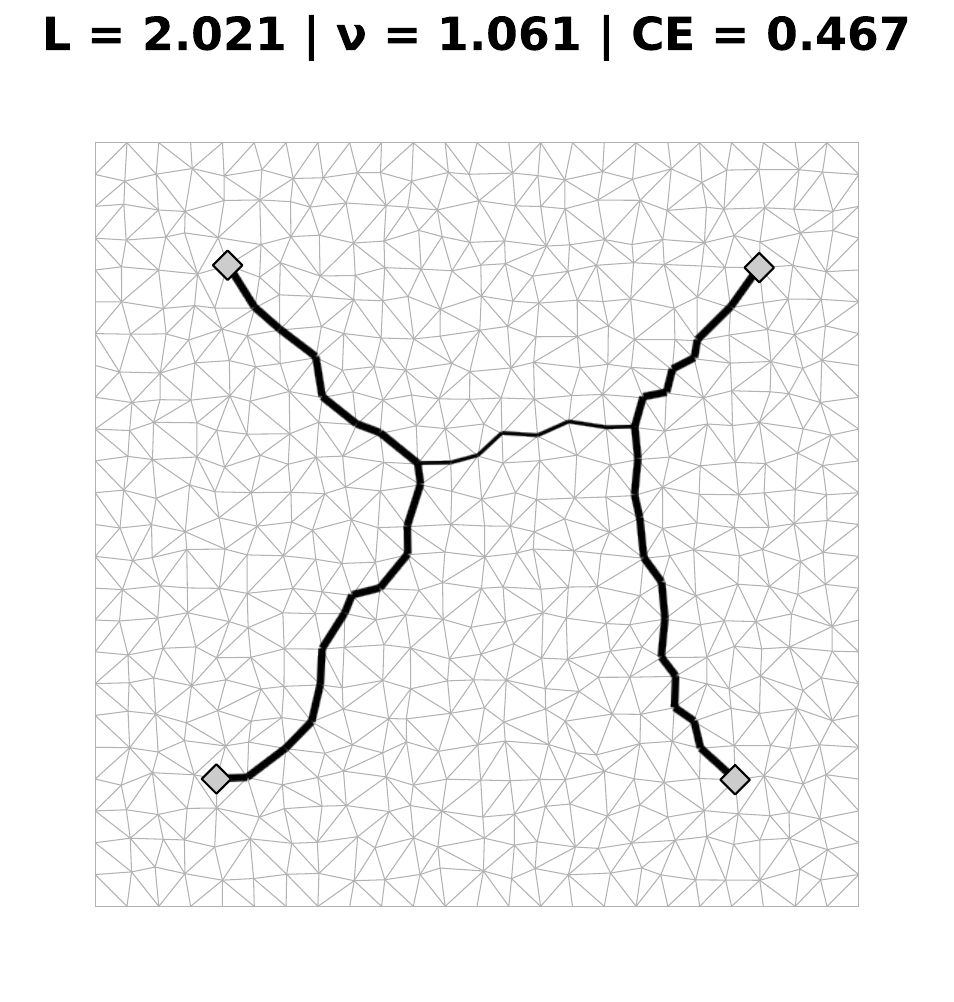}
\caption{Steady state with smallest $L$.}
\label{fig:square_random_half_ss_smallest_L}
\end{subfigure}
\begin{subfigure}[b]{0.32\textwidth}
\centering
\includegraphics[width=\textwidth]{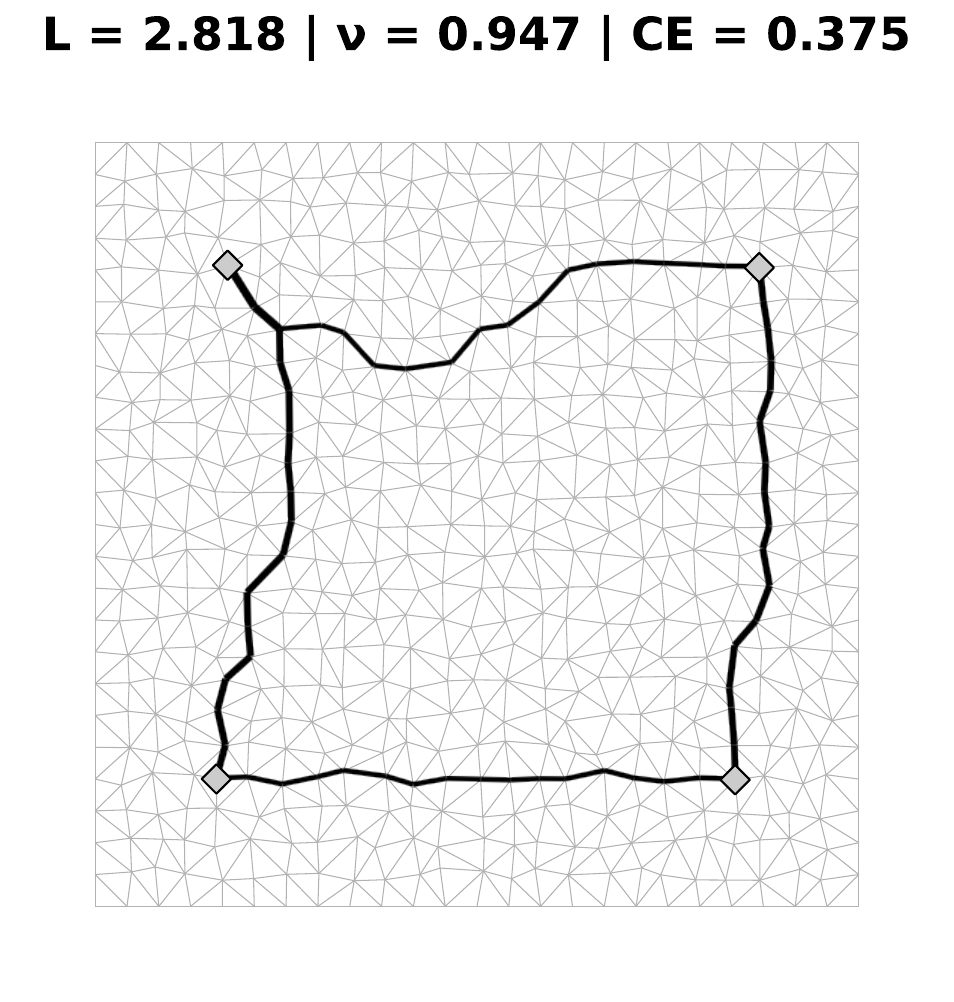}
\caption{Steady state with smallest $\nu$.}
\label{fig:square_random_half_ss_smallest_niu}
\end{subfigure}
\begin{subfigure}[b]{0.32\textwidth}
\centering
\includegraphics[width=\textwidth]{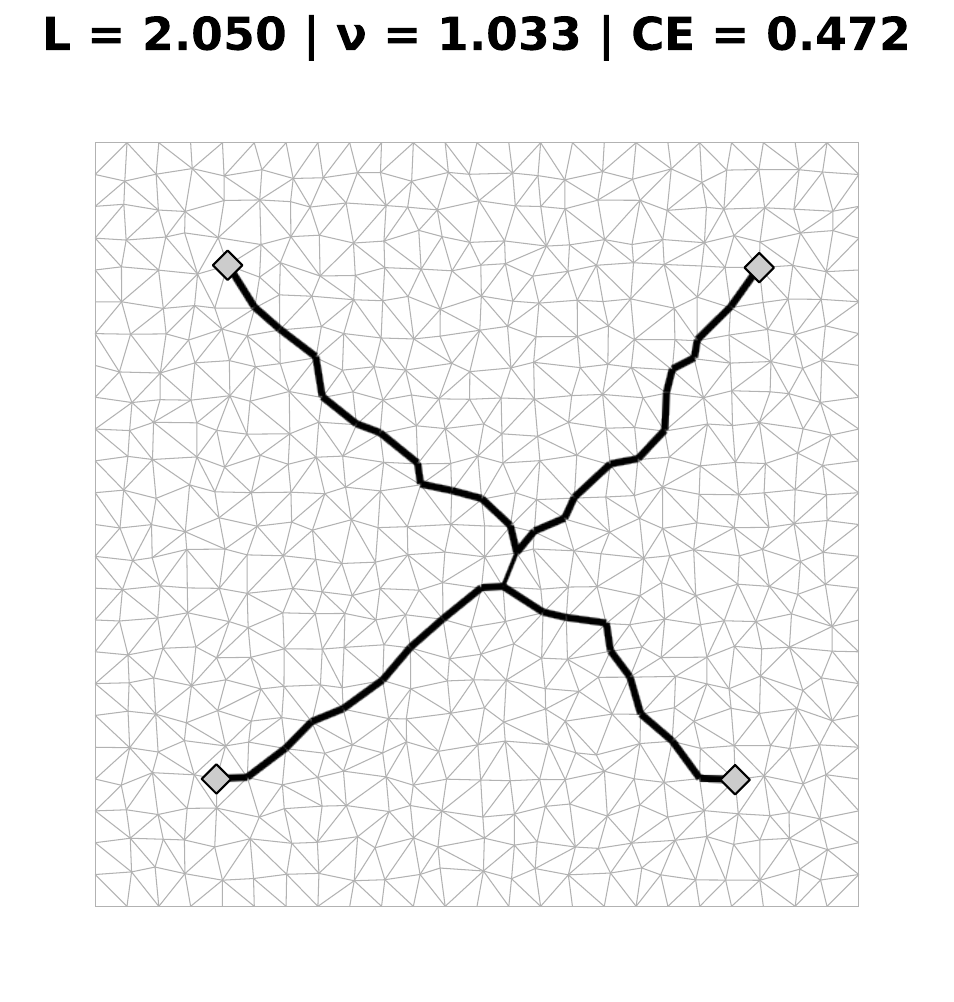}
\caption{Steady state with largest CE.}
\label{fig:square_random_half_ss_largest_CE}
\end{subfigure}
\caption{Best steady states for the square configuration, for the \textbf{Random half} algorithm.}
\label{fig:square_random_half}
\end{figure}

\begin{figure}
\centering
\begin{subfigure}[b]{0.55\textwidth}
\centering
\includegraphics[width=0.73\textwidth]{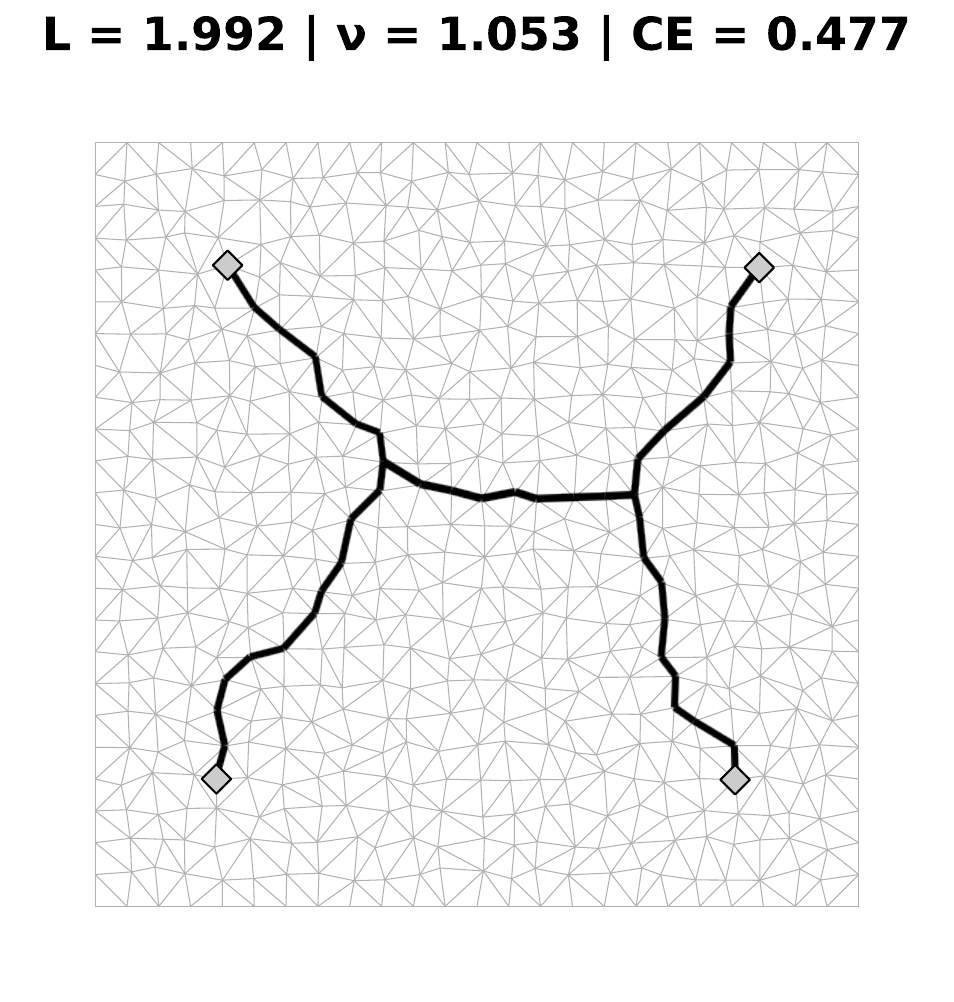}
\caption{Steady state with smallest $L$ and largest CE.}
\label{fig:square_random_source_ss_smallest_L_biggest_CE}
\end{subfigure}
\begin{subfigure}[b]{0.4\textwidth}
\centering
\includegraphics[width=\textwidth]{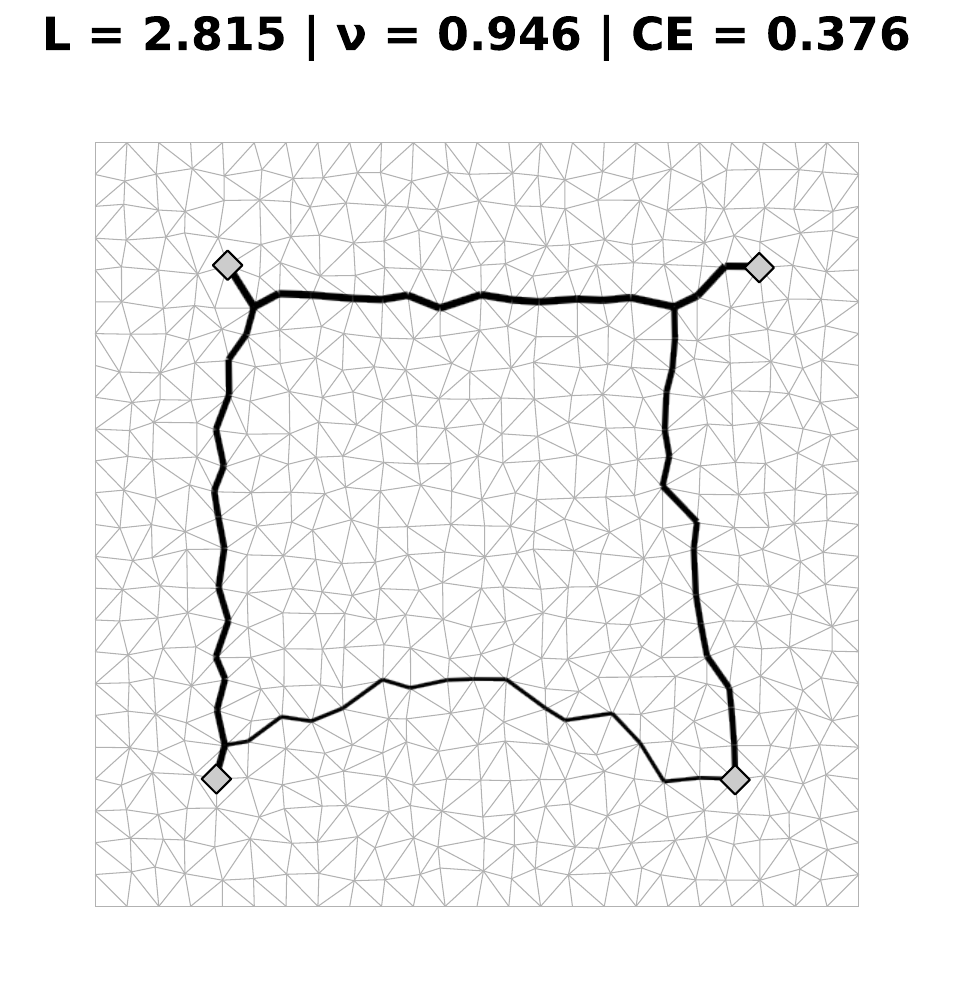}
\caption{Steady state with smallest $\nu$.}
\label{fig:square_random_source_ss_smallest_niu}
\end{subfigure}
\caption{Best steady states for the square configuration, for the \textbf{Random source} algorithm.}
\label{fig:square_random_source}
\end{figure}

\begin{figure}
\centering
\begin{subfigure}[b]{0.55\textwidth}
\centering
\includegraphics[width=0.73\textwidth]{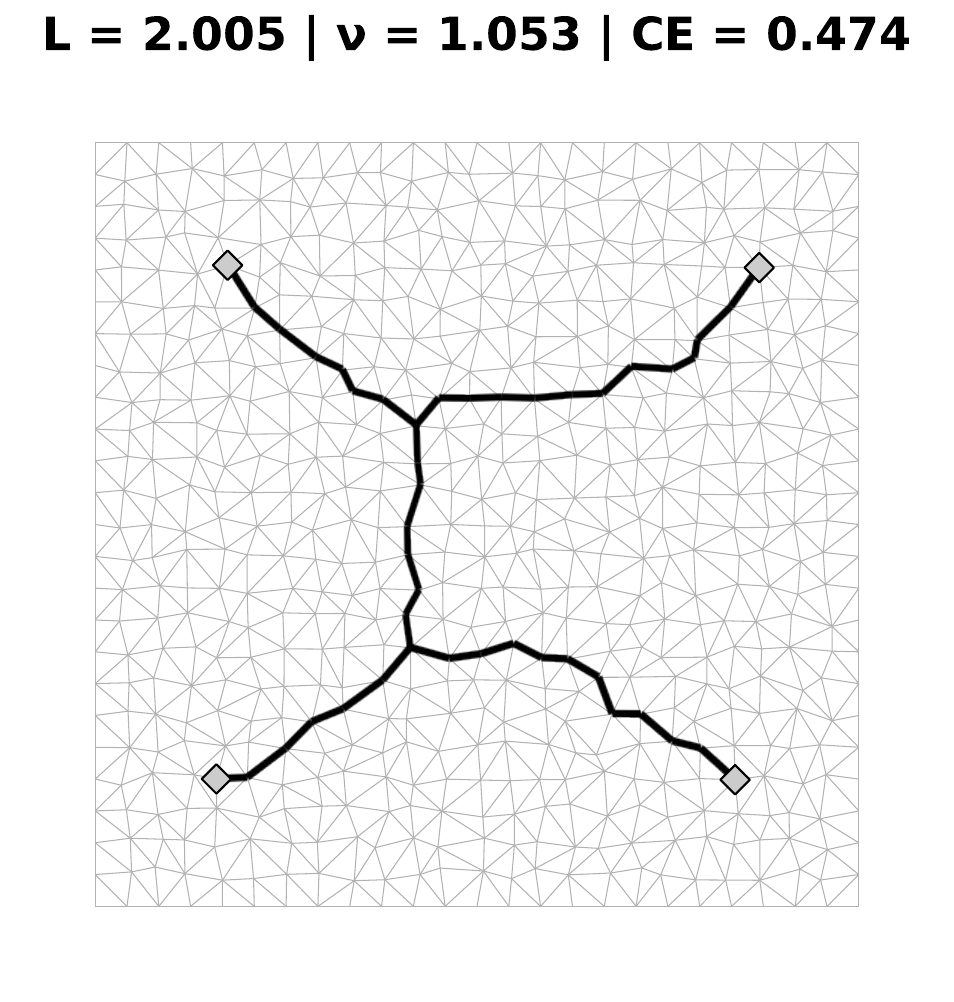}
\caption{Steady state with smallest $L$ and largest CE.}
\label{fig:square_random_fixed_I0_ss_smallest_L_biggest_CE}
\end{subfigure}
\begin{subfigure}[b]{0.4\textwidth}
\centering
\includegraphics[width=\textwidth]{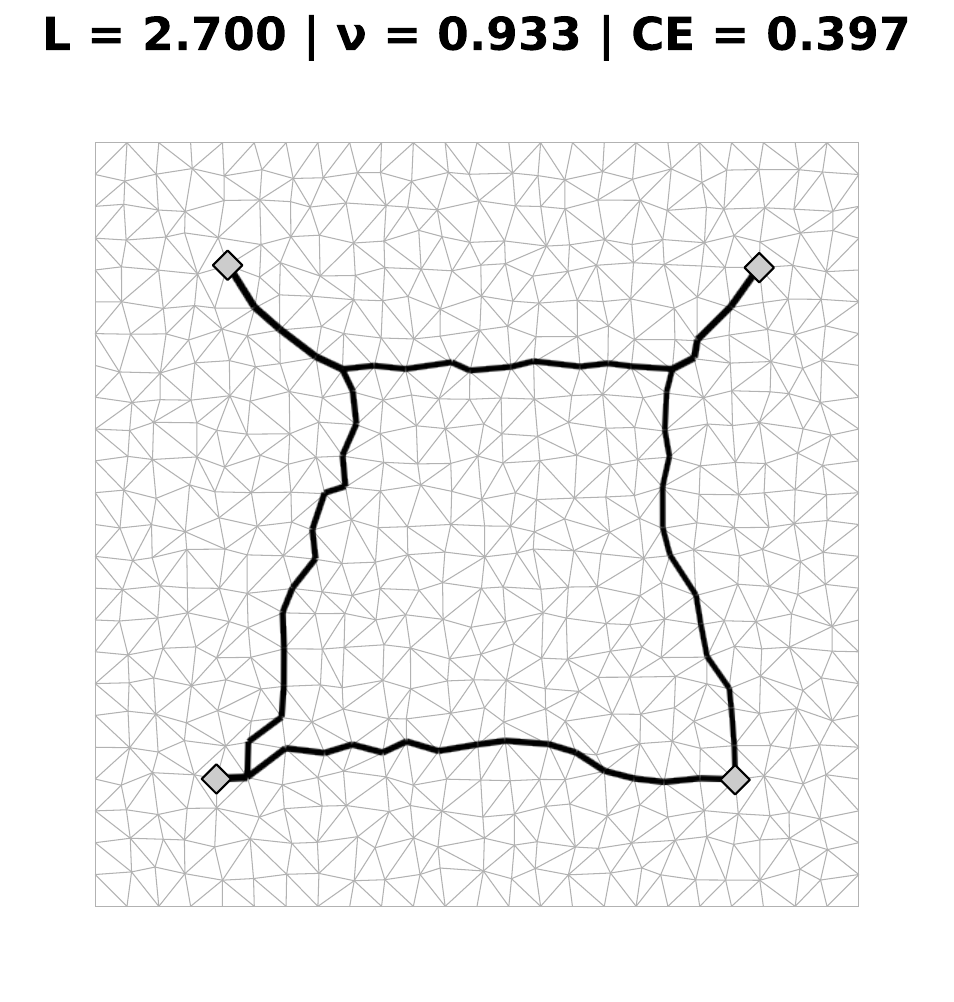}
\caption{Steady state with smallest $\nu$.}
\label{fig:square_random_fixed_I0_ss_smallest_niu}
\end{subfigure}
\caption{Best steady states for the square configuration, for the \textbf{Random random} algorithm.}
\label{fig:square_random_fixed_I0}
\end{figure}

The results for algorithms \texttt{Random pair}, \texttt{Random half}, \texttt{Random source} and \texttt{Random random} are shown in figures \ref{fig:square_random}, \ref{fig:square_random_half}, \ref{fig:square_random_source} and \ref{fig:square_random_fixed_I0} respectively. The results shown are the steady states obtained with the smallest $L$ and $\nu$ and largest CE values, and the graph of $\nu$ as a function of $L$ for all runs for all algorithms can be seen in figure \ref{fig:square_L_niu}. Additional results can be viewed in appendix \ref{appendix:square}.

\begin{figure}
\centering
\begin{subfigure}[b]{0.4\textwidth}
\centering
\includegraphics[width=\textwidth]{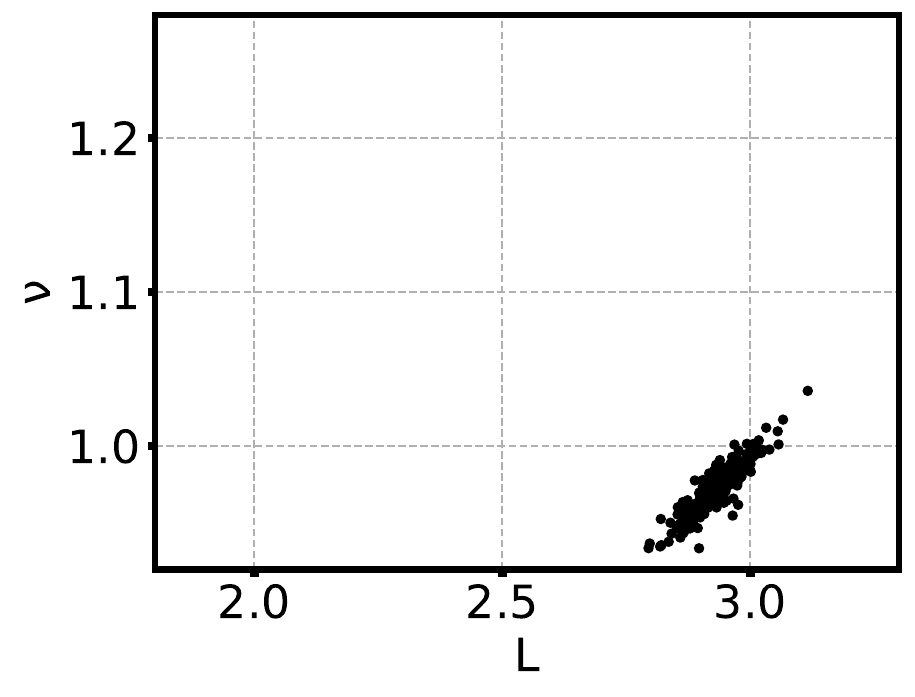}
\caption{\textbf{Random pair}.}
\label{fig:square_random_L_niu}
\end{subfigure}
\begin{subfigure}[b]{0.4\textwidth}
\centering
\includegraphics[width=\textwidth]{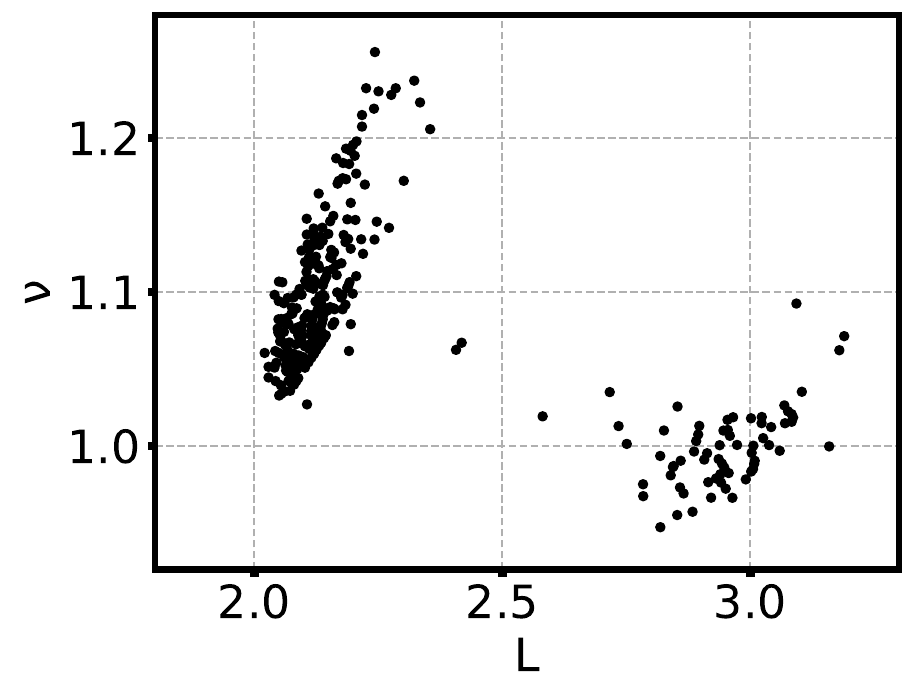}
\caption{\textbf{Random half}.}
\label{fig:square_random_half_L_niu}
\end{subfigure}
\\
\begin{subfigure}[b]{0.4\textwidth}
\centering
\includegraphics[width=\textwidth]{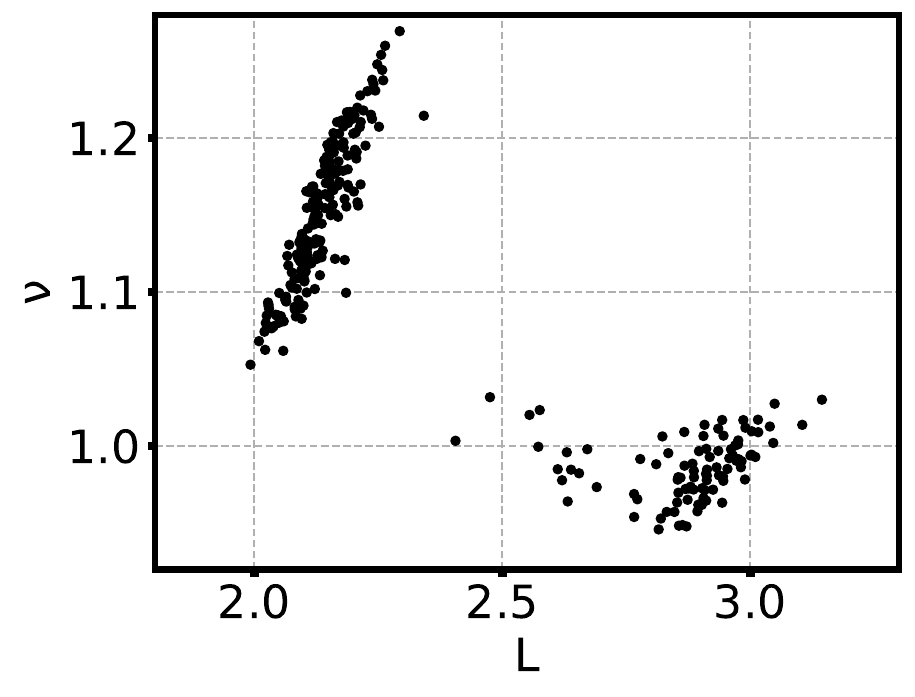}
\caption{\textbf{Random source}.}
\label{fig:square_random_source_L_niu}
\end{subfigure}
\begin{subfigure}[b]{0.4\textwidth}
\centering
\includegraphics[width=\textwidth]{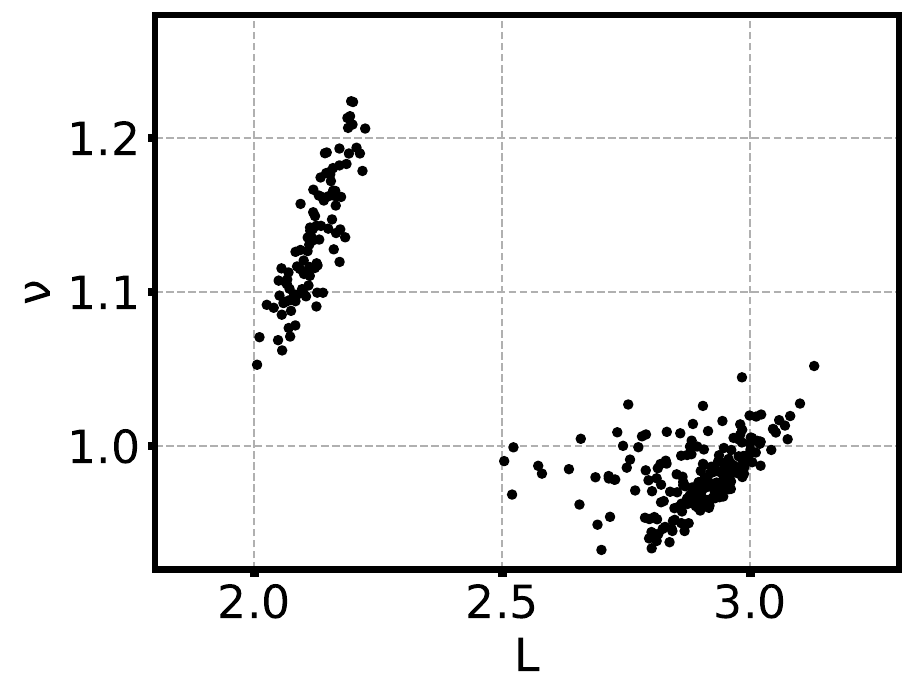}
\caption{\textbf{Random random}.}
\label{fig:square_random_fixed_I0_L_niu}
\end{subfigure}
\caption{Steady state parameters $\nu$ as a function of $L$ for the square configuration, for all algorithms.}
\label{fig:square_L_niu}
\end{figure}

Algorithm \texttt{Random pair} once again did not obtain the minimum Steiner tree for the square. The tree shape obtained was the full perimeter of the square. Once again, the variables $\nu$ and $L$ show a somewhat linear relation, but graph \ref{fig:square_random_L_niu} now displays points located in a region rather than showing a straight line of points.

The remaining algorithms, \texttt{Random half}, \texttt{Random source} and \texttt{Random random}, all obtained a tree very similar to the minimum Steiner tree of the square for the trees with minimum length and maximum CE. The trees with minimum $\nu$ were the (slightly distorted) full perimeter of the square. Once again, the graph of $\nu$ vs. $L$ shows two clusters: one representing trees with large $\nu$ and small $L$ and one representing trees with small $\nu$ and large $L$. Very few runs fall between these two regions (less than 5\% for all these algorithms).

It's relevant to note that the shortest tree is usually the tree with largest cost-efficiency value.

The algorithm that produced the shortest tree and the tree with the largest CE value was \texttt{Random source}. 
The algorithm that most frequently produced short and high CE value trees was \texttt{Random half} (see fig. \ref{fig:square_random_half_L_distribution} and table \ref{tab:square}).

The algorithm that produced the smallest $\nu$ tree was \texttt{Random random}. The algorithm that most frequently produced small $\nu$ value trees was \texttt{Random pair} (see fig. \ref{fig:square_random_niu_distribution} and table \ref{tab:square}).

These solutions could potentially be improved by using the algorithm described in section \ref{sec:steiner_points_calc} to calculate the best position for the Steiner points.

\subsubsection{Pentagonal configuration}

For $n = 5$, the sites approximately form a pentagon of side $0.55$. The theoretical value for the perimeter of the pentagon is $P_{\text{penta}} = 2.802$ (calculated using the coordinates of the sites) and for the minimum Steiner tree length it is $L_{\text{Stei}_{\text{penta}}} = 2.194$ (calculated using the Steiner points locations obtained with the equations of section \ref{sec:steiner_points_calc}). 

\begin{figure}
\centering
\begin{subfigure}[b]{0.6\textwidth}
\centering
\includegraphics[width=0.7\textwidth]{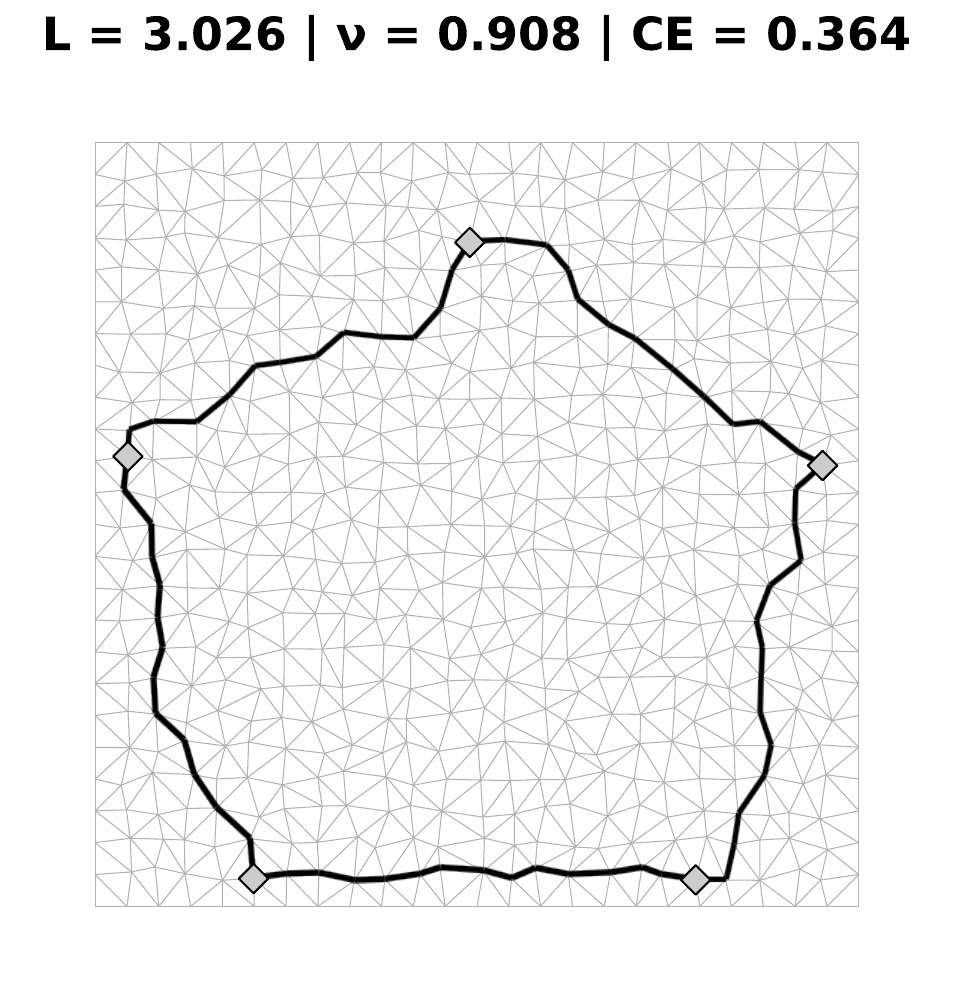}
\caption{Steady state with smallest $L$, smallest $\nu$ and largest CE.}
\label{fig:penta_random_ss_smallest_L_smallest_niu_biggest_CE}
\end{subfigure}
\caption{Best steady states for the pentagonal configuration, for the \textbf{Random pair} algorithm.}
\label{fig:penta_random}
\end{figure}

\begin{figure}
\centering
\begin{subfigure}[b]{0.55\textwidth}
\centering
\includegraphics[width=0.73\textwidth]{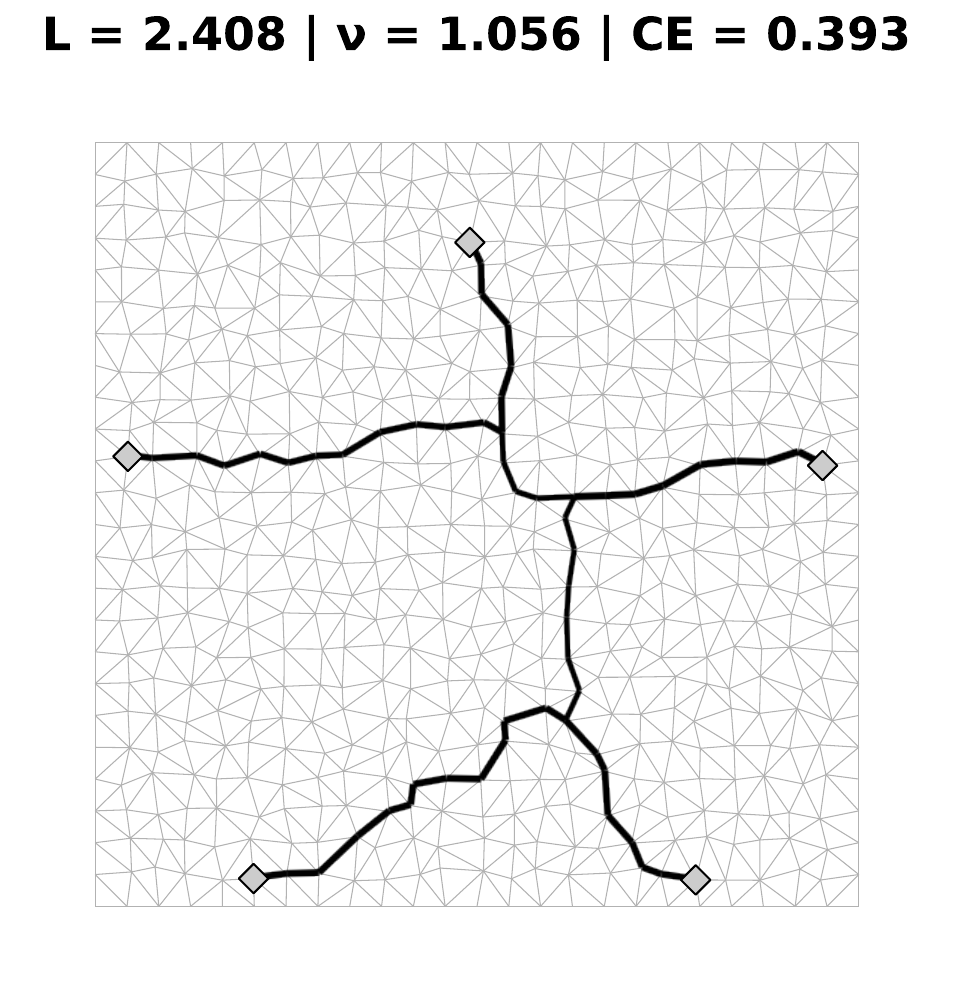}
\caption{Steady state with smallest $L$ and largest CE.}
\label{fig:penta_random_half_ss_smallest_L_largest_CE}
\end{subfigure}
\begin{subfigure}[b]{0.4\textwidth}
\centering
\includegraphics[width=\textwidth]{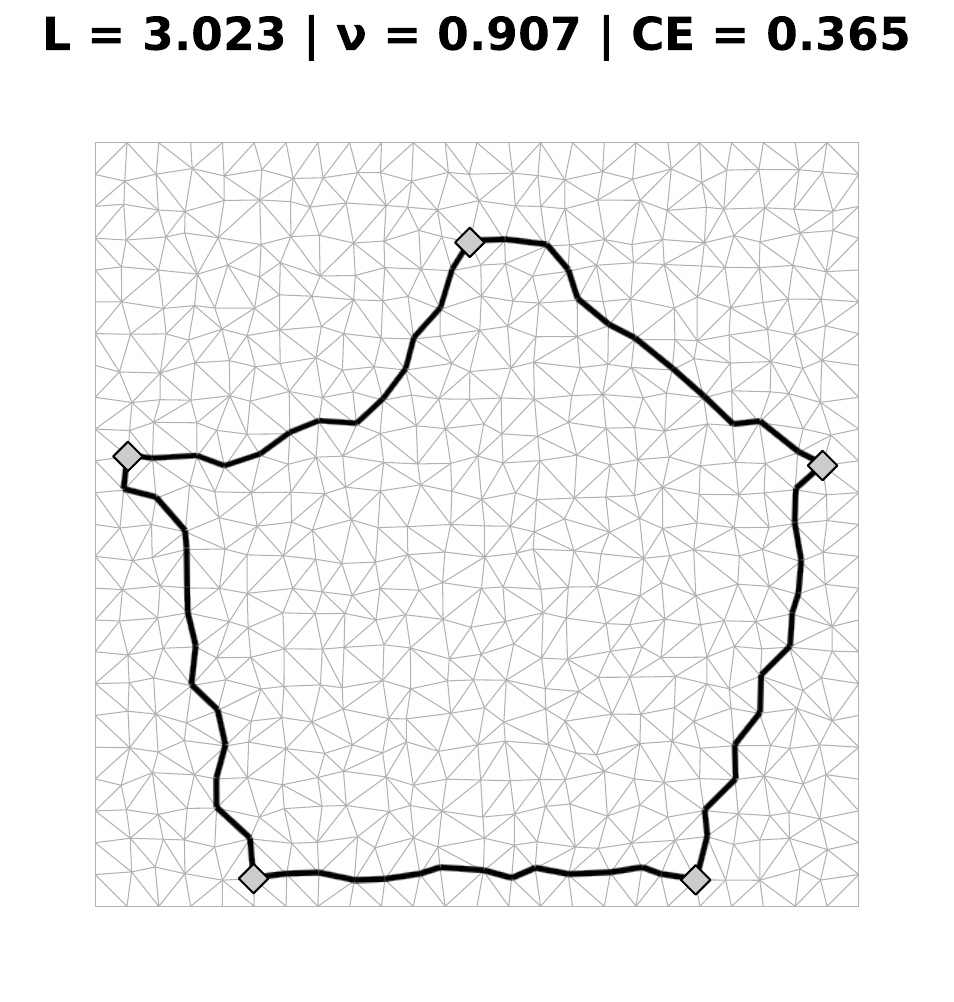}
\caption{Steady state with smallest $\nu$.}
\label{fig:penta_random_half_ss_smallest_niu}
\end{subfigure}
\caption{Best steady states for the pentagonal configuration, for the \textbf{Random half} algorithm.}
\label{fig:penta_random_half}
\end{figure}

\begin{figure}
\centering
\begin{subfigure}[b]{0.55\textwidth}
\centering
\includegraphics[width=0.73\textwidth]{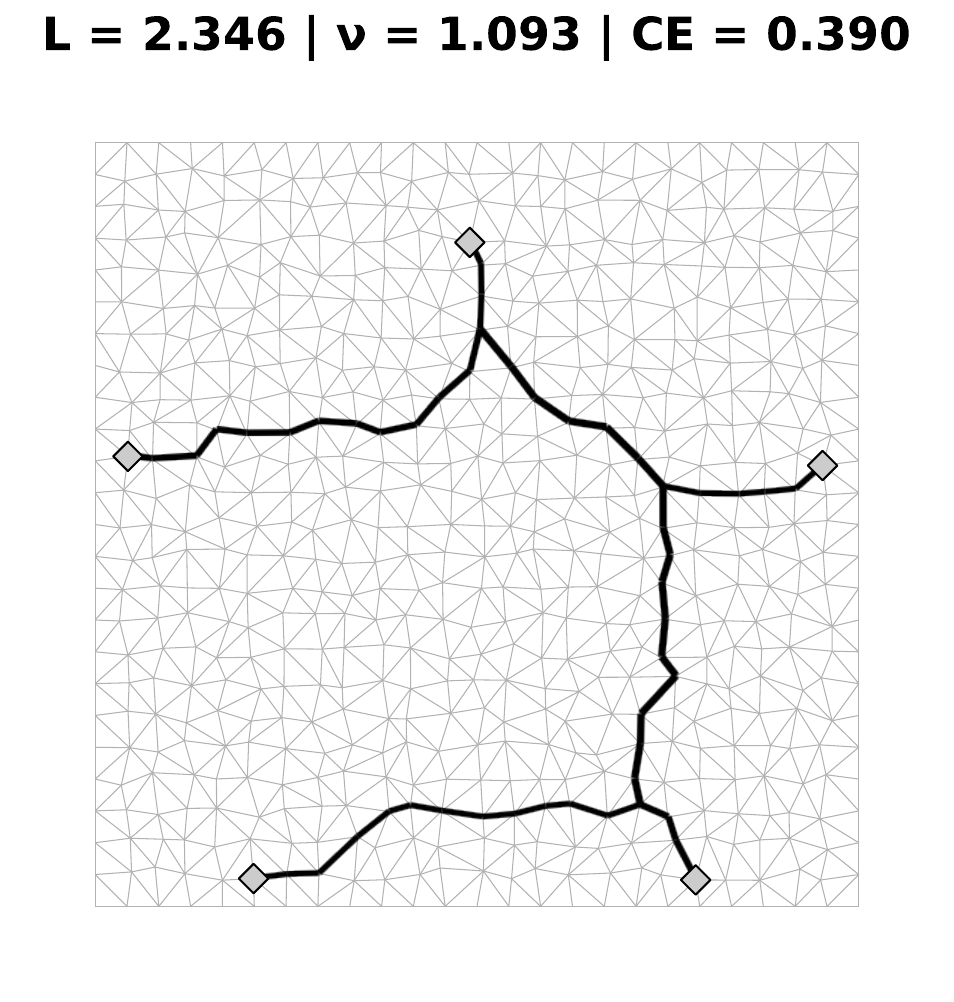}
\caption{Steady state with smallest $L$ and largest CE.}
\label{fig:penta_random_source_ss_smallest_L_largest_CE}
\end{subfigure}
\begin{subfigure}[b]{0.4\textwidth}
\centering
\includegraphics[width=\textwidth]{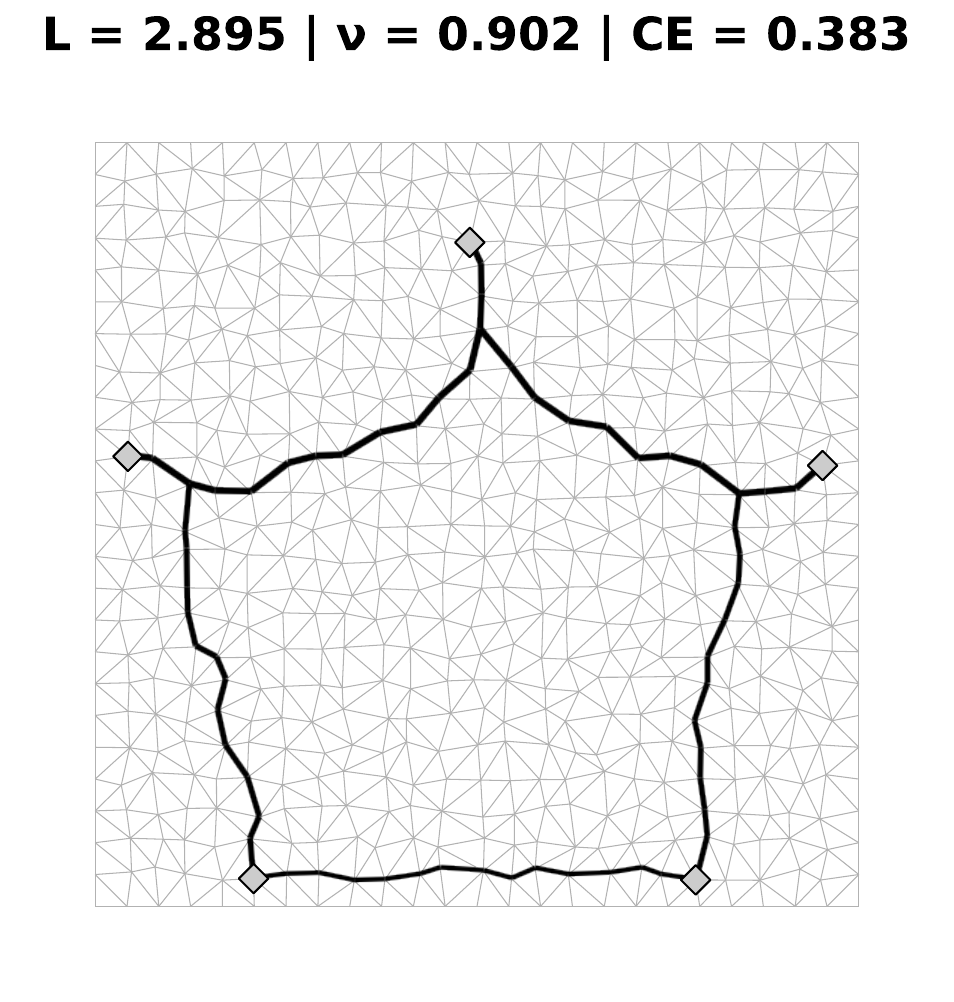}
\caption{Steady state with smallest $\nu$.}
\label{fig:penta_random_source_ss_smallest_niu}
\end{subfigure}
\caption{Best steady states for the pentagonal configuration, for the \textbf{Random source} algorithm.}
\label{fig:penta_random_source}
\end{figure}

\begin{figure}
\centering
\begin{subfigure}[b]{0.32\textwidth}
\centering
\includegraphics[width=\textwidth]{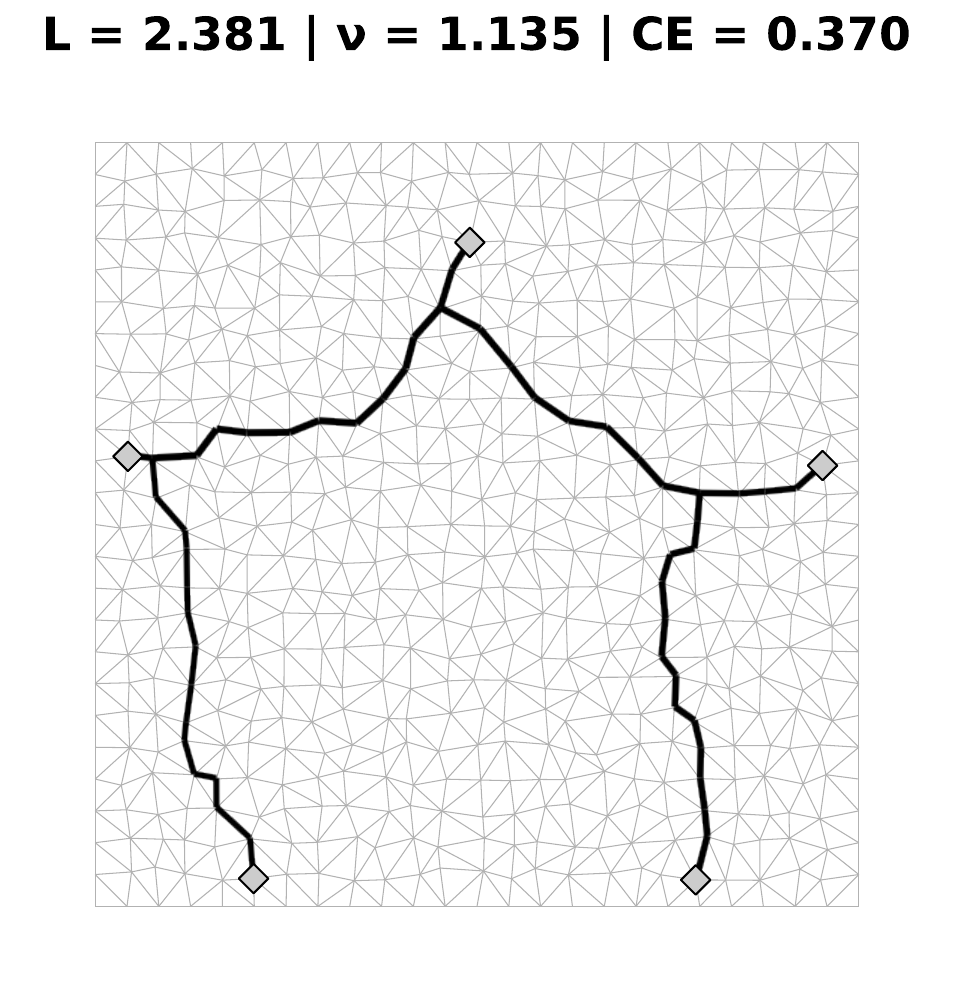}
\caption{Steady state with smallest $L$.}
\label{fig:penta_random_fixed_I0_ss_smallest_L}
\end{subfigure}
\begin{subfigure}[b]{0.32\textwidth}
\centering
\includegraphics[width=\textwidth]{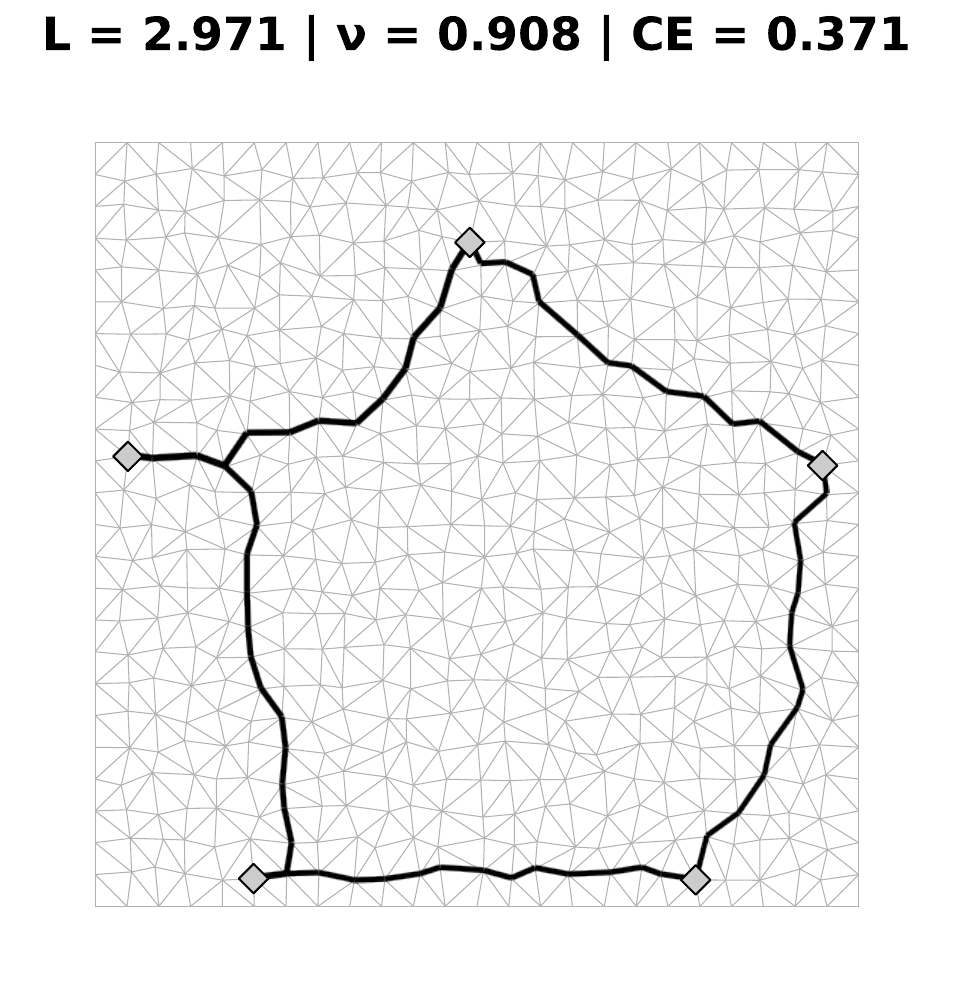}
\caption{Steady state with smallest $\nu$.}
\label{fig:penta_random_fixed_I0_ss_smallest_niu}
\end{subfigure}
\begin{subfigure}[b]{0.32\textwidth}
\centering
\includegraphics[width=\textwidth]{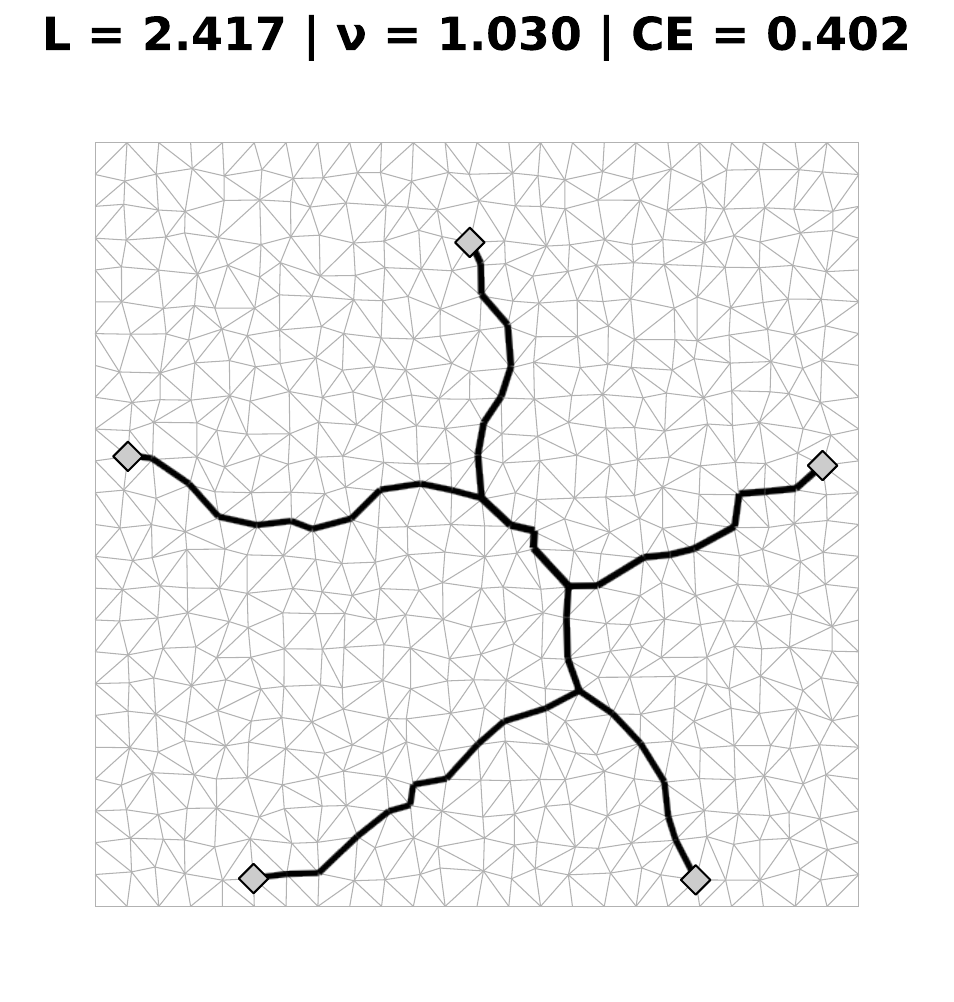}
\caption{Steady state with largest CE.}
\label{fig:penta_random_fixed_I0_ss_largest_CE}
\end{subfigure}
\caption{Best steady states for the pentagonal configuration, for the \textbf{Random random} algorithm.}
\label{fig:penta_random_fixed_I0}
\end{figure}

The results for algorithms \texttt{Random pair}, \texttt{Random half}, \texttt{Random source} and \texttt{Random random} are shown in figures \ref{fig:penta_random}, \ref{fig:penta_random_half}, \ref{fig:penta_random_source} and \ref{fig:penta_random_fixed_I0} respectively. The results shown are the steady states obtained with smallest $L$ and $\nu$ values and largest CE value, and the graph of $\nu$ as a function of $L$ for all runs for all algorithms can be see in figure \ref{fig:penta_L_niu}. Additional results can be viewed in appendix \ref{appendix:penta}.

\begin{figure}
\centering
\begin{subfigure}[b]{0.4\textwidth}
\centering
\includegraphics[width=\textwidth]{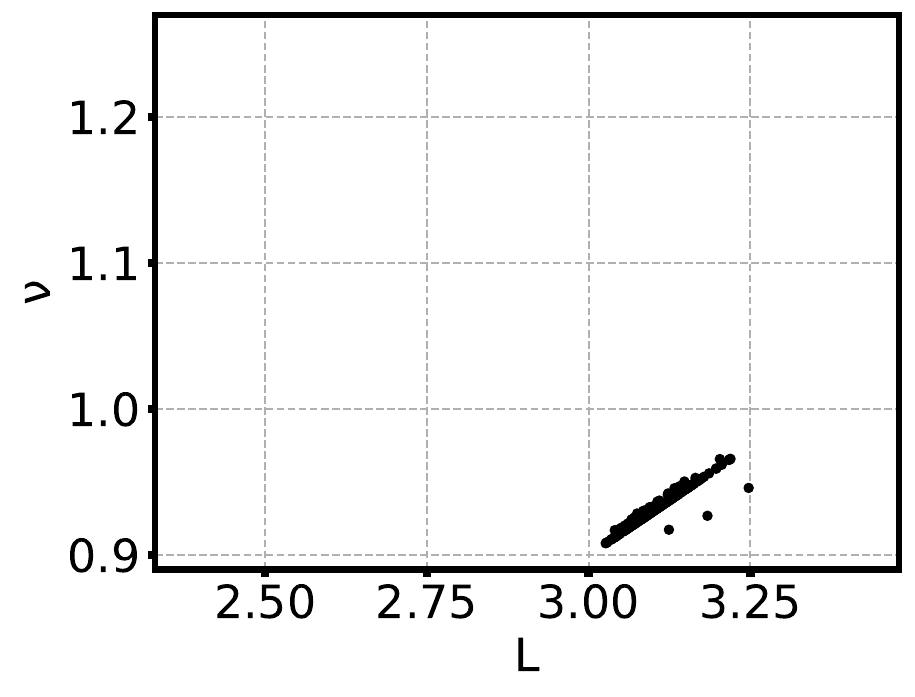}
\caption{\textbf{Random pair}.}
\label{fig:penta_random_L_niu}
\end{subfigure}
\begin{subfigure}[b]{0.4\textwidth}
\centering
\includegraphics[width=\textwidth]{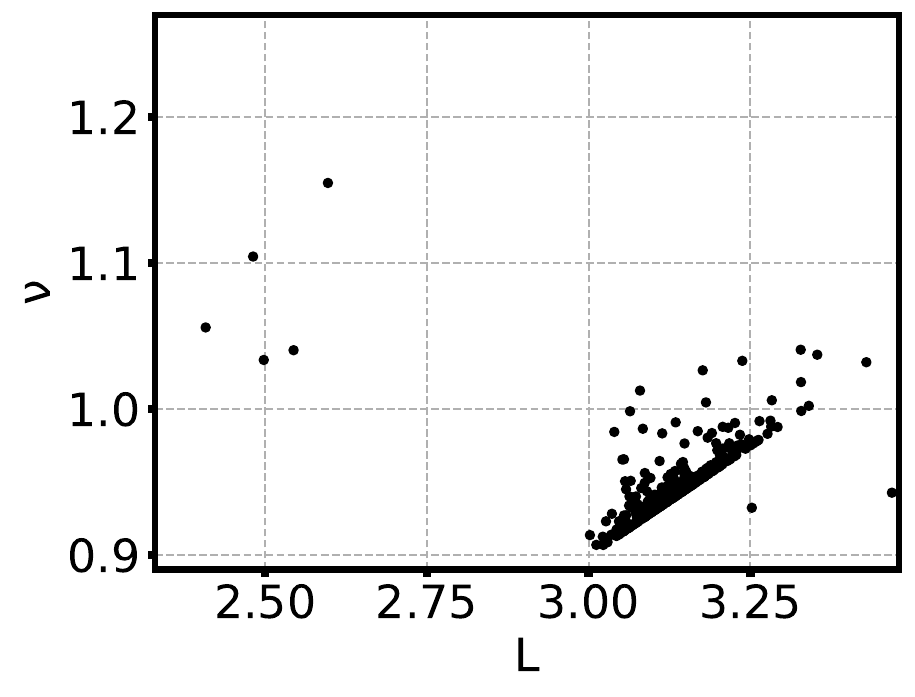}
\caption{\textbf{Random half}.}
\label{fig:penta_random_half_L_niu}
\end{subfigure}
\\
\begin{subfigure}[b]{0.4\textwidth}
\centering
\includegraphics[width=\textwidth]{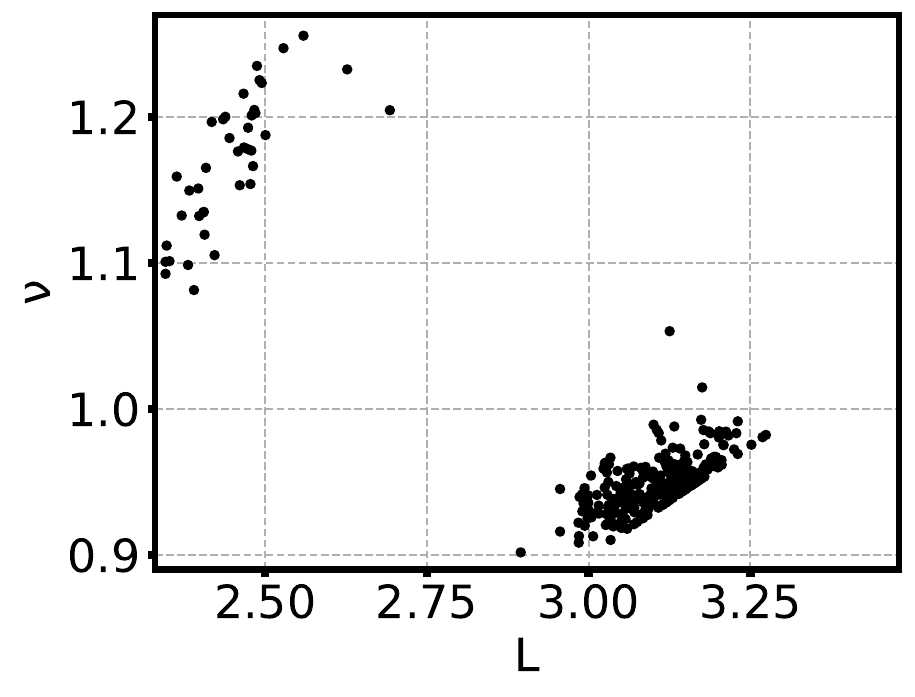}
\caption{\textbf{Random source}.}
\label{fig:penta_random_source_L_niu}
\end{subfigure}
\begin{subfigure}[b]{0.4\textwidth}
\centering
\includegraphics[width=\textwidth]{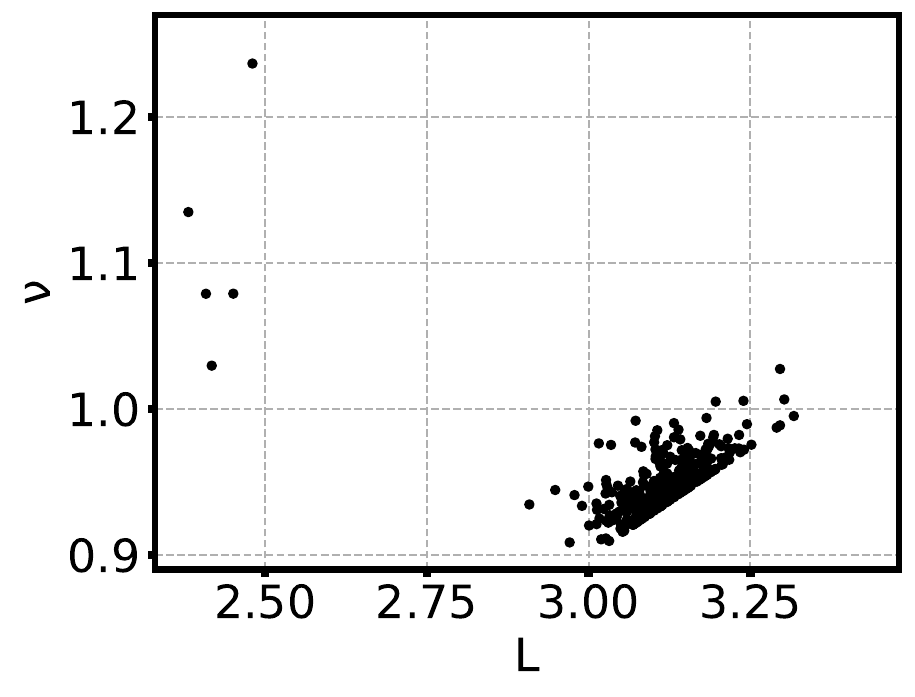}
\caption{\textbf{Random random}.}
\label{fig:penta_random_fixed_I0_L_niu}
\end{subfigure}
\caption{Steady state parameters $\nu$ as a function of $L$ for the pentagonal configuration, for all algorithms.}
\label{fig:penta_L_niu}
\end{figure}

Algorithm \texttt{Random pair} once again did not obtain the minimum Steiner tree for the pentagon. The result obtained was the full perimeter of the pentagon. Once more, the variables $\nu$ and $L$ show a linear relation.

Algorithms \texttt{Random half} and \texttt{Random random} obtained a tree very similar to the minimum Steiner tree of the pentagon for the tree with minimum length for the former and for the tree with maximum CE for the latter, while algorithm \texttt{Random source} obtained a tree which looks like a Steiner tree (right side) conjoined with the perimeter of the pentagon (left side) for the shortest length tree. 
The tree with minimum $\nu$ for algorithm \texttt{Random half} was the (distorted) full perimeter of the square, while for algorithms \texttt{Random source} and \texttt{Random random} it was a mix of the Steiner tree of the pentagon with the perimeter of the pentagon. Once again, the graph of $\nu$ vs. $L$ shows two clusters for all these algorithms: one representing trees with large $\nu$ and small $L$ and one representing trees with small $\nu$ and large $L$.

The algorithm that produced the shortest tree was \texttt{Random source}. The algorithm that most frequently produced short trees was \texttt{Random source} as well (see fig. \ref{fig:penta_random_source_L_distribution} and table \ref{tab:penta}). 

The algorithm that produced the smallest $\nu$ tree was \texttt{Random source}. The algorithm that most frequently produced small $\nu$ value trees was \texttt{Random pair} (see fig. \ref{fig:penta_random_niu_distribution} and table \ref{tab:penta}). 

The algorithm that produced the tree with the largest CE value was \texttt{Random random}. The algorithm that most frequently produced high CE value trees was \texttt{Random pair} (see fig. \ref{fig:penta_random_L_distribution} and table \ref{tab:penta}).

\subsubsection{Remarks about polygonal studies}

Using the adaptive H-P model studied in this thesis, the algorithms studied in section \ref{sec:polygonal_studies} were all able to obtain Steiner minimal trees for all polygonal configurations, with the exception of the \texttt{Random pair} algorithm that was unable to produce a Steiner-like tree.

The tree shape obtained that corresponded to the smallest $\nu$ value, thus the highest efficiency, was mostly the perimeter of the polygon (often including a Steiner point near some of the sites). The tree shape obtained that corresponded to the smallest $L$ value, thus the lowest cost, was mostly the Steiner minimal tree-like result. The tree shape obtained that corresponded to the largest CE value was usually the same as the smallest $L$ value one, that is, the Steiner minimal tree-like result.

The algorithm that produced the shortest trees (smallest $L$ value) was consistently \texttt{Random source} for all configurations. 
The algorithms that produced the most efficient trees (smallest $\nu$ value) were \texttt{Random source} for $n = \{3,5\}$ and \texttt{Random random} for $n = \{4\}$.
The algorithms that produced the trees with the highest cost-efficiency values were \texttt{Random source} for $n = \{3,4\}$ and \texttt{Random random} for $n = \{5\}$.

The algorithms that produced Steiner tree-like shapes more frequently were the \texttt{Random half} algorithm (for the square configuration) and the \texttt{Random source} algorithm (for the pentagonal configuration). 
For the triangular configuration, algorithms \texttt{Random source} and \texttt{Random random} were equally likely to produce Steiner-like trees. 
This is observed by analyzing the length distributions present in appendix \ref{appendix:steiner_geometric}.

\textit{Physarum}, as seen in figure \ref{fig:Nakagaki_physarum_steiner}, is able to produce Steiner minimal tree-like networks, even though it does not produce one every single time. This is exactly what was observed for the simulations of section \ref{sec:polygonal_studies}: while the minimal length steady state trees were obtained a few times for each configuration, they were not obtained every time. This is because the steady state tree shape is heavily dependent on the initial conditions of the conductivities $D_{ij}(t=0)$, as first seen in several simulations of chapter \ref{chapter:properties}. Similarly, \textit{Physarum}'s networks are heavily dependent on external stimuli (like food sources and light sources, for example). Figure \ref{fig:Nakagaki_physarum_steiner} shows a \textit{Physarum} specimen growing on an agar-covered surface; many external stimuli could lead \textit{Physarum} to construct different shapes of its network, like non-uniformities in the agar, for example.

\subsection{Optimizing paths for communication systems}

As an example, the algorithms described will now be used to approximate the mainland Portuguese railroad transport network. This is a continuation of the work started in \cite{rodrigo_tese}, and thus we will use many of the same resources (like the original mesh, railway tree and MST) as the previous work.

\begin{figure}
\centering
\begin{subfigure}[b]{0.37\textwidth}
\centering
\includegraphics[width=\textwidth]{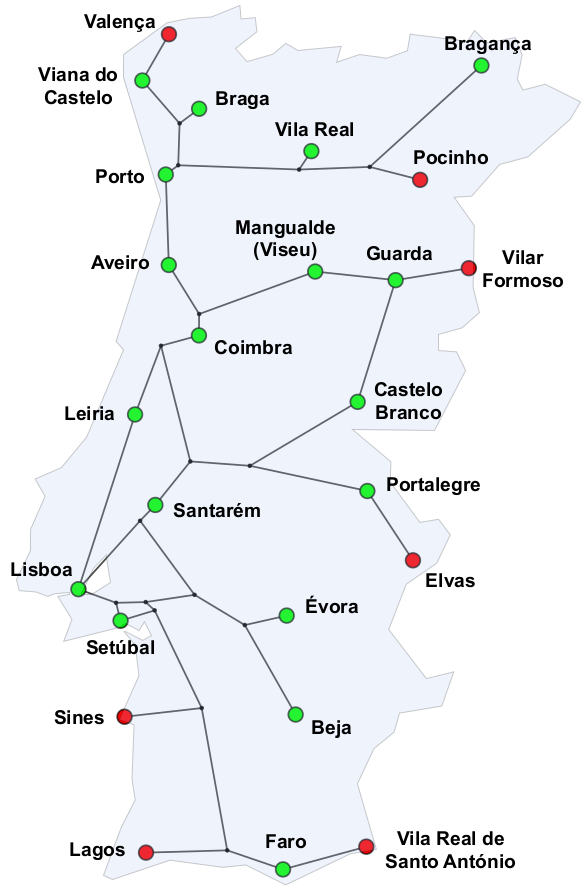}
\caption{Railway with stations.}
\label{fig:Portugal_railway_w_stations}
\end{subfigure}
\begin{subfigure}[b]{0.28\textwidth}
\centering
\includegraphics[width=\textwidth]{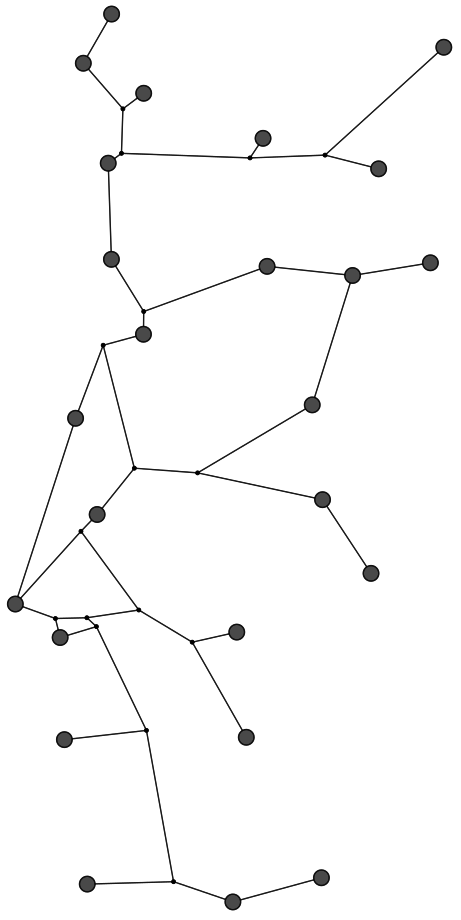}
\caption{Railway.}
\label{fig:Portugal_railway}
\end{subfigure}
\begin{subfigure}[b]{0.28\textwidth}
\centering
\includegraphics[width=\textwidth]{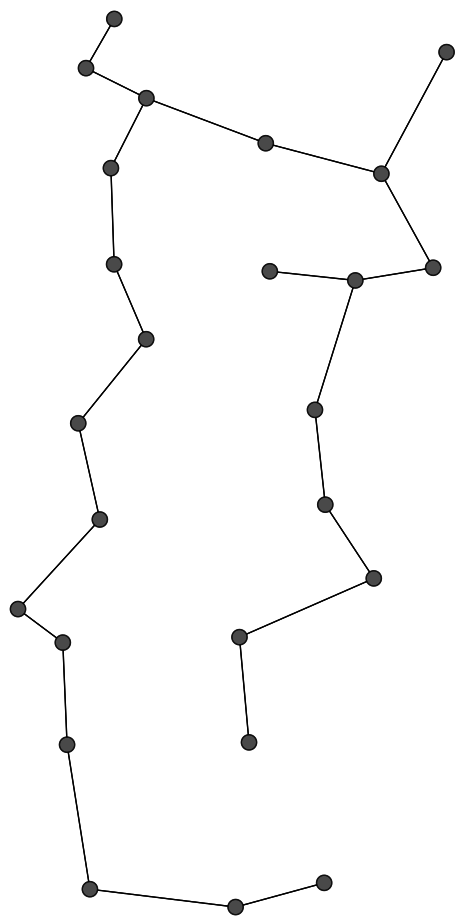}
\caption{MST.}
\label{fig:Portugal_MST}
\end{subfigure}
\caption{\textbf{a)} Approximation of part of the railway of mainland Portugal that shows all 18 district capitals (in green) (note that Viseu's district capital was replaced by Mangualde) and 5 additional cities (in red). Some rail lines are no longer active but are included here for consistency. \textbf{b)} Tree that represents Portugal's railway system. \textbf{c)} Minimum spanning tree that connects all railway sites. Images adapted from \cite{rodrigo_tese}.}
\label{fig:Portugal_railway_MST}
\end{figure}

The mesh used was first described in \cite{rodrigo_tese}. It is shaped like mainland Portugal and is made up of 1005 nodes and 2817 edges resultant of triangulation. We chose 25 sites of interest, which are the 18 Portuguese district capitals and 7 other major cities. These sites and the real railway tree that connects them can be seen in figure \ref{fig:Portugal_railway_MST}, as well as the MST that connects all sites. The length of each edge $L_{ij}$ used was the geodesic distance between nodes $i$ and $j$, which was calculated approximately using the Haversine formula.

The parameters used were: $\beta = 1$, $\sum_{i\in\text{active sources}}S_i = 1$, $V = 100$, $\Delta \tau = 0.1$ and $N_{\text{runs}} = 300$.

\begin{figure}
\centering
\begin{subfigure}[b]{0.32\textwidth}
\centering
\includegraphics[width=\textwidth]{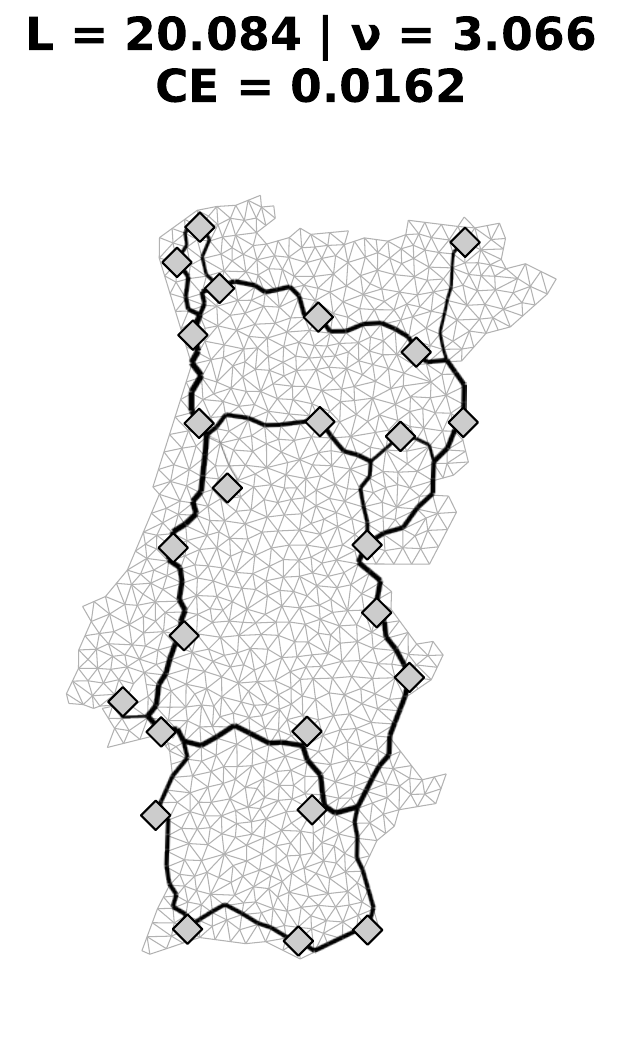}
\caption{Steady state with smallest $L$.}
\label{fig:Portugal_random_ss_smallest_L}
\end{subfigure}
\begin{subfigure}[b]{0.32\textwidth}
\centering
\includegraphics[width=\textwidth]{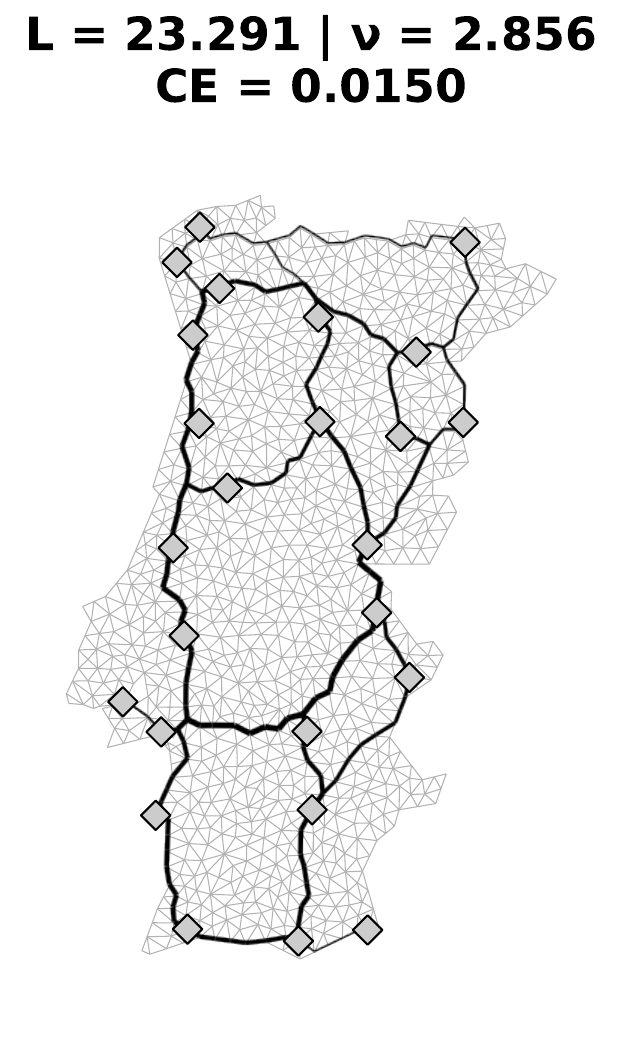}
\caption{Steady state with smallest $\nu$.}
\label{fig:Portugal_random_ss_smallest_niu}
\end{subfigure}
\begin{subfigure}[b]{0.32\textwidth}
\centering
\includegraphics[width=\textwidth]{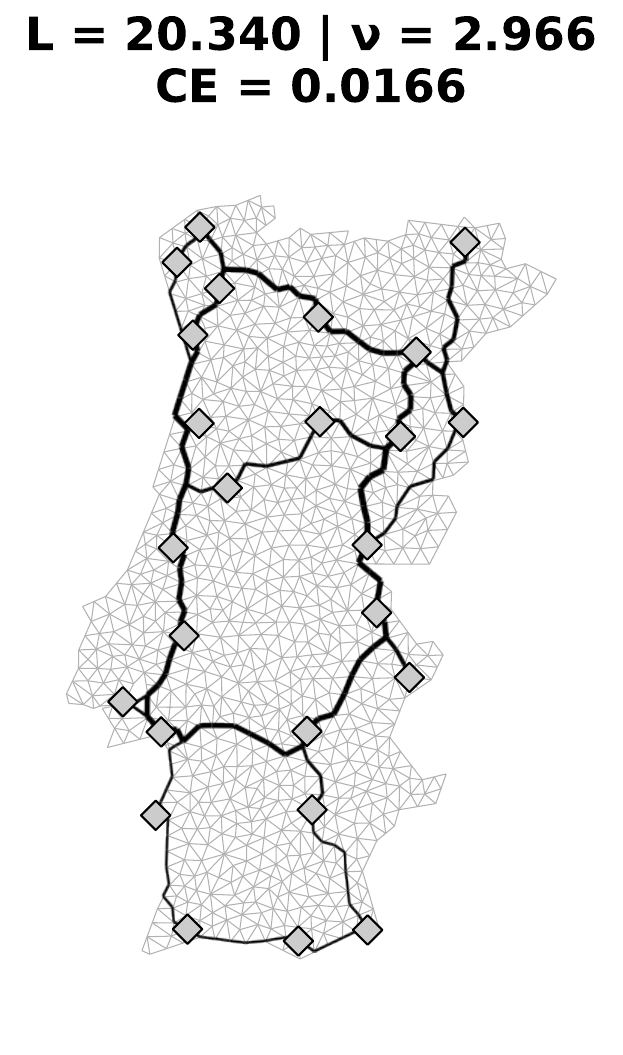}
\caption{Steady state with largest CE.}
\label{fig:Portugal_random_ss_largest_CE}
\end{subfigure}
\caption{Best steady states for the Portugal configuration, for the \textbf{Random pair} algorithm.}
\label{fig:Portugal_random}
\end{figure}

\begin{figure}
\centering
\begin{subfigure}[b]{0.32\textwidth}
\centering
\includegraphics[width=\textwidth]{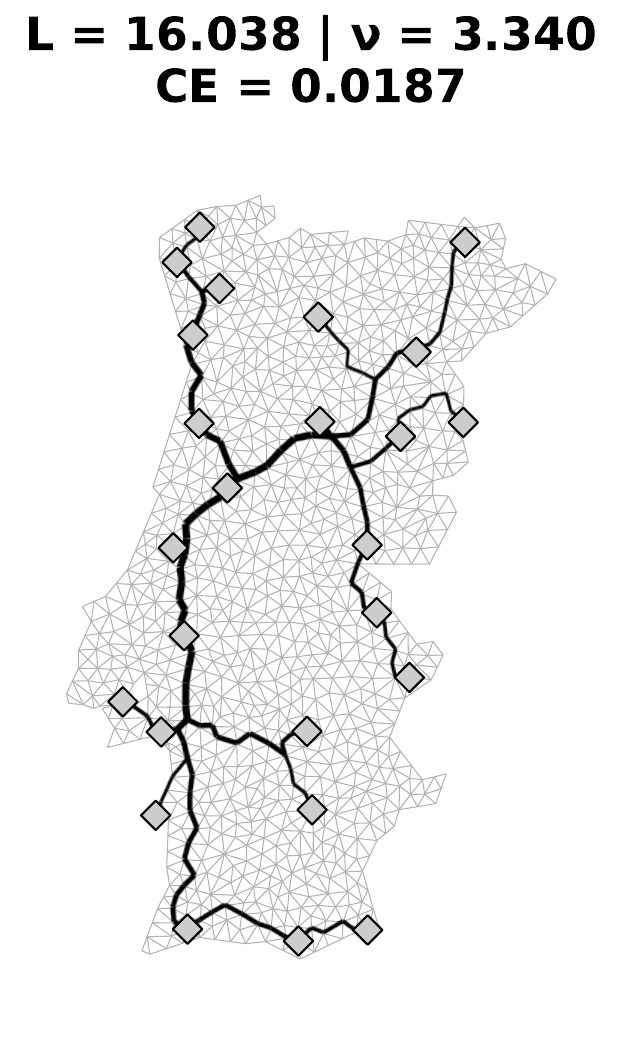}
\caption{Steady state with smallest $L$.}
\label{fig:Portugal_random_source_ss_smallest_L}
\end{subfigure}
\begin{subfigure}[b]{0.32\textwidth}
\centering
\includegraphics[width=\textwidth]{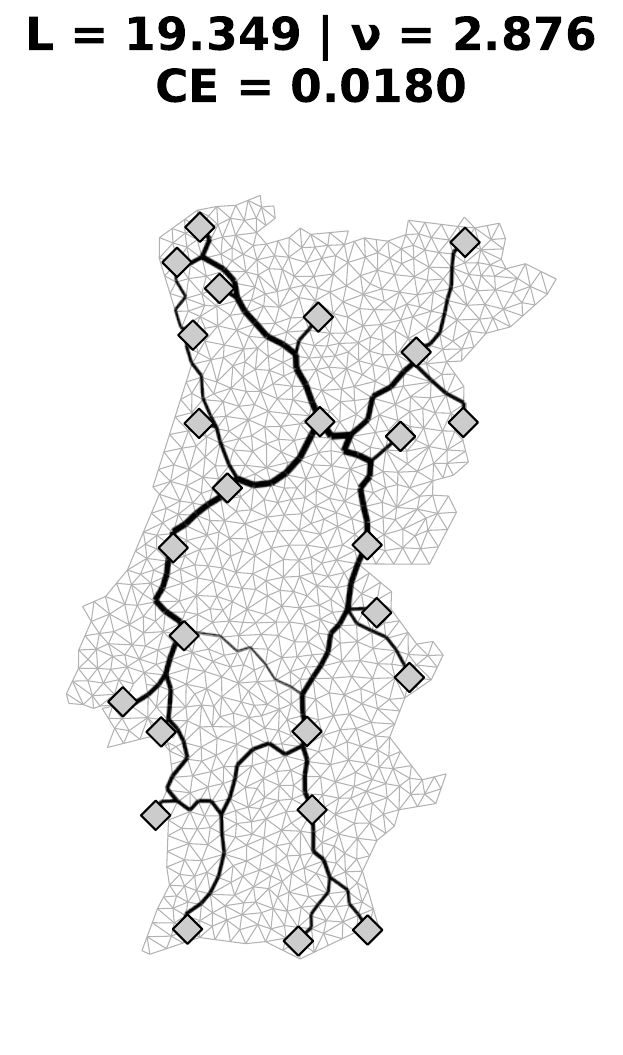}
\caption{Steady state with smallest $\nu$.}
\label{fig:Portugal_random_source_ss_smallest_niu}
\end{subfigure}
\begin{subfigure}[b]{0.32\textwidth}
\centering
\includegraphics[width=\textwidth]{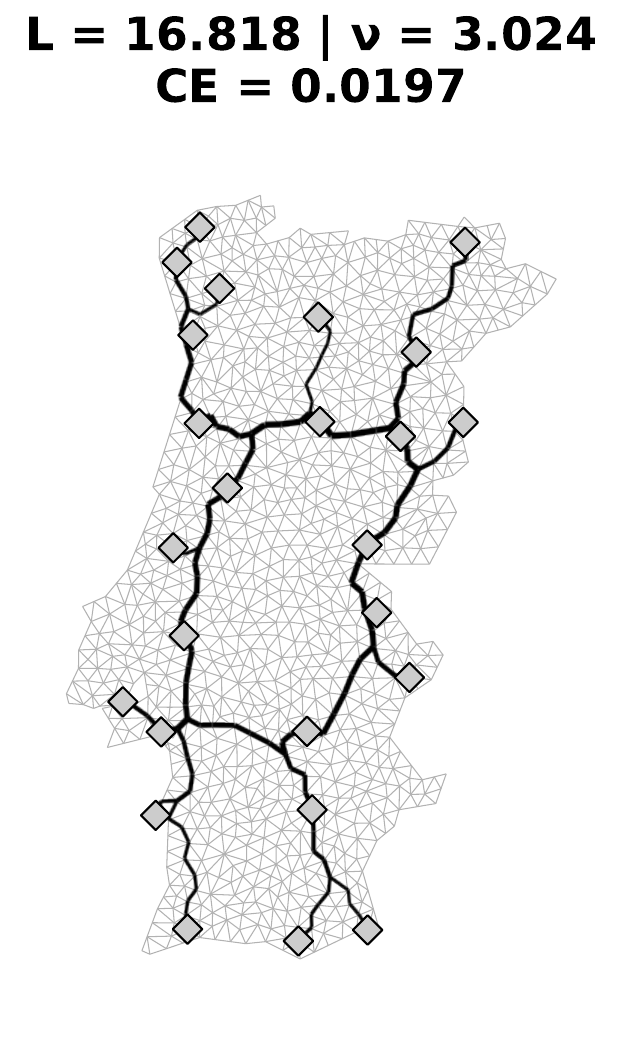}
\caption{Steady state with largest CE.}
\label{fig:Portugal_random_source_ss_largest_CE}
\end{subfigure}
\caption{Best steady states for the Portugal configuration, for the \textbf{Random source} algorithm.}
\label{fig:Portugal_random_source}
\end{figure}

\begin{figure}
\centering
\begin{subfigure}[b]{0.32\textwidth}
\centering
\includegraphics[width=\textwidth]{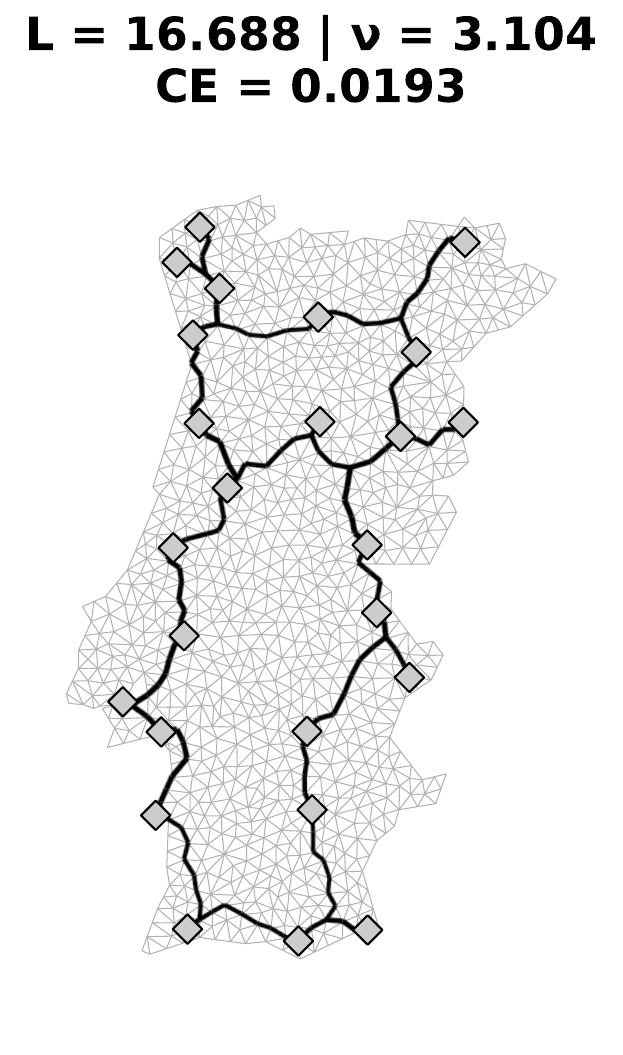}
\caption{Steady state with smallest $L$.}
\label{fig:Portugal_random_half_ss_smallest_L}
\end{subfigure}
\begin{subfigure}[b]{0.32\textwidth}
\centering
\includegraphics[width=\textwidth]{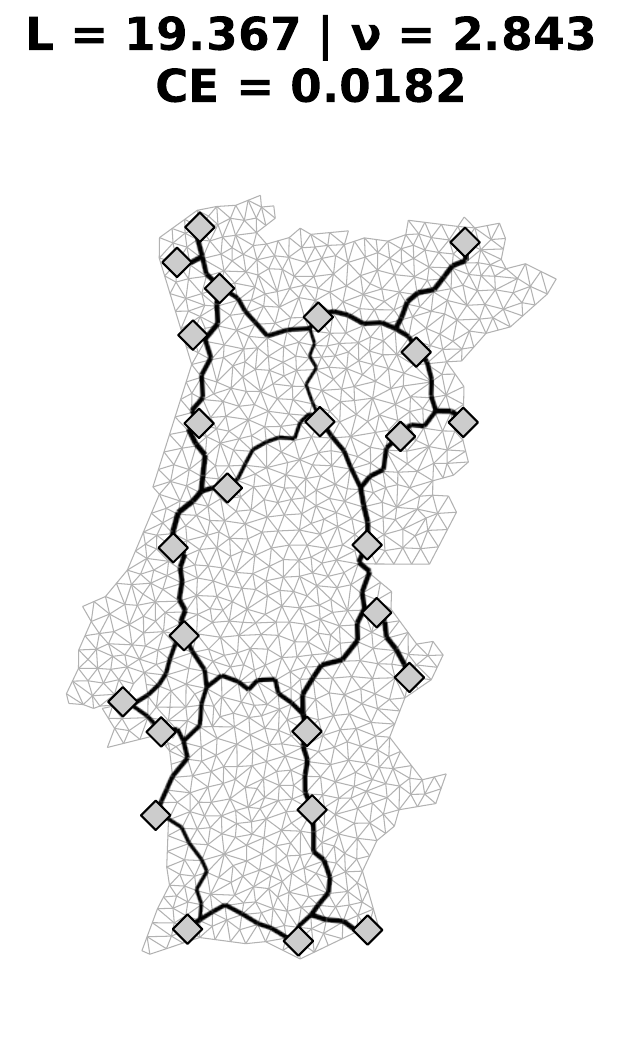}
\caption{Steady state with smallest $\nu$.}
\label{fig:Portugal_random_half_ss_smallest_niu}
\end{subfigure}
\begin{subfigure}[b]{0.32\textwidth}
\centering
\includegraphics[width=\textwidth]{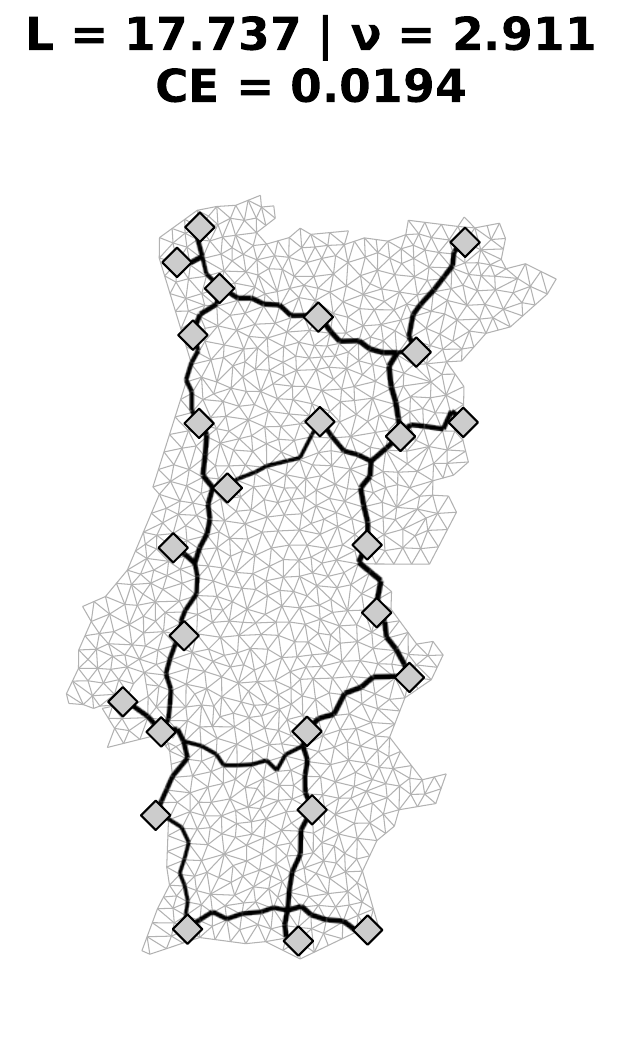}
\caption{Steady state with largest CE.}
\label{fig:Portugal_random_half_ss_largest_CE}
\end{subfigure}
\caption{Best steady states for the Portugal configuration, for the \textbf{Random half} algorithm.}
\label{fig:Portugal_random_half}
\end{figure}

\begin{figure}
\centering
\begin{subfigure}[b]{0.32\textwidth}
\centering
\includegraphics[width=\textwidth]{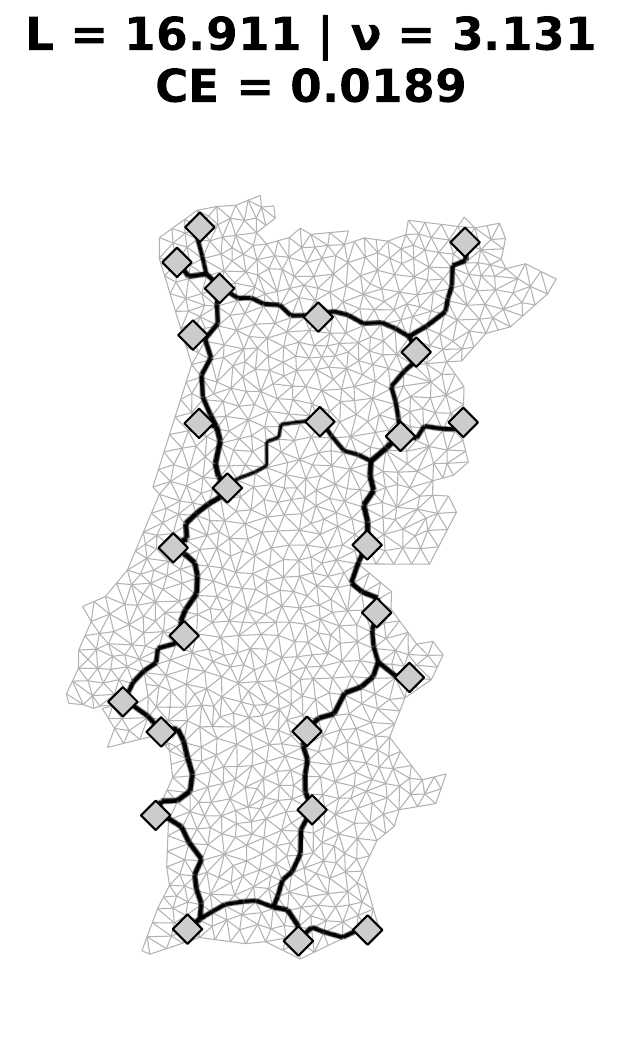}
\caption{Steady state with smallest $L$.}
\label{fig:Portugal_random_fixed_I0_ss_smallest_L}
\end{subfigure}
\begin{subfigure}[b]{0.32\textwidth}
\centering
\includegraphics[width=\textwidth]{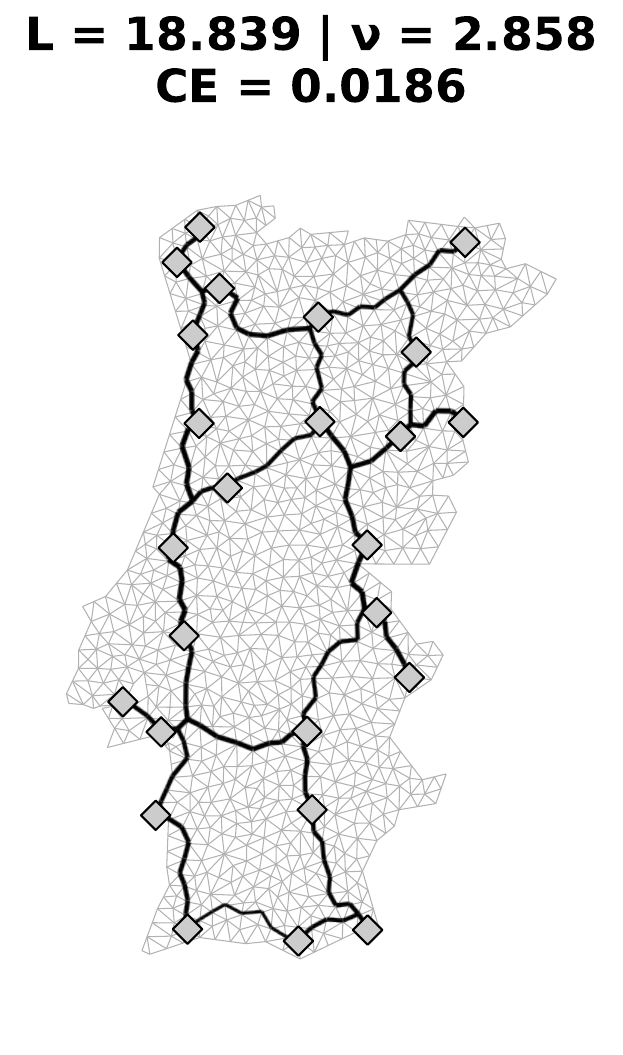}
\caption{Steady state with smallest $\nu$.}
\label{fig:Portugal_random_fixed_I0_ss_smallest_niu}
\end{subfigure}
\begin{subfigure}[b]{0.32\textwidth}
\centering
\includegraphics[width=\textwidth]{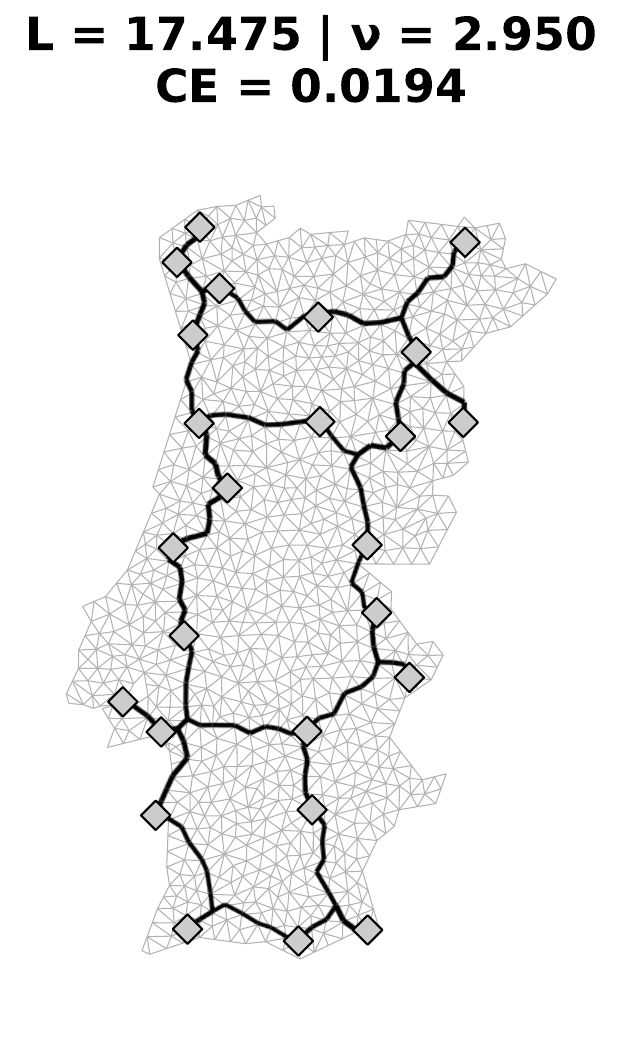}
\caption{Steady state with largest CE.}
\label{fig:Portugal_random_fixed_I0_ss_largest_CE}
\end{subfigure}
\caption{Best steady states for the Portugal configuration, for the \textbf{Random random} algorithm.}
\label{fig:Portugal_random_fixed_I0}
\end{figure}

The results for algorithms \texttt{Random pair}, \texttt{Random half}, \texttt{Random source} and \texttt{Random random} are shown in figures \ref{fig:Portugal_random}, \ref{fig:Portugal_random_half}, \ref{fig:Portugal_random_source} and \ref{fig:Portugal_random_fixed_I0} respectively. The results shown are the steady states obtained with smallest $L$, $\nu$ and CE values, and the graph of steady state parameters $\nu$ as a function of $L$ for all runs (figure \ref{fig:Portugal_L_niu}). Additional results can be seen in appendix \ref{appendix:portugal}.

\begin{figure}
\centering
\begin{subfigure}[b]{0.4\textwidth}
\centering
\includegraphics[width=\textwidth]{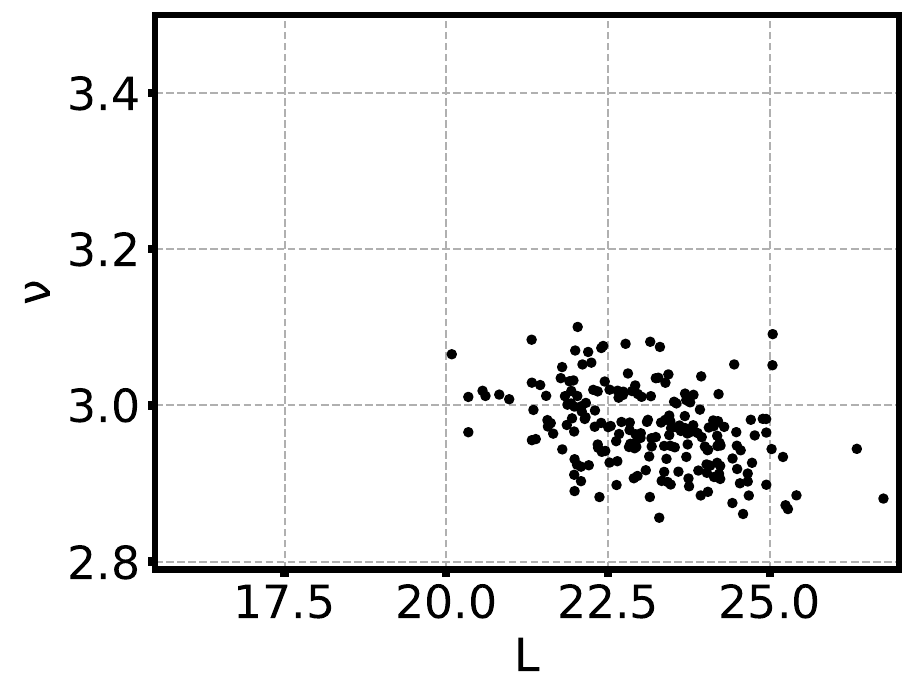}
\caption{\textbf{Random pair}.}
\label{fig:Portugal_random_L_niu}
\end{subfigure}
\begin{subfigure}[b]{0.4\textwidth}
\centering
\includegraphics[width=\textwidth]{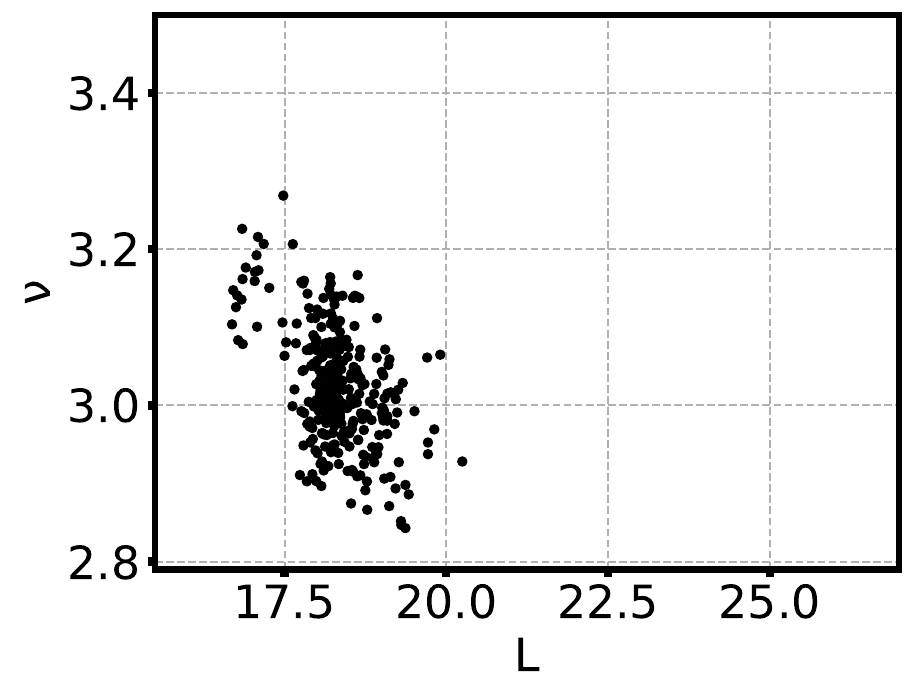}
\caption{\textbf{Random half}.}
\label{fig:Portugal_random_half_L_niu}
\end{subfigure}
\\
\begin{subfigure}[b]{0.4\textwidth}
\centering
\includegraphics[width=\textwidth]{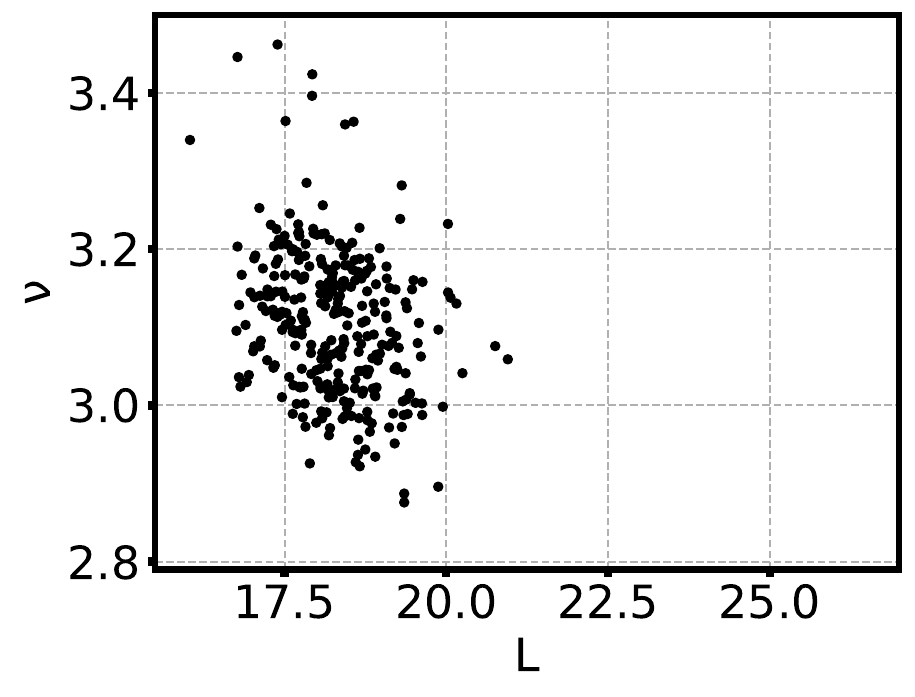}
\caption{\textbf{Random source}.}
\label{fig:Portugal_random_source_L_niu}
\end{subfigure}
\begin{subfigure}[b]{0.4\textwidth}
\centering
\includegraphics[width=\textwidth]{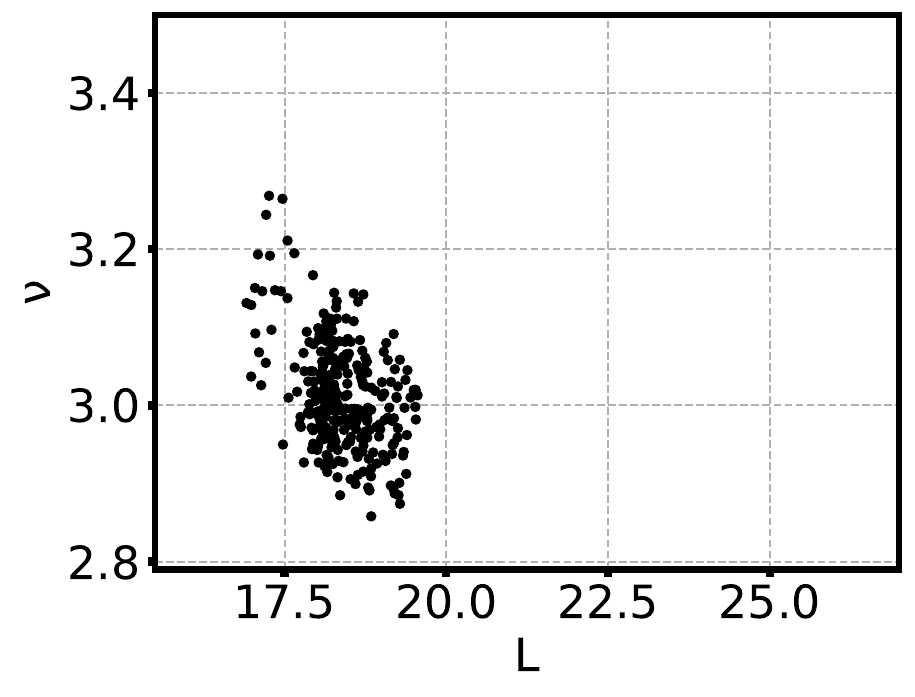}
\caption{\textbf{Random random}.}
\label{fig:Portugal_random_fixed_I0_L_niu}
\end{subfigure}
\caption{Steady state parameters $\nu$ as a function of $L$ for the Portugal configuration, for all algorithms.}
\label{fig:Portugal_L_niu}
\end{figure}

Before any analysis of the results takes place, it is important to refer that algorithm \texttt{Random pair} was not like the others when it comes to running the simulations. It took about 10 times more to run 300 iterations than the other algorithms and only $67\%$ of iterations actually converged (using the stopping criteria of the length of the tree being unchanged for 500 iterations) with a maximum number of iterations of 5000. For this study, only the converged runs were taken into account. This makes \texttt{Random pair} a much less efficient algorithm when compared to the others on a computational level, especially when the number of sites used increases.

First, let us look at the tendencies of each algorithm. Using figure \ref{fig:Portugal_L_niu} and the images and table of appendix \ref{appendix:portugal}, we can see that algorithms \texttt{Random half} and \texttt{Random random} produce very similar results: the trees show a relatively concentrated cluster of runs with low $L$ and (mostly) low $\nu$. We can see that these algorithms show very concentrated $L$ values and much more dispersed $\nu$ values.
On the other hand, algorithm \texttt{Random source} shows a much more dispersed cluster of trees at approximately the same location (with low $L$) but reaching higher $\nu$ values. It shows the most dispersed $\nu$ values of all algorithms and $L$ values slightly more dispersed than the previously mentioned algorithms.
Algorithm \texttt{Random pair} is the most disparate of them all, with a dispersed cluster of trees with high $L$ values and low $\nu$ values. While this cluster is dispersed in $L$ values (unlike the others), it has the lowest standard deviation with regard to $\nu$ values of all the algorithms.

Let us now comment on the shape of the best trees obtained. \texttt{Random pair} shows the trees with the most numerous loops. The smallest $L$ tree, smallest $\nu$ tree, and largest CE tree produced by this algorithm present five, eight, and five loops, respectively. Compared with the other algorithms, \texttt{Random source} presents the fewest loops, with zero, three, and one loop, \texttt{Random half} presents two, five, and three loops, and \texttt{Random random} presents two, four and three loops, respectively. The real railway tree shows four loops, while the MST shows zero loops (see figure \ref{fig:Portugal_railway_MST}). We can see that the most efficient trees present the highest number of loops and the shortest trees the lowest number of loops, with the high CE trees showing a middle ground between the two.

The shortest tree obtained was produced using algorithm \texttt{Random source}. This is also the algorithm that most frequently produces short trees (see fig. \ref{fig:Portugal_random_source_L_distribution} and table \ref{tab:Portugal}).

The tree with lowest $\nu$ value obtained was produced with algorithm \texttt{Random half}. The algorithm that most consistently produces low $\nu$ trees is \texttt{Random pair} (see fig. \ref{fig:Portugal_random_niu_distribution} and table \ref{tab:Portugal}).

The tree with the highest CE score was produced with algorithm \texttt{Random source}. The algorithm that most frequently produces high CE trees is \texttt{Random random} (with \texttt{Random half} very close behind) (see figs. \ref{fig:Portugal_random_fixed_I0_CE_distribution} and \ref{fig:Portugal_random_half_CE_distribution} and table \ref{tab:Portugal}).

Let us now look at table \ref{tab:Portugal} and compare the results obtained for all algorithms with those of the real railway system and with the MST of the sites. 
Algorithms \texttt{Random source}, \texttt{Random half} and \texttt{Random random} all consistently produced trees shorter than the railway, while all the trees produced by \texttt{Random pair} were longer than the railway. The MST is significantly shorter than all the trees obtained by all algorithms.
All trees produced by all algorithms were more efficient than the MST. All algorithms produced trees with an on average larger efficiency (smaller $\nu$ value) than that of the railway. However, only for algorithm \texttt{Random pair} were all trees produced more efficient than the railway.
Algorithms \texttt{Random source}, \texttt{Random half} and \texttt{Random random} all on average produced higher value CE trees than that of the railway or the MST. On the other hand, all trees produced by algorithm \texttt{Random pair} showed CE values lower than that for the railway or MST.

It is important to note that a large CE value does not exactly equate to a better network. After all, the MST has a CE value 3\% larger than the real railway tree, but it has a 31\% larger $\nu$ value than the railway, making it much less efficient. CE was included in this work because it is a commonly used parameter in the literature, and in certain cases can help create a middle ground between construction cost and transport efficiency.

Overall, we can see that algorithm \texttt{Random pair} maximizes transport efficiency (producing small $\nu$ trees), while algorithm \texttt{Random source} minimizes construction costs (producing small $L$ trees). However, \texttt{Random pair} generates exceedingly long trees (its average tree presents a 21\% difference to \texttt{Random source}) and is particularly inefficient computationally as mentioned previously, while \texttt{Random source}'s average tree shows only a 4\% difference in $\nu$ value compared to \texttt{Random pair}. Algorithms \texttt{Random half} and \texttt{Random random} show a middle ground between the two, being able to produce trees as efficient as \texttt{Random pair} and slightly longer than \texttt{Random source} (their shortest trees show a less than 6\% difference to \texttt{Random source}'s shortest tree, and their average trees show a less than 2\% difference to \texttt{Random pair}'s trees).

While all algorithms utilized were successful in producing more efficient trees on average than that of Portugal's mainland's railway system, we would overall recommend algorithm \texttt{Random random} when trying to produce an efficient (and somewhat short) tree, as it was the algorithm that showed, on average, higher CE values (better cost-efficiency) and lower $\nu$ values (higher efficiency), and also presented a smaller standard deviation with respect to $\nu$, thus more consistently produced efficient trees. This was similarly the same result obtained in \cite{rodrigo_tese}.







\chapter{Concluding remarks}
\label{chapter:conclusion}

\section{Achievements}

In this thesis, we aimed to test the adaptive Hagen-Poiseuille flow model (described in section \ref{sec:adaptive-model}) as a way to both mimic \textit{Physarum polycephalum}'s network adaptation dynamics and to be able to produce efficient networks. 

We started by making a slight change to the model's implementation that allowed us to control the volume of fluid in each simulation, which is a key parameter that must be regulated to permit comparing between different runs of the simulation.

To determine whether or not the model studied produced realistic results and was worth studying, we tested certain key physical characteristics of the trees produced by it. We showed that the model tested was able to produce results that were consistent with physical observations, namely the tendency of the simulations to always choose the shortest path, especially as the continuous limit was approached. Additionally. 
Murray's law (which is experimentally observed in several biological networks) was verified for bifurcation nodes in steady state trees, but it was not verified dynamically for the time of the simulation leading up to steady state. 
By studying the results of the model applied to simple one-channel and two-channel networks, we observed that the conductivities and flows of said network's channels behaved like other biological contractile veins.

The model was also tested with regards to how well it could mimic \textit{Physarum polycephalum}'s network topology, using both fixed sources and sinks and non-static sources and sinks. Both cases were compared to \textit{Physarum}'s radial growth pattern, and both cases presented a key characteristic present in \textit{Physarum} networks: showing fewer, thicker veins closer to the center of the network that branch out into many thinner veins as one gets closer to the periphery. However, the non-static sources and sinks steady states showed more accurate resemblance, as they also displayed other characteristics: connections between the sites on the periphery of the network (even though they were scarce compared to real \textit{Physarum} networks) and thus the presence of loops in the network.
As such, it was determined that the model represents quite accurately \textit{Physarum}-like network adaptation patterns, that is, it produces network shapes that resemble those of "older" (\textit{Physarum} grows radially outward, so closer to the center of the network), more optimized parts of \textit{Physarum}'s networks. However, it is not successful at accurately describing the peripheral, foraging, "most recent" parts of \textit{Physarum}'s networks, that are highly connected by numerous thin veins and connect them using numerous loops.

The non-static sources and sinks configuration were also tested to determine whether or not they showed two key physical characteristics of \textit{Physarum}: shuttle streaming and peristalsis. While this model does not aim to describe the chemical or biological dynamics of \textit{Physarum}, it describes the basic hydrodynamics involved in the movement of cytoplasm inside its body. As this model has at its core the adaptation of vein walls of the network, it describes the contraction of the walls that is thought to contribute to shuttle streaming in \textit{Physarum}. 
It was shown that the simulations produced by the model presented peristalsis (oscillatory contraction and relaxation of the vein walls) on all edges tested. 
It was also determined that the model did show shuttle streaming. This shuttle streaming was present in peripheral parts of the network (channels connecting sinks at the edge) and in parts of the network that were part of a loop, before a steady state was reached. As such, it is possible that the presence of loops in \textit{Physarum}'s trees may also contribute to the shuttle streaming phenomenon and the efficient transport of nutrients and resources throughout its cell.
Peristalsis and shuttle streaming were not observed for static sites configurations, and were only seen when using asynchronous sites. We can conclude that shuttle streaming and peristalsis are most likely consequences of the asynchronous resource consumption that occurs in \textit{Physarum}, as mentioned previously.

Then, this thesis intended to determine if this model could be used to produce efficient networks, like \textit{Physarum} has been witnessed to do; by efficient, one means networks with short distances between relevant sites. This was contrasted with determining whether or not the model was able to create short networks, like Steiner minimum trees. 
Transport efficiency and network cost are two defining characteristics of a network that are heavily studied and weighed before a transport network is constructed. This biologically-inspired algorithm could be advantageous in constructing cost-efficient networks if proven to be successful. 
The implementation of the model involves a stochastic algorithm, as the final shape of a network produced by this model is heavily influenced by the starting conditions of conductivities, which are set to be random; like \textit{Physarum}, this model can produce many different types of networks, depending on its environment. The choice of sources and sinks was stochastic as well and followed one of four different tested algorithms. The results were tested on two different types of networks: first, connecting $n=\{3,4,5\}$ sites in a regular polygon configuration; then, connecting $n=25$ sites in a configuration that is meant to mirror mainland Portugal, thus creating a realistic transport network.
All algorithms collectively were shown to be able to produce short trees, efficient trees, and cost-effective trees, and all produced trees that were more efficient than the studied Portuguese railway network. Algorithm \texttt{Random pair} produced overall long, quite efficient trees (about 20\% longer than \texttt{Random source} for the Portuguese configuration), and was the most computationally demanding one. Algorithm \texttt{Random source} produced mainly short, less efficient trees (yet showing only about a 4\% difference in efficiency to \texttt{Random pair}). The recommended algorithm to use to produce cost-efficient trees would be \texttt{Random random}, as it presented the overall highest efficiency values (behind \texttt{Random pair}) with reduced length values and adequate computation time.

\section{Future work}

The model studied in this work does not fully mirror \textit{Physarum}'s network characteristics; namely, it doesn't mimic the peripheral, "more recent" parts of the trees that often present numerous thin veins connected by many loops. However, the model produces networks that share similarities with the "older", more optimized parts of \textit{Physarum}'s network. Thus, the model should be extended to include a description of the growth and foraging mechanisms of slime mold that can accurately reproduce network patterns.

While shuttle streaming was observed using this model, it wasn't observed at steady state or in all edges tested. The model used does not explicitly incorporate peristaltic mechanisms \cite{alim_peristalsis} or other mechanisms that are theorized to be responsible for shuttle streaming like the transport of signaling molecules released in regions that come in contact with stimuli \cite{alim_signal_prop}. Including such mechanisms in the model could result in the active presence of shuttle streaming and more accurate network patterns.

The general form of the adaptation equation (eq. \eqref{eq:Dij_adapt}) allows for any choice of $g$ function. In this work, only the $g$ function described in eq. \eqref{eq:g_23} was used, as it minimizes the dissipated power. It would be interesting to study the results of using a different $g$ function, particularly one that saturates for large flux values to mimic what is observed in real biological networks (like blood vessels \cite{rubinow}).

The study of the efficiency of the networks created using this model could be extended by using different $g$ functions (like $g = |Q_{ij}|^\gamma$ for different $\gamma$ values, like as seen in \cite{rodrigo_tese}) to create them, as this could lead to different topologies and efficiencies. This study could also be significantly improved by optimizing the model's computational implementation to reduce run times; as this is a stochastic algorithm that relies on running the simulations numerous times, optimizing the run time of the algorithm (by, for example, parallelizing the code) is essential.

There's still much work to be done in modeling all of \textit{Physarum polycephalum}'s complex behavior. After all, we still lack a full understanding of the mechanisms behind its cytoplasmic movement, among other aspects. Because of this, we have yet to find a consolidated model that can mimic \textit{Physarum}'s network dynamics and intelligent behavior. This is an ambitious open-ended problem that could have exciting and beneficial consequences for a large range of fields.


\phantomsection
\addcontentsline{toc}{chapter}{\bibname}


\printbibliography

\appendix

\chapter{Steiner trees additional images}
\label{appendix:steiner_geometric}

\section{Triangular configuration} \label{appendix:tri}

\begin{figure}[H]
\centering
\begin{subfigure}[b]{0.4\textwidth}
 \centering
 \includegraphics[width=0.8\textwidth]{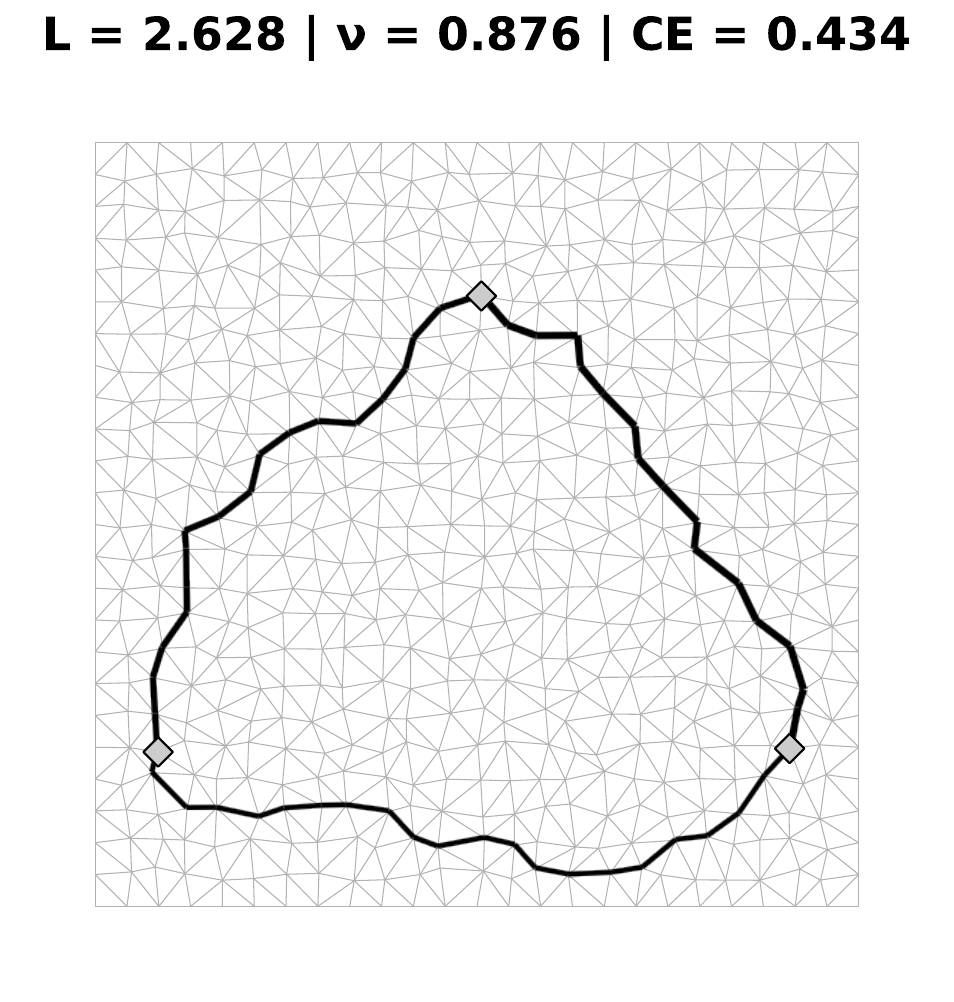}
 \caption{Steady state with biggest $L$ and smallest CE.}
 \label{fig:tri_random_ss_biggest_L_smallest_CE}
\end{subfigure}
\begin{subfigure}[b]{0.4\textwidth}
 \centering
 \includegraphics[width=0.8\textwidth]{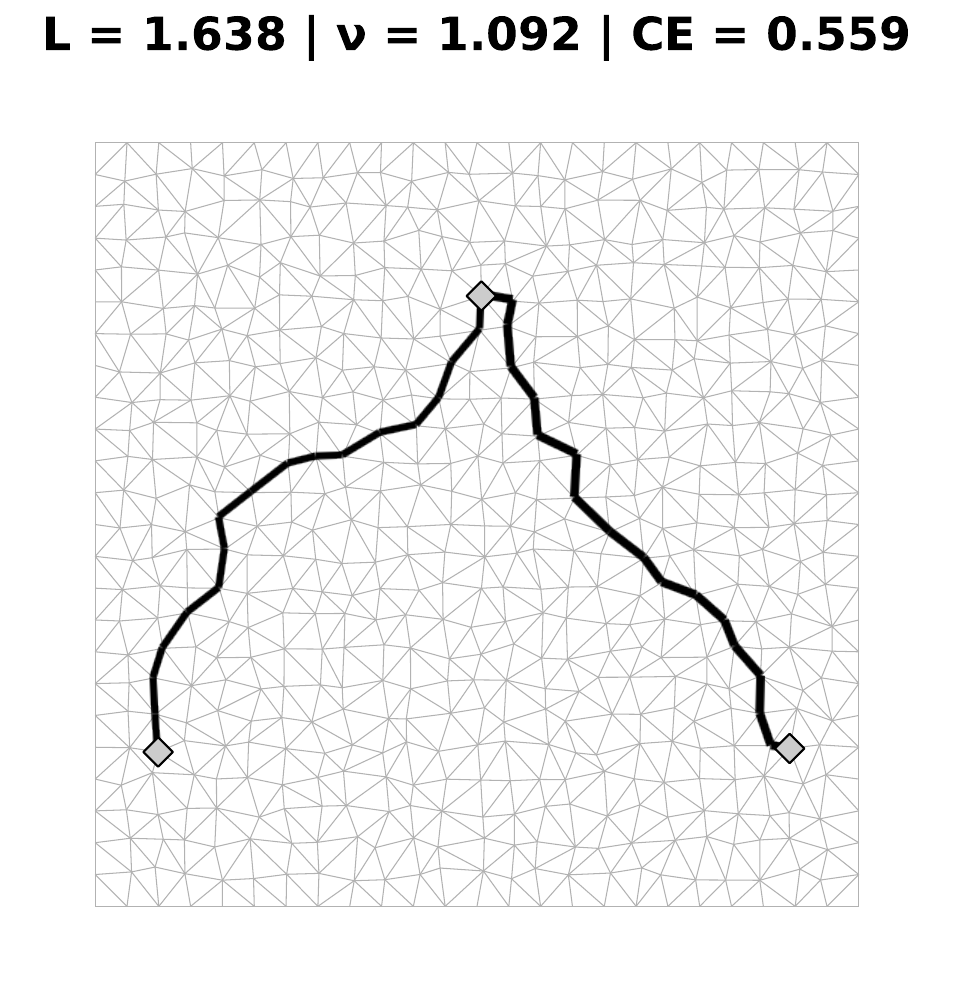}
 \caption{Steady state with biggest $\nu$.}
 \label{fig:tri_random_ss_biggest_niu}
\end{subfigure}
\\
\begin{subfigure}[b]{0.32\textwidth}
\centering
\includegraphics[width=\textwidth]{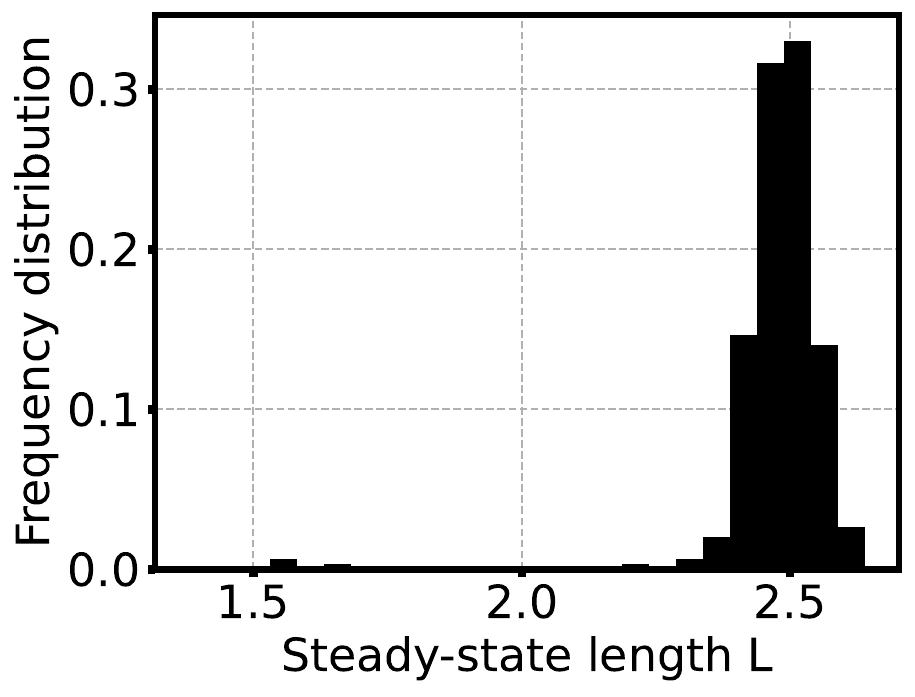}
\caption{$L$ distribution.}
\label{fig:tri_random_L_distribution}
\end{subfigure}
\hfill
\begin{subfigure}[b]{0.32\textwidth}
\centering
\includegraphics[width=\textwidth]{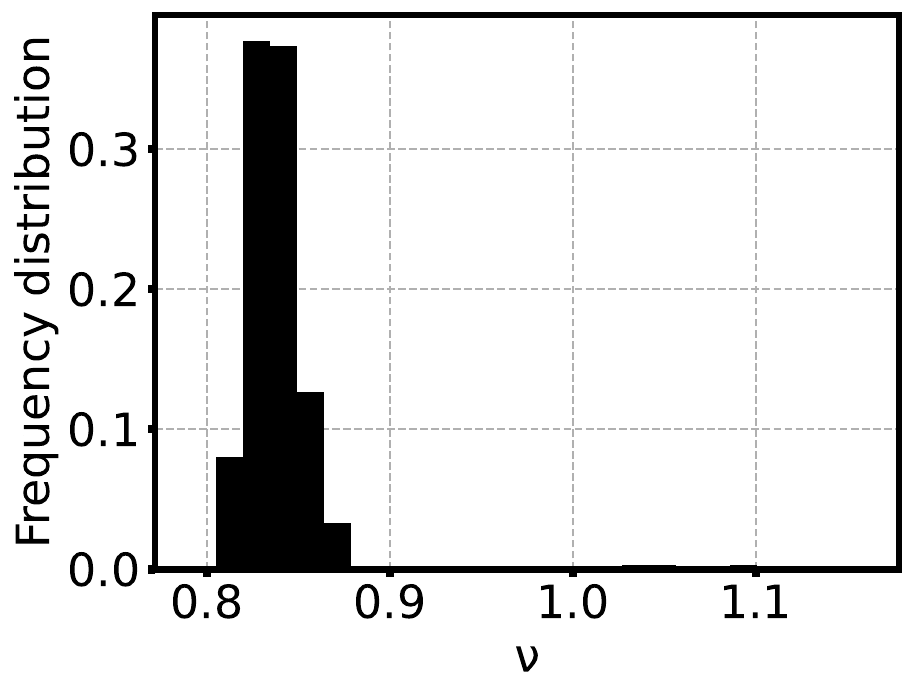}
\caption{$\nu$ distribution.}
\label{fig:tri_random_niu_distribution}
\end{subfigure}
\hfill
\begin{subfigure}[b]{0.32\textwidth}
\centering
\includegraphics[width=\textwidth]{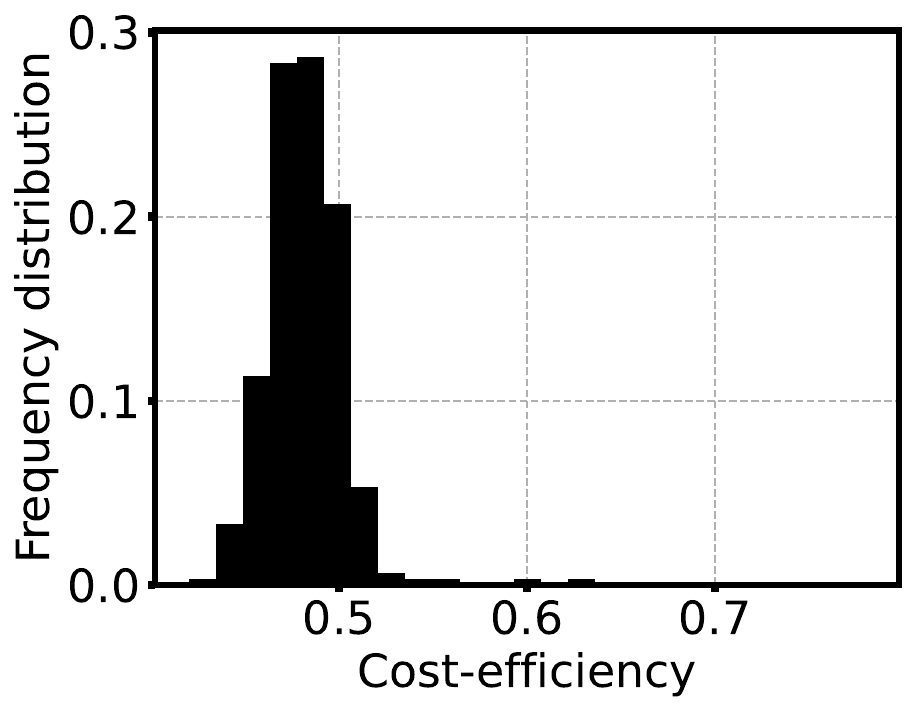}
\caption{CE distribution.}
\label{fig:tri_random_CE_distribution}
\end{subfigure}
\caption{Additional results for the triangular configuration, for the \textbf{Random pair} algorithm.}
\label{fig:tri_random_add}
\end{figure}

\begin{figure}[H]
\centering
\begin{subfigure}[b]{0.6\textwidth}
 \centering
 \includegraphics[width=0.55\textwidth]{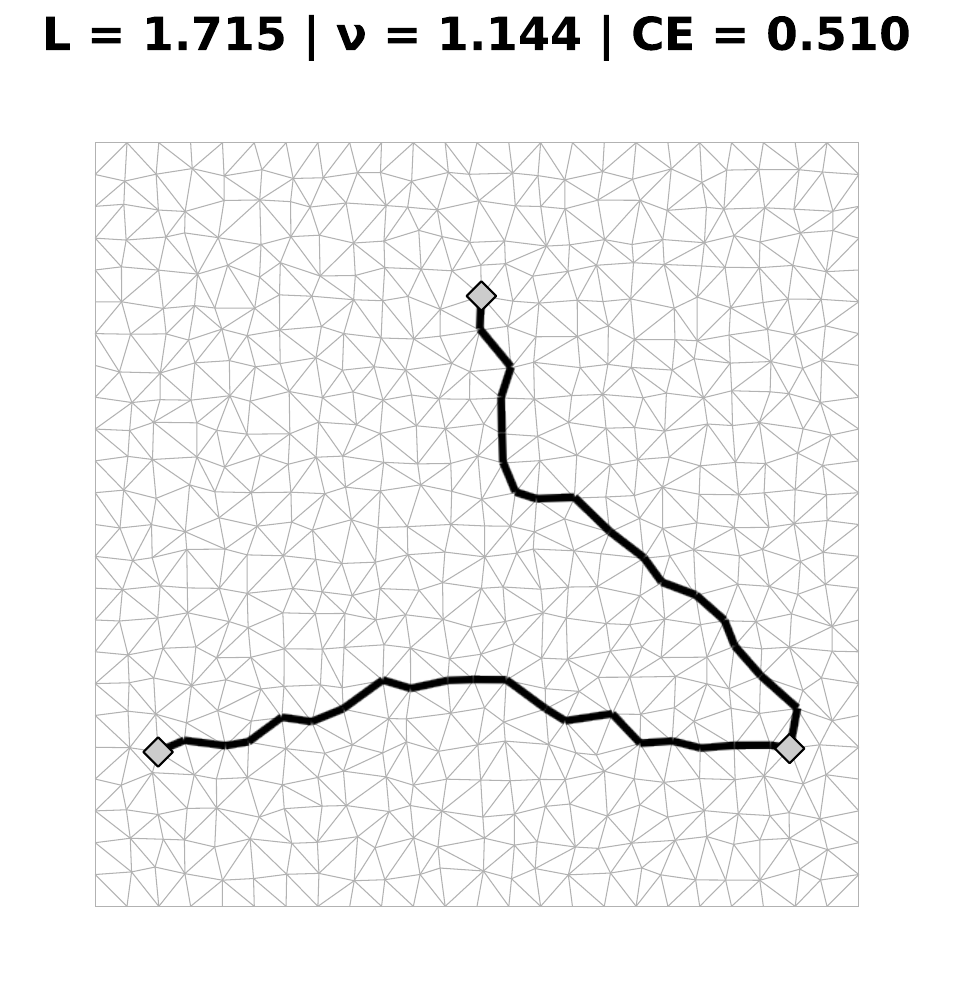}
 \caption{Steady state with largest $L$, largest $\nu$ and smallest CE.}
 \label{fig:tri_random_source_ss_biggest_L_biggest_nu_smallest_CE}
\end{subfigure}
\\
\begin{subfigure}[b]{0.32\textwidth}
\centering
\includegraphics[width=\textwidth]{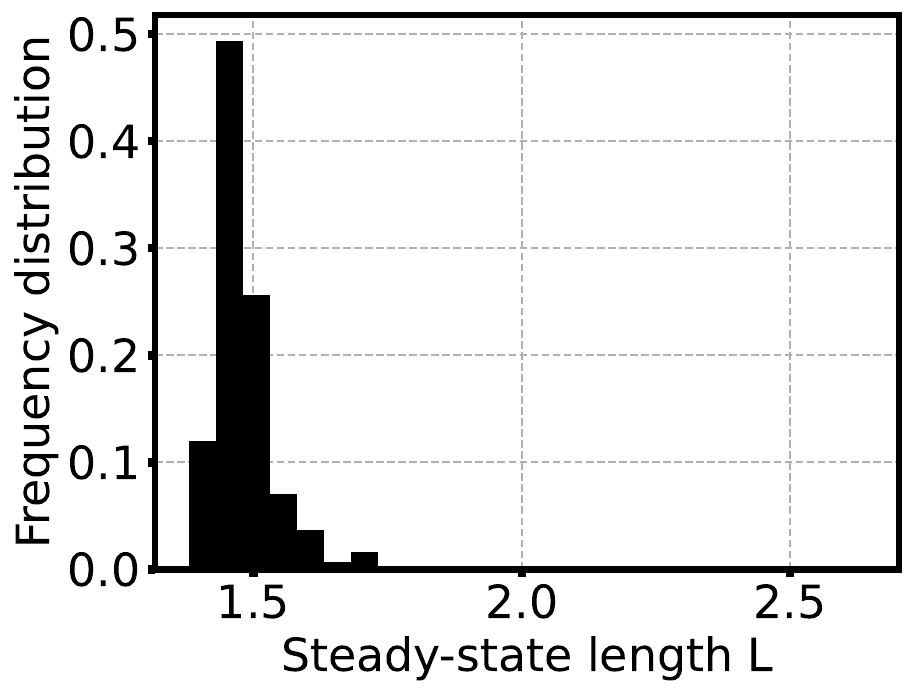}
\caption{$L$ distribution.}
\label{fig:tri_random_source_L_distribution}
\end{subfigure}
\hfill
\begin{subfigure}[b]{0.32\textwidth}
\centering
\includegraphics[width=\textwidth]{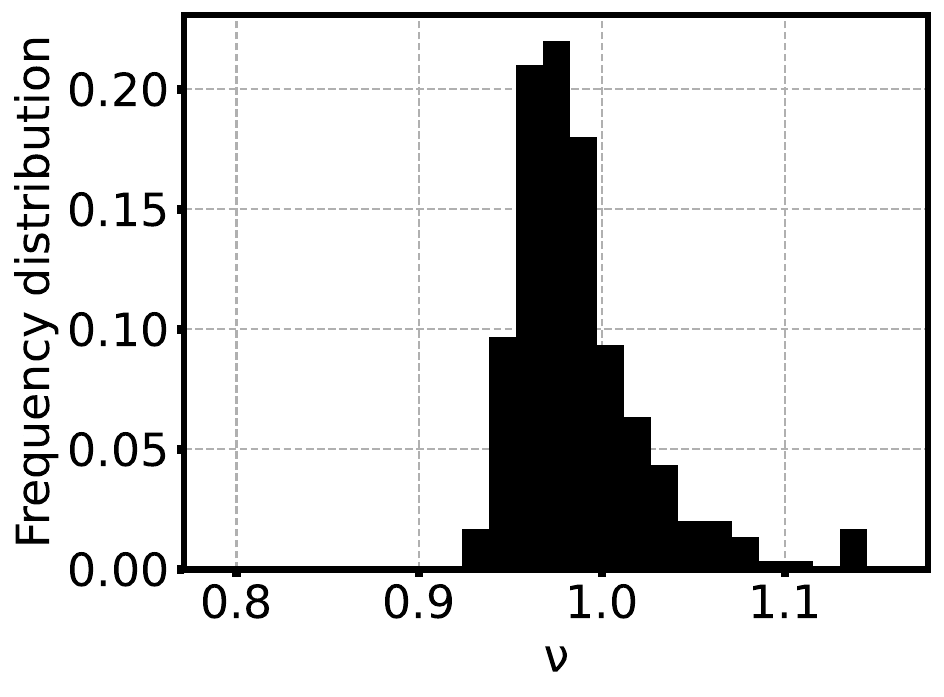}
\caption{$\nu$ distribution.}
\label{fig:tri_random_source_niu_distribution}
\end{subfigure}
\hfill
\begin{subfigure}[b]{0.32\textwidth}
\centering
\includegraphics[width=\textwidth]{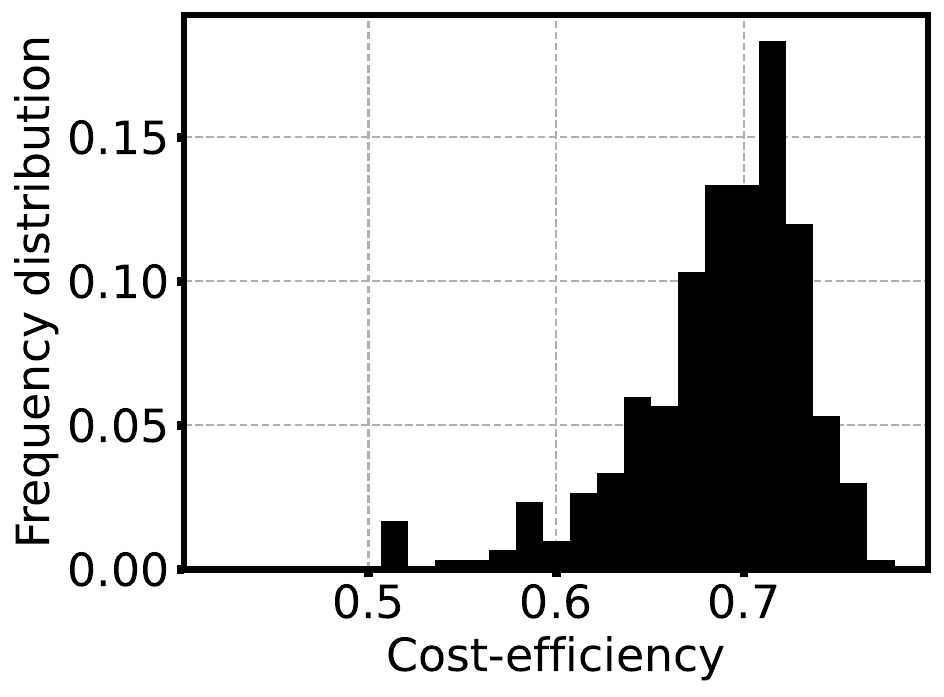}
\caption{CE distribution.}
\label{fig:tri_random_source_CE_distribution}
\end{subfigure}
\caption{Additional results for the triangular configuration, for the \textbf{Random source} algorithm.}
\label{fig:tri_random_source_add}
\end{figure}

\begin{figure}[H]
\centering
\begin{subfigure}[b]{0.6\textwidth}
\centering
\includegraphics[width=0.55\textwidth]{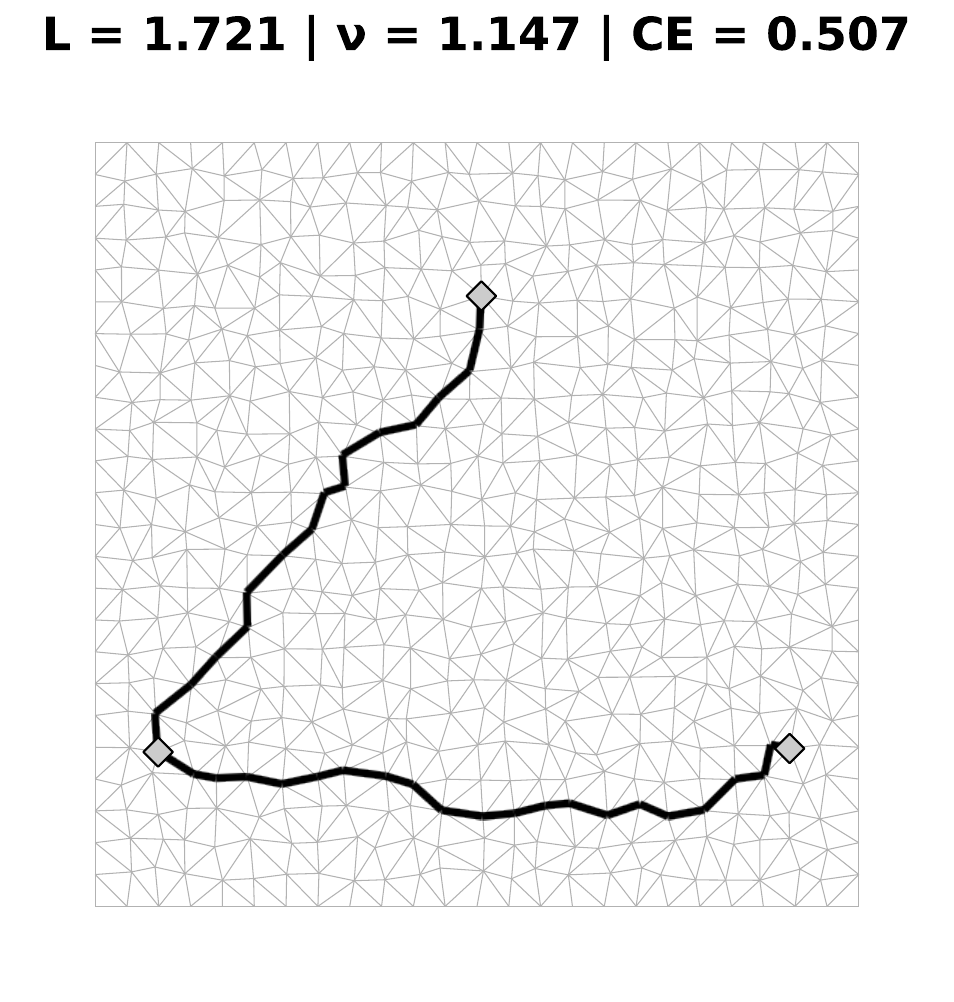}
\caption{Steady state with largest $L$, largest $\nu$ and smallest CE.}
\label{fig:tri_random_fixed_I0_ss_biggest_L_biggest_niu_smallest_CE}
\end{subfigure}
\\
\begin{subfigure}[b]{0.32\textwidth}
\centering
\includegraphics[width=\textwidth]{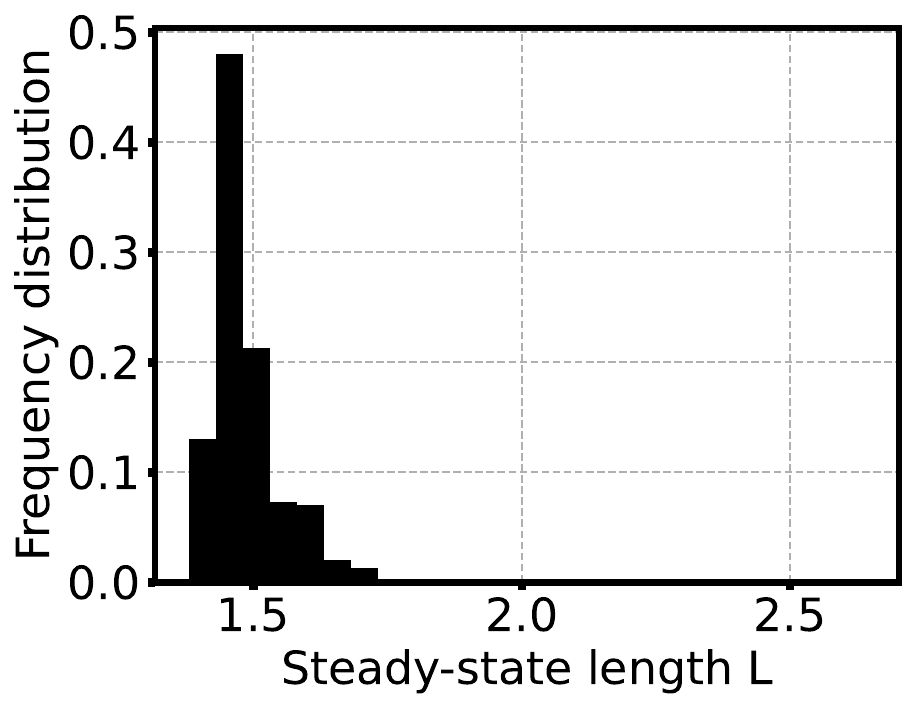}
\caption{$L$ distribution.}
\label{fig:tri_random_fixed_I0_L_distribution}
\end{subfigure}
\hfill
\begin{subfigure}[b]{0.32\textwidth}
\centering
\includegraphics[width=\textwidth]{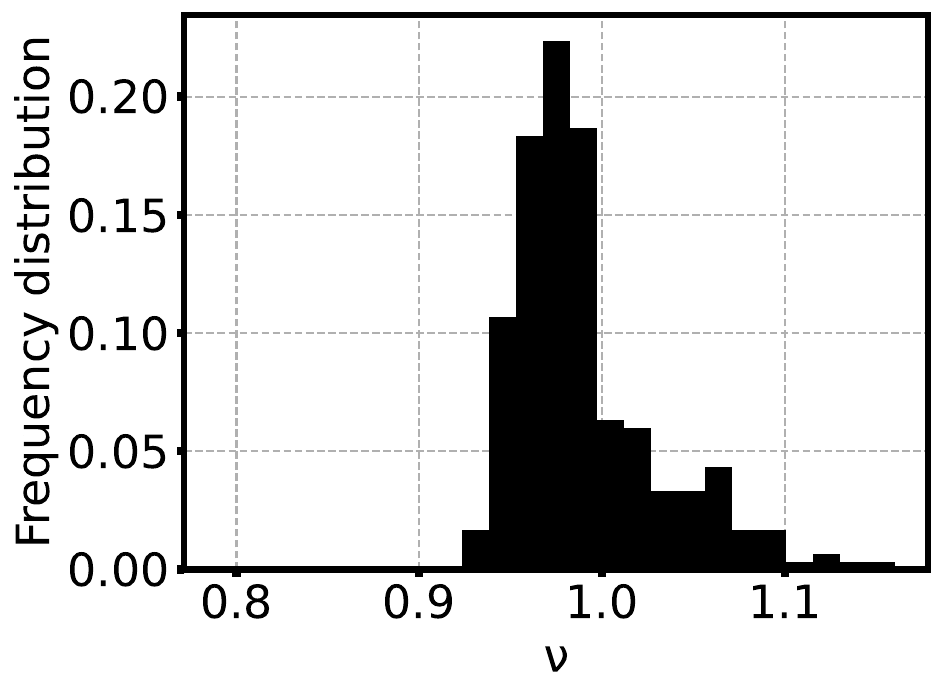}
\caption{$\nu$ distribution.}
\label{fig:tri_random_fixed_I0_niu_distribution}
\end{subfigure}
\hfill
\begin{subfigure}[b]{0.32\textwidth}
\centering
\includegraphics[width=\textwidth]{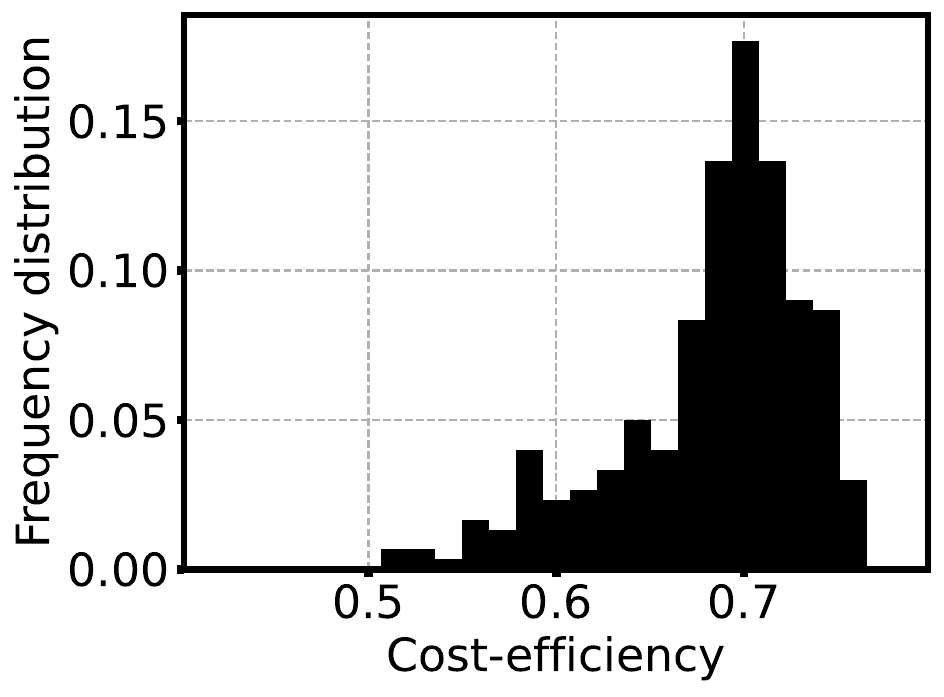}
\caption{CE distribution.}
\label{fig:tri_random_fixed_I0_CE_distribution}
\end{subfigure}
\caption{Additional results for the triangular configuration, for the \textbf{Random random} algorithm.}
\label{fig:tri_random_fixed_I0_add}
\end{figure}

\begin{table}[H]
\centering
\begin{tabular}{c|c|c|c}
\toprule
Algorithm & $L$ & $\nu$ & CE \\
\midrule
\texttt{Random pair} & 2.48 $\pm$ 0.11 & 0.839 $\pm$ 0.026 & 0.482 $\pm$ 0.021 \\
\texttt{Random source} & 1.481 $\pm$ 0.055 & 0.987 $\pm$ 0.037 & 0.687 $\pm$ 0.047 \\
\texttt{Random random} & 1.485 $\pm$ 0.059 & 0.990 $\pm$ 0.040 & 0.683 $\pm$ 0.051 \\
\midrule
Perimeter & 2.275 & 0.758 & 0.595 \\
SMT & 1.313 & 0.876 & 0.521 \\
\bottomrule
\end{tabular}
\caption{Average values and standard deviation of $L$, $\nu$ and CE for the triangular configuration, for all algorithms, and comparison for the theoretical values for the perimeter and SMT of the triangle (calculated using the coordinates of the sites and the Steiner point coordinates obtained in section \ref{sec:steiner_points_calc}). Note that \texttt{Random pair} is equivalent to \texttt{Random half} for this configuration.}
\label{tab:tri}
\end{table}

\section{Square configuration} \label{appendix:square}

\begin{figure}[H]
\centering
\begin{subfigure}[b]{0.6\textwidth}
\centering
\includegraphics[width=0.55\textwidth]{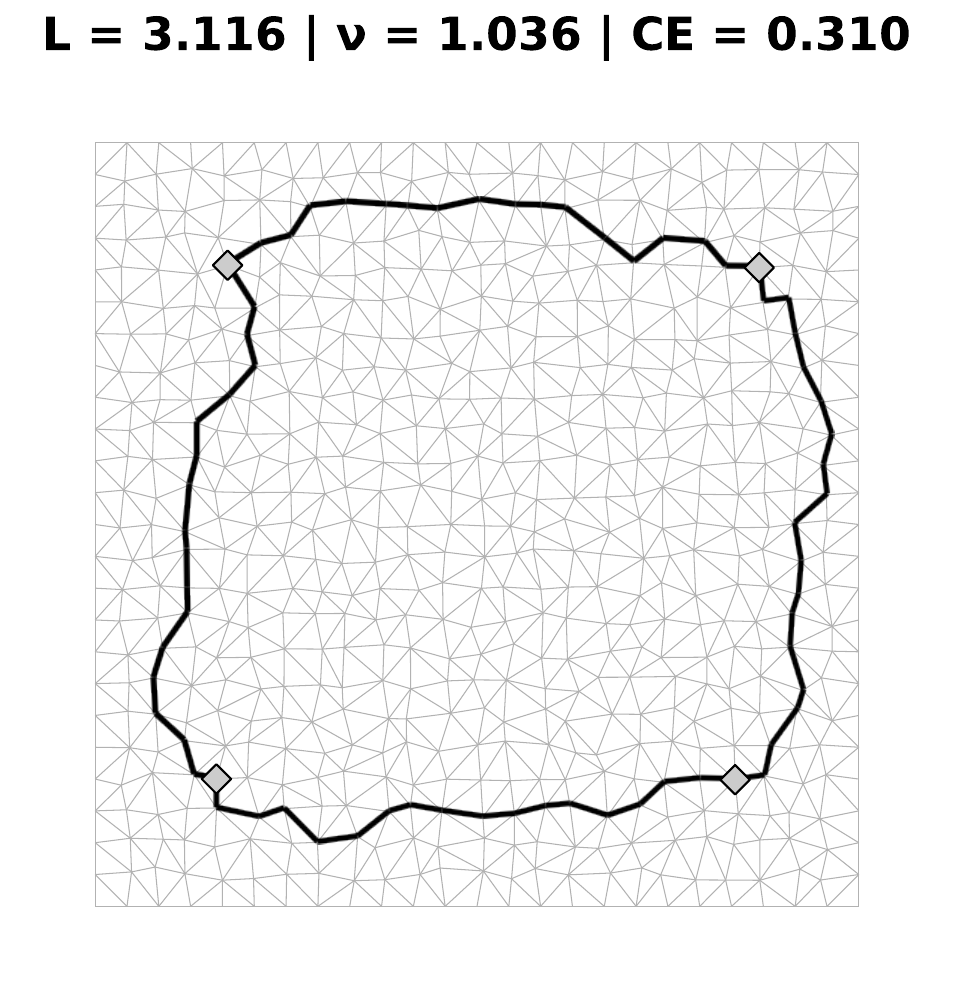}
\caption{Steady state with largest $L$, largest $\nu$ and smallest CE.}
\label{fig:square_random_ss_biggest_L_biggest_niu_smallest_CE}
\end{subfigure}
\\
\begin{subfigure}[b]{0.32\textwidth}
\centering
\includegraphics[width=\textwidth]{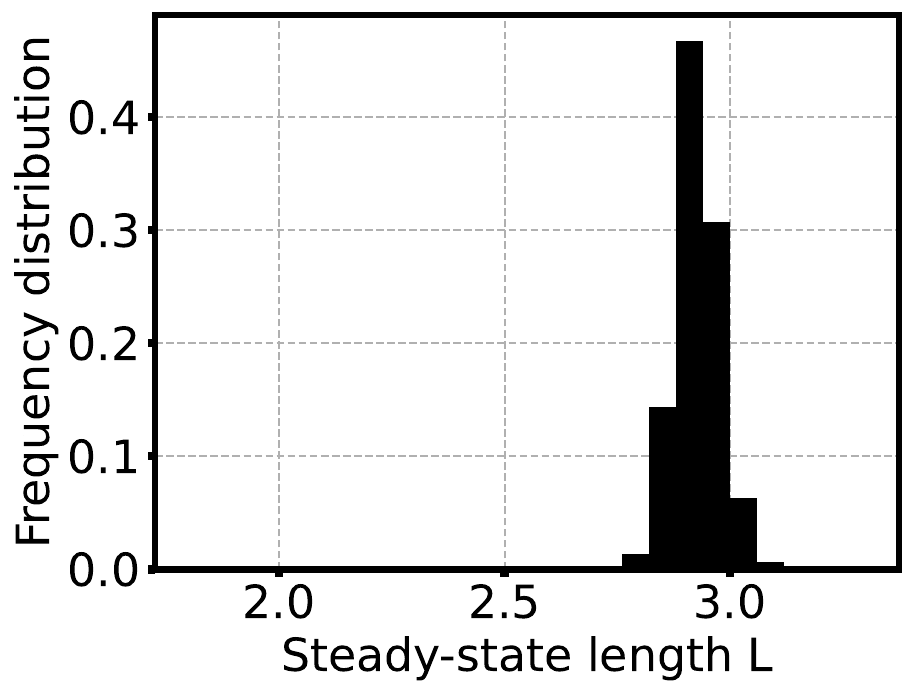}
\caption{$L$ distribution.}
\label{fig:square_random_L_distribution}
\end{subfigure}
\hfill
\begin{subfigure}[b]{0.32\textwidth}
\centering
\includegraphics[width=\textwidth]{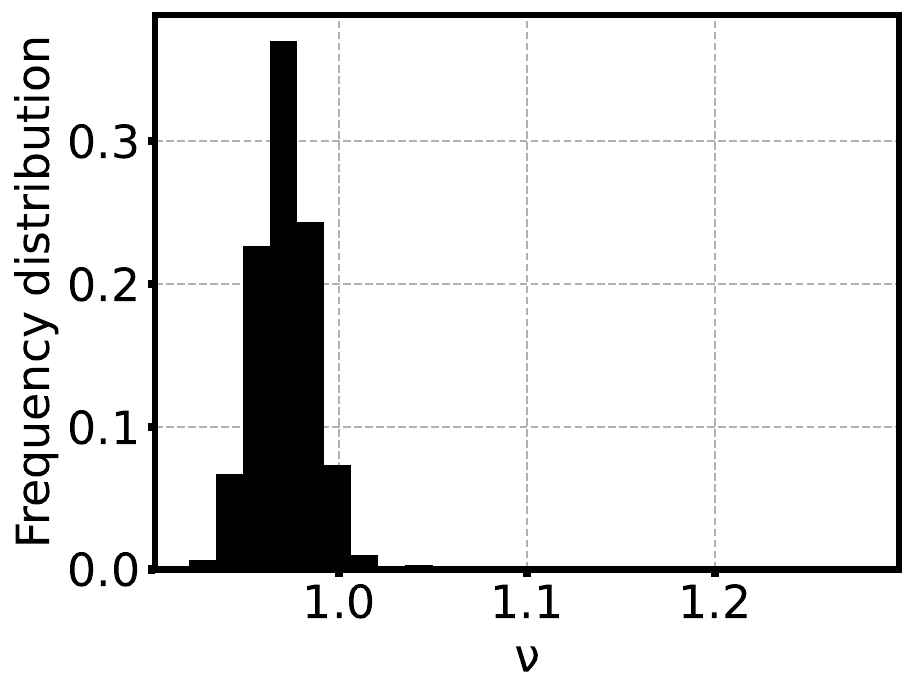}
\caption{$\nu$ distribution.}
\label{fig:square_random_niu_distribution}
\end{subfigure}
\hfill
\begin{subfigure}[b]{0.32\textwidth}
\centering
\includegraphics[width=\textwidth]{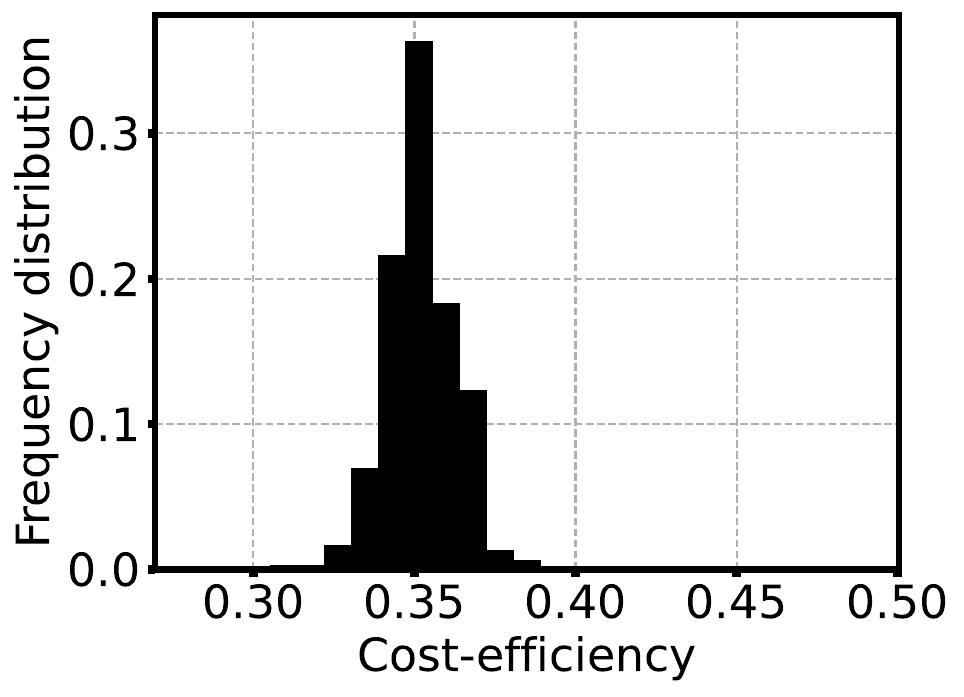}
\caption{CE distribution.}
\label{fig:square_random_CE_distribution}
\end{subfigure}
\caption{Additional results for the square configuration, for the \textbf{Random pair} algorithm.}
\label{fig:square_random_add}
\end{figure}

\begin{figure}[H]
\centering
\begin{subfigure}[b]{0.4\textwidth}
 \centering
 \includegraphics[width=0.8\textwidth]{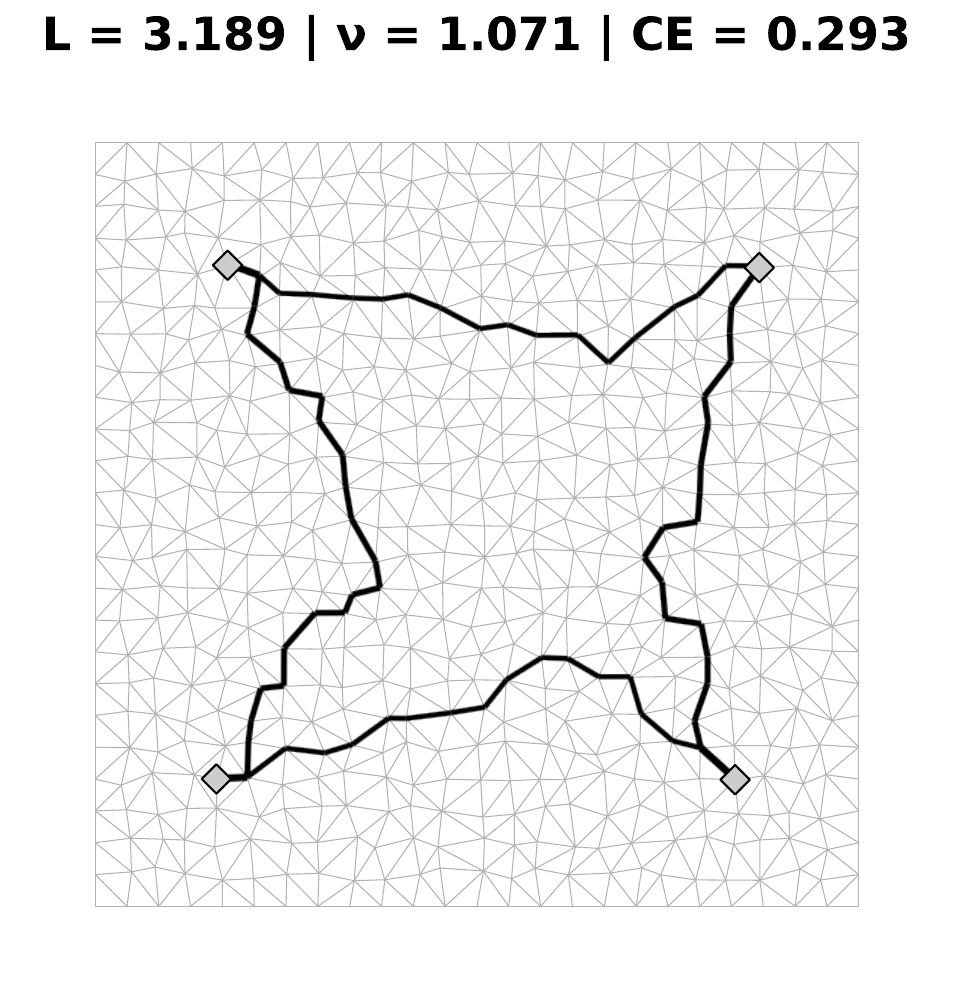}
 \caption{Steady state with largest $L$ and smallest CE.}
 \label{fig:square_random_half_ss_biggest_L_smallest_CE}
\end{subfigure}
\begin{subfigure}[b]{0.4\textwidth}
 \centering
 \includegraphics[width=0.8\textwidth]{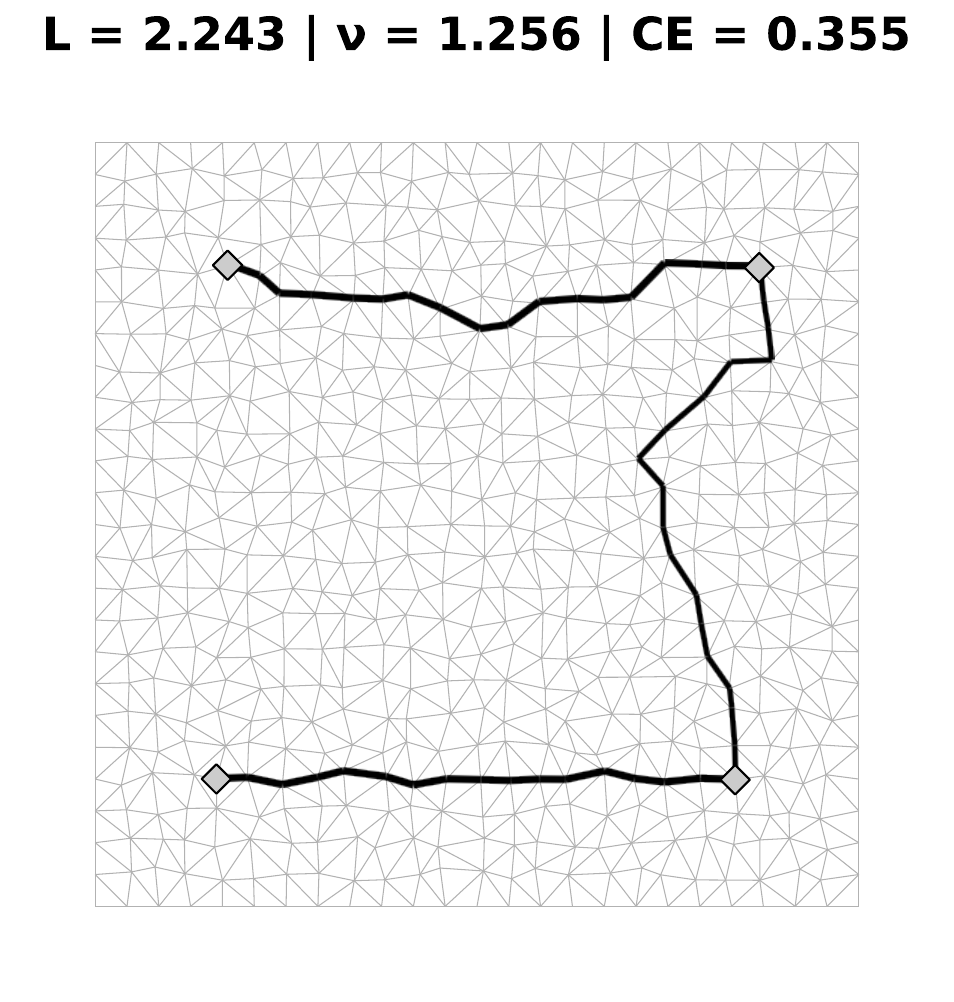}
 \caption{Steady state with largest $\nu$.}
 \label{fig:square_random_half_ss_biggest_niu}
\end{subfigure}
\\
\begin{subfigure}[b]{0.32\textwidth}
\centering
\includegraphics[width=\textwidth]{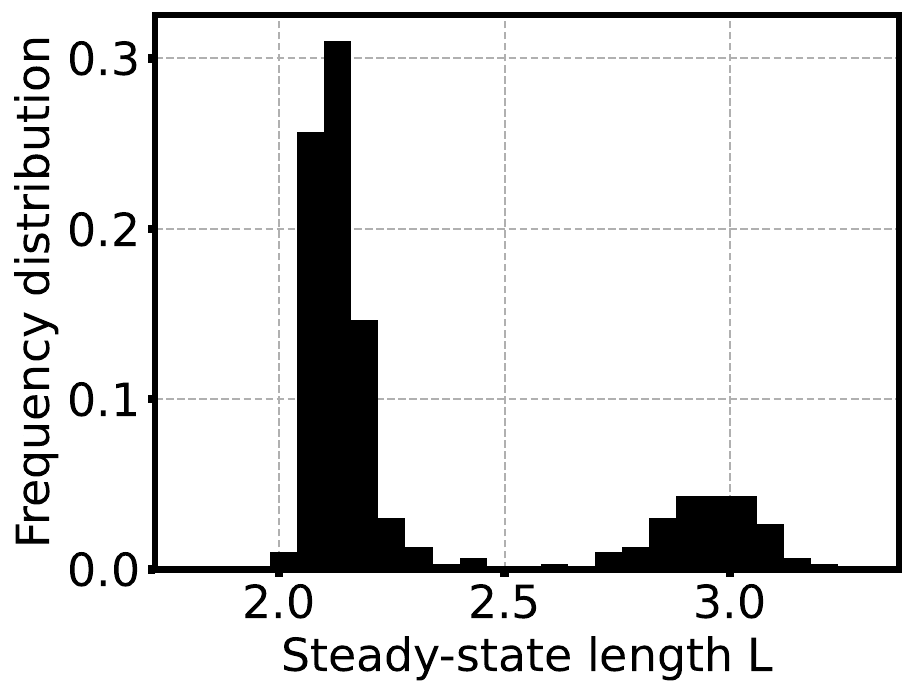}
\caption{$L$ distribution.}
\label{fig:square_random_half_L_distribution}
\end{subfigure}
\hfill
\begin{subfigure}[b]{0.32\textwidth}
\centering
\includegraphics[width=\textwidth]{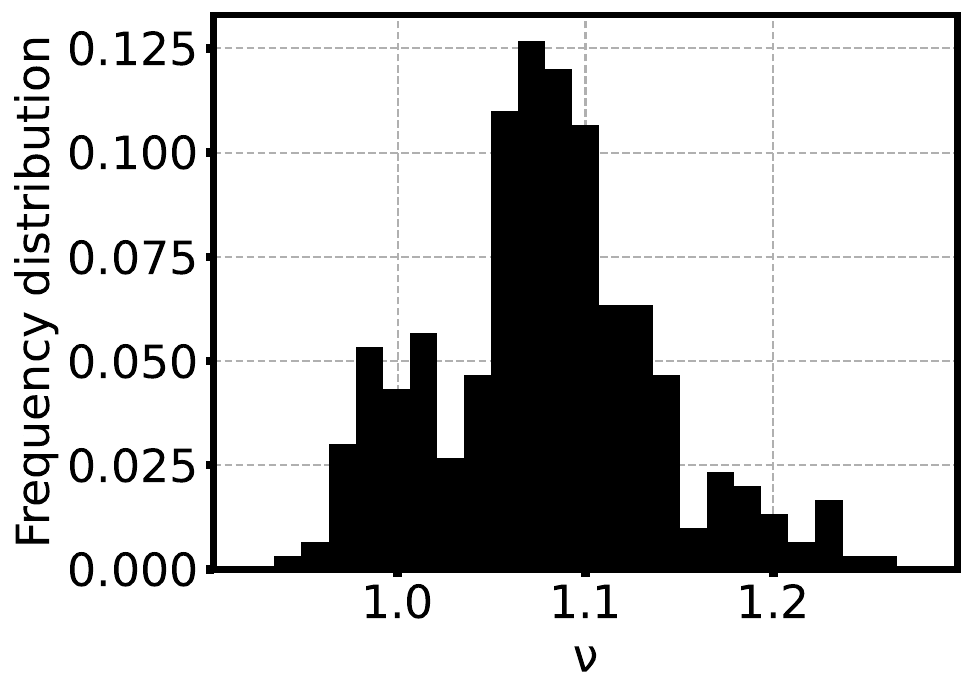}
\caption{$\nu$ distribution.}
\label{fig:square_random_half_niu_distribution}
\end{subfigure}
\hfill
\begin{subfigure}[b]{0.32\textwidth}
\centering
\includegraphics[width=\textwidth]{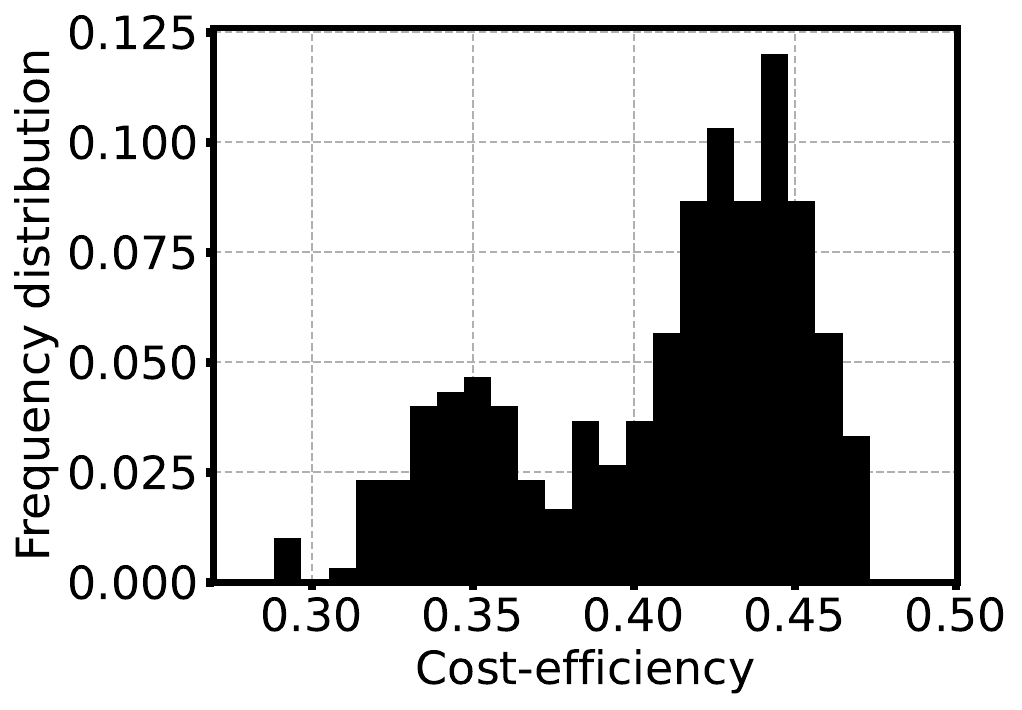}
\caption{CE distribution.}
\label{fig:square_random_half_CE_distribution}
\end{subfigure}
\caption{Additional results for the square configuration, for the \textbf{Random half} algorithm.}
\label{fig:square_random_half_add}
\end{figure}

\begin{figure}[H]
\centering
\begin{subfigure}[b]{0.4\textwidth}
 \centering
 \includegraphics[width=0.8\textwidth]{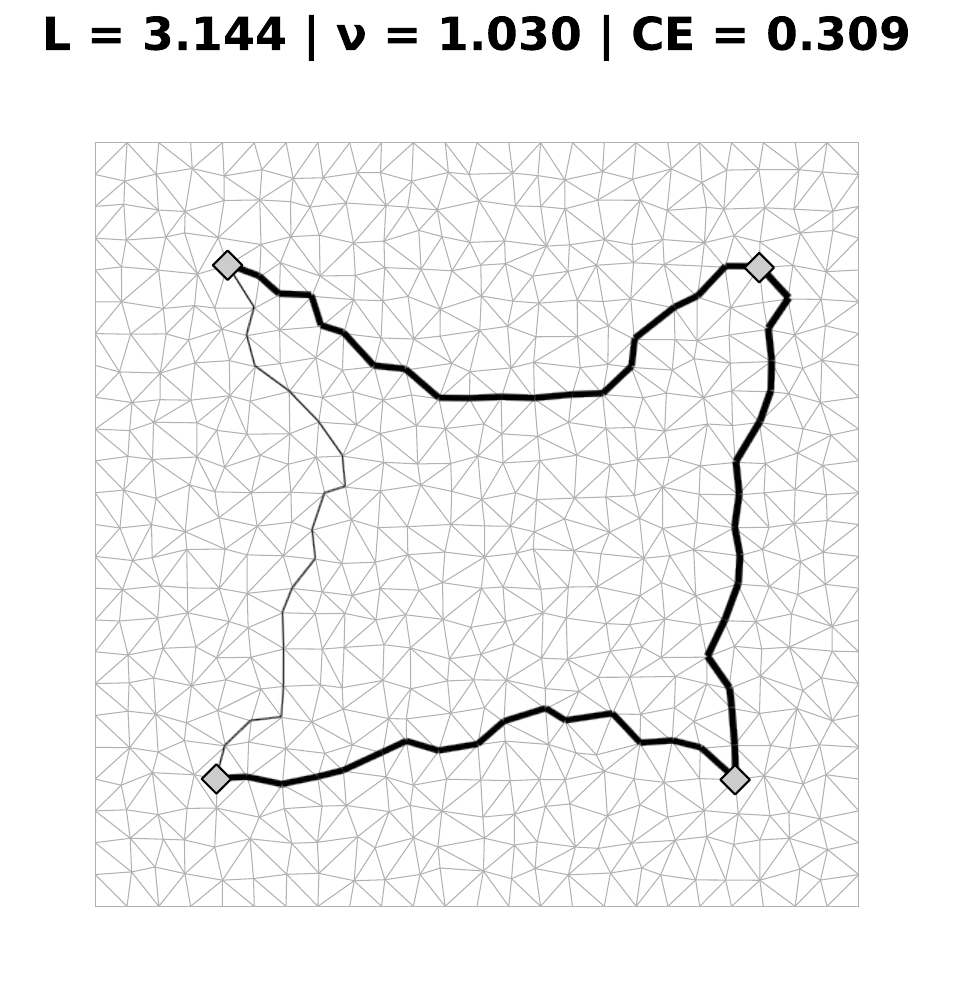}
 \caption{Steady state with largest $L$ and smallest CE.}
 \label{fig:square_random_source_ss_biggest_L}
\end{subfigure}
\begin{subfigure}[b]{0.4\textwidth}
 \centering
 \includegraphics[width=0.8\textwidth]{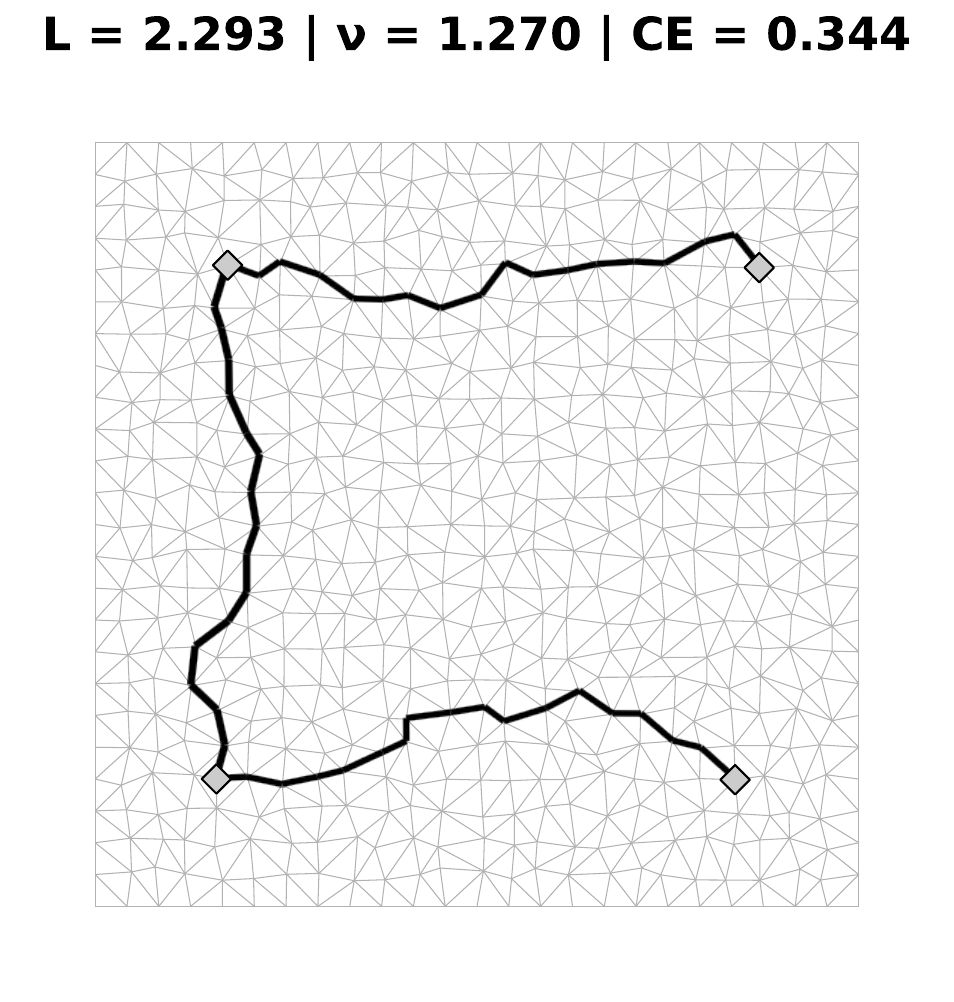}
 \caption{Steady state with largest $\nu$.}
 \label{fig:square_random_source_ss_biggest_niu}
\end{subfigure}
\\
\begin{subfigure}[b]{0.32\textwidth}
\centering
\includegraphics[width=\textwidth]{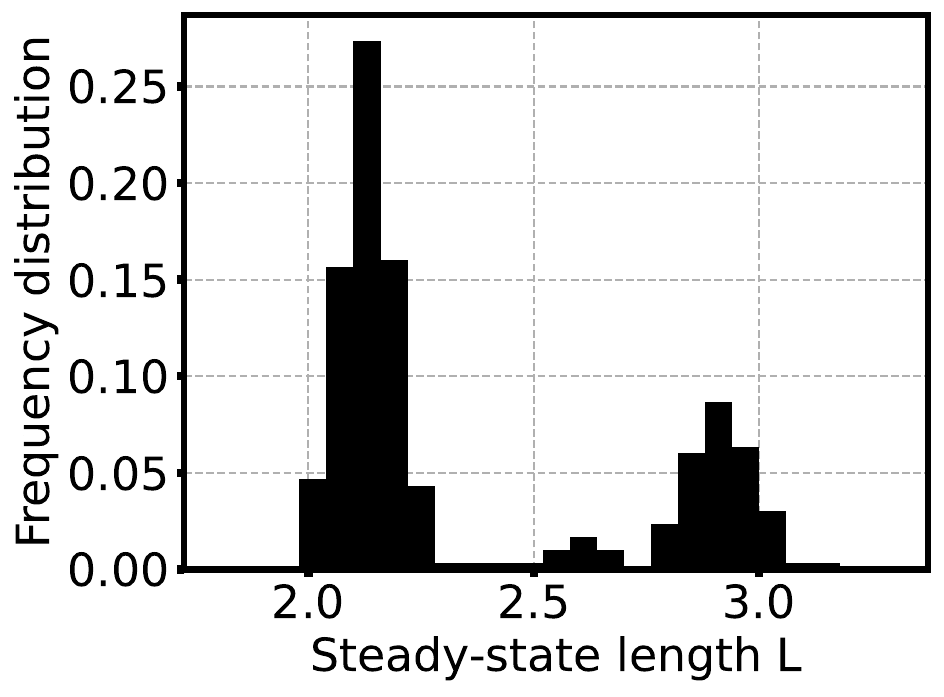}
\caption{$L$ distribution.}
\label{fig:square_random_source_L_distribution}
\end{subfigure}
\hfill
\begin{subfigure}[b]{0.32\textwidth}
\centering
\includegraphics[width=\textwidth]{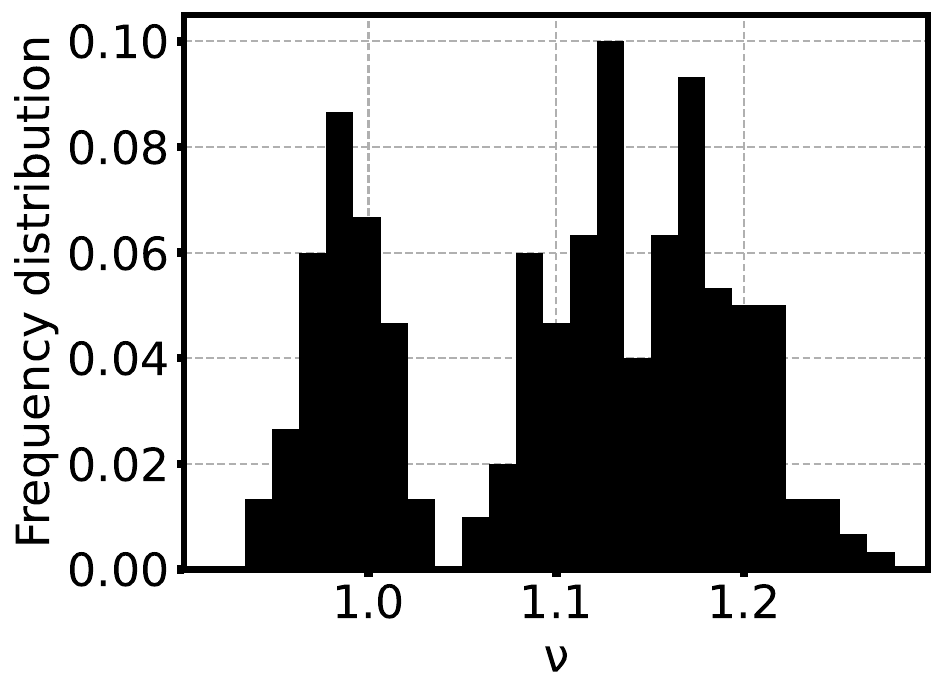}
\caption{$\nu$ distribution.}
\label{fig:square_random_source_niu_distribution}
\end{subfigure}
\hfill
\begin{subfigure}[b]{0.32\textwidth}
\centering
\includegraphics[width=\textwidth]{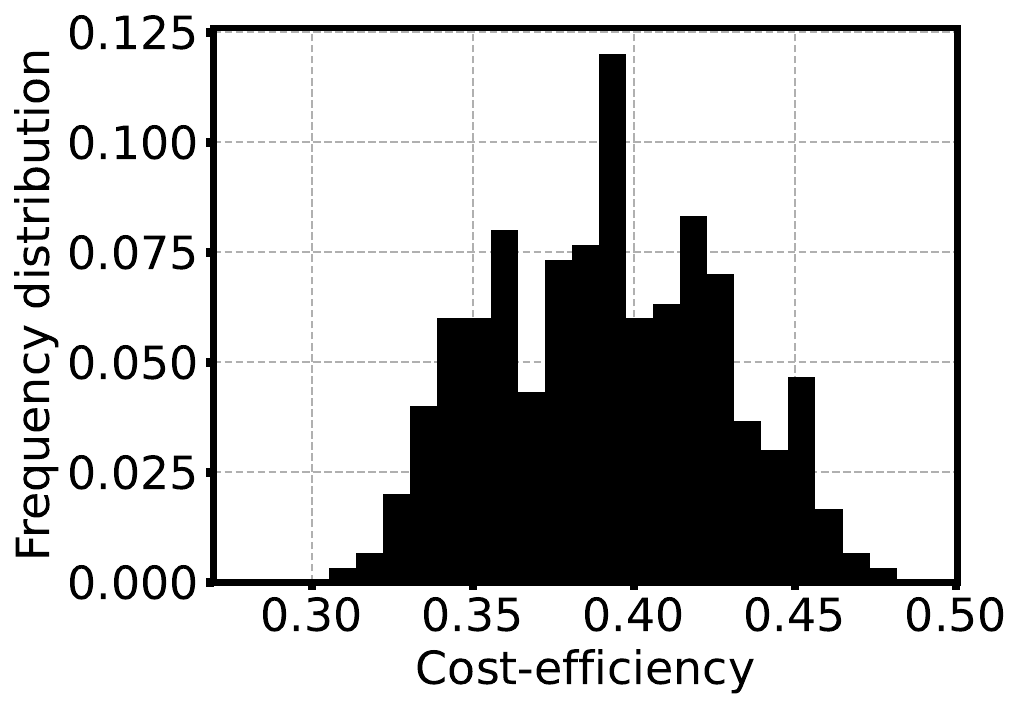}
\caption{CE distribution.}
\label{fig:square_random_source_CE_distribution}
\end{subfigure}
\caption{Additional results for the square configuration, for the \textbf{Random source} algorithm.}
\label{fig:square_random_source_add}
\end{figure}

\begin{figure}[H]
\centering
\begin{subfigure}[b]{0.4\textwidth}
 \centering
 \includegraphics[width=0.8\textwidth]{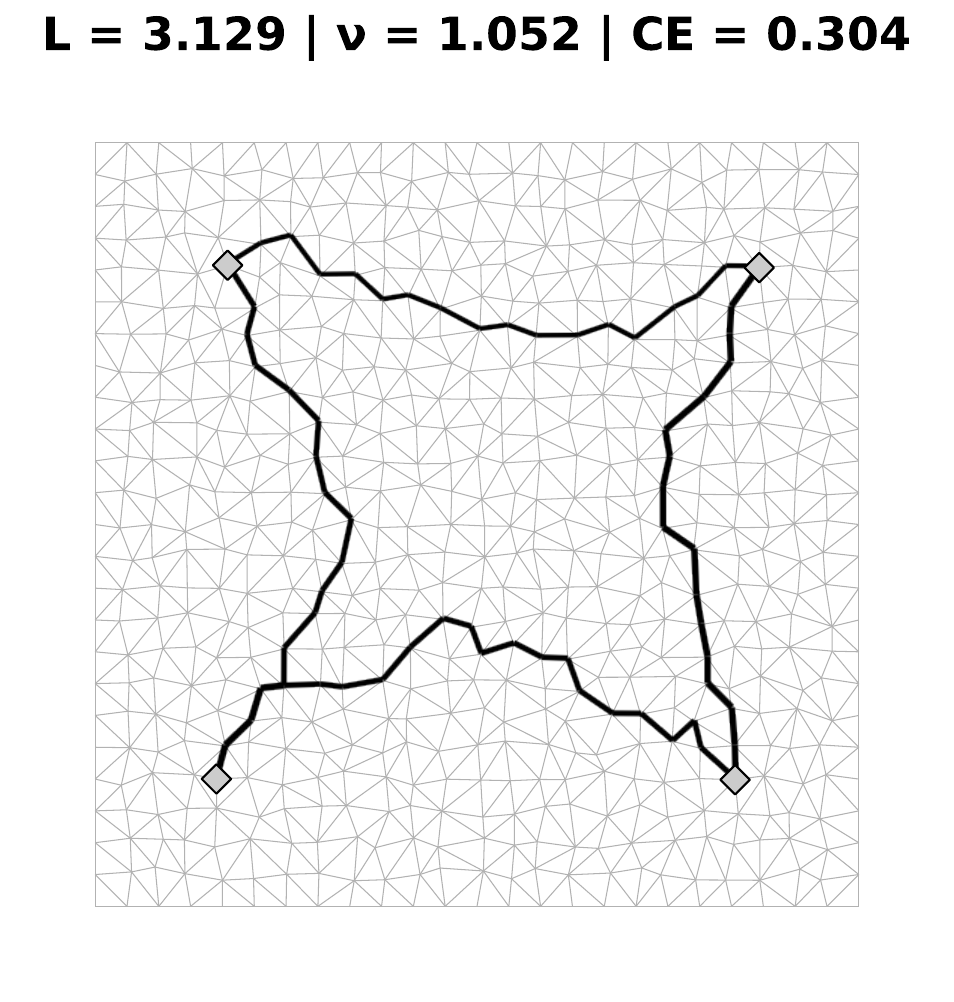}
 \caption{Steady state with largest $L$ and smallest CE.}
 \label{fig:square_random_fixed_I0_ss_biggest_L_smallest_CE}
\end{subfigure}
\begin{subfigure}[b]{0.4\textwidth}
 \centering
 \includegraphics[width=0.8\textwidth]{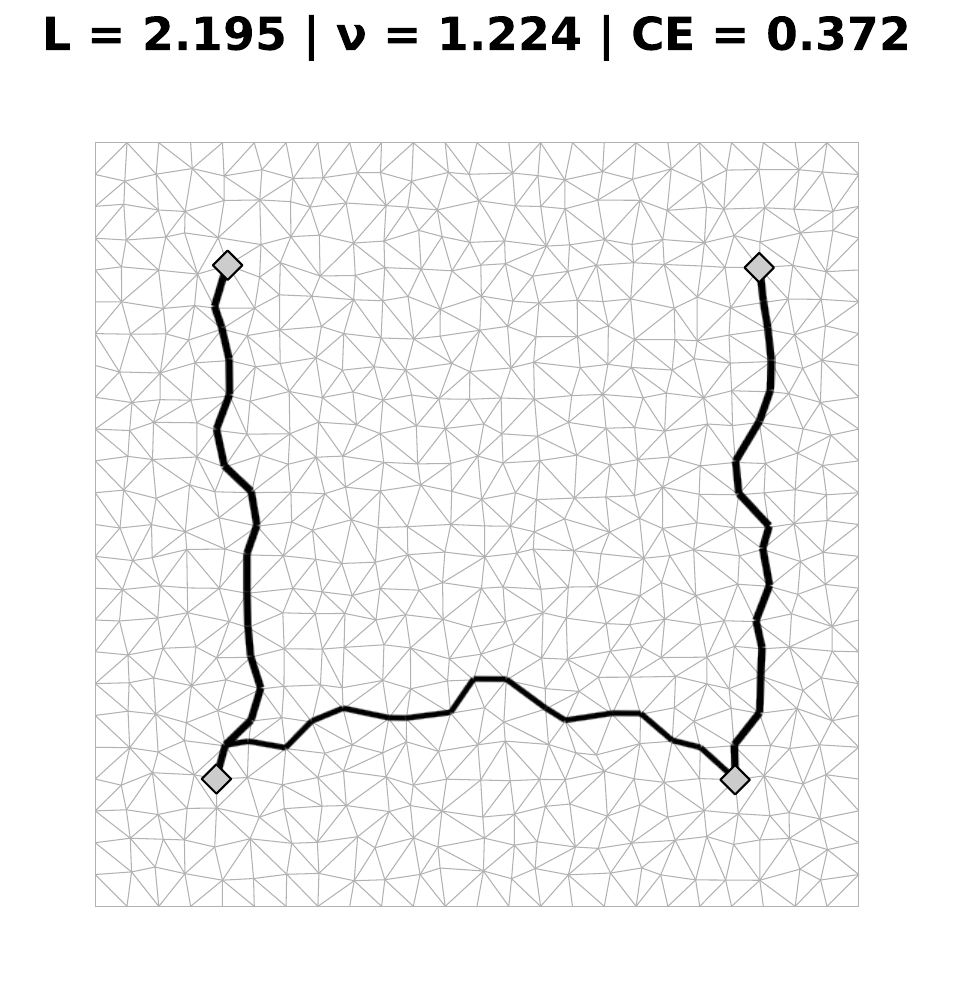}
 \caption{Steady state with largest $\nu$.}
 \label{fig:square_random_fixed_I0_ss_biggest_niu}
\end{subfigure}
\\
\begin{subfigure}[b]{0.32\textwidth}
\centering
\includegraphics[width=\textwidth]{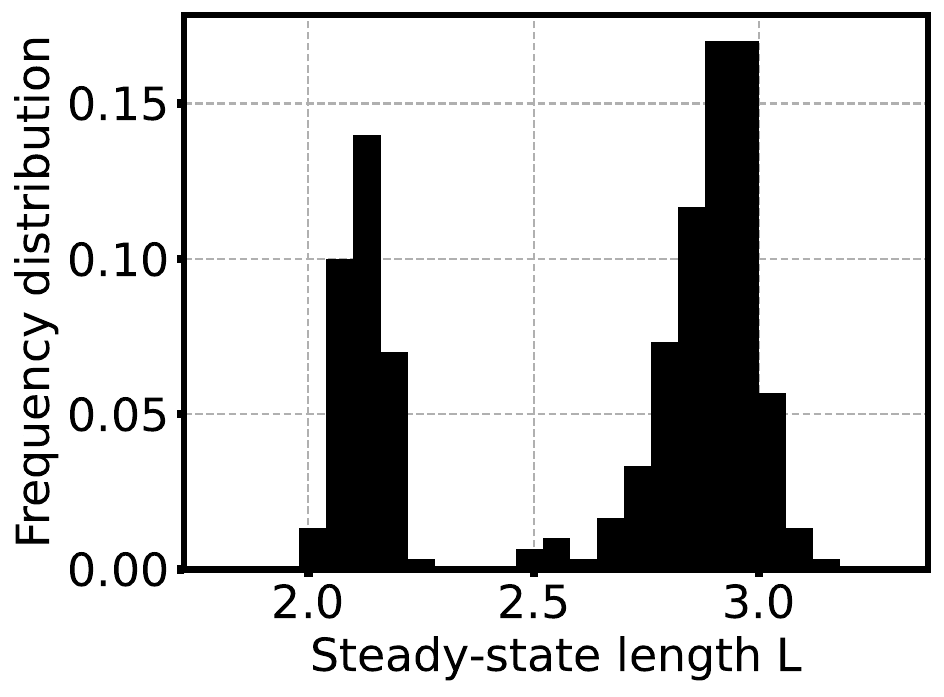}
\caption{$L$ distribution.}
\label{fig:square_random_fixed_I0_L_distribution}
\end{subfigure}
\hfill
\begin{subfigure}[b]{0.32\textwidth}
\centering
\includegraphics[width=\textwidth]{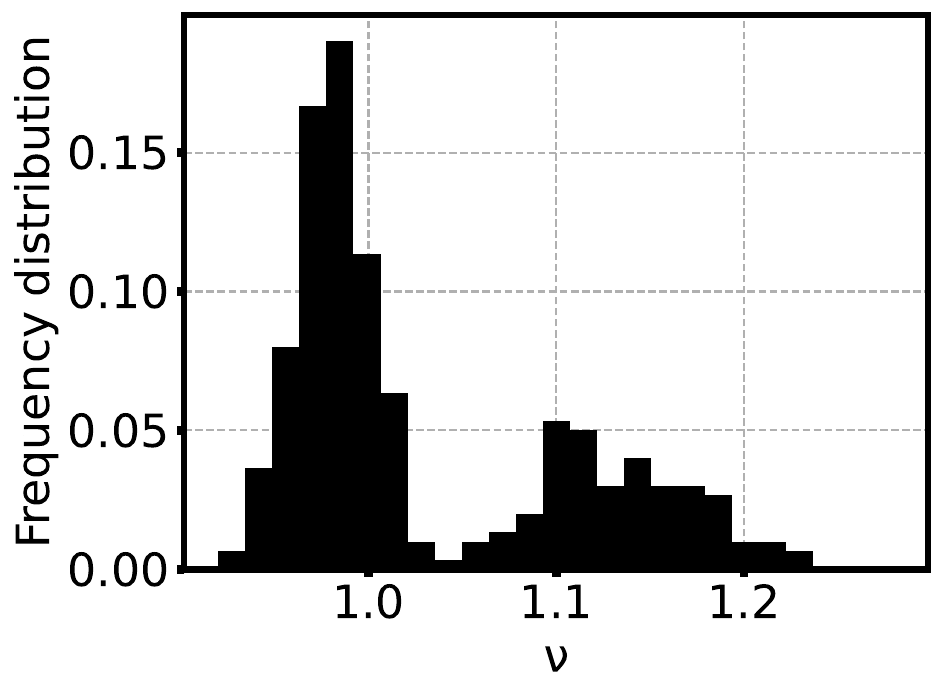}
\caption{$\nu$ distribution.}
\label{fig:square_random_fixed_I0_niu_distribution}
\end{subfigure}
\hfill
\begin{subfigure}[b]{0.32\textwidth}
\centering
\includegraphics[width=\textwidth]{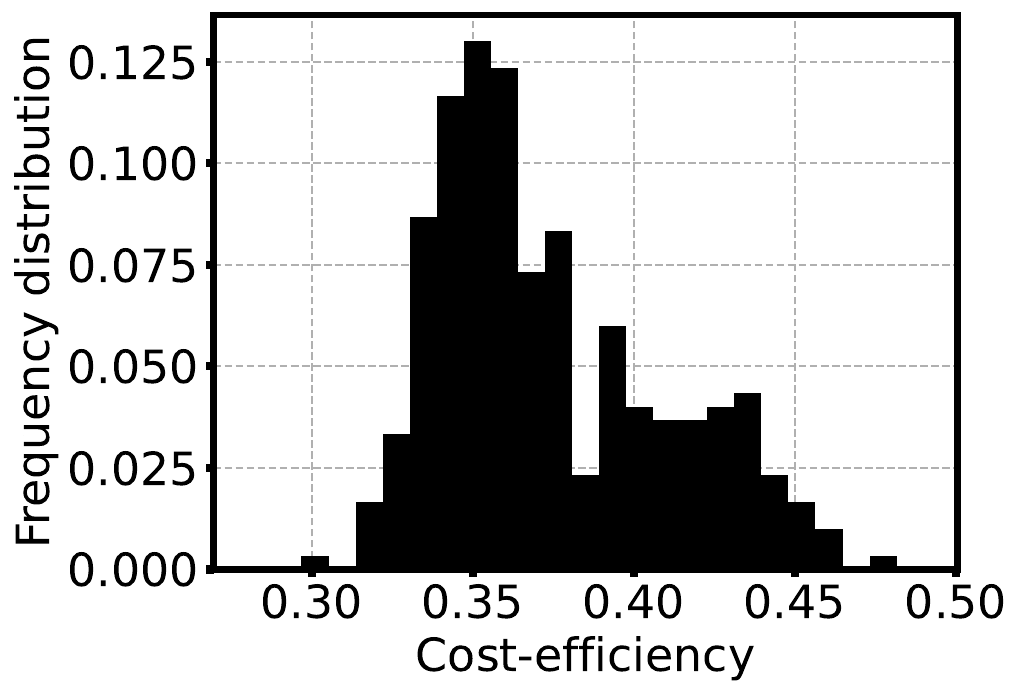}
\caption{CE distribution.}
\label{fig:square_random_fixed_I0_CE_distribution}
\end{subfigure}
\caption{Additional results for the square configuration, for the \textbf{Random random} algorithm.}
\label{fig:square_random_fixed_I0_add}
\end{figure}

\begin{table}[H]
\centering
\begin{tabular}{c|c|c|c}
\toprule
Algorithm & $L$ & $\nu$ & CE \\
\midrule
\texttt{Random pair} & 2.928 $\pm$ 0.047 & 0.971 $\pm$ 0.015 & 0.352 $\pm$ 0.011 \\
\texttt{Random source} & 2.37 $\pm$ 0.35 & 1.099 $\pm$ 0.086 & 0.391 $\pm$ 0.036 \\
\texttt{Random half} & 2.31 $\pm$ 0.35 & 1.079 $\pm$ 0.060 & 0.408 $\pm$ 0.044 \\
\texttt{Random random} & 2.64 $\pm$ 0.37 & 1.032 $\pm$ 0.078 & 0.374 $\pm$ 0.036 \\
\midrule
Perimeter & 2.715 & 1.107 & 0.383 \\
SMT & 1.851 & 0.983 & 0.514 \\
\bottomrule
\end{tabular}
\caption{Average values and standard deviation of $L$, $\nu$ and CE for the square configuration, for all algorithms, and comparison for the theoretical values for the perimeter and SMT of the square (calculated using the coordinates of the sites and the Steiner point coordinates obtained in section \ref{sec:steiner_points_calc}).}
\label{tab:square}
\end{table}

\section{Pentagonal configuration} \label{appendix:penta}

\begin{figure}[H]
\centering
\begin{subfigure}[b]{0.4\textwidth}
 \centering
 \includegraphics[width=0.8\textwidth]{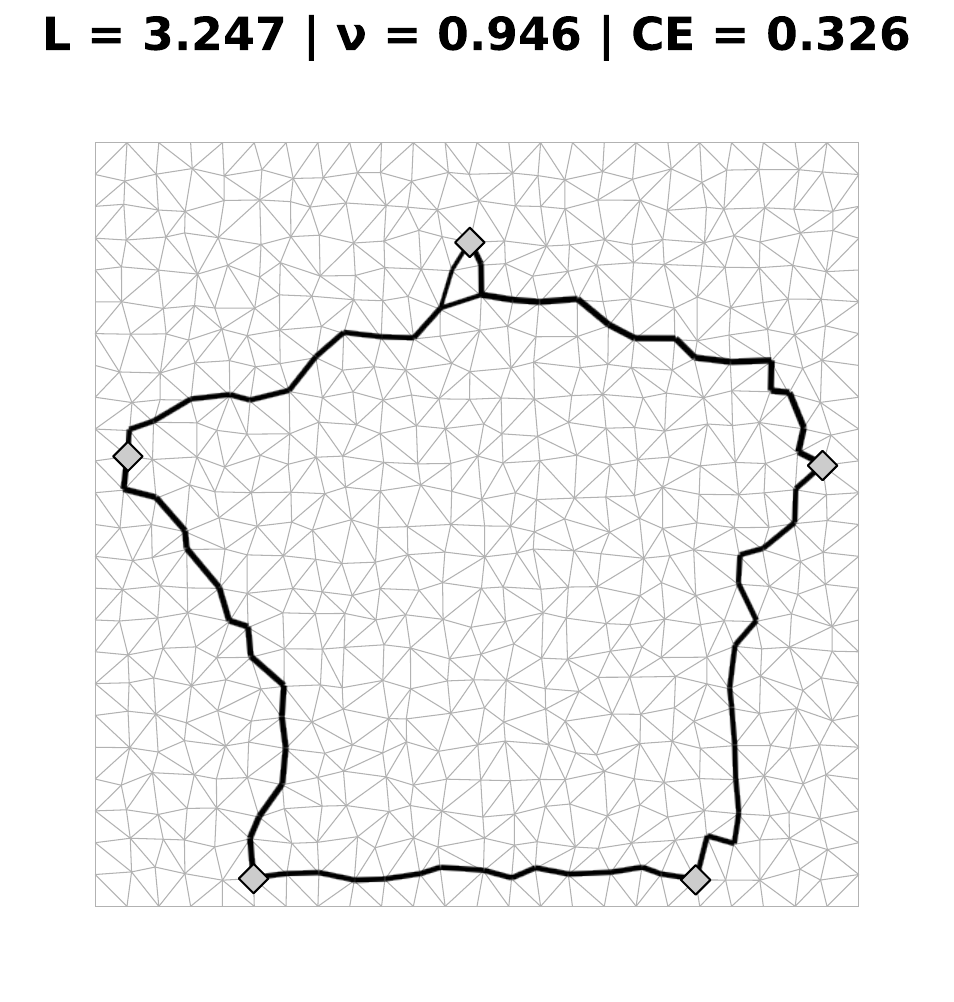}
 \caption{Steady state with largest $L$.}
 \label{fig:penta_random_ss_biggest_L}
\end{subfigure}
\begin{subfigure}[b]{0.4\textwidth}
 \centering
 \includegraphics[width=0.8\textwidth]{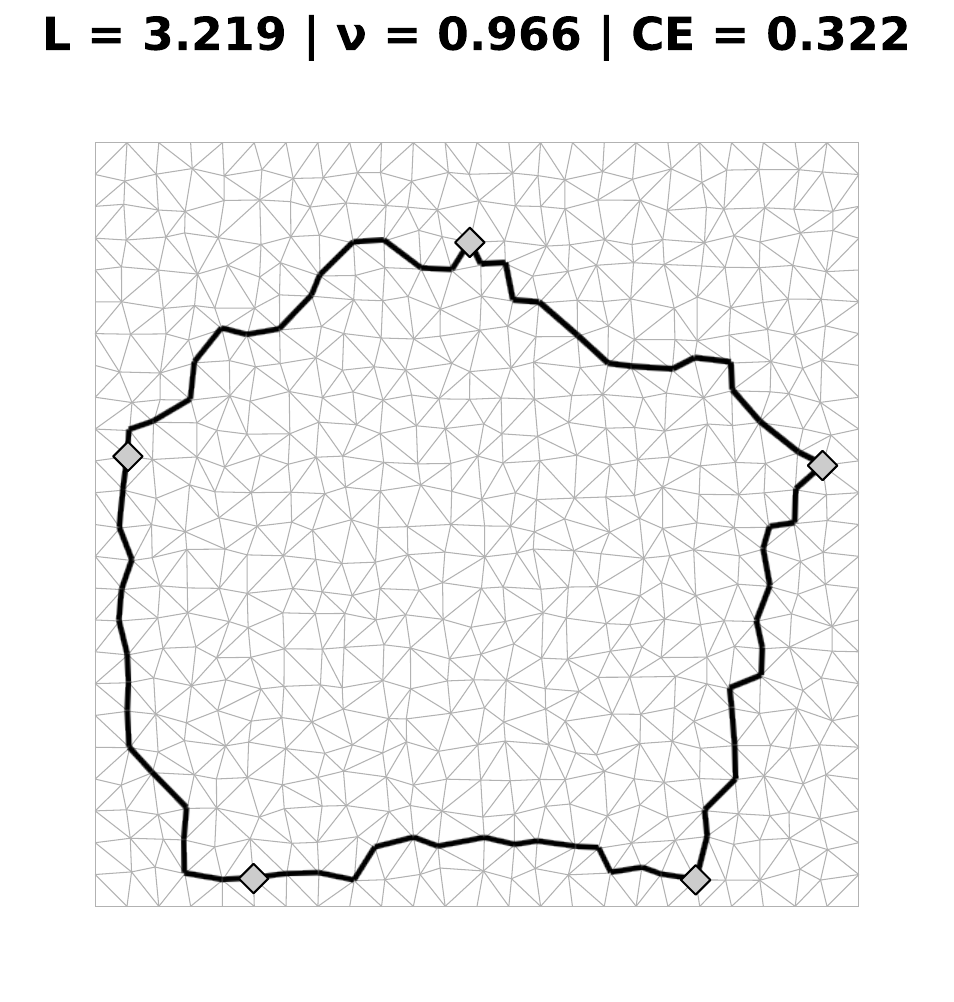}
 \caption{Steady state with largest $\nu$ and smallest CE.}
 \label{fig:penta_random_ss_biggest_niu}
\end{subfigure}
\\
\begin{subfigure}[b]{0.32\textwidth}
\centering
\includegraphics[width=\textwidth]{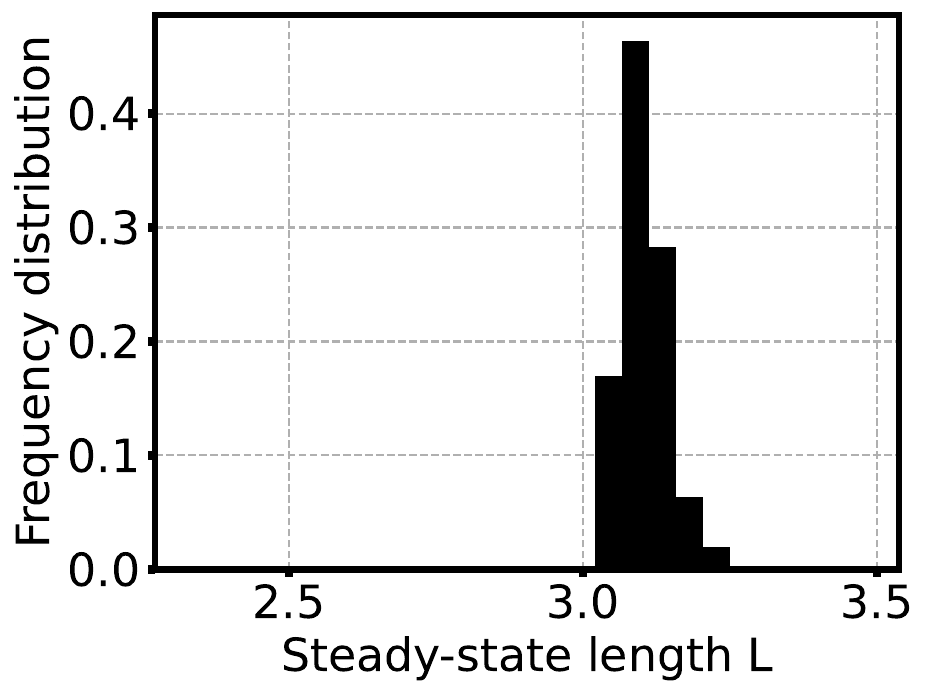}
\caption{$L$ distribution.}
\label{fig:penta_random_L_distribution}
\end{subfigure}
\hfill
\begin{subfigure}[b]{0.32\textwidth}
\centering
\includegraphics[width=\textwidth]{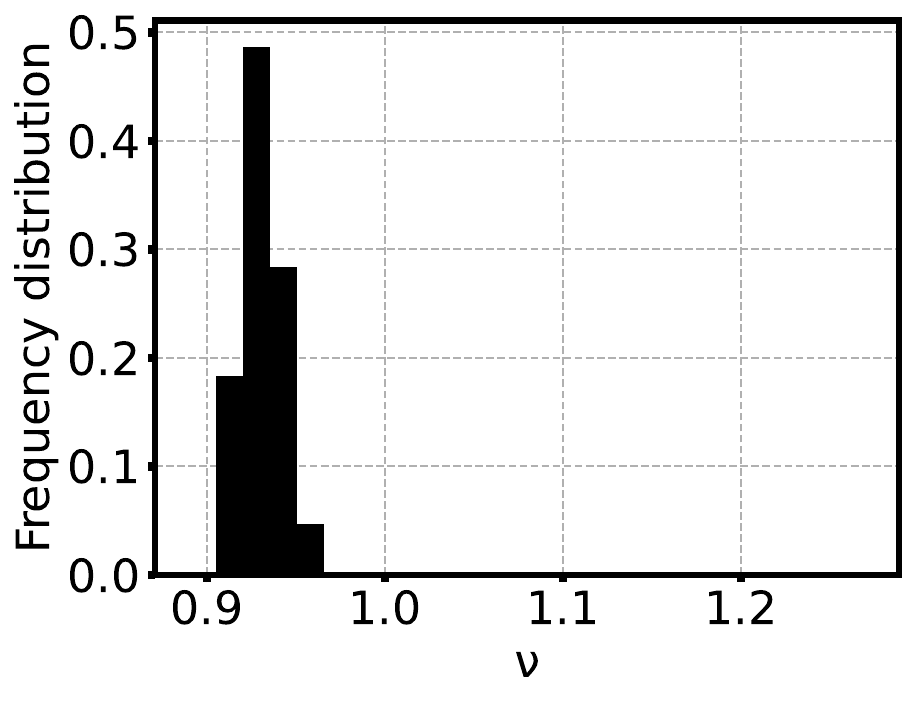}
\caption{$\nu$ distribution.}
\label{fig:penta_random_niu_distribution}
\end{subfigure}
\hfill
\begin{subfigure}[b]{0.32\textwidth}
\centering
\includegraphics[width=\textwidth]{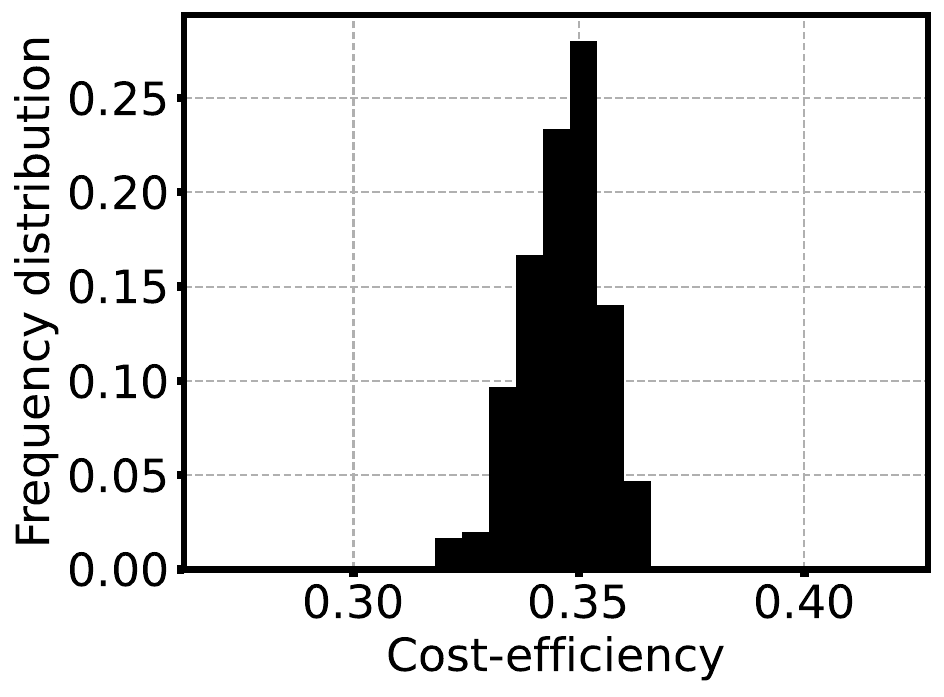}
\caption{CE distribution.}
\label{fig:penta_random_CE_distribution}
\end{subfigure}
\caption{Additional results for the pentagonal configuration, for the \textbf{Random pair} algorithm.}
\label{fig:penta_random_add}
\end{figure}

\begin{figure}[H]
\centering
\begin{subfigure}[b]{0.32\textwidth}
 \centering
 \includegraphics[width=\textwidth]{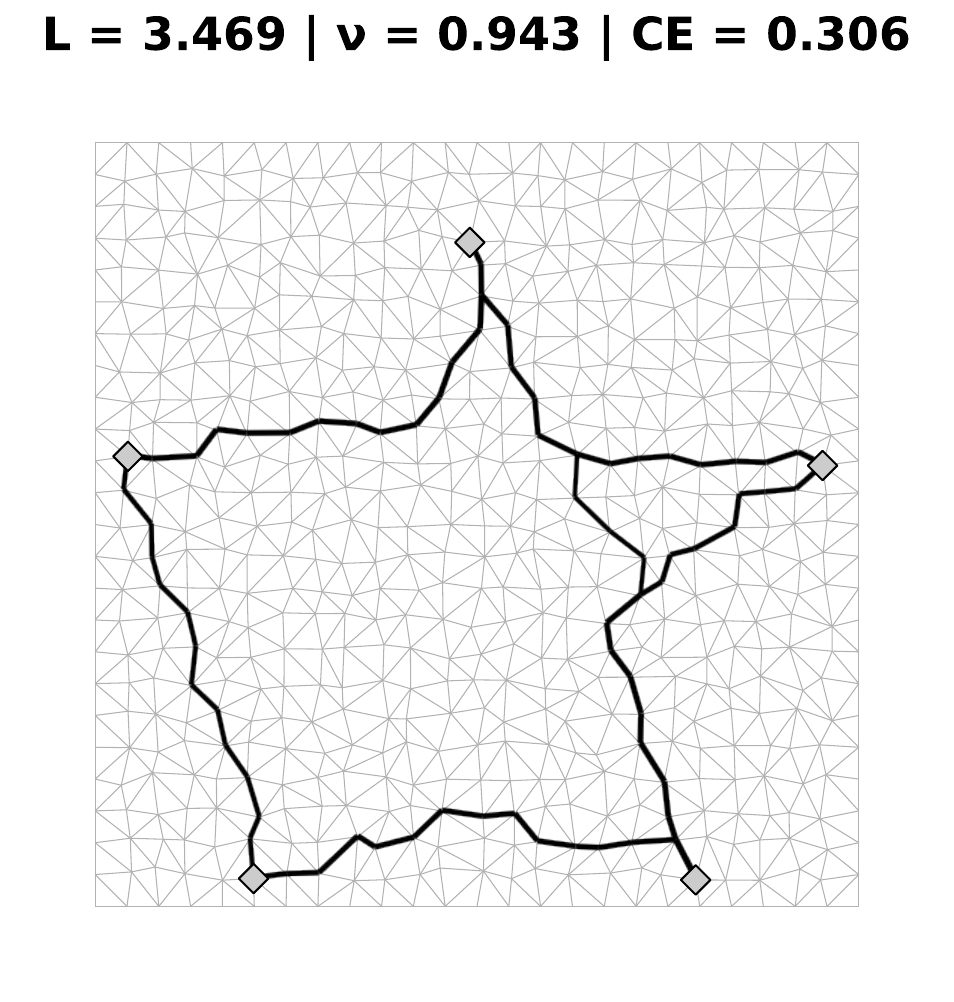}
 \caption{Steady state with largest $L$.}
 \label{fig:penta_random_half_ss_biggest_L}
\end{subfigure}
\begin{subfigure}[b]{0.32\textwidth}
 \centering
 \includegraphics[width=\textwidth]{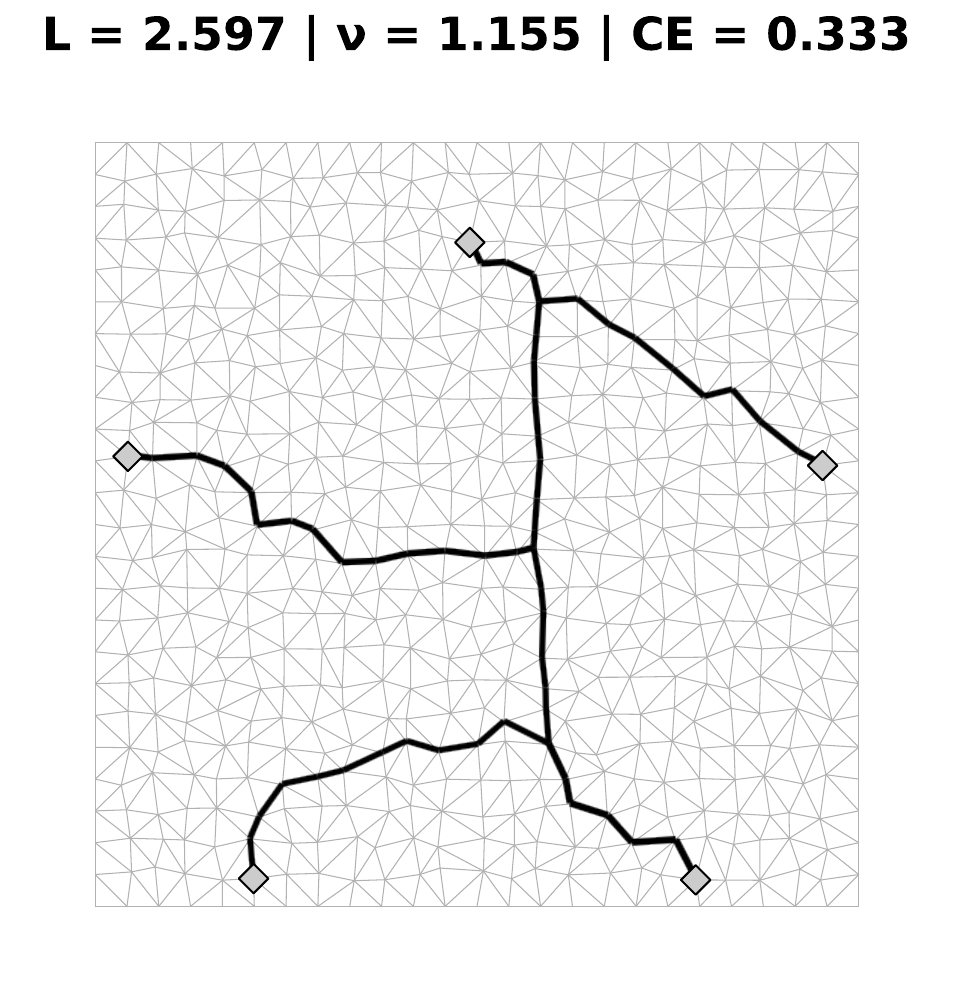}
 \caption{Steady state with largest $\nu$.}
 \label{fig:penta_random_half_ss_biggest_niu}
\end{subfigure}
\begin{subfigure}[b]{0.32\textwidth}
 \centering
 \includegraphics[width=\textwidth]{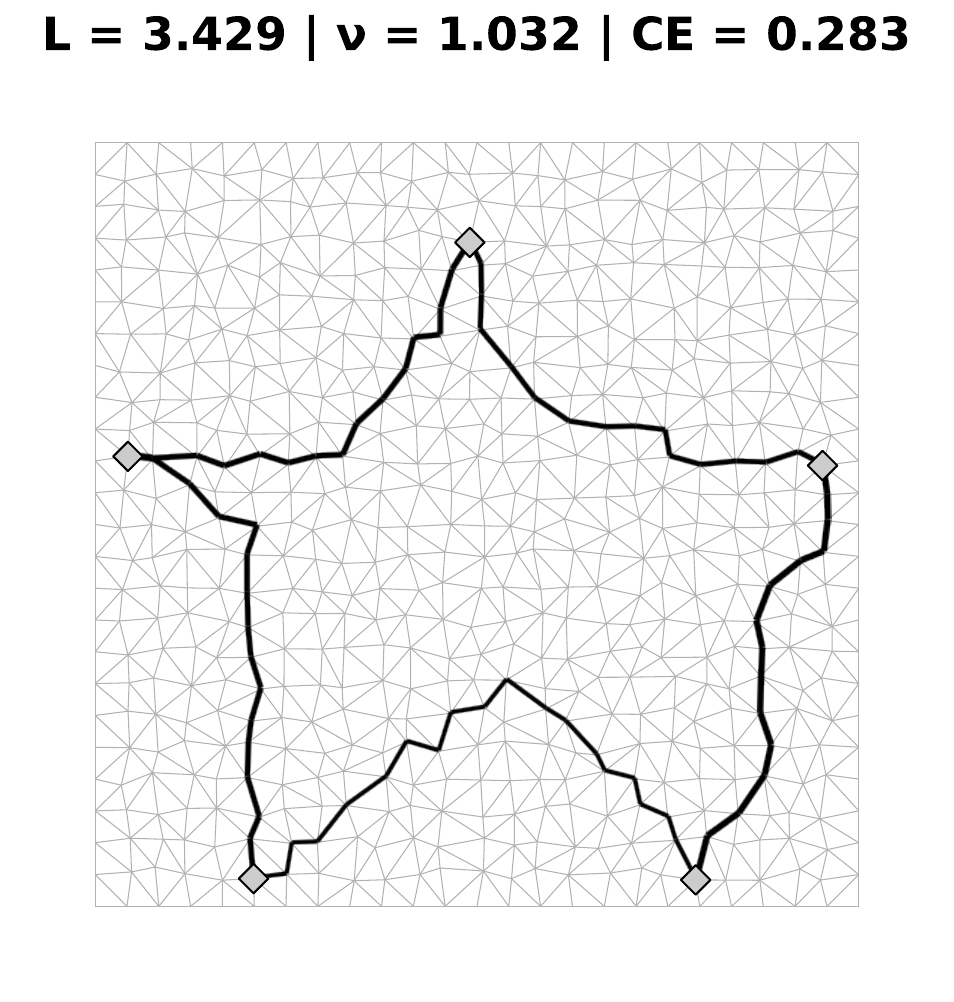}
 \caption{Steady state with smallest CE.}
 \label{fig:penta_random_half_ss_smallest_CE}
\end{subfigure}
\\
\begin{subfigure}[b]{0.32\textwidth}
\centering
\includegraphics[width=\textwidth]{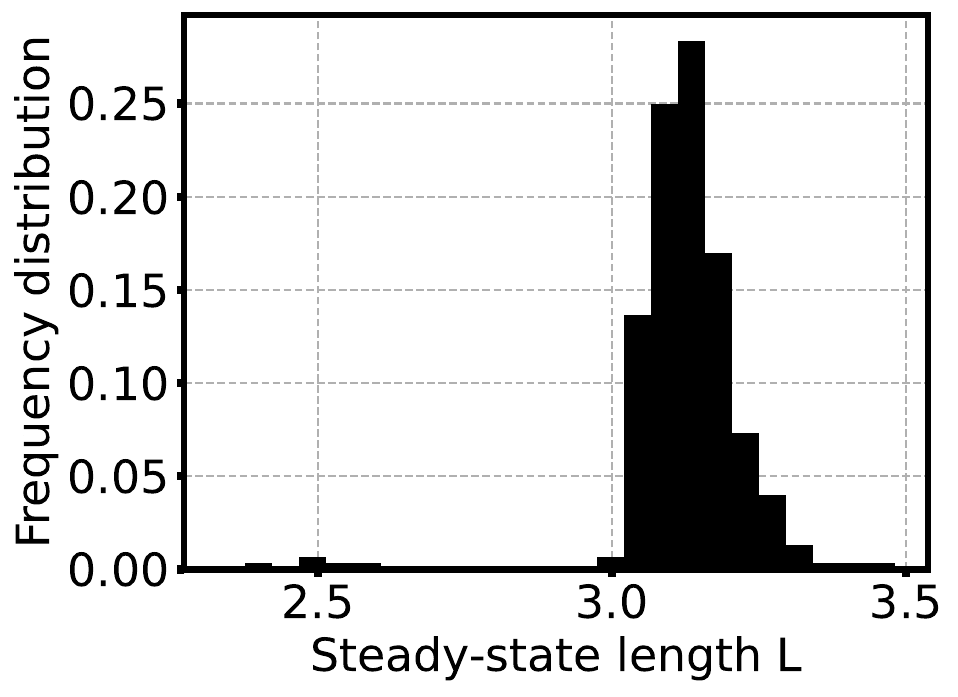}
\caption{$L$ distribution.}
\label{fig:penta_random_half_L_distribution}
\end{subfigure}
\hfill
\begin{subfigure}[b]{0.32\textwidth}
\centering
\includegraphics[width=\textwidth]{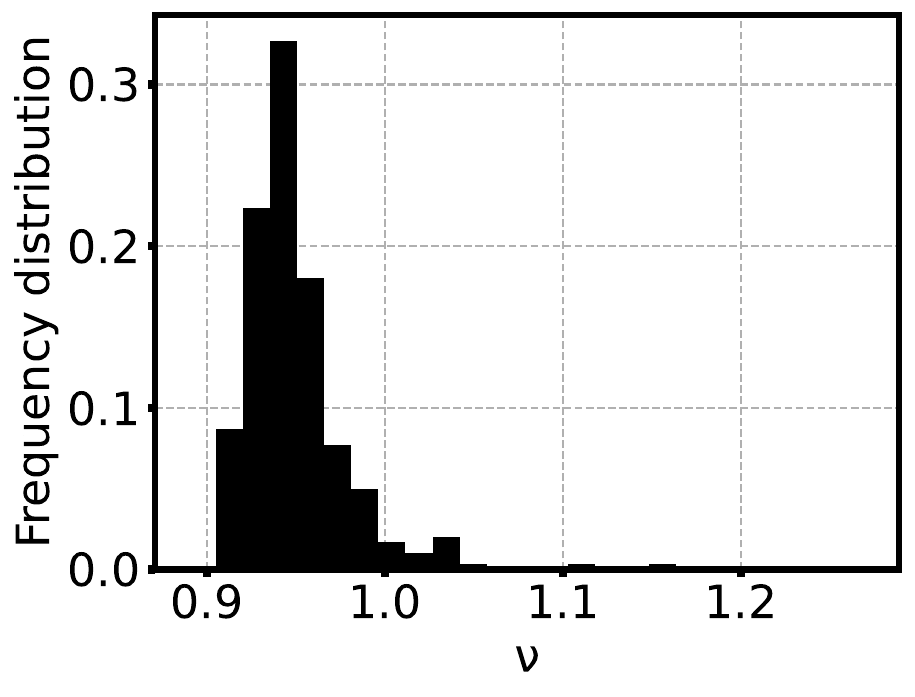}
\caption{$\nu$ distribution.}
\label{fig:penta_random_half_niu_distribution}
\end{subfigure}
\hfill
\begin{subfigure}[b]{0.32\textwidth}
\centering
\includegraphics[width=\textwidth]{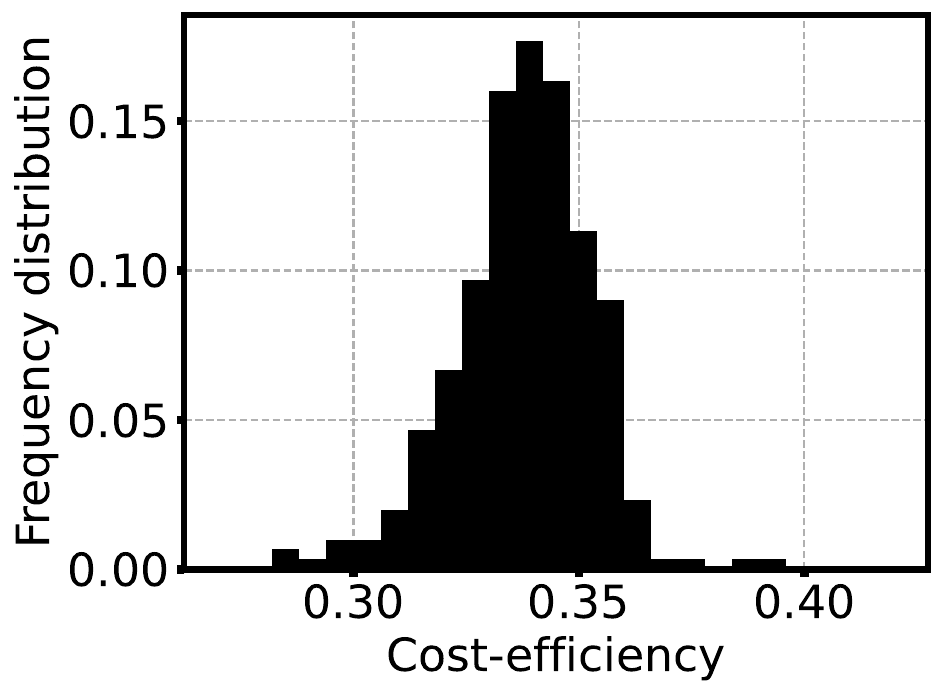}
\caption{CE distribution.}
\label{fig:penta_random_half_CE_distribution}
\end{subfigure}
\caption{Additional results for the pentagonal configuration, for the \textbf{Random half} algorithm.}
\label{fig:penta_random_half_add}
\end{figure}

\begin{figure}[H]
\centering
\begin{subfigure}[b]{0.32\textwidth}
 \centering
 \includegraphics[width=\textwidth]{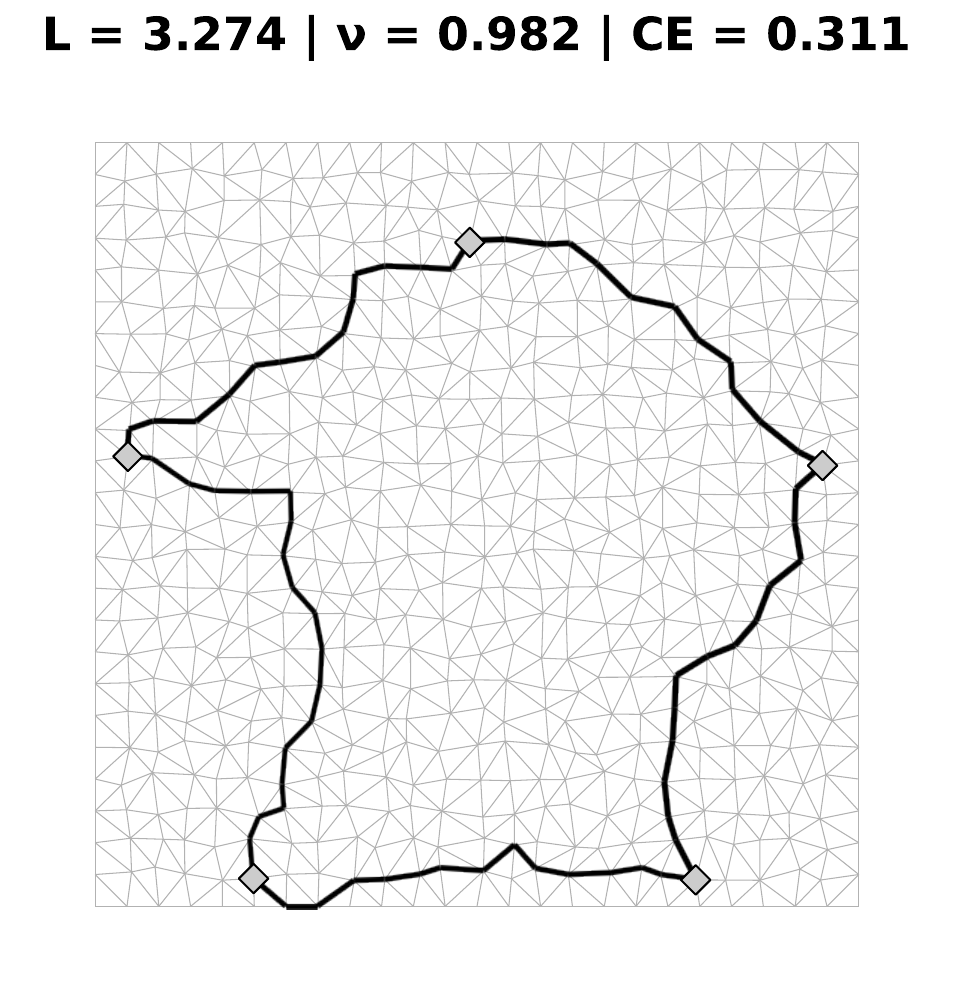}
 \caption{Steady state with largest $L$.}
 \label{fig:penta_random_source_ss_biggest_L}
\end{subfigure}
\begin{subfigure}[b]{0.32\textwidth}
 \centering
 \includegraphics[width=\textwidth]{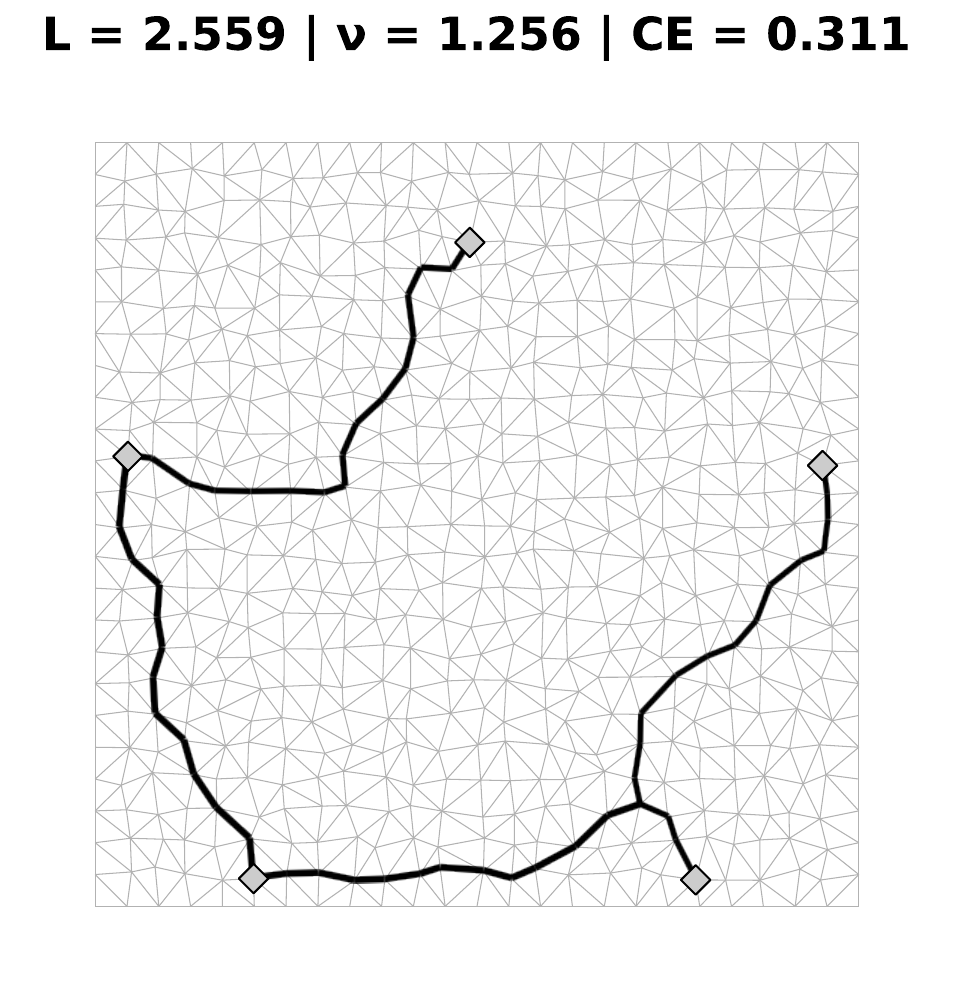}
 \caption{Steady state with largest $\nu$.}
 \label{fig:penta_random_source_ss_biggest_niu}
\end{subfigure}
\begin{subfigure}[b]{0.32\textwidth}
 \centering
 \includegraphics[width=\textwidth]{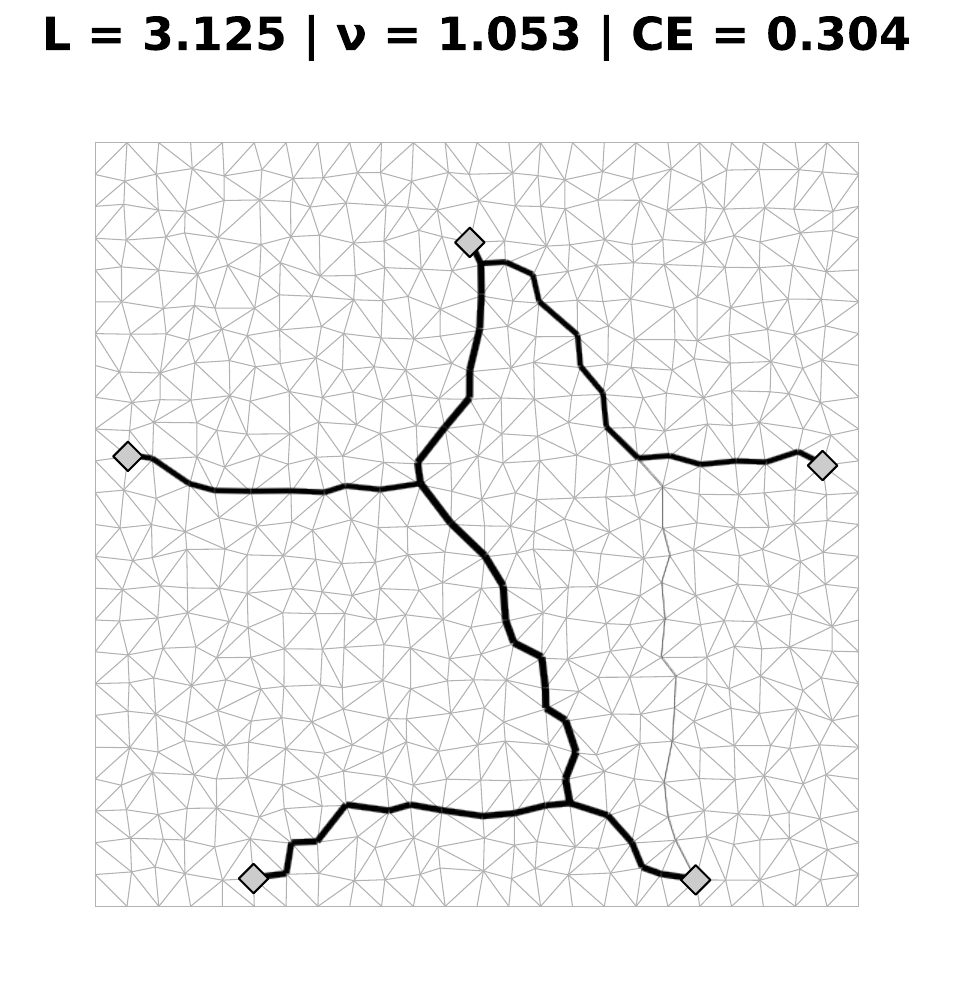}
 \caption{Steady state with smallest CE.}
 \label{fig:penta_random_source_ss_smallest_CE}
\end{subfigure}
\\
\begin{subfigure}[b]{0.32\textwidth}
\centering
\includegraphics[width=\textwidth]{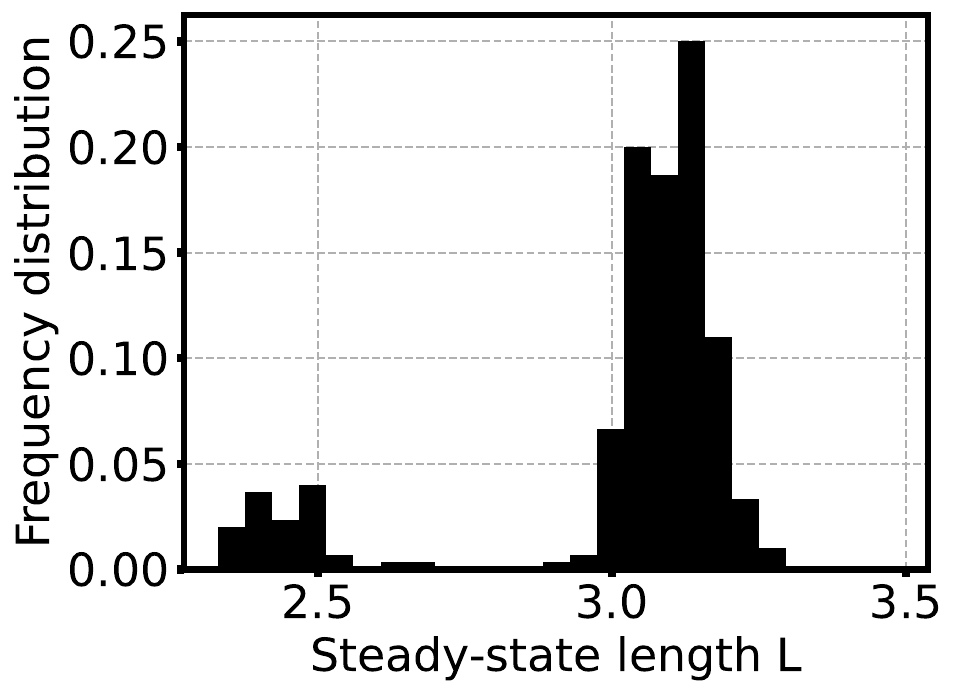}
\caption{$L$ distribution.}
\label{fig:penta_random_source_L_distribution}
\end{subfigure}
\hfill
\begin{subfigure}[b]{0.32\textwidth}
\centering
\includegraphics[width=\textwidth]{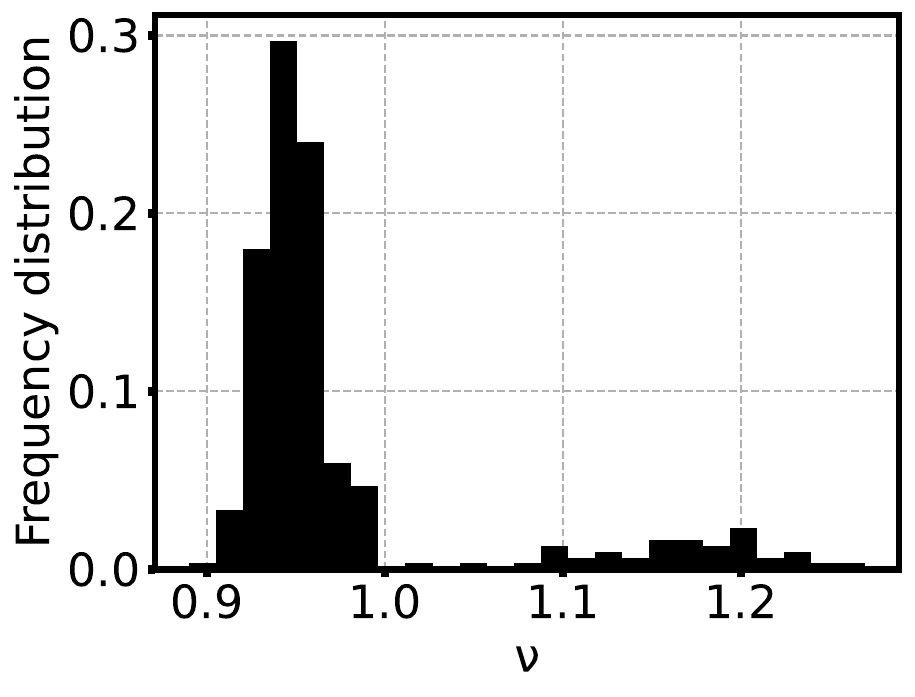}
\caption{$\nu$ distribution.}
\label{fig:penta_random_source_niu_distribution}
\end{subfigure}
\hfill
\begin{subfigure}[b]{0.32\textwidth}
\centering
\includegraphics[width=\textwidth]{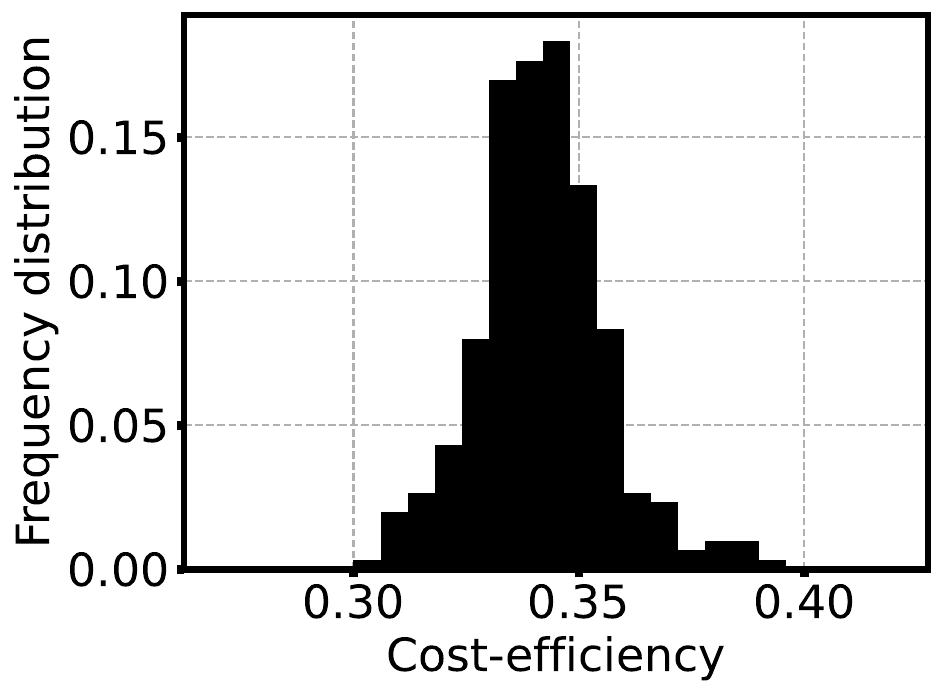}
\caption{CE distribution.}
\label{fig:penta_random_source_CE_distribution}
\end{subfigure}
\caption{Additional results for the pentagonal configuration, for the \textbf{Random source} algorithm.}
\label{fig:penta_random_source_add}
\end{figure}

\begin{figure}[H]
\centering
\begin{subfigure}[b]{0.32\textwidth}
 \centering
 \includegraphics[width=\textwidth]{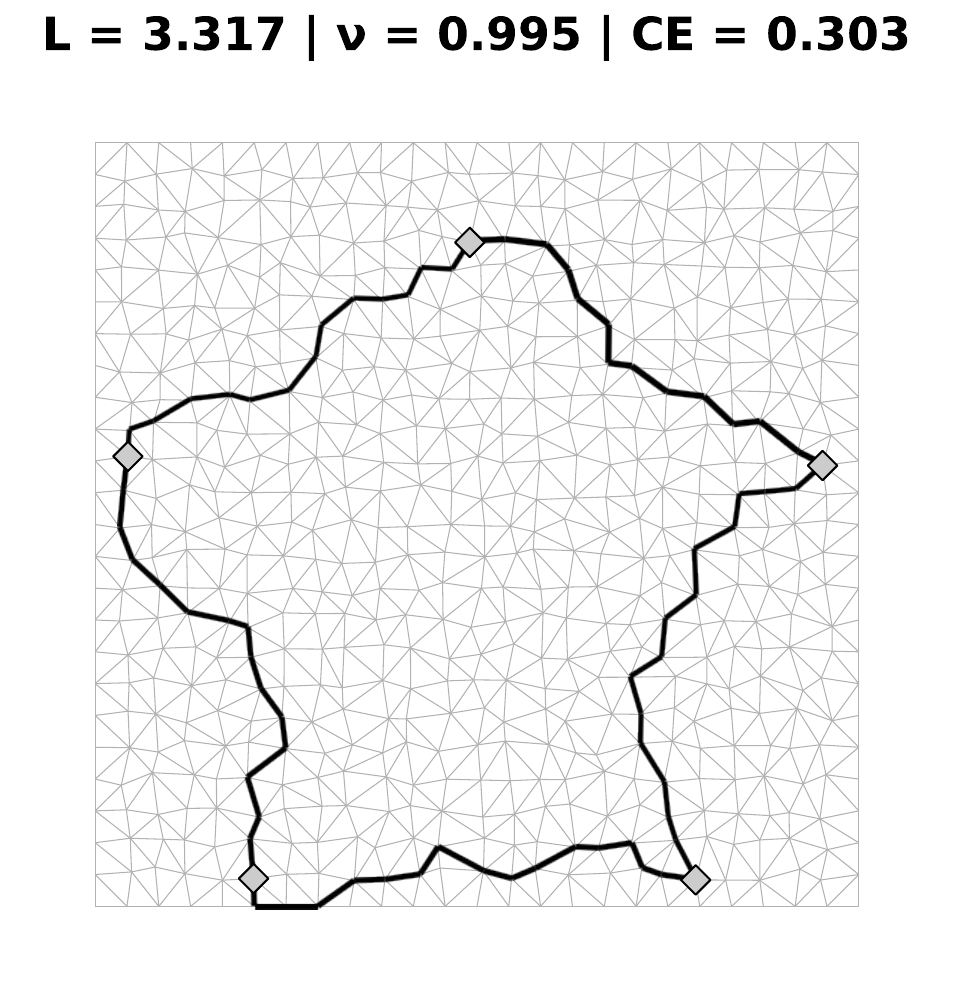}
 \caption{Steady state with largest $L$.}
 \label{fig:penta_random_fixed_I0_ss_biggest_L}
\end{subfigure}
\begin{subfigure}[b]{0.32\textwidth}
 \centering
 \includegraphics[width=\textwidth]{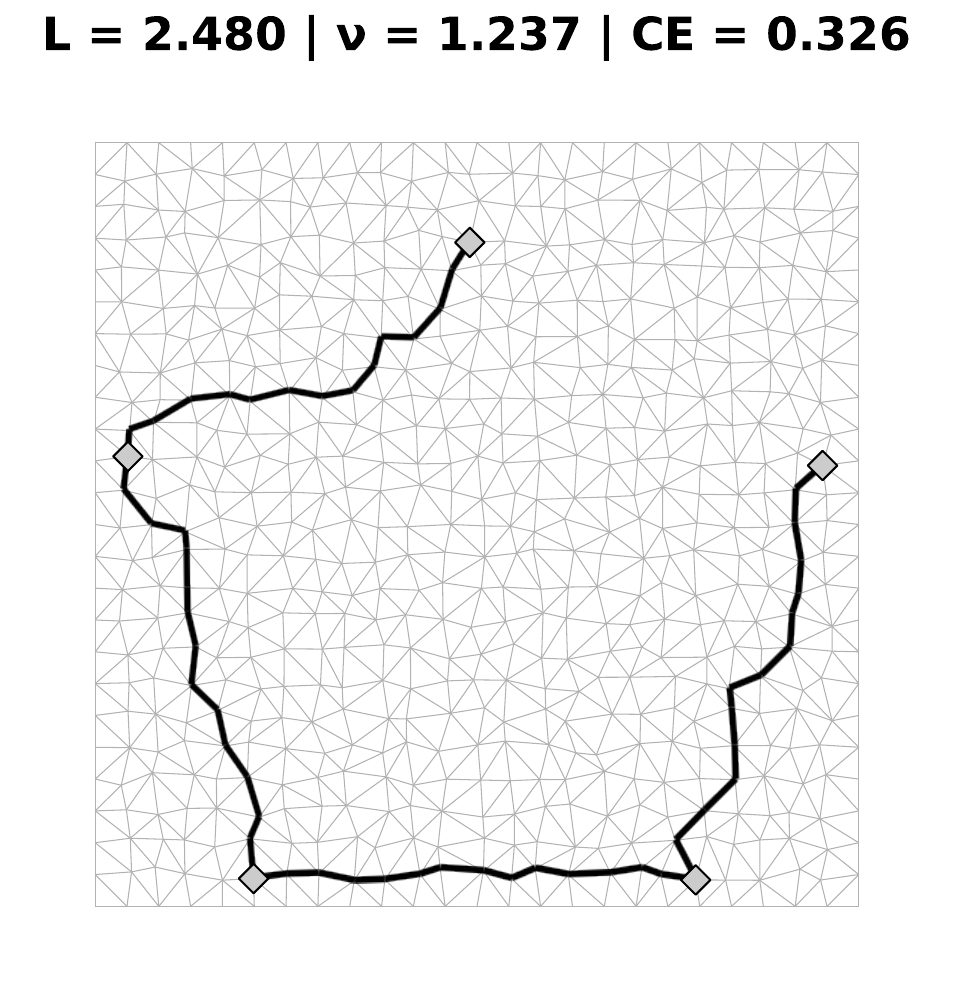}
 \caption{Steady state with largest $\nu$.}
 \label{fig:penta_random_fixed_I0_ss_biggest_niu}
\end{subfigure}
\begin{subfigure}[b]{0.32\textwidth}
 \centering
 \includegraphics[width=\textwidth]{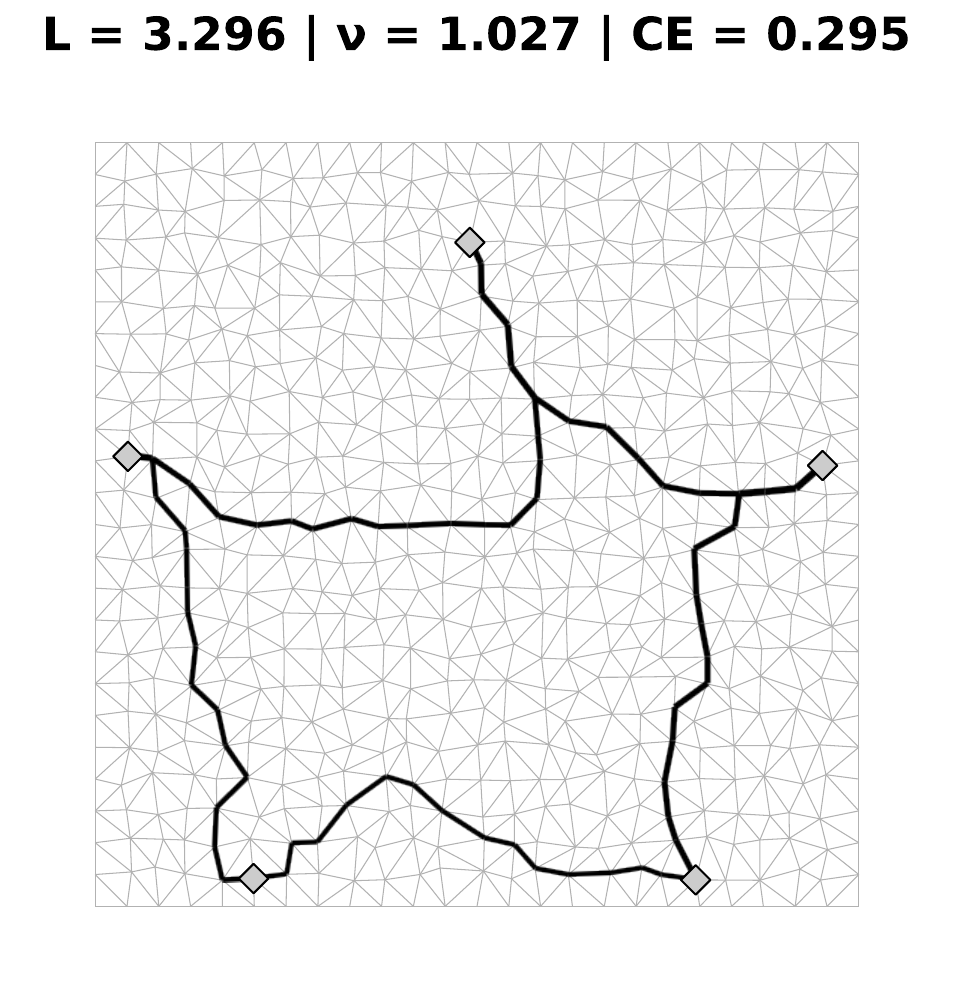}
 \caption{Steady state with smallest CE.}
 \label{fig:penta_random_fixed_I0_ss_smallest_CE}
\end{subfigure}
\\
\begin{subfigure}[b]{0.32\textwidth}
\centering
\includegraphics[width=\textwidth]{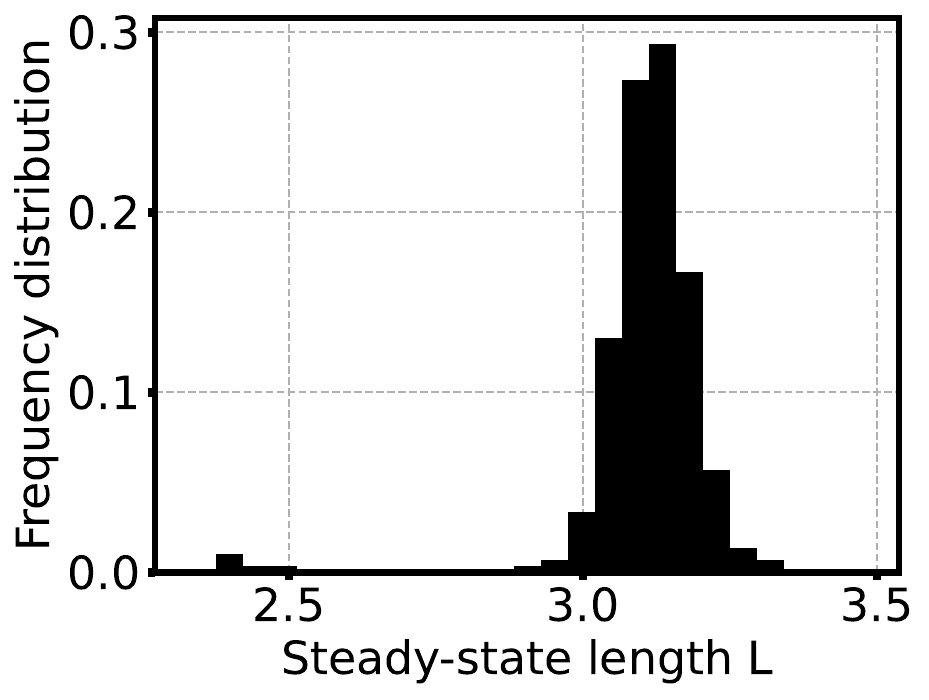}
\caption{$L$ distribution.}
\label{fig:penta_random_fixed_I0_L_distribution}
\end{subfigure}
\hfill
\begin{subfigure}[b]{0.32\textwidth}
\centering
\includegraphics[width=\textwidth]{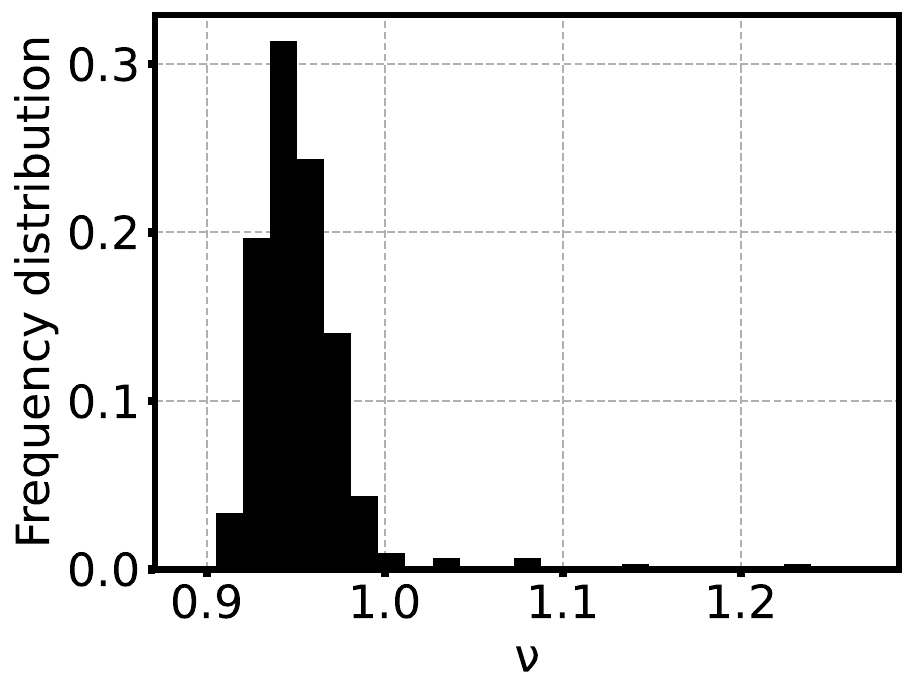}
\caption{$\nu$ distribution.}
\label{fig:penta_random_fixed_I0_niu_distribution}
\end{subfigure}
\hfill
\begin{subfigure}[b]{0.32\textwidth}
\centering
\includegraphics[width=\textwidth]{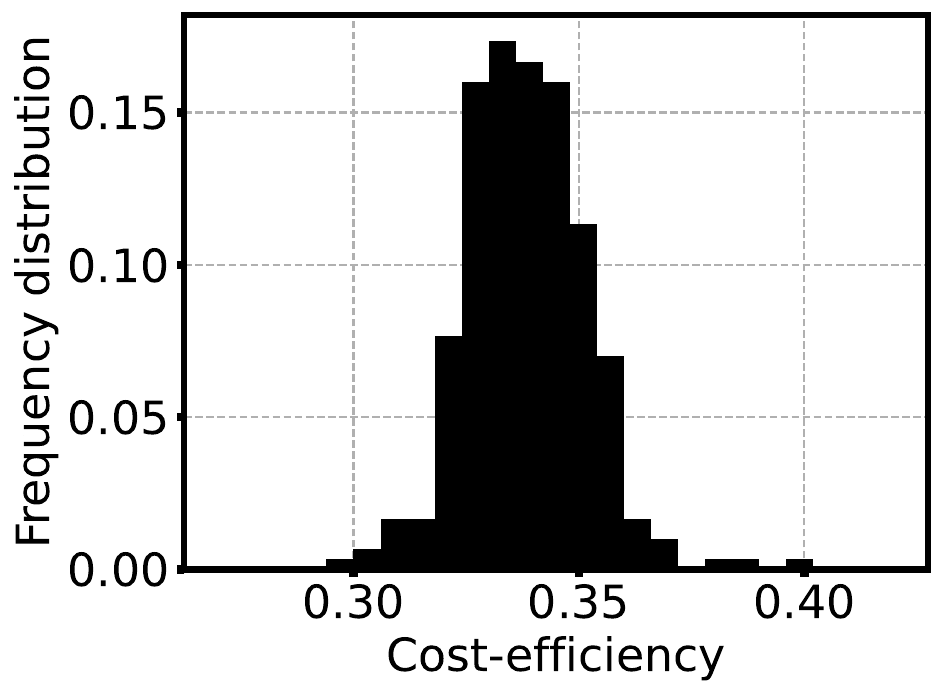}
\caption{CE distribution.}
\label{fig:penta_random_fixed_I0_CE_distribution}
\end{subfigure}
\caption{Additional results for the pentagonal configuration, for the \textbf{Random random} algorithm.}
\label{fig:penta_random_fixed_I0_add}
\end{figure}

\begin{table}[H]
\centering
\begin{tabular}{c|c|c|c}
\toprule
Algorithm & $L$ & $\nu$ & CE \\
\midrule
\texttt{Random pair} & 3.103 $\pm$ 0.039 & 0.931 $\pm$ 0.012 & 0.3463 $\pm$ 0.0085 \\
\texttt{Random source} & 3.01 $\pm$ 0.23 & 0.978 $\pm$ 0.079 & 0.342 $\pm$ 0.014 \\
\texttt{Random half} & 3.13 $\pm$ 0.11 & 0.950 $\pm$ 0.029 & 0.338 $\pm$ 0.015 \\
\texttt{Random random} & 3.11 $\pm$ 0.11 & 0.953 $\pm$ 0.029 & 0.338 $\pm$ 0.013 \\
\midrule
Perimeter & 2.802 & 0.841 & 0.441 \\
SMT & 2.194 & 0.992 & 0.474 \\
\bottomrule
\end{tabular}
\caption{Average values and standard deviation of $L$, $\nu$ and CE for the pentagonal configuration, for all algorithms, and comparison for the theoretical values for the perimeter and SMT of the pentagon (calculated using the coordinates of the sites and the Steiner point coordinates obtained in section \ref{sec:steiner_points_calc}).}
\label{tab:penta}
\end{table}

\section{Mainland Portugal's configuration} \label{appendix:portugal}

\begin{figure}[H]
\centering
\begin{subfigure}[b]{0.32\textwidth}
 \centering
 \includegraphics[width=\textwidth]{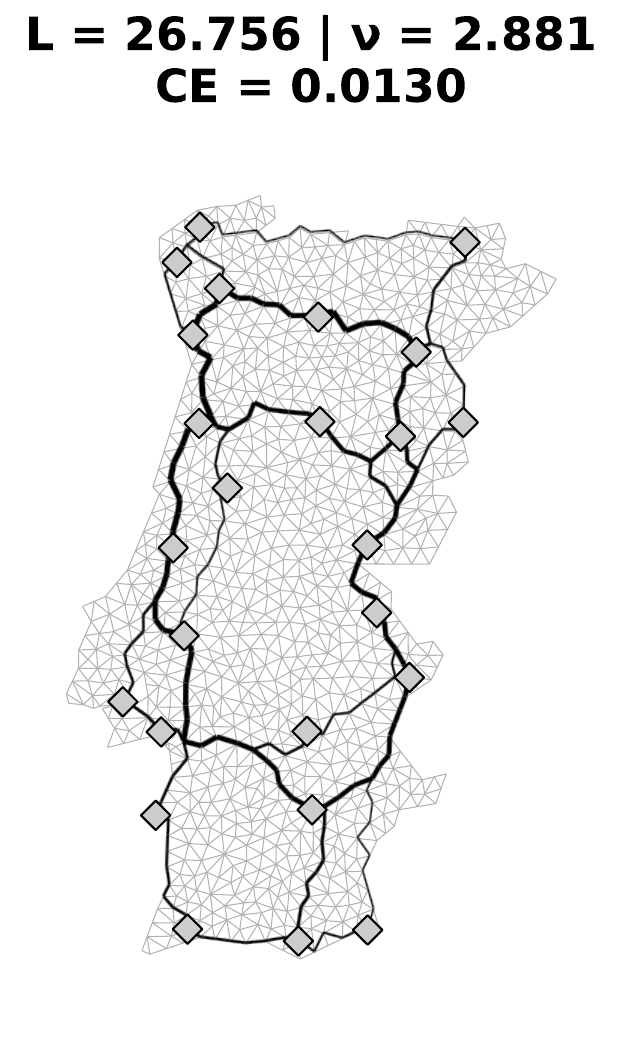}
 \caption{Steady state with largest $L$.}
 \label{fig:Portugal_random_ss_biggest_L}
\end{subfigure}
\begin{subfigure}[b]{0.32\textwidth}
 \centering
 \includegraphics[width=\textwidth]{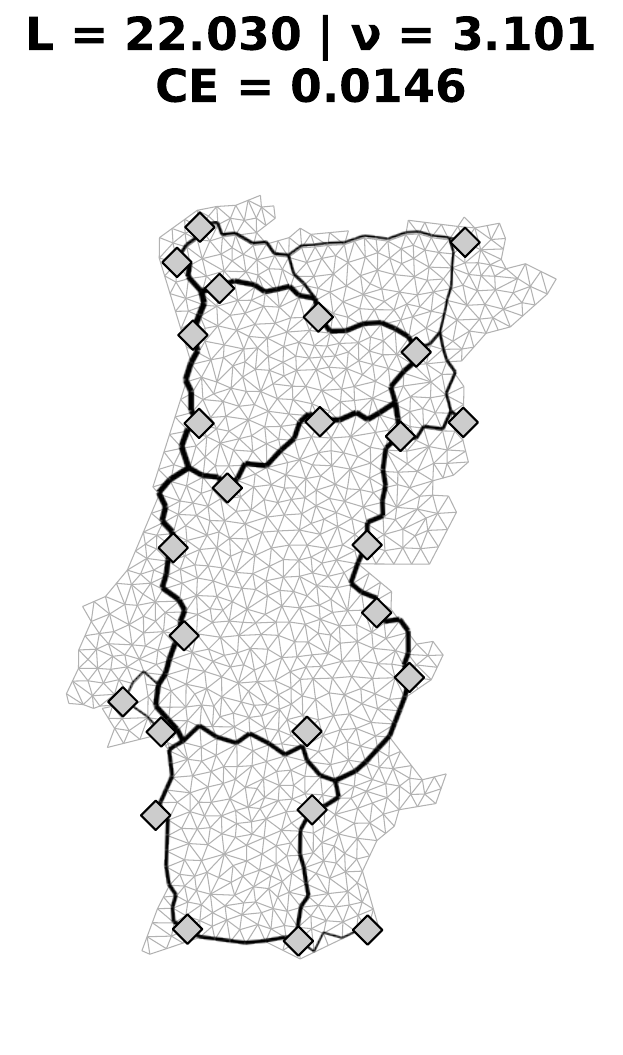}
 \caption{Steady state with largest $\nu$.}
 \label{fig:Portugal_random_ss_biggest_niu}
\end{subfigure}
\begin{subfigure}[b]{0.32\textwidth}
 \centering
 \includegraphics[width=\textwidth]{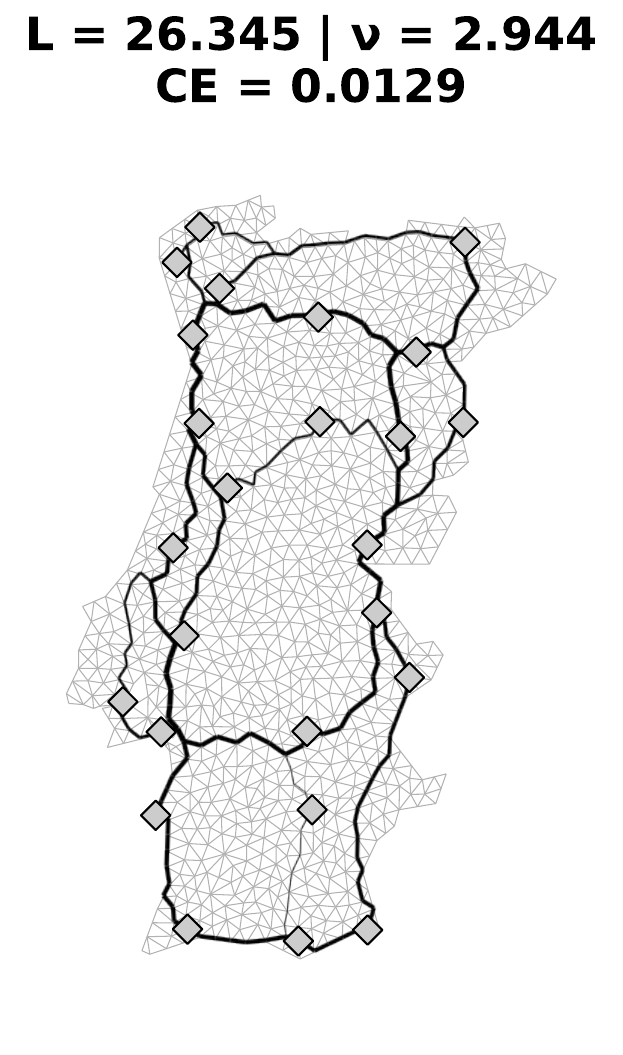}
 \caption{Steady state with smallest CE.}
 \label{fig:Portugal_random_ss_smallest_CE}
\end{subfigure}
\\
\begin{subfigure}[b]{0.32\textwidth}
\centering
\includegraphics[width=\textwidth]{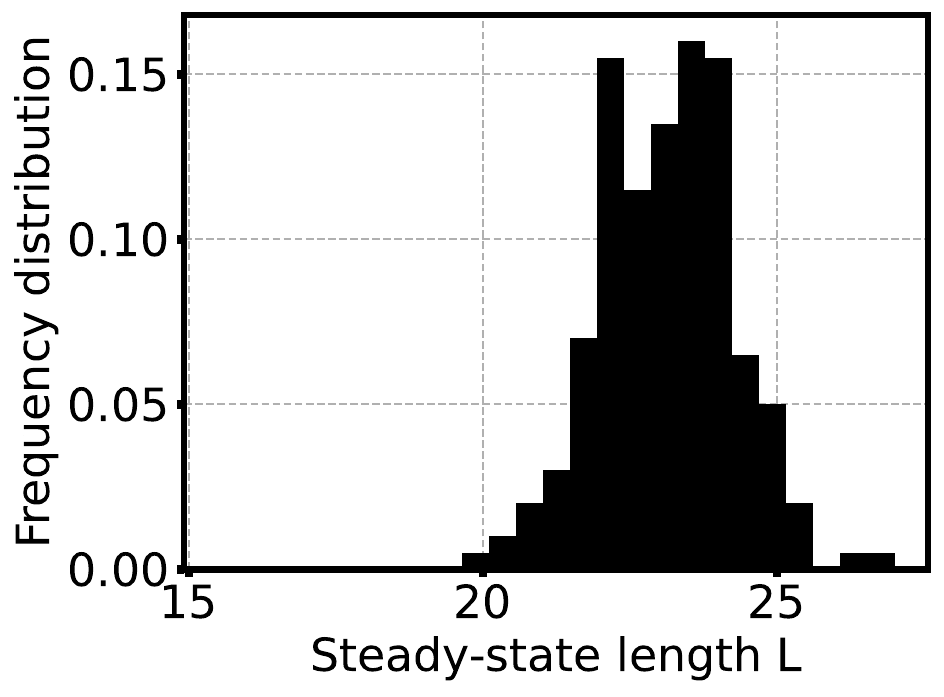}
\caption{$L$ distribution.}
\label{fig:Portugal_random_L_distribution}
\end{subfigure}
\hfill
\begin{subfigure}[b]{0.32\textwidth}
\centering
\includegraphics[width=\textwidth]{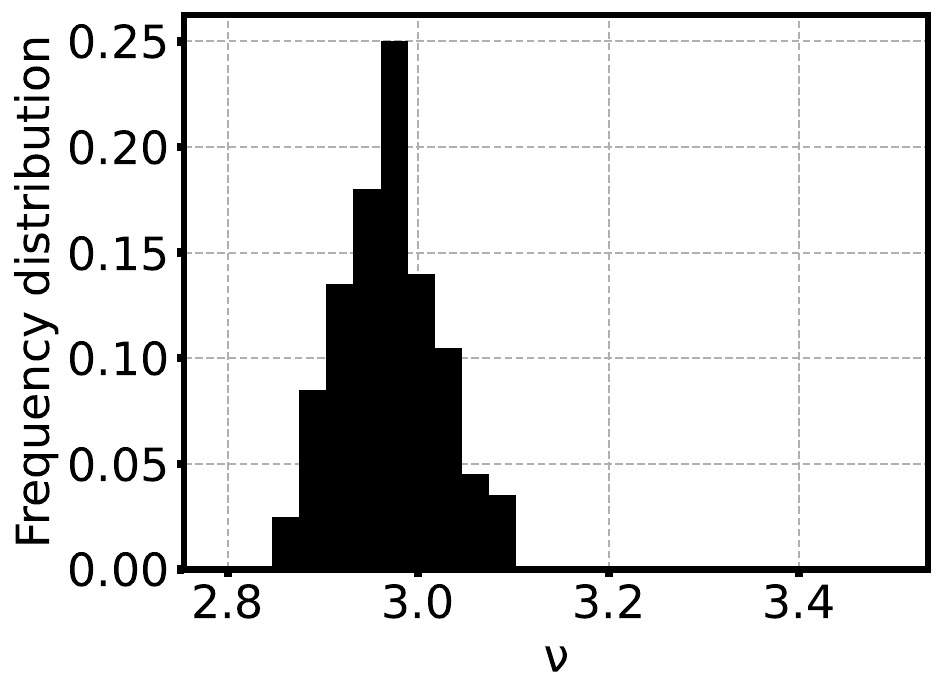}
\caption{$\nu$ distribution.}
\label{fig:Portugal_random_niu_distribution}
\end{subfigure}
\hfill
\begin{subfigure}[b]{0.32\textwidth}
\centering
\includegraphics[width=\textwidth]{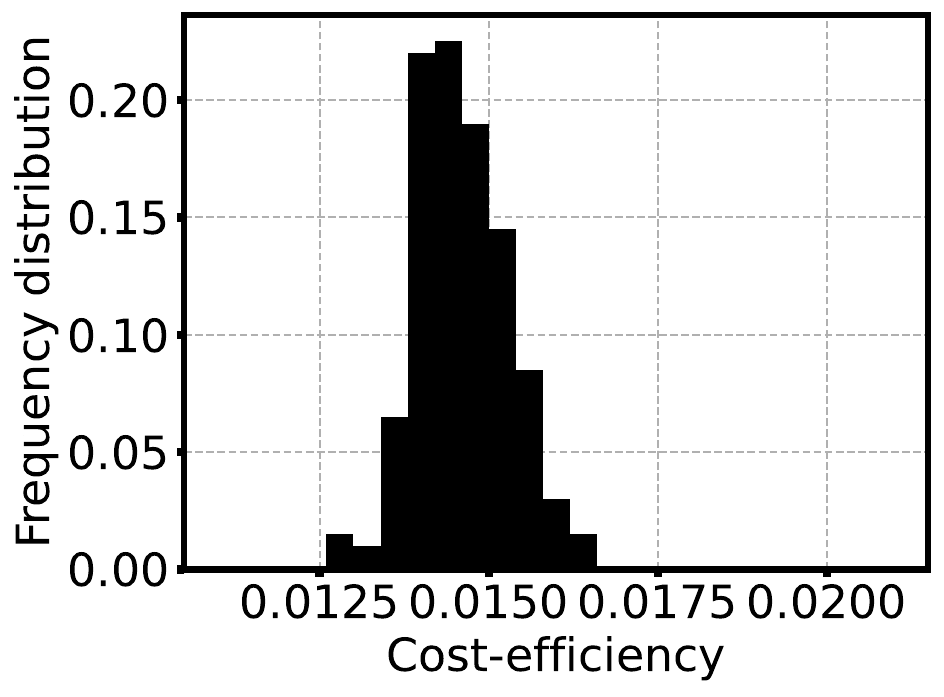}
\caption{CE distribution.}
\label{fig:Portugal_random_CE_distribution}
\end{subfigure}
\caption{Additional results for the Portugal configuration, for the \textbf{Random pair} algorithm.}
\label{fig:Portugal_random_add}
\end{figure}

\begin{figure}[H]
\centering
\begin{subfigure}[b]{0.32\textwidth}
 \centering
 \includegraphics[width=\textwidth]{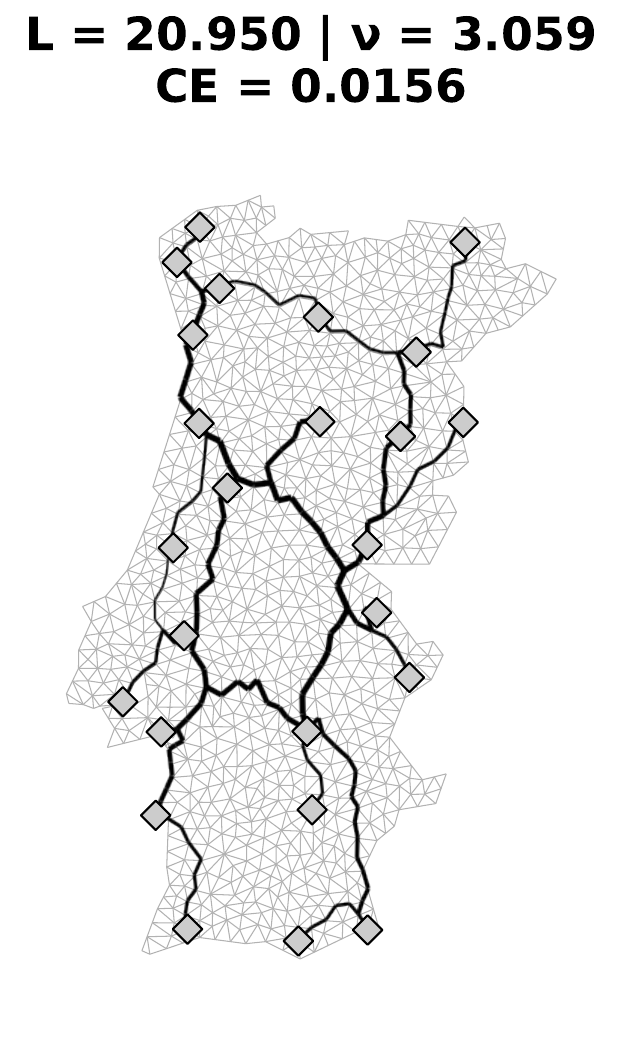}
 \caption{Steady state with largest $L$.}
 \label{fig:Portugal_random_source_ss_biggest_L}
\end{subfigure}
\begin{subfigure}[b]{0.32\textwidth}
 \centering
 \includegraphics[width=\textwidth]{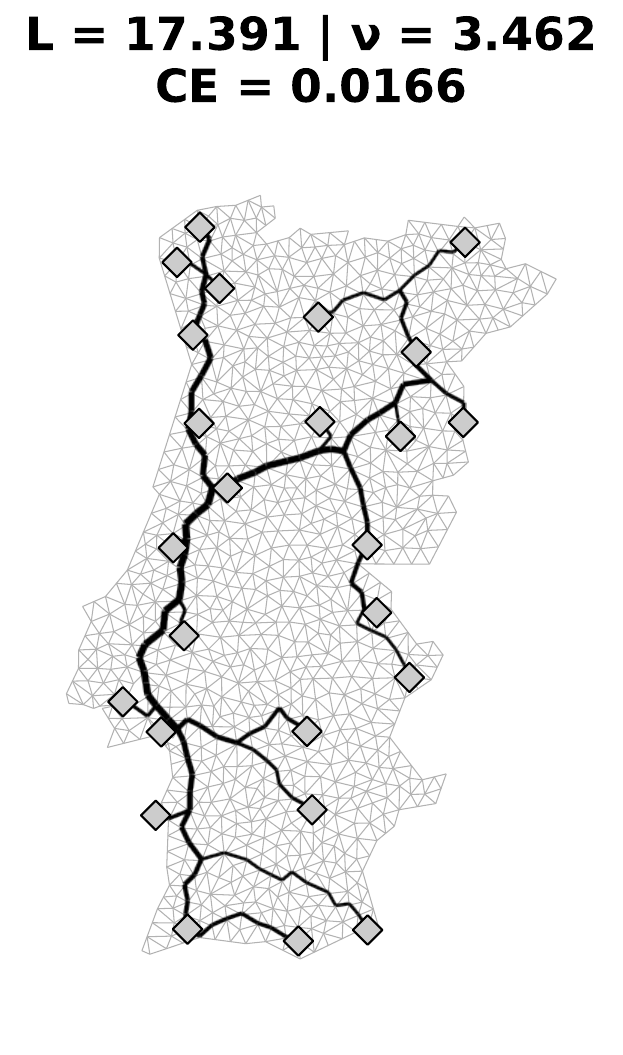}
 \caption{Steady state with largest $\nu$.}
 \label{fig:Portugal_random_source_ss_biggest_niu}
\end{subfigure}
\begin{subfigure}[b]{0.32\textwidth}
 \centering
 \includegraphics[width=\textwidth]{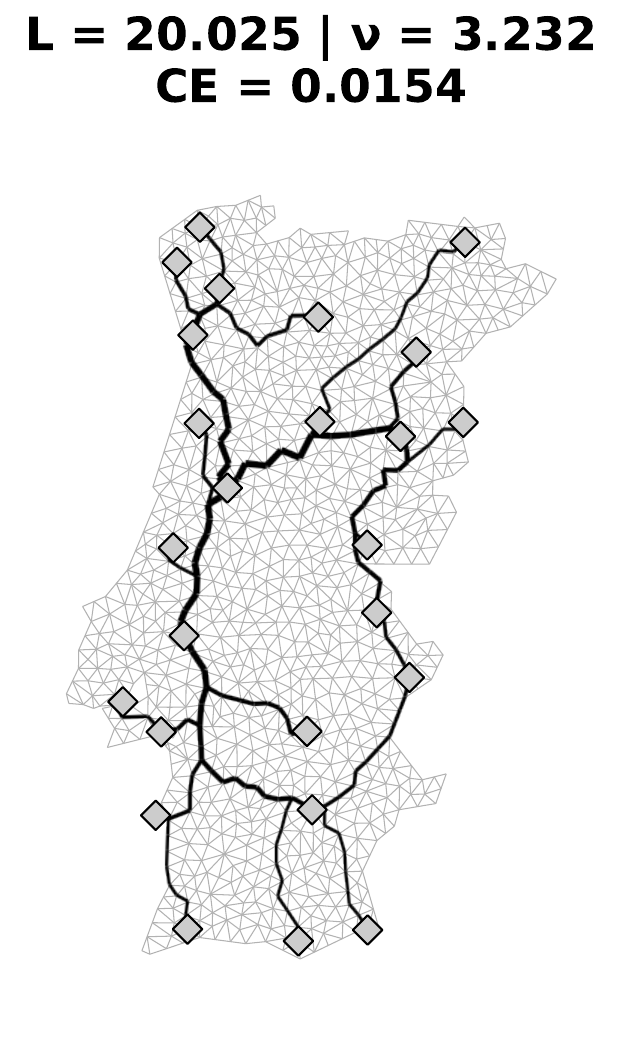}
 \caption{Steady state with smallest CE.}
 \label{fig:Portugal_random_source_ss_smallest_CE}
\end{subfigure}
\\
\begin{subfigure}[b]{0.32\textwidth}
\centering
\includegraphics[width=\textwidth]{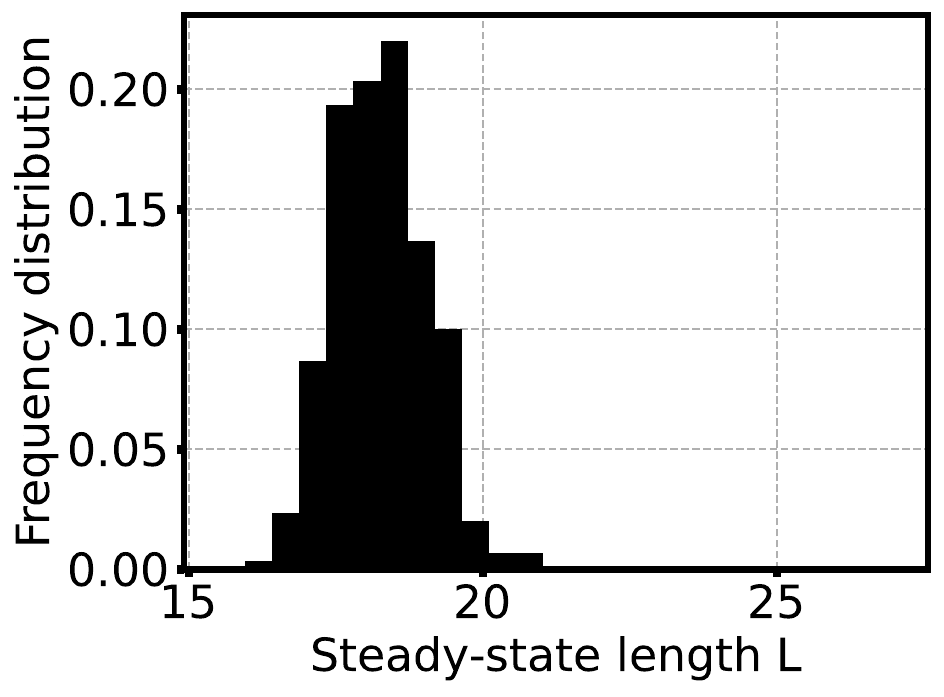}
\caption{$L$ distribution.}
\label{fig:Portugal_random_source_L_distribution}
\end{subfigure}
\hfill
\begin{subfigure}[b]{0.32\textwidth}
\centering
\includegraphics[width=\textwidth]{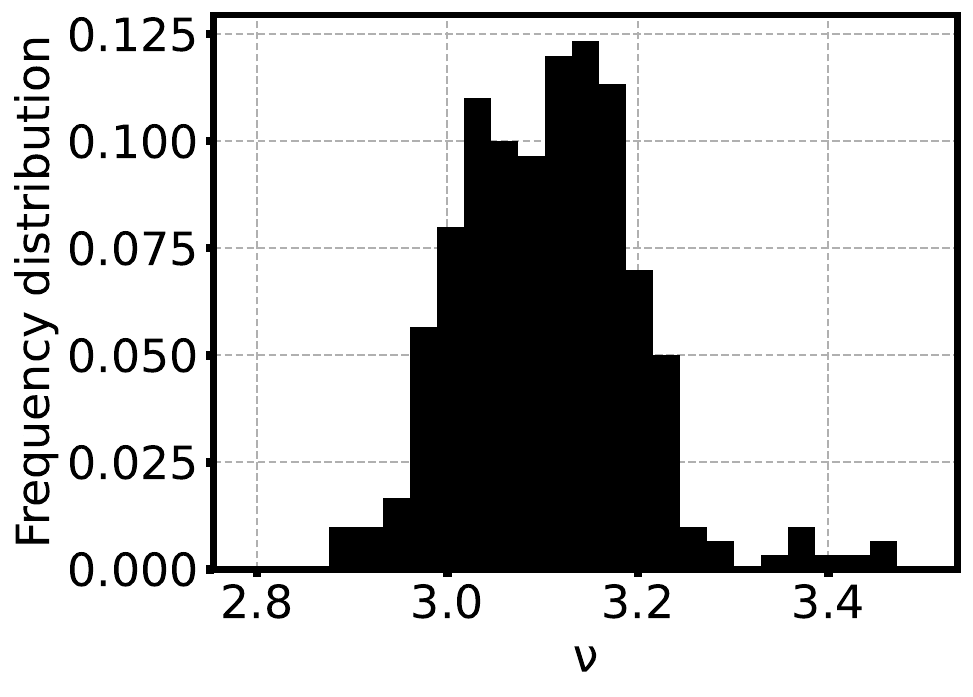}
\caption{$\nu$ distribution.}
\label{fig:Portugal_random_source_niu_distribution}
\end{subfigure}
\hfill
\begin{subfigure}[b]{0.32\textwidth}
\centering
\includegraphics[width=\textwidth]{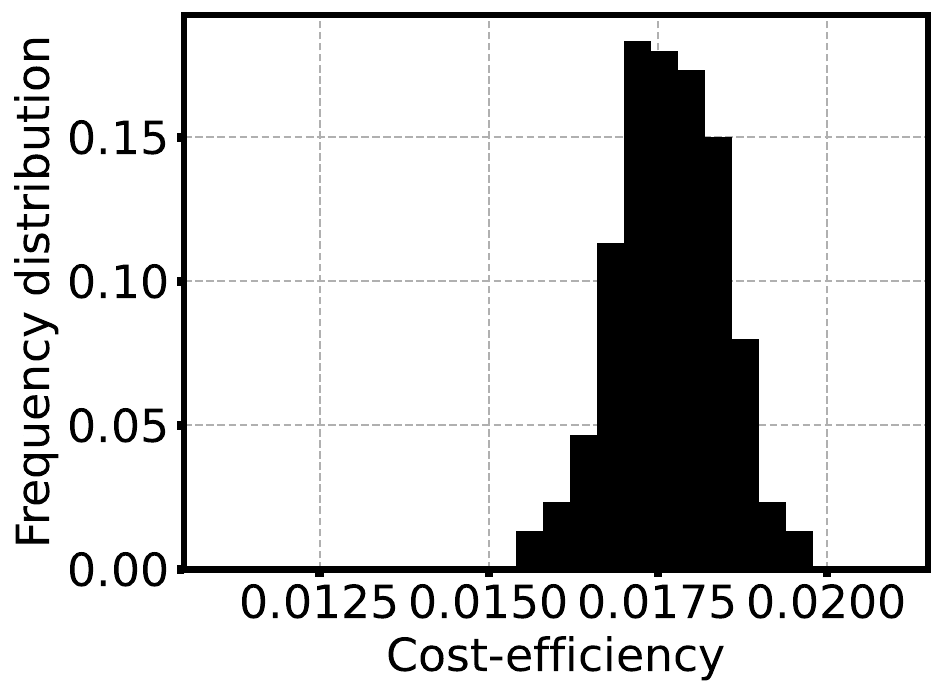}
\caption{CE distribution.}
\label{fig:Portugal_random_source_CE_distribution}
\end{subfigure}
\caption{Additional results for the Portugal configuration, for the \textbf{Random source} algorithm.}
\label{fig:Portugal_random_source_add}
\end{figure}

\begin{figure}[H]
\centering
\begin{subfigure}[b]{0.32\textwidth}
 \centering
 \includegraphics[width=\textwidth]{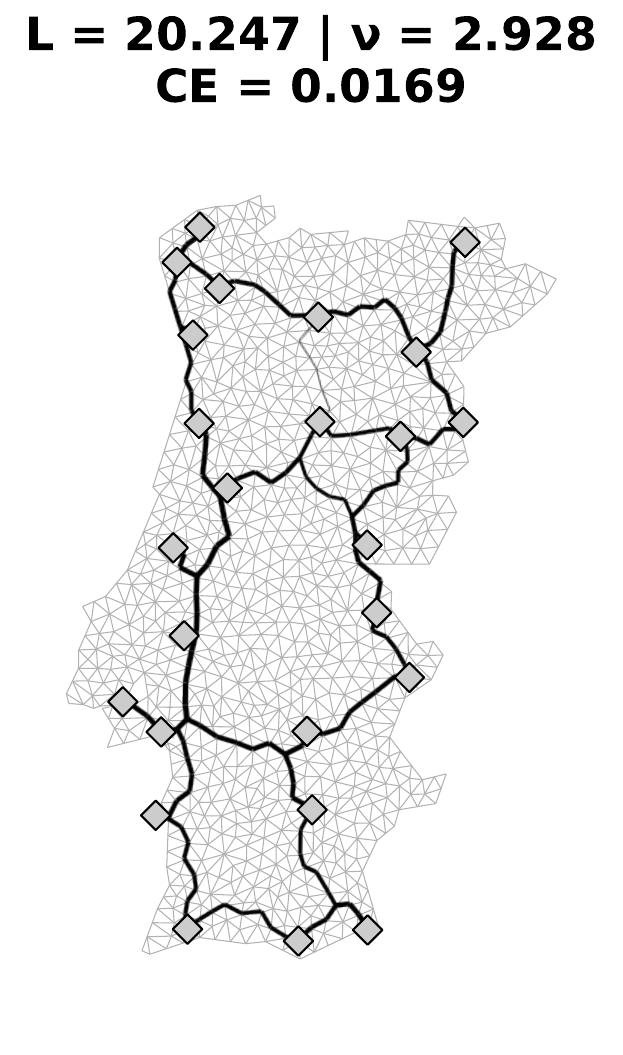}
 \caption{Steady state with largest $L$.}
 \label{fig:Portugal_random_half_ss_biggest_L}
\end{subfigure}
\begin{subfigure}[b]{0.32\textwidth}
 \centering
 \includegraphics[width=\textwidth]{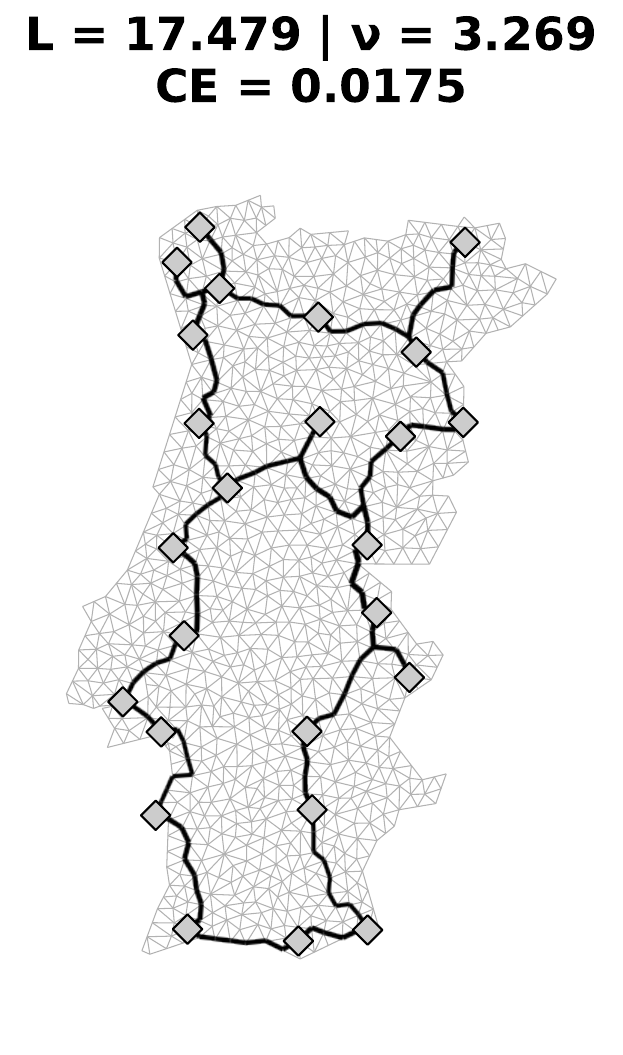}
 \caption{Steady state with largest $\nu$.}
 \label{fig:Portugal_random_half_ss_biggest_niu}
\end{subfigure}
\begin{subfigure}[b]{0.32\textwidth}
 \centering
 \includegraphics[width=\textwidth]{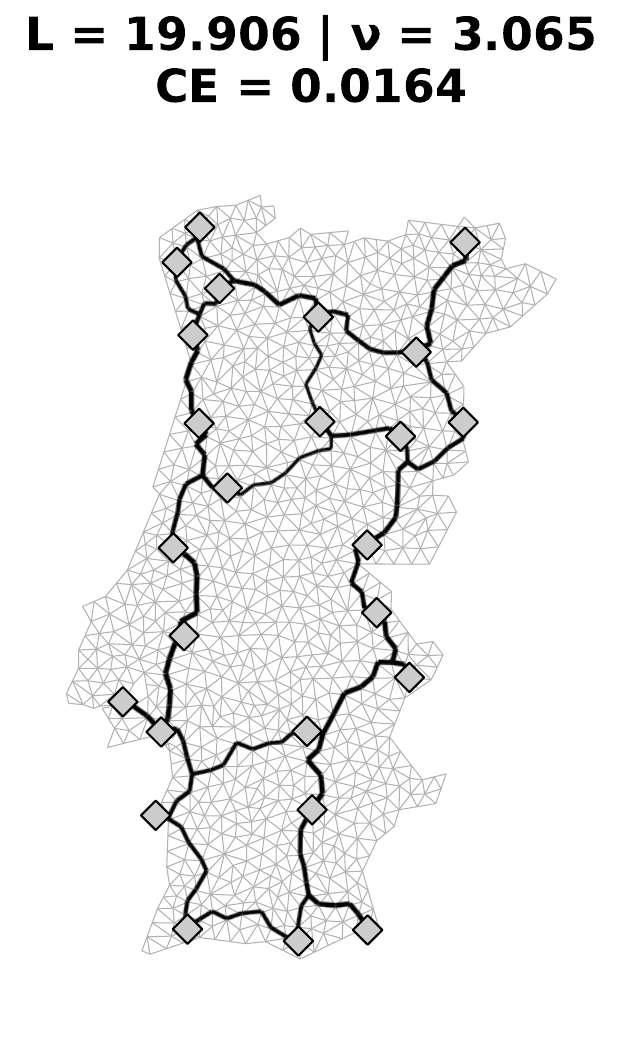}
 \caption{Steady state with smallest CE.}
 \label{fig:Portugal_random_half_ss_smallest_CE}
\end{subfigure}
\\
\begin{subfigure}[b]{0.32\textwidth}
\centering
\includegraphics[width=\textwidth]{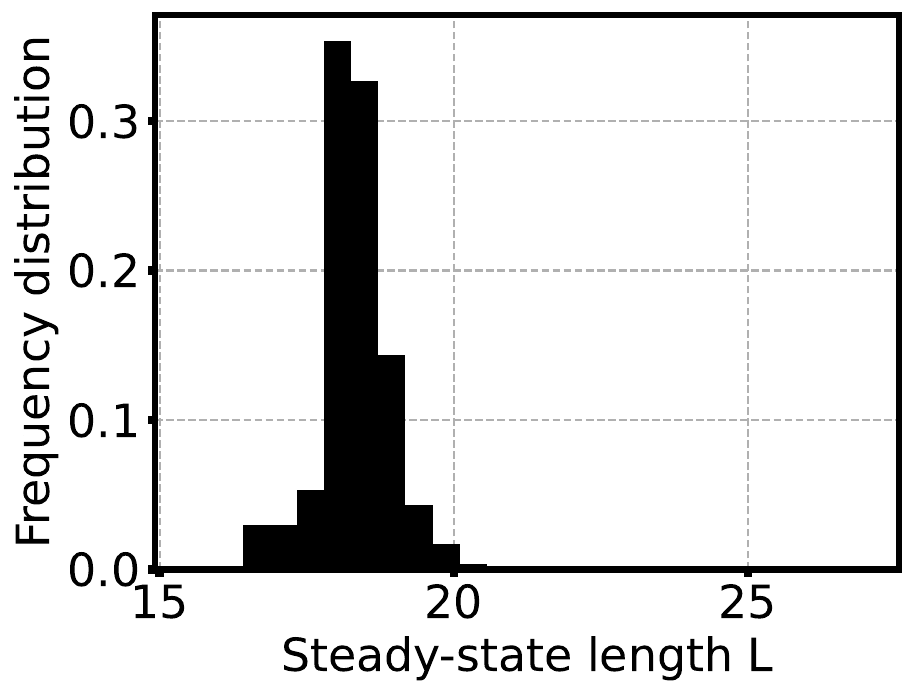}
\caption{$L$ distribution.}
\label{fig:Portugal_random_half_L_distribution}
\end{subfigure}
\hfill
\begin{subfigure}[b]{0.32\textwidth}
\centering
\includegraphics[width=\textwidth]{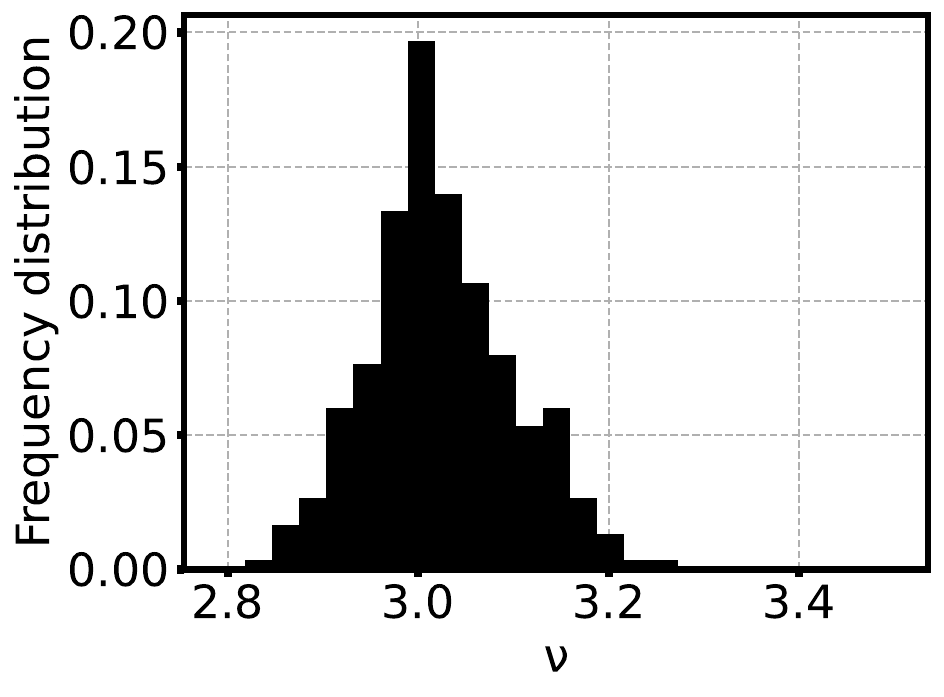}
\caption{$\nu$ distribution.}
\label{fig:Portugal_random_half_niu_distribution}
\end{subfigure}
\hfill
\begin{subfigure}[b]{0.32\textwidth}
\centering
\includegraphics[width=\textwidth]{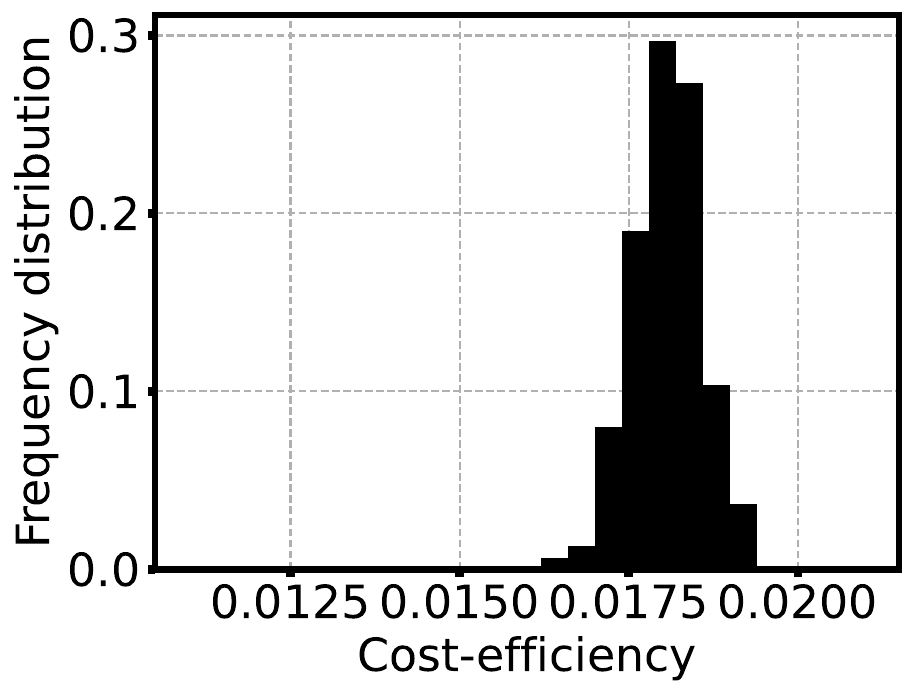}
\caption{CE distribution.}
\label{fig:Portugal_random_half_CE_distribution}
\end{subfigure}
\caption{Additional results for the Portugal configuration, for the \textbf{Random half} algorithm.}
\label{fig:Portugal_random_half_add}
\end{figure}

\begin{figure}[H]
\centering
\begin{subfigure}[b]{0.32\textwidth}
 \centering
 \includegraphics[width=\textwidth]{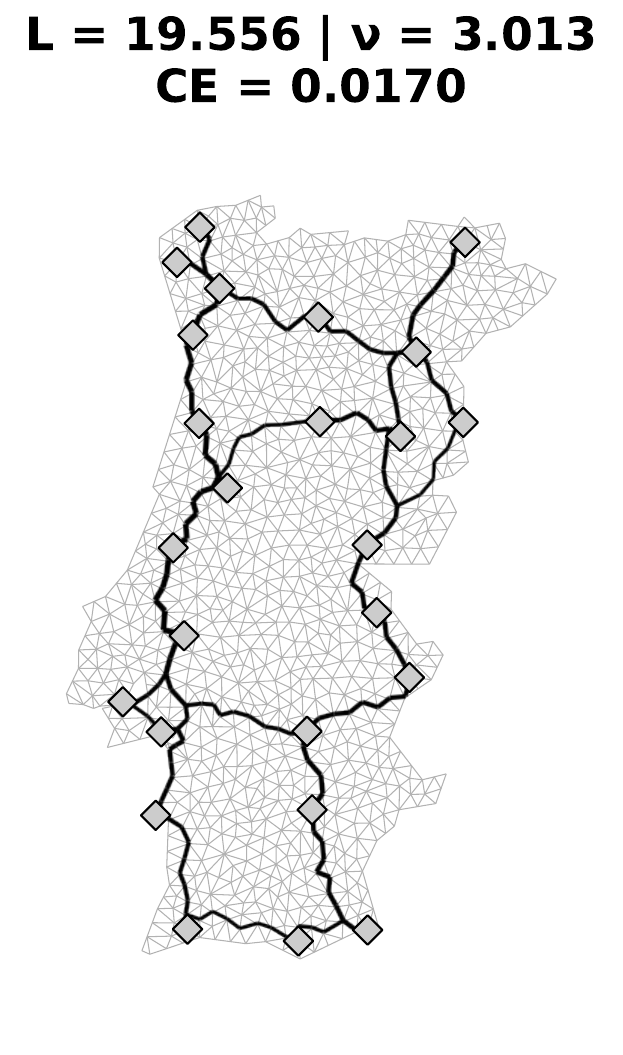}
 \caption{Steady state with largest $L$.}
 \label{fig:Portugal_random_fixed_I0_ss_biggest_L}
\end{subfigure}
\begin{subfigure}[b]{0.32\textwidth}
 \centering
 \includegraphics[width=\textwidth]{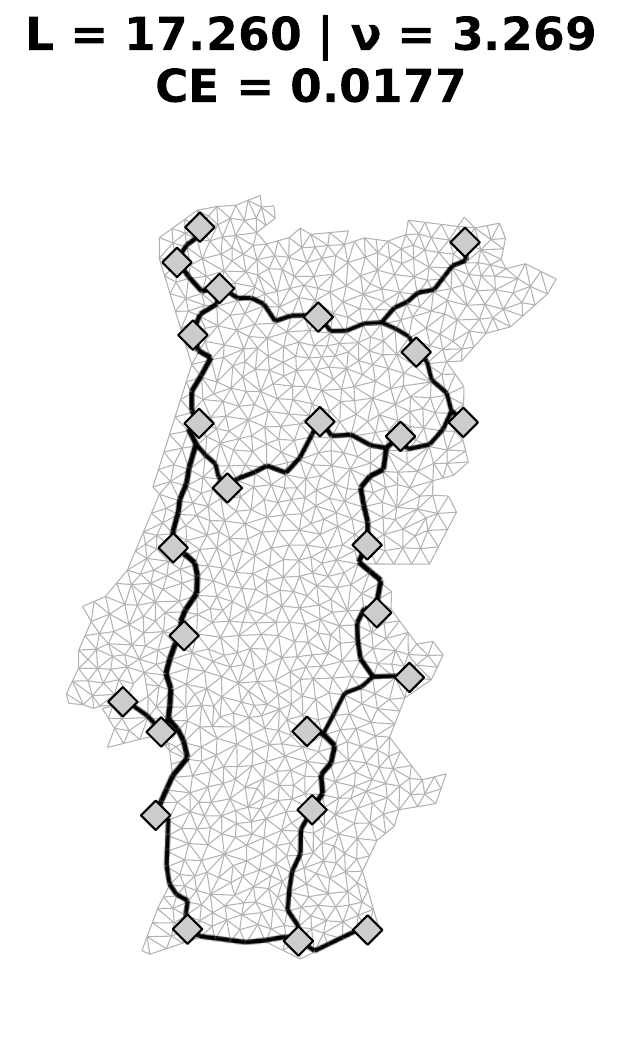}
 \caption{Steady state with largest $\nu$.}
 \label{fig:Portugal_random_fixed_I0_ss_biggest_niu}
\end{subfigure}
\begin{subfigure}[b]{0.32\textwidth}
 \centering
 \includegraphics[width=\textwidth]{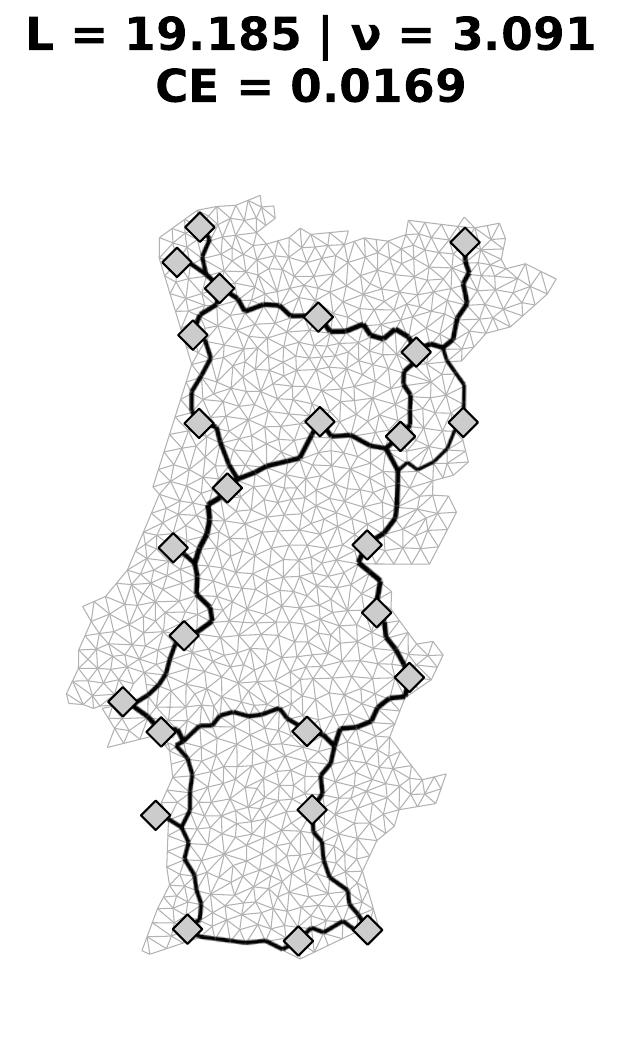}
 \caption{Steady state with smallest CE.}
 \label{fig:Portugal_random_fixed_I0_ss_smallest_CE}
\end{subfigure}
\\
\begin{subfigure}[b]{0.32\textwidth}
\centering
\includegraphics[width=\textwidth]{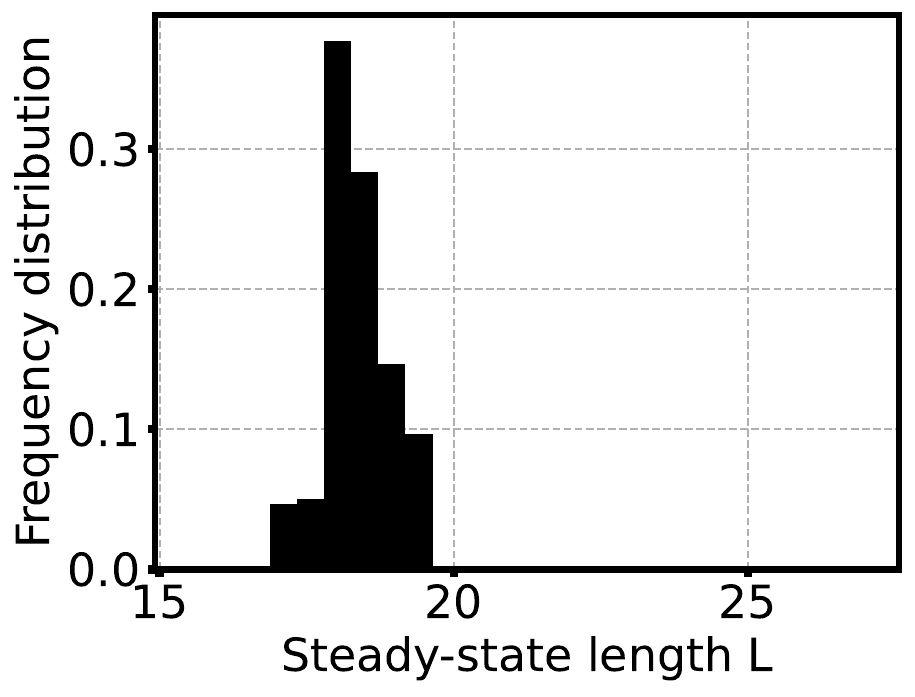}
\caption{$L$ distribution.}
\label{fig:Portugal_random_fixed_I0_L_distribution}
\end{subfigure}
\hfill
\begin{subfigure}[b]{0.32\textwidth}
\centering
\includegraphics[width=\textwidth]{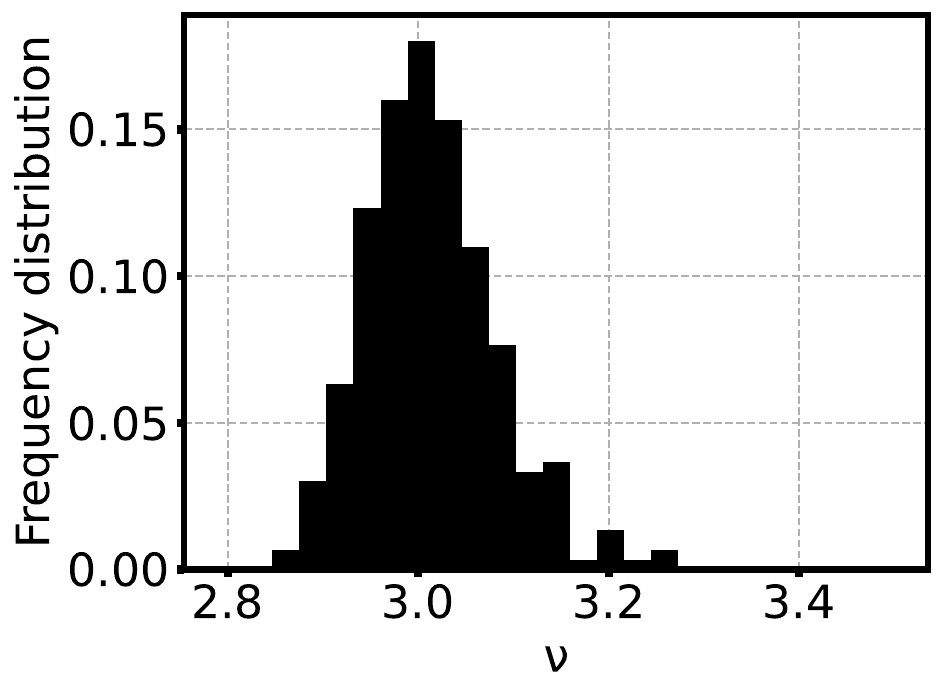}
\caption{$\nu$ distribution.}
\label{fig:Portugal_random_fixed_I0_niu_distribution}
\end{subfigure}
\hfill
\begin{subfigure}[b]{0.32\textwidth}
\centering
\includegraphics[width=\textwidth]{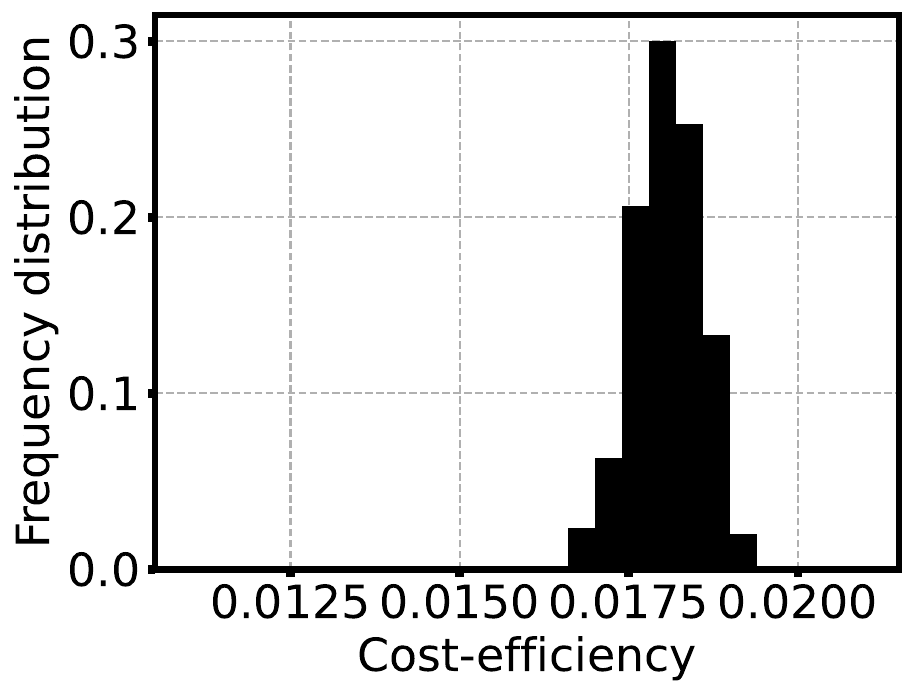}
\caption{CE distribution.}
\label{fig:Portugal_random_fixed_I0_CE_distribution}
\end{subfigure}
\caption{Additional results for the Portugal configuration, for the \textbf{Random random} algorithm.}
\label{fig:Portugal_random_fixed_I0_add}
\end{figure}

\begin{table}[H]
\centering
\begin{tabular}{c|c|c|c}
\toprule
Algorithm & $L$ & $\nu$ & CE \\
\midrule
\texttt{Random pair} & 23.1 $\pm$ 1.1 & 2.970 $\pm$ 0.051 & 0.01458 $\pm$ 0.00067 \\
\texttt{Random source} & 18.27 $\pm$ 0.78 & 3.105 $\pm$ 0.093 & 0.01767 $\pm$ 0.00078 \\
\texttt{Random half} & 18.32 $\pm$ 0.57 & 3.024 $\pm$ 0.075 & 0.01808 $\pm$ 0.00052 \\
\texttt{Random random} & 18.37 $\pm$ 0.53 & 3.014 $\pm$ 0.069 & 0.01808 $\pm$ 0.00051 \\
\midrule
Railway & 18.49 & 3.154 & 0.01715 \\
MST & 13.85 & 4.118 & 0.01754 \\
\bottomrule
\end{tabular}
\caption{Average values and standard deviation of $L$, $\nu$ and CE for the Portuguese configuration, for all algorithms, and comparison with value for mainland Portugal's railway system and with the theoretical value for the MST (last two rows obtained from \cite{rodrigo_tese}).}
\label{tab:Portugal}
\end{table}

\end{document}